\documentclass{article}%
\usepackage{amsfonts}
\usepackage{amsmath}
\usepackage{amssymb}
\usepackage{charter}
\usepackage{graphicx}%
\providecommand{\U}[1]{\protect \rule{.1in}{.1in}}

\providecommand{\U}[1]{\protect \rule{.1in}{.1in}}
\begin{document}

\title{Quantum Optical Version of Classical Optical Transformations and Beyond }
\author{Hong-yi Fan$^{1}$ and Li-yun Hu$^{2}$\thanks{{\small Corresponding author.
E-mail address: hlyun2008@126.com (L.Y. Hu)}}\\$^{1}${\small Department of Physics, Shanghai Jiao Tong University, Shanghai
200030, China; }\\{\small Department of Material Science and Engineering, University of Science
and Technology of China, Hefei, Anhui 230026, China}\\$^{2}${\small College of Physics \& Communication Electronics, Jiangxi Normal
University, Nanchang 330022, China}}
\maketitle

\begin{abstract}
{\small By virtue of the newly developed technique of integration within an
ordered product (IWOP) of operators, we explore quantum optical version of
classical optical transformations such as optical Fresnel transform, Hankel
transform, fractional Fourier transform, Wigner transform, wavelet transform
and Fresnel-Hadmard combinatorial transform etc. In this way one may gain
benefit for developing classical optics theory from the research in quantum
optics, or vice-versa. We can not only find some new quantum mechanical
unitary operators which correspond to the known optical transformations,
deriving a new theorem for calculating quantum tomogram of density operators,
but also can reveal some new classical optical transformations. For examples,
we find the generalized Fresnel operator (GFO) to correspond to the
generalized Fresnel transform (GFT) in classical optics. We derive GFO's
normal product form and its canonical coherent state representation and find
that GFO is the loyal representation of symplectic group multiplication rule.
We show that GFT is just the transformation matrix element of GFO in the
coordinate representation such that two successive GFTs is still a GFT. The
ABCD rule of the Gaussian beam propagation is directly demonstrated in the
context of quantum optics. Especially, the introduction of quantum mechanical
entangled state representations opens up a new area to finding new classical
optical transformations. The complex wavelet transform and the condition of
mother wavelet are studied in the context of quantum optics too. Throughout
our discussions, the coherent state, the entangled state representation of the
two-mode squeezing operators and the technique of integration within an
ordered product (IWOP) of operators are fully used. All these confirms Dirac's
assertion: \textquotedblleft \ }$...${\small for a quantum dynamic system that
has a classical analogue, unitary transformation in the quantum theory is the
analogue of contact transformation in the classical theory".}

{\small Keywords: Dirac's symbolic method; IWOP technique; entangled state of
continuum variables; entangled Fresnel transform; Collins formula; Generalized
Fresnel operator; complex wavelet transform; complex Wigner transform; complex
fractional Fourier transform; symplectic wavelet transform; entangled
symplectic wavelet transform; Symplectic-dilation mixed wavelet transform;
fractional Radon transform; new eigenmodes of fractional Fourier transform}

\end{abstract}
\tableofcontents

\section{Introduction}

The history of quantum mechanics records that from the very beginning the
founders of the quantum theory realized that there might exist formal
connection between classical optics and quantum mechanics. For example,
Schr\"{o}dinger considered that classical dynamics of a point particle should
be the \textquotedblleft geometrical optics\textquotedblright \ approximation
of a linear wave equation, in the same way as ray optics is a limiting
approximation of wave optics; Schr\"{o}dinger also searched for some quantum
mechanical state which behaves like a classical `particle', and this state was
later recognized as the coherent state \cite{Glauber,Klauder,Schro}, which
plays an essential role in quantum optics theory and laser physics; As Dirac
wrote in his famous book
$<$%
Principles of Quantum Mechanics%
$>$%
\cite{Dirac}: \textquotedblleft$\cdot \cdot \cdot$ \textbf{for a quantum dynamic
system that has a classical analogue, unitary transformation in the quantum
theory is the analogue of contact transformation in the classical theory}".
According to Dirac, there should exist a formal correspondence between quantum
optics unitary-transform operators and classical optics transformations.
Indeed, in the last century physicists also found some rigorous mathematical
analogies between classical optics and quantum mechanics, i.e. the similarity
between the optical Helmholtz equation and the time-independent
Schr\"{o}dinger equation; Since 1960s, the advent of a laser and the
appearance of coherent state theory of radiation field
\cite{Glauber,Klauder,glau1}, quantum optics has experienced rapid development
and achieved great success in revealing and explaining the quantum mechanical
features of optical field and non-classical behavior (for instance,
Hanbury-Brown-Twiss effect, photon antibunching, squeezing, sub-Poissonian
photon statistics) of photons in various photon-atom interactions
\cite{mandel}. The relationship between classical and quantum coherence has
been discussed in the book of Mandel and Wolf \cite{mandel}; The Hermite-Gauss
or the Laguerre-Gauss modes of a laser beam are described using the bosonic
operator algebra by Nienhuis and Allen \cite{Nienhuis}. In addition, displaced
light beams refracted by lenses according to the law of geometrical optics,
were found to be the paraxial optics analog of a coherent state. Besides,
phase space correspondence between classical optics and quantum mechanics, say
for example, the Wigner function theory, is inspected in the literature
\cite{Dragoman}.

On the other hand, classical optics, which tackles vast majority of
physical-optics experiments and is based on Maxwell's equations, has never
ceased its own evolving steps, physicists have endeavored to develop various
optical transforms in light propagation through lens systems and various
continuous media. The two research fields, quantum optics and classical
optics, have their own physical objects and conceptions. From the point of
view of mathematics, classical optics is framed in the group transform and
associated representations on appropriate function space, while quantum optics
deals with operators and state vectors, and their overlap seems little at
first glance. It seems to us that if one wants to further relate them to each
other, one needs some new theoretical method to "bridge" them. For example,
what is the quantum mechanical unitary operator corresponding to the Fresnel
transform in Fourier optics? Is there any so-called Fresnel operator as the
image of classical generalized Fresnel transform? Since generalized Fresnel
transforms are very popularly used in optical instrument design and optical
propagation through lenses and various media, it is worth of studying these
transforms in the context of quantum optics theory, especially based on
coherent state, squeezed state \cite{squeezed1,squeezed2} and the newly
invented entangled state theory \cite{entangle,entangle1,entangle2,entangle3}.

Fortunately, the recently developed technique of integration within an ordered
product (IWOP) of operators \cite{IWOP1,IWOP2,IWOP3} is of great aid to
studying quantum optical version of classical optical transformations. Using
the IWOP technique one may gain benefit for classical optics from quantum
optics' research, or vice-versa.

Our present Review is arranged as follows: in section 2 we briefly recall the
classical diffraction theory \cite{Born,Goodman}, this is preparing for later
sections in which we shall show that most frequently employed classical
optical transforms have their counterparts in quantum optics theory. In
section 3 we introduce the IWOP technique and demonstrate that the
completeness relation of fundamental quantum mechanical representations can be
recast into normally ordered Gaussian operator form. Using the IWOP technique
we can directly perform the asymmetric ket-bra integration $\mu^{-1/2}%
\int_{-\infty}^{\infty}dq\left \vert q/\mu \right \rangle \left \langle
q\right \vert $ in the coordinate representation, which leads to the normally
ordered single-mode squeezing operator, this seems to be a direct way to
understanding the squeezing mechanism as a mapping from the classical scaling
$q\rightarrow q/\mu$. In section 4 with the help of IWOP technique and based
on the concept of quantum entanglement of Einstein-Podolsky-Rosen \cite{EPR}
we construct two mutually conjugate entangled states of continuum variables,
$\left \vert \eta \right \rangle $ versus $\left \vert \xi \right \rangle ,$ and
their \textit{deduced entangled states (or named }%
\textbf{correlated-amplitude---number-difference entangled states}), they are
all qualified to make up quantum mechanical representations. It is remarkable
that using the IWOP technique to performing the asymmetric ket-bra integration
$\mu^{-1}\int d^{2}\eta \left \vert \eta/\mu \right \rangle \left \langle
\eta \right \vert $ leads to the two-mode normally ordered two-mode squeezing
operator, this implies that the two-mode squeezed state is simultaneously an
entangled state. We point out that the entangled state $\left \vert
\eta \right \rangle $ also embodies entanglement in the aspect of correlative
amplitude and the phase. We are also encouraged that the overlap between two
mutually conjugate \textit{deduced entangled states} is just the Bessel
function--- the optical Hankel transform kernel \cite{Fanpla1}$,$ which again
shows that the new representations in the context of physics theory match
beautiful mathematical formalism exactly. We then employ the \textit{deduced}
entangled states to derive \textbf{quantum optical version of classical
circular harmonic correlation. }Section 5 is devoted to finding a quantum
operator which corresponds to the optical Fresnel transform, with use of the
coherent state representation and by projecting the classical sympletic
transform $z\rightarrow sz-rz^{\ast}$ ($\left \vert s\right \vert ^{2}%
-\left \vert r\right \vert ^{2}=1)$ in phase space onto the quantum mechanical
Hilbert space, we are able to recognize which operator is the single-mode
Fresnel operator (FO). It turns out that the 1-dimensional optical Fresnel
transform is just the matrix element of the Fresnel operator $F$ in the
coordinate eigenstates. Besides, the coherent state projection operator
representation of FO constitutes a loyal realization of symplectic group,
which coincides with the fact that two successive optical Fresnel transforms
make up a new Fresnel transform. Then in Section 6 based on the coherent state
projection representation of FO, we prove $ABCD$ rule for optical propagation
in the context of quantum optics. In section 7 the quadratic operator form of
FO is also presented and the four fundamental optical operators are derived by
decomposing the FO. In section 8 we discuss how to apply the Fresnel operator
to quantum tomography theory, by introducing the Fresnel quadrature phase
$FXF^{\dagger}=X_{F},$ we point out that Wigner operator's Radon
transformation is just the pure state projection operator $\left \vert
x\right \rangle _{s,rs,r}$\ $\left \langle x\right \vert $, where $\left \vert
x\right \rangle _{s,r}=F\left \vert x\right \rangle $ and $\left \vert
x\right \rangle $ is the position eigenstate, so the probability distribution
for the Fresnel quadrature phase is the Radon transform of the Wigner
function. Moreover, the tomogram of quantum state $\left \vert \psi
\right \rangle $ is just the squared modulus of the wave function
$_{s,r}\left \langle x\right \vert \left.  \psi \right \rangle .$ This new
relation between quantum tomography and optical Fresnel transform may provide
experimentalists to figure out new approach for testing tomography. In
addition, we propose another new theorem for calculating tomogram, i.e., the
tomogram of a density operator $\rho$ is equal to the marginal integration of
the classical Weyl correspondence function of $F^{\dagger}\rho F$. In section
9 by virtue of the coherent state and IWOP method we propose two-mode
generalized Fresnel operator (GFO), in this case we employ the entangled state
representation to relate the 2-mode GFO to classical transforms, since the
2-mode GFO is not simply the direct product of two 1-mode GFOs. The
corresponding quantum optics $ABCD$ rule for two-mode case is also proved. The
2-mode GFO can also be expressed in quadratic operators form in entangled way.
The relation between optical FT and quantum tomography in two-mode case is
also revealed. In section 10 we propose a kind of integration transformation,
$\iint_{-\infty}^{\infty}\frac{dpdq}{\pi}e^{2i\left(  p-x\right)  \left(
q-y\right)  }h(p,q)\equiv f\left(  x,y\right)  ,$ which is invertible and
obeys Parseval theorem. Remarkably, it can convert chirplet function to the
kernel of fractional Fourier transform (FrFT). This transformation can also
serve for solving some operator ordering problems. In section 11 we employ the
entangled state representation to introduce the complex FrFT (CFrFT), which is
not the direct product of two independent 1-dimensional FrFT transform. The
eigenmodes on CFrFT is derived. New eigen-modes for light propagation in
graded-index medium and the fractional Hankel transform are presented. The
Wigner transform theory is extended to the complex form and its relation to
CFrFT is shown; The integration transformation in section 10 is also extended
to the entangled case. In section 12 we shall treat the adaption problem of
Collins diffraction formula to the CFrFT with the use of two-mode (3
parameters) squeezing operator and in the entangled state representation of
continuous variables, in so doing the quantum mechanical version of associated
theory of classical diffraction and classical CFrFT is obtained, which
connects classical optics and quantum optics in this aspect. In section 13 we
introduce a convenient way for constructing the fractional Radon transform.
the complex fractional Randon transform is also proposed; In sections 14 and
15 we discuss quantum optical version of classical wavelet transforms (WTs),
including how to recast the condition of\ mother wavelet into the context of
quantum optics; how to introduce complex wavelet transform with use of the
entangled state representations. Some properties, such as Parseval theorem,
Inversion formula, and orthogonal property, the relation between WT and
Wigner-Husimi distribution function are also discussed. In section 16, we
generalize the usual wavelet transform to symplectic wavelet transformation
(SWT) by using the coherent state representation and making transformation
$z\rightarrow s\left(  z-\kappa \right)  -r\left(  z^{\ast}-\kappa^{\ast
}\right)  $ ($\left \vert s\right \vert ^{2}-\left \vert r\right \vert ^{2}=1)$ in
phase space. The relation between SWT and optical Fresnel transformation is
revealed. Then the SWT is extended to the entangled case by mapping the
classical mixed transformation $\left(  z,z^{\prime}\right)  \rightarrow
\left(  sz+rz^{\prime \ast},sz^{\prime}+rz^{\ast}\right)  $ in 2-mode coherent
state $\left \vert z,z^{\prime}\right \rangle $ representation. At the end of
this section, we introduce a new symplectic-dilation mixed WT by employing a
new entangled-coherent state representation $\left \vert \alpha,x\right \rangle
$. The corresponding classical optical transform is also presented. In the
last section, we introduce the Fresnel-Hadamard combinatorial operator by
virtue of the IWOP technique and $\left \vert \alpha,x\right \rangle $. This
unitary operator plays the role of both Fresnel transformation for mode
$\frac{a_{1}-a_{2}}{\sqrt{2}}$ and Hadamard transformation for mode
$\frac{a_{1}+a_{2}}{\sqrt{2}},$ respectively, and the two transformations are
combinatorial. All these sections are used to prove the existence of a
one-to-one correspondence between quantum optical operators that transform
state vectors in Hilbert space and the classical optical transforms that
change the distribution of optical field.

\section{Some typical classical optical transformations}

Here we briefly review some typical optical transforms based on light
diffraction theory. These transformations, as one can see in later sections,
are just the correspondence of some representation transformations between
certain quantum mechanical states of which some are newly constructed.

It was Huygens who gave a first illustrative explanation to wave theory by
proposing every point in the propagating space as a sub-excitation source of a
new sub-wave. An intuitive theory mathematically supporting Huygens' principle
is the scalar diffraction approximation, so named because optical fields
(electromagnetic fields) actually are vector fields, whereby the theory is
valid approximately. This theory is based on the superposition of the combined
radiation field of multiple re-emission sources initiated by Huygens. Light
diffraction phenomena has played an important role in the development of the
wave theory of light, and now underlies the Fourier optics and information
optics. The formulation of a diffraction problem essentially considers an
incident free-space wave whose propagation is interrupted by an obstacle or
mask which changes the phase and/or amplitude of the wave locally by a well
determined factor \cite{opticalPhyiscs}. A more rigorous, but still in the
scheme of scalar wave, derivation has been given by Kirchhoff who reformulated
the diffraction problem as a boundary-value problem, which essentially
justifies the use of Huygens principle. The Fresnel-Kirchhoff (or
Rayleigh-Sommerfeld) diffraction formula is practically reduced to the Fresnel
integral formula in paraxial and far-field approximation \cite{Born,Goodman}
that reads:
\begin{equation}
U_{2}\left(  x_{2},y_{2}\right)  =\frac{\exp \left(  ikz\right)  }{i\lambda
z}\int \int_{-\infty}^{\infty}U_{1}\left(  x_{1},y_{1}\right)  \exp \left \{
i\frac{k}{2z}\left[  \left(  x_{2}-x_{1}\right)  ^{2}+\left(  y_{2}%
-y_{1}\right)  ^{2}\right]  \right \}  dx_{1}dy_{1}, \label{2.1}%
\end{equation}
where $U_{1}\left(  x_{1},y_{1}\right)  $ is the optical distribution of a
2-dimensional light source and $U_{2}\left(  x_{2},y_{2}\right)  $ is its
image on the observation plane, $\lambda$ is the optical wavelength,
$k=\frac{2\pi}{\lambda}$ is the wave number in the vacuum and $z$ is the
propagation distance. When
\begin{equation}
z^{2}\gg \frac{k}{2}\left(  x_{1}^{2}+y_{1}^{2}\right)  _{\max}, \label{2.2}%
\end{equation}
is satisfied, Eq. (\ref{2.1}) reduces to
\begin{align}
U_{2}\left(  x_{2},y_{2}\right)   &  =\frac{\exp \left(  ikz\right)
\exp \left[  i\frac{k}{2z}\left(  x_{1}^{2}+y_{1}^{2}\right)  \right]
}{i\lambda z}\nonumber \\
&  \times \int \int_{-\infty}^{\infty}U_{1}\left(  x_{1},y_{1}\right)
\exp \left[  -i\frac{2\pi}{\lambda z}\left(  x_{1}x_{2}+y_{1}y_{2}\right)
\right]  dx_{1}dy_{1}, \label{2.3}%
\end{align}
which is named the Fraunhofer diffraction formula.

The Fresnel integral is closely related to the fractional Fourier transform
(FrFT), actually, it has been proved that the Fresnel transform can be
interpreted as a scaled FrFT with a residual phase curvature \cite{Torre}. The
FrFT is a very useful tool in Fourier optics and information optics. This
concept was firstly introduced in 1980 by Namias \cite{Namias} but not brought
enough attention until FrFT was defined physically, based on propagation in
quadratic graded-index media (GRIN media). Mendlovic and Ozaktas
\cite{Mendlovic,Ozakatas} defined the $\alpha$th FrFT as follows: Let the
original function be input from one side of quadratic GRIN medium, at $z=0$.
Then, the light distribution observed at the plane $z=z_{0}$ corresponds to
the $\alpha$ equal to the ($z_{0}/L$)th fractional Fourier transform of the
input fraction, where $L\equiv(\pi/2)(n_{1}/n_{2})^{1/2}$ is a characteristic
distance. The FrFT can also be implemented by lenses. Another approach for
introducing FrFT was made by Lohmann who pointed out the algorithmic
isomorphism among image rotation, rotation of the Wigner distribution function
\cite{Wigner}, and fractional Fourier transforming \cite{Lohmann}. Lohmann
proposed the FrFT as the transform performed on a function that leads to a
rotation with an angle of the associated Wigner distribution function, in this
sense, the FrFT bridges the gap between classical optics and optical Wigner
distribution theory. Recently, the FrFT has been paid more and more attention
within different contexts of both mathematics and physics
\cite{Namias,Mendlovic,Ozakatas,Lohmann,Bernardo}. The FrFT is defined as
\begin{align}
&  \mathcal{F}_{\alpha}\left[  U_{1}\right]  \left(  x_{2},y_{2}\right)
=\frac{e^{i(1-\alpha)\frac{\pi}{2}}}{2\sin \left(  \frac{\pi}{2}\alpha \right)
}\exp \left[  -\frac{i\left(  x_{2}^{2}+y_{2}^{2}\right)  }{2\tan \left(
\frac{\pi}{2}\alpha \right)  }\right] \nonumber \\
&  \times \int \int_{-\infty}^{\infty}\frac{dx_{1}dy_{1}}{\pi}\exp \left[
-\frac{i\left(  x_{1}^{2}+y_{1}^{2}\right)  }{2\tan \left(  \frac{\pi}{2}%
\alpha \right)  }\right]  \exp \left[  \frac{i\left(  x_{2}x_{1}+y_{2}%
y_{1}\right)  }{\sin \left(  \frac{\pi}{2}\alpha \right)  }\right]  U_{1}\left(
x_{1},y_{1}\right)  . \label{2.4}%
\end{align}
We can see that $F_{0}$ is the identity operator and $F_{\pi/2}$ is just the
Fourier transform. The most important property of FRFT is that $F_{\alpha}$
obeys the semigroup property, i.e. two successive FrFTs of order $\alpha$\ and
$\beta$ \ makes up the FrFT of order $\alpha+\beta.$ A more general form
describing the light propagation in an optical system characterized by the
$\left[  A,B;C,D\right]  $ ray transfer matrix is the Collins diffraction
integral formula \cite{Collins1},
\begin{align}
U_{2}\left(  x_{2},y_{2}\right)   &  =\frac{k\exp \left(  ikz\right)  }{2\pi
Bi}\int \int_{-\infty}^{\infty}dx_{1}dy_{1}U_{1}\left(  x_{1},y_{1}\right)
\nonumber \\
&  \times \exp \left \{  \frac{ik}{2B}\left[  A\left(  x_{1}^{2}+y_{1}%
^{2}\right)  -2\left(  x_{1}x_{2}+y_{1}y_{2}\right)  +D\left(  x_{2}^{2}%
+y_{2}^{2}\right)  \right]  \right \}  dx_{1}dy_{1}, \label{2.5}%
\end{align}
where $AD-BC=1$ if the system is lossless. One can easily find the similarity
between Collins formula and the FrFT by some scaling transform and relating
the $\left[  A,B,C,D\right]  $ matrix to $\alpha$ in the FrFT
\cite{MatrixOptics}. Note that $M=\left(
\begin{array}
[c]{cc}%
A & B\\
C & D
\end{array}
\right)  $ is a ray transfer matrix describing optical systems belonging to
the unimodular symplectic group. When treating the light propagation in
optical elements in near-axis approximation, matrices $M$ representing linear
transformations are a convenient mathematical tool for calculating the
fundamental properties of optical systems, which is the origin of the name of
matrix optics. In cylindrical coordinates the Collins formula is expressed as
\cite{Collins1,Collins2}
\begin{equation}
U_{2}\left(  r_{2},\varphi \right)  =\frac{i}{\lambda B}{\int}_{0}^{\infty
}{\int}_{0}^{2\pi}\exp \left \{  -\frac{i\pi}{\lambda B}\left[  Ar_{1}%
^{2}+Dr_{2}^{2}-2r_{1}r_{2}\cos \left(  \theta-\varphi \right)  \right]
\right \}  U_{1}\left(  r_{1},\theta \right)  r_{1}dr_{1}d\theta \label{2.6}%
\end{equation}
where $x_{1}=r_{1}\cos \theta,$ $y_{1}=r_{1}\sin \theta,$ $x_{2}=r_{2}%
\cos \varphi$ and $y_{2}=r_{2}\sin \varphi.$ When $U_{1}\left(  r_{1}%
,\theta \right)  $ has rotational symmetry
\begin{equation}
U_{1}\left(  r_{1},\theta \right)  =u_{1}\left(  r_{1}\right)  \exp \left(
im\theta \right)  ,\text{ }U_{2}\left(  r_{2},\varphi \right)  =u_{2}\left(
r_{2}\right)  \exp \left(  im\varphi \right)  , \label{2.7}%
\end{equation}
then (\ref{2.6}) becomes
\begin{equation}
u_{2}\left(  r_{2}\right)  =\frac{2\pi}{\lambda B}\exp \left[  i\left(
1+m\right)  \frac{\pi}{2}\right]  {\int}_{0}^{\infty}\exp \left[  -\frac{i\pi
}{\lambda B}\left[  Ar_{1}^{2}+Dr_{2}^{2}\right]  \right]  J_{m}\left(
\frac{2\pi r_{1}r_{2}}{\lambda B}\right)  u_{1}\left(  r_{1}\right)
r_{1}dr_{1}, \label{2.8}%
\end{equation}
where we have used the $m$-order Bessel function
\begin{equation}
J_{m}\left(  x\right)  =\frac{1}{2\pi}{\int}_{0}^{2\pi}\exp \left[
ix\cos \theta+im\left(  \theta-\frac{\pi}{2}\right)  \right]  d\theta.
\label{2.9}%
\end{equation}
When $A=0$, (\ref{2.8}) reduces to the standard Hankel transform (up to a
phase factor)%
\begin{equation}
u_{2}\left(  r_{2}\right)  \rightarrow \frac{2\pi}{\lambda B}{\int}_{0}%
^{\infty}J_{m}\left(  \frac{2\pi r_{1}r_{2}}{\lambda B}\right)  u_{1}\left(
r_{1}\right)  r_{1}dr_{1}, \label{2.10}%
\end{equation}
The compact form of one-dimensional Collins formula is
\begin{equation}
g\left(  x_{2}\right)  =\int_{-\infty}^{\infty}\mathcal{K}^{M}\left(
x_{2},x_{1}\right)  f\left(  x_{1}\right)  dx_{1}, \label{2.11}%
\end{equation}
where the transform kernel is
\begin{equation}
\mathcal{K}^{M}\left(  x_{2},x_{1}\right)  =\frac{1}{\sqrt{2\pi iB}}%
\exp \left[  \frac{i}{2B}\left(  Ax_{1}^{2}-2x_{2}x_{1}+Dx_{2}^{2}\right)
\right]  , \label{2.12}%
\end{equation}
$M$ is the parameter matrix $\left[  A,B,C,D\right]  $. Eq. (\ref{2.12}) is
called generalized Fresnel transform \cite{GFT,GFT1,Alieva1,agarwal}. In the
following sections we will show how we find the quantum optical counterpart
for those transformations of classical optics. For this purpose in the next
chapter we introduce the IWOP technique to demonstrate how Dirac's symbolic
method can be developed and be applied to quantum optics theory. Also, we
briefly review some properties of the entangled state
\cite{entangle,entangle1,entangle2,entangle3} and reveal the connection
between the mutual transform generated by these entangled states and the
Hankel transform in classical optics.

\section{The IWOP technique and two mutually conjugate entangled states}

\subsection{The IWOP technique}

The history of mathematics tells us that whenever there appears a new
important mathematical symbol, there coexists certain operational rules for
it, the quantum mechanical operators in ket-bra projective form (the core of
Dirac's symbolic method) also need their own operational rules. The
terminology \textquotedblleft symbolic method\textquotedblright \ was first
shown in the preface of Dirac's book
$<$%
The Principle of Quantum Mechanics%
$>$%
: \textquotedblleft \textit{The symbolic method, which deals directly in an
abstract way with the quantities of fundamental importance}$\cdot \cdot \cdot
$\textit{, however, seems to go more deeply into the nature of things. It
enables one to express the physical law in a neat and concise way,} and will
probably be increasingly used in the future as it becomes better understood
and its own special mathematics gets developed\textbf{.\textquotedblright}
\cite{Dirac} Then two questions naturally arise: How to better understand the
\textit{symbolic method? How to develop Dirac's symbolic method, especially
its mathematics? We noticed that }\textbf{Newton-Leibniz integration rule only
applies to commuting functions of continuum variables, while operators made of
Dirac's symbols\ (ket versus bra, e.g., }$\left \vert q/\mu \right \rangle
\left \langle q\right \vert $\textbf{\ of continuous parameter }$q$\textbf{) in
quantum mechanics are usually not commutative. Therefore integrations over the
operators of type }$\left \vert \  \right \rangle \left \langle \  \right \vert $
(\textbf{where ket- and bra- state vectors need not to be Hermitian-conjugate
to each other) can not be directly performed by the Newton-Leibniz rule. Thus
we invented an innovative technique of integration within an ordered product
(IWOP) of operators that made the integration of non-commutative operators
possible. The core of IWOP technique is to arrange non-commutable quantum
operators within an ordered product (say, normal ordering) in a way that they
become commutable, in this sense the gap between q-numbers and c-numbers is
"narrowed". However, the nature of operators which which are within : : is not
changed, they are still q-numbers, not c-numbers. After the integration over
c-numbers within ordered product is performed, we can get rid of the normal
ordering symbol after putting the integration result in normal ordering.
\cite{Weyl}.The IWOP technique thus bridges this mathematical gap between
classical mechanics and quantum mechanics, and further reveals the beauty and
elegance of Dirac's symbolic method and transformation theory. This technique
develops symbolic method significantly, i. e. makes Dirac's representation
theory and the transformation theory more plentiful, and consequently to be
better understood. The beauty and elegance of Dirac's symbolic method are
further revealed. Various applications of the IWOP technique, including
constructing the entangled state, developing \textbf{the} nonlinear coherent
state theory, Wigner function theory, etc. are found; many new unitary
operators and operator-identities as well as new quantum mechanical
representations can be derived too, which are partly} summarized in the Review
Articles \cite{entangle1}.

We begin with listing some properties of normal product of operators which
means all the bosonic creation operators $a^{\dagger}$ are standing on the
left of annihilation operators $a$ in a monomial of $a^{\dagger}$ and $a$.

1. The order of Bose operators $a$ and $a^{\dagger}$ within a normally ordered
product can be permuted. That is to say, even though $\left[  a,a^{\dagger
}\right]  =1$, we can have $\colon aa^{\dagger}\colon=\colon a^{\dagger
}a\colon=a^{\dagger}a,$ where $:$ $:$ denotes normal ordering.

2. $c$-numbers can be taken out of the symbol $:$ $\colon$ as one wishes.

3. The symbol $:$ : which is within another symbol $:$ $\colon$ can be deleted.

4. The vacuum projection operator $|0\rangle \langle0|$ has the normal product
form
\begin{equation}
|0\rangle \langle0|=\colon e^{-a^{\dagger}a}\colon. \label{3.1}%
\end{equation}

5. A normally ordered product can be integrated or differentiated with respect
to a $c$-number provided the integration is convergent.

\subsection{The IWOP technique for deriving normally ordered Gaussian form of
the completeness relations of fundamental quantum mechanical representations}

As an application of IWOP, (in the following, unless particularly mentioned,
we take $\hbar=\omega=m=1$ for convenience.) Using \ the Fock representation
of the coordinate eigenvector $Q|q\rangle=q|q\rangle,$ ($Q=(a+a^{\dagger
})/\sqrt{2}$)%
\begin{equation}
|q\rangle=\pi^{-1/4}e^{-\frac{q^{2}}{2}+\sqrt{2}qa^{\dagger}-\frac
{a^{\dagger2}}{2}}|0\rangle, \label{3.7}%
\end{equation}
we perform the integration below
\begin{align}
S_{1}  &  \equiv \int_{-\infty}^{\infty}\frac{dq}{\sqrt{\mu}}|\frac{q}{\mu
}\rangle \langle q|\nonumber \\
&  =\int_{-\infty}^{\infty}\frac{dq}{\sqrt{\pi \mu}}e^{-\frac{q^{2}}{2\mu^{2}%
}+\sqrt{2}\frac{q}{\mu}a^{\dagger}-\frac{a^{\dagger2}}{2}}|0\rangle
\langle0|e^{-\frac{q^{2}}{2}+\sqrt{2}qa-\frac{a^{2}}{2}}. \label{3.8a}%
\end{align}
Substituting (\ref{3.1}) into (\ref{3.8a}) we see
\begin{equation}
S_{1}=\int_{-\infty}^{\infty}\frac{dq}{\sqrt{\pi \mu}}e^{-\frac{q^{2}}{2\mu
^{2}}+\sqrt{2}\frac{q}{\mu}a^{\dagger}-\frac{a^{\dagger2}}{2}}\colon
e^{-a^{\dagger}a}\colon e^{-\frac{q^{2}}{2}+\sqrt{2}qa-\frac{a^{2}}{2}}.
\label{3.9}%
\end{equation}
Note that on the left of $\colon e^{-a^{+}a}\colon$ are all creation
operators, while on its right are all annihilation operators, so the whole
integral is in normal ordering, thus using property 1 we have
\begin{equation}
S_{1}=\int_{-\infty}^{\infty}\frac{dq}{\sqrt{\pi \mu}}\colon e^{-\frac{q^{2}%
}{2}(1+\frac{1}{\mu^{2}})+\sqrt{2}q(\frac{a^{\dagger}}{\mu}+a)-\frac{1}%
{2}(a+a^{\dagger})^{2}}\colon. \label{3.10}%
\end{equation}
As $a$ commutes with $a^{\dagger \text{ }}$ within $:$ $:$, so $a^{\dagger
\text{ }}$and $a$ can be considered as if they were parameters while the
integration is performing. Therefore, by setting $\mu=e^{\lambda}$,
sech$\lambda=\frac{2\mu}{1+\mu^{2}},$ tanh$\lambda=\frac{\mu^{2}+1}{\mu^{2}%
-1},$ we are able to perform the integration and obtain%
\begin{align}
S_{1}  &  =\sqrt{\frac{2\mu}{1+\mu^{2}}}\colon \exp \left \{  \frac{\left(
\frac{a^{\dagger}}{\mu}+a\right)  ^{2}}{1+\frac{1}{\mu^{2}}}-\frac{1}%
{2}\left(  a+a^{\dagger}\right)  ^{2}\right \}  \colon \nonumber \\
&  =\left(  \operatorname*{sech}\lambda \right)  ^{1/2}e^{-\frac{a^{\dagger2}%
}{2}\tanh \lambda}\colon e^{\left(  \operatorname*{sech}\lambda-1\right)
a^{\dagger}a}\colon e^{\frac{a^{2}}{2}\tanh \lambda}, \label{3.01}%
\end{align}
which is just the single-mode squeezing operator in normal ordering appearing
in many references. It is worth mentioning that we have not used the SU(1,1)
Lie algebra method in the derivation. The integration automatically arranges
the squeezing operator in normal ordering. Using
\begin{align}
e^{\lambda a^{\dagger}a}  &  =\sum_{n=0}^{\infty}e^{\lambda n}|n\rangle \langle
n|=\sum_{n=0}^{\infty}e^{\lambda n}\frac{a^{\dagger n}}{n!}\colon
e^{-a^{\dagger}a}\colon a^{n}\nonumber \\
&  =\colon \exp[\left(  e^{\lambda}-1\right)  a^{\dagger}a]\colon, \label{3.02}%
\end{align}
Eq. (\ref{3.01}) becomes
\begin{equation}
\int_{-\infty}^{\infty}\frac{dq}{\sqrt{\mu}}|\frac{q}{\mu}\rangle \langle
q|=e^{-\frac{a^{+2}}{2}\tanh \lambda}e^{(a^{+}a+\frac{1}{2})\ln
\operatorname*{sech}\lambda}e^{\frac{a^{2}}{2}\tanh \lambda}. \label{3.03}%
\end{equation}
This shows the classical dilation $q\rightarrow \frac{q}{\mu}$ maps into the
normally ordered squeezing operator manifestly. It also exhibits that the
fundamental representation theory can be formulated in not so abstract way, as
we can now directly perform the integral over ket-bra projection operators.
Moreover, the IWOP\ technique can be employed to perform many complicated
integrations for ket-bra projection operators.

There is a deep ditch between quantum mechanical operators ($q$-numbers)
theory and classical numbers ($c$-numbers) theory. The IWOP technique arranges
non-commutable operators within an ordered product symbol in a way that they
become commutable, in this sense the `ditch' between $q$-numbers and
$c$-numbers is \textquotedblright shoaled\textquotedblright. However, the
nature of operators are not changed, they are still $q-$numbers, not
$c$-numbers. After the integration over $c$-numbers within ordered product is
performed, we can finally get rid of the normal ordering symbol by using
(\ref{3.02}).

When $\mu=1,$ Eq. (\ref{3.03}) becomes
\begin{align}
\int_{-\infty}^{\infty}dq|q\rangle \langle q|  &  =\int_{-\infty}^{\infty}%
\frac{dq}{\sqrt{\pi}}\colon e^{-q^{2}+2q(\frac{a+a^{\dagger}}{\sqrt{2}}%
)-\frac{1}{2}(a+a^{\dagger})^{2}}\colon \nonumber \\
&  =\int_{-\infty}^{\infty}\frac{dq}{\sqrt{\pi}}\colon e^{-\left(  q-Q\right)
^{2}}\colon=1,\; \; \label{3.04}%
\end{align}
a\ real simple Gaussian\ integration! This immediately leads us to put the
completeness relation of the momentum representation into the normally ordered
Gaussian form
\begin{equation}
\int_{-\infty}^{\infty}dp|p\rangle \langle p|=\int_{-\infty}^{\infty}\frac
{dp}{\sqrt{\pi}}\colon e^{-\left(  p-P\right)  ^{2}}\colon=1, \label{3.05}%
\end{equation}
where $P=\left(  a-a^{\dagger}\right)  /(i\sqrt{2}),$ and $\left \vert
p\right \rangle $ is the momentum eigenvector $P\left \vert p\right \rangle
=p\left \vert p\right \rangle $,
\begin{equation}
\left \vert p\right \rangle =\pi^{-\frac{1}{4}}\exp \left[  -\frac{1}{2}%
p^{2}+i\sqrt{2}pa^{\dagger}+\frac{1}{2}a^{\dagger2}\right]  \left \vert
0\right \rangle . \label{3.8}%
\end{equation}
In addition, we should notice that $\left \vert q\right \rangle $ and
$\left \vert p\right \rangle $ are related by the Fourier transform (FT), i.e.
$\left \langle p\right \vert \left.  q\right \rangle =\frac{1}{\sqrt{2\pi}}%
\exp \left(  -iqp\right)  ,$ the integral kernel of the Fraunhofer diffraction
formula in 1-dimensional is such a FT, so FT in classical optics has its
correspondence in quantum mechanical representations' transform. This
enlightens us that in order to find more general analogy between unitary
operators in quantum optics and transformations in classical optics we should
construct new representations for quantum optics theory, and these are the
bi-partite entangled state and many-particle entangled state. These ideal
states can be implemented by optical devices and optical network \cite{JPA1}.
In the following we focus on the bi-partite entangled state.

\subsection{Single-mode Wigner operator}

When we combine (\ref{3.04}) and (\ref{3.05}) we can obtain
\begin{equation}
\pi^{-1}\colon e^{-\left(  q-Q\right)  ^{2}-\left(  p-P\right)  ^{2}}%
\colon \equiv \Delta \left(  q,p\right)  , \label{3.06}%
\end{equation}
which is just the normally ordered Wigner operator since its marginal
integration gives $|q\rangle \langle q|$ and $|p\rangle \langle p|$
respectively, i.e.,
\begin{align}
\int_{-\infty}^{\infty}dq\Delta \left(  q,p\right)   &  =\frac{1}{\sqrt{\pi}%
}\colon e^{-\left(  p-P\right)  ^{2}}\colon=|p\rangle \langle p|,\label{3.07}\\
\int_{-\infty}^{\infty}dp\Delta \left(  q,p\right)   &  =\frac{1}{\sqrt{\pi}%
}\colon e^{-\left(  q-Q\right)  ^{2}}\colon=|q\rangle \langle q|. \label{3.08}%
\end{align}
Thus the Wigner function of quantum state $\rho$ can be calculated as
$W(q,p)=$Tr$[\rho \Delta \left(  q,p\right)  ]$. On the other hand, the Wigner
operator (\ref{3.06}) can be recast into the coherent state representation,
\begin{equation}
\Delta \left(  q,p\right)  \rightarrow \Delta \left(  \alpha,\alpha^{\ast
}\right)  =\int \frac{d^{2}z}{\pi}\left \vert \alpha+z\right \rangle \left \langle
\alpha-z\right \vert e^{\alpha z^{\ast}-\alpha^{\ast}z}, \label{3.09}%
\end{equation}
where $\left \vert z\right \rangle $ is a coherent state. In fact, using the
IWOP technique we can obtain
\begin{align}
\Delta \left(  \alpha,\alpha^{\ast}\right)   &  =\int \frac{d^{2}z}{\pi}%
\colon \exp \{-\left \vert z\right \vert ^{2}+\left(  \alpha+z\right)  a^{\dag
}+\left(  \alpha^{\ast}-z^{\ast}\right)  a\nonumber \\
&  +\alpha z^{\ast}-\alpha^{\ast}z-\left \vert \alpha \right \vert ^{2}\}
\colon \nonumber \\
&  =\frac{1}{\pi}\colon \exp \left \{  -2\left(  a-\alpha \right)  \left(
a^{\dag}-\alpha^{\ast}\right)  \right \}  \colon, \label{3.010}%
\end{align}
which is the same as (\ref{3.06}).

\subsection{Entangled state $\left \vert \eta \right \rangle $ and its Fourier
transform in complex form}

The concept of quantum entanglement was first employed by Einstein, Rosen and
Poldosky (EPR) to challenge that quantum mechanics is incomplete when they
observed that two particles' relative position $Q_{1}-Q_{2}$ and the total
momentum $P_{1}+P_{2}$ are commutable. Hinted by EPR, the bipartite entangled
state $\left \vert \eta \right \rangle $ is introduced as \cite{fank,fanyue}
\begin{equation}
\left \vert \eta \right \rangle =\exp \left[  -\frac{1}{2}\left \vert
\eta \right \vert ^{2}+\eta a_{1}^{\dagger}-\eta^{\ast}a_{2}^{\dagger}%
+a_{1}^{\dagger}a_{2}^{\dagger}\right]  \left \vert 00\right \rangle .
\label{3.11}%
\end{equation}
$\left \vert \eta=\eta_{1}+\mathtt{i}\eta_{2}\right \rangle $ is the common
eigenstate of relative coordinate $Q_{1}-Q_{2}$ and the total momentum
$P_{1}+P_{2}$,
\begin{equation}
\left(  Q_{1}-Q_{2}\right)  \left \vert \eta \right \rangle =\sqrt{2}\eta
_{1}\left \vert \eta \right \rangle ,\text{ }\, \text{\ }\left(  P_{1}%
+P_{2}\right)  \left \vert \eta \right \rangle =\sqrt{2}\eta_{2}\left \vert
\eta \right \rangle , \label{3.12}%
\end{equation}
where $Q_{i}=(a_{j}+a_{j}^{\dagger})/\sqrt{2},\ P_{j}=(a_{j}-a_{j}^{\dagger
})/(\mathtt{i}\sqrt{2}),$ $j=1,2.$ Using the IWOP technique, we can
immediately prove that $\left \vert \eta \right \rangle $ possesses the
completeness relation
\begin{equation}
\int \frac{d^{2}\eta}{\pi}\left \vert \eta \right \rangle \left \langle
\eta \right \vert =\int \frac{d^{2}{\eta}}{\pi}\colon e^{-\left[  \eta^{\ast
}-(a_{1}^{\dagger}-a_{2})\right]  \left[  \eta-(a_{1}-a_{2}^{\dagger})\right]
}\colon=1,\,d^{2}\eta=d\eta_{1}d\eta_{2}, \label{3.13}%
\end{equation}
and orthonormal relation%
\begin{equation}
\left \langle \eta \right \vert \left.  \eta^{\prime}\right \rangle =\pi
\delta(\eta_{1}-\eta_{1}^{\prime})\delta(\eta_{2}-\eta_{2}^{\prime}).
\label{3.14}%
\end{equation}
The Schmidt decomposition of $\left \vert \eta \right \rangle $ is
\begin{equation}
\left \vert \eta \right \rangle =e^{-i\eta_{2}\eta_{1}}\int_{-\infty}^{\infty
}dx\left \vert q\right \rangle _{1}\otimes \left \vert q-\sqrt{2}\eta
_{1}\right \rangle _{2}e^{i\sqrt{2}\eta_{2}x}, \label{3.21}%
\end{equation}
The $\left \vert \eta \right \rangle $ state can also be Schmidt-decomposed in
momentum eigenvector space as
\begin{equation}
\left \vert \eta \right \rangle =e^{i\eta_{1}\eta_{2}}\int_{-\infty}^{\infty
}dp\left \vert p\right \rangle _{1}\otimes \left \vert \sqrt{2}\eta_{2}%
-p\right \rangle _{2}e^{-i\sqrt{2}\eta_{1}p}. \label{3.22}%
\end{equation}
The $\left \vert \eta \right \rangle $ is physically appealing in quantum optics
theory, because the two-mode squeezing operator has its natural representation
on $\left \langle \eta \right \vert $ basis \cite{fanyue}
\begin{equation}%
{\displaystyle \int}
\frac{d^{2}\eta}{\pi \mu}\left \vert \eta/\mu \right \rangle \left \langle
\eta \right \vert =e^{a_{1}^{^{\dagger}}a_{2}^{^{\dagger}}\tanh \lambda}%
e^{(a_{1}^{^{\dagger}}a_{1}+a_{2}^{^{\dagger}}a_{2}+1)\ln \operatorname*{sech}%
\lambda}e^{-a_{1}a_{2}\tanh \lambda},\; \mu=e^{\lambda}, \label{3.23}%
\end{equation}
The proof of (\ref{3.23}) is proceeded by virtue of the IWOP technique%
\begin{align}%
{\displaystyle \int}
\frac{d^{2}\eta}{\pi \mu}\left \vert \eta/\mu \right \rangle \left \langle
\eta \right \vert  &  =%
{\displaystyle \int}
\frac{d^{2}\eta}{\pi \mu}\colon \exp \left \{  -\frac{|\eta|^{2}}{2}\left(
1+\frac{1}{\mu^{2}}\right)  +\eta \left(  \frac{a_{1}^{^{\dagger}}}{\mu}%
-a_{2}\right)  \right. \nonumber \\
&  +\left.  \eta^{\ast}\left(  a_{1}-\frac{a_{2}^{^{\dagger}}}{\mu}\right)
+a_{1}^{\dagger}a_{2}^{^{\dagger}}+a_{1}a_{2}-a_{1}^{\dagger}a_{1}%
-a_{2}^{\dagger}a_{2}\right \} \nonumber \\
&  =\frac{2\mu}{1+\mu^{2}}\colon \exp \left \{  \frac{\mu^{2}}{1+\mu^{2}}\left(
\frac{a_{1}^{^{\dagger}}}{\mu}-a_{2}\right)  \left(  a_{1}-\frac
{a_{2}^{^{\dagger}}}{\mu}\right)  -\left(  a_{1}-a_{2}^{^{\dagger}}\right)
\left(  a_{1}^{^{\dagger}}-a_{2}\right)  \right \}  \colon \nonumber \\
&  =e^{a_{1}^{^{\dagger}}a_{2}^{^{\dagger}}\tanh \lambda}e^{(a_{1}^{^{\dagger}%
}a_{1}+a_{2}^{^{\dagger}}a_{2}+1)\ln \operatorname*{sech}\lambda}e^{-a_{1}%
a_{2}\tanh \lambda}\equiv S_{2}, \label{3.24}%
\end{align}
so the necessity of introducing $\left \vert \eta \right \rangle $ into quantum
optics is clear. $S_{2}$ squeezes $\left \vert \eta \right \rangle $ in the
manifest way
\begin{equation}
S_{2}\left \vert \eta \right \rangle =\frac{1}{\mu}\left \vert \eta/\mu
\right \rangle ,\text{ \ }\mu=e^{\lambda}\ , \label{3.15}%
\end{equation}
and the two-mode squeezed state itself is an entangled state which entangles
the idle mode and signal mode as an outcome of a parametric-down conversion
process \cite{PDC}.

We can also introduce the conjugate state of $\left \vert \eta \right \rangle $
\cite{PRA},
\begin{equation}
\left \vert \xi \right \rangle =\exp \left[  -\frac{1}{2}\left \vert \xi \right \vert
^{2}+\xi a_{1}^{\dagger}+\xi^{\ast}a_{2}^{\dagger}-a_{1}^{\dagger}%
a_{2}^{\dagger}\right]  \left \vert 00\right \rangle ,\text{ }\xi=\xi_{1}%
+i\xi_{2}, \label{3.17}%
\end{equation}
which obeys the eigen-equations
\begin{equation}
\left(  Q_{1}+Q_{2}\right)  \left \vert \xi \right \rangle =\sqrt{2}\xi
_{1}\left \vert \xi \right \rangle ,\, \left(  P_{1}-P_{2}\right)  \left \vert
\xi \right \rangle =\sqrt{2}\xi_{2}\left \vert \xi \right \rangle . \label{3.18}%
\end{equation}
Because $\left[  \left(  Q_{1}-Q_{2}\right)  ,\left(  P_{1}-P_{2}\right)
\right]  =2\mathtt{i},$ so we name the conjugacy between $\left \vert
\xi \right \rangle $ and $\left \vert \eta \right \rangle .$ The completeness and
orthonormal relations of $\left \vert \xi \right \rangle $ are%
\begin{align}
\int \frac{d^{2}\xi}{\pi}\left \vert \xi \right \rangle \left \langle
\xi \right \vert  &  =\int \frac{d^{2}{\xi}}{\pi}\colon e^{-\left[  \xi^{\ast
}-(a_{1}^{\dagger}+a_{2})\right]  \left[  \xi-(a_{1}+a_{2}^{\dagger})\right]
}\colon=1,\label{3.19a}\\
\left \langle \xi \right \vert \left.  \xi^{\prime}\right \rangle  &  =\pi
\delta(\xi_{1}-\xi_{1}^{\prime})\delta(\xi_{2}-\xi_{2}^{\prime}),\text{ }%
d^{2}\xi=d\xi_{1}d\xi_{2}, \label{3.19}%
\end{align}
respectively. $\left \vert \eta \right \rangle $ and $\left \vert \xi \right \rangle
$ can be related to each other by
\begin{equation}
\left \langle \eta|\xi \right \rangle =\frac{1}{2}\exp \left(  \frac{\xi \eta
^{\ast}-\xi^{\ast}\eta}{2}\right)  , \label{3.20}%
\end{equation}
since $\xi^{\ast}\eta-\xi \eta^{\ast}$ is a pure imaginary number, Eq.
(\ref{3.20}) is the Fourier transform kernel in complex form (or named
entangled Fourier transform, this concept should also be extended to
multipartite entangled states.) It will be shown in later sections that
departing from entangled states $\left \vert \eta \right \rangle $ and
$\left \vert \xi \right \rangle $ and the generalized Fresnel operator a new
entangled Fresnel transforms in classical optics can be found.

\subsection{Two-mode Wigner operator in the $\left \vert \eta \right \rangle $
representation}

Combining (\ref{3.13}) and (\ref{3.19a}) we can construct the following
operator
\begin{align}
&  \frac{1}{\pi^{2}}\colon e^{-\left[  \sigma^{\ast}-(a_{1}^{\dagger}%
-a_{2})\right]  \left[  \sigma-(a_{1}-a_{2}^{\dagger})\right]  -\left[
\gamma^{\ast}-(a_{1}^{\dagger}+a_{2})\right]  \left[  \gamma-(a_{1}%
+a_{2}^{\dagger})\right]  }\colon \nonumber \\
&  =\Delta \left(  \alpha,\alpha^{\ast}\right)  \otimes \Delta \left(
\beta,\beta^{\ast}\right)  \equiv \Delta \left(  \sigma,\gamma \right)  ,
\label{3.25}%
\end{align}
where%
\begin{equation}
\sigma=\alpha-\beta^{\ast},\; \gamma=\alpha+\beta^{\ast}. \label{3.26}%
\end{equation}
Eq. (\ref{3.25}) is just equal to the direct product of two single-mode Wigner
operators. It is convenient to express the Wigner operator in the $\left \vert
\eta \right \rangle $ representation as \cite{R1}
\begin{equation}
\Delta \left(  \sigma,\gamma \right)  =\int \frac{d^{2}\eta}{\pi^{3}}\left \vert
\sigma-\eta \right \rangle \left \langle \sigma+\eta \right \vert e^{\eta
\gamma^{\ast}-\eta^{\ast}\gamma}. \label{3.27}%
\end{equation}
For two-mode correlated system, it prefers to using $\Delta \left(
\sigma,\gamma \right)  $ to calculate quantum states' Wigner function. For
example, noticing $\left \langle \eta \right \vert \left.  00\right \rangle
=\exp \{-\left \vert \eta \right \vert ^{2}/2\},$ the two-mode squeezed states'
Wigner function is
\begin{align}
&  \left \langle 00\right \vert S_{2}^{\dag}\left(  \mu \right)  \Delta \left(
\sigma,\gamma \right)  S_{2}\left(  \mu \right)  \left \vert 00\right \rangle
\nonumber \\
&  =\left \langle 00\right \vert \mu^{2}\int \frac{d^{2}\eta}{\pi^{3}}\left \vert
\mu \left(  \sigma-\eta \right)  \right \rangle \left \langle \mu \left(
\sigma+\eta \right)  \right \vert e^{\eta \gamma^{\ast}-\eta^{\ast}\gamma
}\left \vert 00\right \rangle \nonumber \\
&  =\pi^{-2}\exp \left[  -\mu^{2}\left \vert \sigma \right \vert ^{2}-\left \vert
\gamma \right \vert ^{2}/\mu^{2}\right]  . \label{3.28}%
\end{align}

\section{Two deduced entangled state representations and Hankel transform}

\subsection{Deduced entangled states}

Starting from the entangled state $\left \vert \eta=re^{i\theta}\right \rangle $
and introducing an integer $m$, we can deduce new states \cite{fanzou},
\begin{equation}
\left \vert m,r\right \rangle =\frac{1}{2\pi}\int_{0}^{2\pi}d\theta \left \vert
\eta=re^{i\theta}\right \rangle e^{-im\theta}, \label{4.1}%
\end{equation}
which is worth of paying attention because when we operate the
number-difference operator,
\begin{equation}
D\equiv a_{1}^{\dagger}a_{1}-a_{2}^{\dagger}a_{2} \label{4.2}%
\end{equation}
on $\left \vert \eta \right \rangle ,$ using Eq.(\ref{3.11}) we see%
\begin{equation}
D\left \vert \eta \right \rangle =\left(  \eta a_{1}^{^{\dagger}}+\eta^{\ast
}a_{2}^{^{\dagger}}\right)  \left \vert \eta \right \rangle =-i\frac{\partial
}{\partial \theta}\left \vert \eta \right \rangle ,\; \eta=|\eta|e^{i\theta},
\label{4.3}%
\end{equation}
so the number-difference operator corresponds to a differential operation
$i\frac{\partial}{\partial \theta}$ in the $\left \langle \eta \right \vert
\;$representation, this is a remarkable property of $\left \langle
\eta \right \vert $. It then follows%
\begin{equation}
D\left \vert m,r\right \rangle =\int_{0}^{2\pi}\frac{d\theta}{2\pi}e^{-im\theta
}\left(  -i\frac{\partial}{\partial \theta}\left \vert \eta=re^{i\theta
}\right \rangle \right)  =m\left \vert m,r\right \rangle . \label{4.4}%
\end{equation}
On the other hand, by defining
\begin{equation}
K\equiv(a_{1}-a_{2}^{\dagger})(a_{1}^{\dagger}-a_{2}), \label{4.5}%
\end{equation}
we see $\left[  D,K\right]  =0,$ and $\left \vert m,r\right \rangle $ is its
eigenstate,
\begin{equation}
\ K\left \vert m,r\right \rangle =r^{2}\left \vert m,r\right \rangle , \label{4.6}%
\end{equation}
where $K$ is named correlated-amplitude operator since $K\left \vert
\eta \right \rangle =|\eta|^{2}\left \vert \eta \right \rangle .$ Thus we name
$\left \vert m,r\right \rangle $ correlated-amplitude---number-difference
entangled states. It is not difficult to prove completeness and orthonormal
property of $\left \vert m,r\right \rangle $,%
\begin{equation}
\sum_{m=-\infty}^{\infty}\int_{0}^{\infty}d\left(  r^{2}\right)  \left \vert
m,r\right \rangle \left \langle m,r\right \vert =1,\; \label{4.7}%
\end{equation}%
\begin{equation}
\left \langle m,r\right \vert \left.  m^{\prime},r^{\prime}\right \rangle
=\delta_{m,m^{\prime}}\frac{1}{2r}\delta \left(  r-r^{\prime}\right)  .
\label{4.8}%
\end{equation}
On the other hand, from $\left \vert \xi \right \rangle $ we can derive another
state
\begin{equation}
\left \vert s,r^{\prime}\right \rangle =\frac{1}{2\pi}\int_{0}^{2\pi}%
d\varphi \left \vert \xi=r^{\prime}e^{i\varphi}\right \rangle e^{-is\varphi},
\label{4.9}%
\end{equation}
which satisfies%
\begin{equation}
D\left \vert \xi \right \rangle =\left(  a_{1}^{\dagger}\xi-a_{2}^{\dagger}%
\xi^{\ast}\right)  \left \vert \xi \right \rangle =-i\frac{\partial}%
{\partial \varphi}\left \vert \xi=r^{\prime}e^{i\varphi}\right \rangle .
\label{4.10}%
\end{equation}
So $D$ in $\left \langle \xi=r^{\prime}e^{i\varphi}\right \vert $ representation
is equal to $i\frac{\partial}{\partial \varphi}$. Consequently,
\begin{equation}
D\left \vert s,r^{\prime}\right \rangle =\int_{0}^{2\pi}\frac{d\theta}{2\pi
}e^{-is\theta}\left(  -i\frac{\partial}{\partial \theta}\left \vert
\xi=r^{\prime}e^{i\theta}\right \rangle \right)  =s\left \vert s,r^{\prime
}\right \rangle . \label{4.11}%
\end{equation}
Note $\left[  D,(a_{1}^{\dagger}+a_{2})(a_{1}+a_{2}^{\dagger})\right]  =0$ and%
\begin{equation}
(a_{1}^{\dagger}+a_{2})(a_{1}+a_{2}^{\dagger})\left \vert s,r^{\prime
}\right \rangle =r^{\prime2}\left \vert s,r^{\prime}\right \rangle . \label{4.12}%
\end{equation}
$\left \vert s,r^{\prime}\right \rangle $\ is qualified to be a new
representation since
\begin{equation}
\sum_{s=-\infty}^{\infty}\int_{0}^{\infty}d\left(  r^{\prime2}\right)
\left \vert s,r^{\prime}\right \rangle \left \langle s,r^{\prime}\right \vert
=1,\text{ }\left \langle s,r^{\prime}\right \vert \left.  s^{\prime}%
,r^{\prime \prime}\right \rangle =\delta_{s,s^{\prime}}\frac{1}{2r^{\prime}%
}\delta \left(  r^{\prime}-r^{\prime \prime}\right)  . \label{4.13}%
\end{equation}

\subsection{Hankel transform between two deduced entangled state
representations}

Since $\left \vert \xi \right \rangle $ and $\left \vert \eta \right \rangle $ are
mutual conjugate, $\left \vert s,r^{\prime}\right \rangle $ is the conjugate
state of $\left \vert m,r\right \rangle $. From the definition of $\left \vert
m,r\right \rangle $ and $\left \vert s,r^{\prime}\right \rangle $ and
(\ref{3.20}) we calculate the overlap \cite{Fanpla1}%
\begin{align}
\left \langle s,r^{\prime}\right \vert \left.  q,r\right \rangle  &  =\frac
{1}{4\pi^{2}}\int_{0}^{2\pi}d\varphi e^{is\varphi}\left \langle \xi=r^{\prime
}e^{i\varphi}\right \vert \int_{0}^{2\pi}d\theta \left \vert \eta=re^{i\theta
}\right \rangle e^{-im\theta}\nonumber \\
&  =\frac{1}{8\pi^{2}}\int_{0}^{2\pi}\int_{0}^{2\pi}e^{is\varphi-im\theta}%
\exp \left[  irr^{\prime}\sin \left(  \theta-\varphi \right)  \right]  d\theta
d\varphi \nonumber \\
&  =\frac{1}{8\pi^{2}}\int_{0}^{2\pi}\int_{0}^{2\pi}e^{is\varphi-im\theta}%
\sum_{l=-\infty}^{\infty}J_{l}\left(  rr^{\prime}\right)  e^{il\left(
\theta-\varphi \right)  }\nonumber \\
&  =\frac{1}{2}\sum_{l=-\infty}^{\infty}\delta_{l,m}\delta_{l,s}J_{l}\left(
rr^{\prime}\right)  =\frac{1}{2}\delta_{s,m}J_{s}\left(  rr^{\prime}\right)  ,
\label{4.14}%
\end{align}
where we have identified the generating function of the s-order Bessel
function $J_{l},$%
\begin{equation}
e^{ix\sin t}=\sum_{l=-\infty}^{\infty}J_{l}\left(  x\right)  e^{ilt},\;
\label{4.15}%
\end{equation}
and%
\begin{equation}
J_{l}\left(  x\right)  =\sum_{k=0}^{\infty}\frac{\left(  -1\right)  ^{l}%
}{k!\left(  l+k\right)  !}\left(  \frac{x}{2}\right)  ^{l+2k}. \label{4.16}%
\end{equation}
Eq. (\ref{4.14}) is remarkable, because $J_{s}\left(  rr^{\prime}\right)  $ is
just the integral kernel of Hankel transform. In fact, if we define
\begin{equation}
\left \langle m,r\right \vert \left.  g\right \rangle \equiv g\left(  m,r\right)
,\text{ }\left \langle s,r^{\prime}\right \vert \left.  g\right \rangle
\equiv \mathcal{G}\left(  s,r^{\prime}\right)  , \label{4.17}%
\end{equation}
and use (\ref{4.7}) as well as (\ref{4.14}), we obtain%
\begin{align}
\mathcal{G}\left(  s,r^{\prime}\right)   &  =\sum_{m=-\infty}^{\infty}\int
_{0}^{\infty}d\left(  r^{2}\right)  \left \langle s,r^{\prime}\right \vert
\left.  m,r\right \rangle \left \langle m,r\right \vert \left.  g\right \rangle
\nonumber \\
&  =\frac{1}{2}\int_{0}^{\infty}d\left(  r^{2}\right)  J_{s}\left(
rr^{\prime}\right)  g\left(  s,r\right)  \equiv \mathcal{H}\left[  g\left(
s,r\right)  \right]  , \label{4.18}%
\end{align}
which is just the Hankel transform of $g\left(  m,r\right)  $ (or it can be
regarded as a simplified form of the Collins formula in cylindrical
coordinate, see (\ref{2.10})). The inverse transform of (\ref{4.18}) is%
\begin{align}
g\left(  m,r\right)   &  =\left \langle m,\mathfrak{r}\right \vert
\sum_{s=-\infty}^{\infty}\int_{0}^{\infty}d\left(  r^{\prime2}\right)
\left \vert s,r^{\prime}\right \rangle \left \langle s,r^{\prime}\right \vert
\left.  g\right \rangle \nonumber \\
&  =\frac{1}{2}\int_{0}^{\infty}d\left(  r^{\prime2}\right)  J_{q}\left(
rr^{\prime}\right)  \mathcal{G}\left(  m,r^{\prime}\right)  \equiv
\mathcal{H}^{-1}\left[  \mathcal{G}\left(  m,r^{\prime}\right)  \right]  .
\label{4.19}%
\end{align}
Now we know that the quantum optical image of classical Hankel transform just
corresponds to the representation transformation between two mutually
conjugate entangled states $\left \langle s,r^{\prime}\right \vert $ and
$\left \vert m,r\right \rangle ,$ this is like the case that the Fourier
transform kernel is just the matrix element between the coordinate state and
the momentum state, a wonderful result unnoticed before. Therefore the
bipartite entangled state representations' transforms, which can lead us to
the Hankel transform, was proposed first in classical optics, can find their
way back in quantum optics.

\subsection{Quantum optical version of classical circular harmonic
correlation}

From Eq.(\ref{4.1}) we can see that its reciprocal relation is the circular
harmonic expansion,%
\begin{equation}
\left \vert \eta=re^{i\theta}\right \rangle =\sum_{m=-\infty}^{\infty}\left \vert
m,r\right \rangle e^{im\theta}, \label{4.23}%
\end{equation}
or correlated-amplitude---number-difference entangled state $\left \vert
m,r\right \rangle $ can be considered as circular harmonic decomposition of
$\left \vert \eta=re^{i\theta}\right \rangle .$ Let $g\left(  r,\theta \right)
$, a general 2-dimensional function expressed in polar coordinates, be
periodic in the variable $\theta,$ it can be looked as the wavefunction of the
state vector $\left \vert g\right \rangle $ in the $\left \langle \eta
=re^{i\theta}\right \vert $ representation
\begin{equation}
g\left(  r,\theta \right)  =\left \langle \eta=re^{i\theta}\right.  \left \vert
g\right \rangle ,\text{ }\text{\ } \label{4.24}%
\end{equation}
using (\ref{4.23}) we have%
\begin{equation}
g\left(  r,\theta \right)  =\sum_{m=-\infty}^{\infty}g_{m}\left(  r\right)
e^{-im\theta},\text{ \ }g_{m}\left(  r\right)  =\left \langle m,r\right.
\left \vert g\right \rangle , \label{4.25}%
\end{equation}
$g_{m}\left(  r\right)  $ is the wavefunction of $\left \vert g\right \rangle $
in $\left \langle m,r\right \vert $ representation. By using\ (\ref{4.7}) and
noticing that It then follows from (\ref{4.3})%
\begin{equation}
e^{-i\alpha(a_{1}^{^{\dagger}}a_{1}-a_{2}^{^{\dagger}}a_{2})}\left \vert
\eta=re^{i\theta}\right \rangle =e^{-\alpha \frac{\partial}{\partial \theta}%
}\left \vert \eta=re^{i\theta}\right \rangle =\left \vert \eta=re^{i\left(
\theta-\alpha \right)  }\right \rangle , \label{4.26}%
\end{equation}
so $e^{-i\alpha \left(  a_{1}^{^{\dagger}}a_{1}-a_{2}^{^{\dagger}}a_{2}\right)
}$ behaves a rotation operator in $\left \vert \eta \right \rangle $
representation, we see that the expectation value of $e^{-i\alpha \left(
a_{1}^{^{\dagger}}a_{1}-a_{2}^{^{\dagger}}a_{2}\right)  }$ in $\left \vert
g\right \rangle $ is%
\begin{align}
\pi \left \langle g\right \vert e^{-i\alpha(a_{1}^{^{\dagger}}a_{1}-a_{2}%
^{^{\dag}}a_{2})}\left \vert g\right \rangle  &  =\pi \left \langle g\right \vert
\int \frac{d^{2}\eta}{\pi}e^{-i\alpha(a_{1}^{^{\dagger}}a_{1}-a_{2}^{^{\dagger
}}a_{2})}\left \vert \eta \right \rangle \left \langle \eta \right.  \left \vert
g\right \rangle \nonumber \\
&  =\int_{0}^{\infty}rdrd\theta \int_{0}^{2\pi}\left \langle g\right.
\left \vert \eta^{\prime}=re^{i\left(  \theta-\alpha \right)  }\right \rangle
\left \langle \eta=re^{i\theta}\right.  \left \vert g\right \rangle \nonumber \\
&  =\int_{0}^{\infty}rdr\int_{0}^{2\pi}g^{\ast}\left(  r,\theta-\alpha \right)
g\left(  r,\theta \right)  d\theta \equiv R_{\alpha}, \label{4.27}%
\end{align}
which is just the cross-correlation between $g\left(  r,\theta \right)  $ and
an angularly rotated version of the same function, $g^{\ast}\left(
r,\theta-\alpha \right)  $. On the other hand, using (\ref{4.7}) we have%
\begin{equation}
\left \vert g\right \rangle =\sum_{m=-\infty}^{\infty}\int_{0}^{\infty}d\left(
r^{2}\right)  \left \vert m,r\right \rangle \left \langle m,r\right \vert \left.
g\right \rangle =\sum_{m=-\infty}^{\infty}\int_{0}^{\infty}d\left(
r^{2}\right)  \left \vert m,r\right \rangle g_{m}\left(  r\right)  .
\label{4.28}%
\end{equation}
Substituting (\ref{4.28}) into (\ref{4.27}) and using the eigenvector equation
(\ref{4.4}) as well as (\ref{4.8}) we obtain%
\begin{align}
R_{\alpha}  &  =\pi \sum_{m^{\prime}=-\infty}^{\infty}\int_{0}^{\infty}d\left(
r^{\prime2}\right)  \left \langle m^{\prime},r^{\prime}\right \vert
g_{m^{\prime}}^{\ast}\left(  r^{\prime}\right)  e^{-i\alpha \left(
a_{1}^{^{\dagger}}a_{1}-a_{2}^{^{\dagger}}a_{2}\right)  }\sum_{m=-\infty
}^{\infty}\int_{0}^{\infty}d\left(  r^{2}\right)  \left \vert m,r\right \rangle
g_{m}\left(  r\right) \nonumber \\
&  =\pi \sum_{m^{\prime}=-\infty}^{\infty}\sum_{m=-\infty}^{\infty}\int
_{0}^{\infty}d\left(  r^{\prime2}\right)  g_{m^{\prime}}^{\ast}\left(
r^{\prime}\right)  e^{-im\alpha}\int_{0}^{\infty}d\left(  r^{2}\right)
g_{m}\left(  r\right)  \delta_{m,m^{\prime}}\frac{1}{2r}\delta \left(
r-r^{\prime}\right) \nonumber \\
&  =2\pi \sum_{m=-\infty}^{\infty}e^{-im\alpha}\int_{0}^{\infty}r|g_{m}\left(
r\right)  |^{2}dr, \label{4.29}%
\end{align}
from which we see that each of the circular harmonic components of the
crosscorrelation undergoes a different phase shift $-m\alpha,$ so $R_{\alpha}$
is not rotation invariant. However, when we consider only one harmonic
component
\begin{equation}
R_{\alpha,M}=2\pi e^{-iM\alpha}\int_{0}^{\infty}r|g_{M}\left(  r\right)
|^{2}dr, \label{4.30}%
\end{equation}
is extracted digitally, then from the phase associated with this component it
is possible to determine the angular shift that one version of the object has
undergone. When an optical filter that is matched to $R_{\alpha,M}$ of a
particular object is constructed, then if that some object is entered as an
input to the system with any angular rotation, a correlation peak of strength
proportional to $\int_{0}^{\infty}r|g_{M}\left(  r\right)  |^{2}dr$ will be
produced, independent of rotation. Hence an optical correlator can be
constructed that will recognize that object independent of rotation
\cite{Goodman}.

So far we have studied the circular harmonic correlation in the context of
quantum optics, we have endowed the crosscorrelation $R_{\alpha}$ with a
definite quantum mechanical meaning, i.e. the overlap between $\left \langle
g\right \vert $ and the rotated state $e^{i\alpha \left(  a_{1}^{^{\dagger}%
}a_{1}-a_{2}^{^{\dagger}}a_{2}\right)  }\left \vert g\right \rangle ,$ in the
entangled state representation.

Note that Fourier-based correlators is also very sensitive to magnification,
however, the magnitude of Mellin transform is independent of scale-size
changes in the input \cite{Goodman}. Now we examine when $\left \vert
g\right \rangle $ is both rotated and squeezed (by\ a two-mode squeezing
operator $S_{2}\left(  \lambda \right)  =\exp[\lambda(a_{1}^{\dagger}%
a_{2}^{\dagger}-a_{1}a_{2})])$, then from (\ref{3.15}) and (\ref{4.24}) we
have%
\begin{equation}
S_{2}\left(  \lambda \right)  \left \vert g\right \rangle =\int \frac{d^{2}\eta
}{\pi \mu}\left \vert \eta/\mu \right \rangle \left \langle \eta \right.  \left \vert
g\right \rangle =\int \frac{d^{2}\eta}{\pi \mu}\left \vert \eta/\mu \right \rangle
g\left(  r,\theta \right)  , \label{4.31}%
\end{equation}
it follows the overlap between $\left \langle g\right \vert $ and the state
$e^{-i\alpha(a_{1}^{^{\dagger}}a_{1}-a_{2}^{^{\dagger}}a_{2})}S\left(
\lambda \right)  \left \vert g\right \rangle ,$
\begin{align}
W_{\alpha,\lambda}  &  \equiv \pi \left \langle g\right \vert e^{-i\alpha \left(
a_{1}^{^{\dagger}}a_{1}-a_{2}^{^{\dagger}}a_{2}\right)  }S\left(
\lambda \right)  \left \vert g\right \rangle \nonumber \\
&  =\left \langle g\right \vert \int \frac{d^{2}\eta}{\mu}e^{-i\alpha \left(
a_{1}^{^{\dagger}}a_{1}-a_{2}^{^{\dagger}}a_{2}\right)  }\left \vert \eta
/\mu \right \rangle g\left(  r,\theta \right) \nonumber \\
&  =\int_{0}^{\infty}\frac{rdr}{\mu}\int_{0}^{2\pi}\left \langle g\right.
\left \vert \eta^{\prime}=e^{i\left(  \theta-\alpha \right)  }r/\mu \right \rangle
g\left(  r,\theta \right)  d\theta \nonumber \\
&  =\int_{0}^{\infty}\frac{rdr}{\mu}\int_{0}^{2\pi}g^{\ast}\left(
r/\mu,\theta-\alpha \right)  g\left(  r,\theta \right)  d\theta, \label{4.32}%
\end{align}
which corresponds to the crosscorrelation arising from combination of
squeezing and rotation (joint transform correlator). On the other hand, from
(\ref{4.1}) and (\ref{3.15}) we see
\begin{equation}
S_{2}\left(  \lambda \right)  \left \vert m,r\right \rangle =\frac{1}{2\pi \mu
}\int_{0}^{2\pi}d\theta \left \vert \eta=\frac{r}{\mu}e^{i\theta}\right \rangle
e^{-im\theta}=\frac{1}{\mu}\left \vert m,\frac{r}{\mu}\right \rangle ,
\label{4.33}%
\end{equation}
and therefore%
\begin{align}
W_{\alpha,\lambda}  &  =\frac{\pi}{\mu}\sum_{m^{\prime}=-\infty}^{\infty}%
\int_{0}^{\infty}d\left(  r^{\prime2}\right)  \left \langle m^{\prime
},r^{\prime}\right \vert g_{m^{\prime}}^{\ast}\left(  r^{\prime}\right)
e^{-i\alpha \left(  a_{1}^{^{\dagger}}a_{1}-a_{2}^{^{\dagger}}a_{2}\right)
}\sum_{m=-\infty}^{\infty}\int_{0}^{\infty}d\left(  r^{2}\right)  \left \vert
m,\frac{r}{\mu}\right \rangle g_{m}\left(  r\right) \nonumber \\
&  =\frac{\pi}{\mu}\sum_{m^{\prime}=-\infty}^{\infty}\sum_{m=-\infty}^{\infty
}\int_{0}^{\infty}d\left(  r^{\prime2}\right)  g_{m^{\prime}}^{\ast}\left(
r^{\prime}\right)  e^{-im\alpha}\int_{0}^{\infty}d\left(  r^{2}\right)
g_{m}\left(  r\right)  \delta_{m,m^{\prime}}\frac{1}{2r^{\prime}}\delta \left(
\frac{r}{\mu}-r^{\prime}\right) \nonumber \\
&  =\frac{2\pi}{\mu}\sum_{m=-\infty}^{\infty}e^{-im\alpha}\int_{0}^{\infty
}rg_{m}\left(  r\right)  g_{m^{\prime}}^{\ast}\left(  re^{-\lambda}\right)
dr,\text{ \ } \label{4.34}%
\end{align}
from which one can see that to achieve simultaneous scale and rotation
invariance, a two-dimensional object $g\left(  r,\theta \right)  $ should be
entered into the optical system in a distorted polar coordinate system, the
distortation arising from the fact that the radial coordinate is stretched by
a logarithmic transformation $\left(  \lambda=-\ln \mu \right)  $, which
coincides with Ref. \cite{Casasent}. The quantum optical version is thus
established which is a new tie connecting Fourier optics and quantum optics
\cite{Fan OC}.

At the end of this section, using the two-variable Hermite polynomials'
definition \cite{Erd}
\begin{equation}
H_{m,n}\left(  \xi,\xi^{\ast}\right)  =\sum_{l=0}^{\min \left(  m,n\right)
}\frac{m!n!\left(  -1\right)  ^{l}\xi^{m-l}\xi^{\ast n-l}}{l!\left(
m-l\right)  !\left(  n-l\right)  !}, \label{4.20}%
\end{equation}
which is quite different from the product of two single-variable Hermite
polynomials, and its generating function formula is
\begin{equation}
\sum_{m,n=0}^{\infty}\frac{t^{m}t^{\prime n}}{m!n!}H_{m,n}\left(  \xi
,\xi^{\ast}\right)  =\exp \left[  -tt^{\prime}+t\xi+t^{\prime}\xi^{\ast
}\right]  , \label{4.21}%
\end{equation}
and noting $H_{m,n}(\xi,\xi^{\ast})=e^{i(m-n)\varphi}H_{m,n}(r^{\prime
},r^{\prime}),$ we can directly perform the integral in (\ref{4.9}) and derive
the explicit form of $\left \vert s,r^{\prime}\right \rangle ,$%
\begin{align}
\left \vert s,r^{\prime}\right \rangle  &  =\frac{1}{2\pi}\int_{0}^{2\pi
}d\varphi \exp \{-r^{\prime2}/2+\xi a_{1}^{\dagger}+\xi^{\ast}a_{2}^{\dagger
}-is\varphi-a_{1}^{\dagger}a_{2}^{\dagger}\} \left \vert 00\right \rangle
\nonumber \\
&  =\frac{1}{2\pi}e^{-r^{\prime2}/2}\int_{0}^{2\pi}d\varphi \sum \limits_{m,n=0}%
^{\infty}\frac{a_{1}^{\dagger m}a_{2}^{\dagger n}}{m!n!}H_{m,n}(\xi,\xi^{\ast
})e^{-is\varphi}\left \vert 00\right \rangle \nonumber \\
&  =\frac{1}{2\pi}e^{-r^{\prime2}/2}\int_{0}^{2\pi}d\varphi \sum \limits_{m,n=0}%
^{\infty}\frac{1}{\sqrt{m!n!}}H_{m,n}(r^{\prime},r^{\prime})e^{i\varphi \left(
m-n-s\right)  }\left \vert m,n\right \rangle \nonumber \\
&  =e^{-r^{\prime2}/2}\sum \limits_{n=0}^{\infty}\frac{1}{\sqrt{\left(
n+s\right)  !n!}}H_{n+s,n}(r^{\prime},r^{\prime})\left \vert n+s,n\right \rangle
, \label{4.22}%
\end{align}
which is really an entangled state in two-mode Fock space. Eqs. (\ref{4.21})
and (\ref{4.22}) will be often used in the following discussions.

In the following we concentrate on finding the generalized Fresnel operators
in both one- and two- mode cases with use of the IWOP technique.

\section{Single-mode Fresnel operator as the image of the classical Optical
Fresnel Transform}

In this section we shall mainly introduce so-called generalized Fresnel
operators (GFO) (in one-mode and two-mode cases both) \cite{PLAFAN} and some
appropriate quantum optical representations (e.g. coherent state
representation and entangled state representation) to manifestly link the
formalisms in quantum optics to those in classical optics. In so doing, we
find that the various transforms in classical optics are just the result of
generalized Fresnel operators inducing transforms on appropriate quantum state
vectors, i.e. classical optical Fresnel transforms have their counterpart in
quantum optics. Besides, we can study the important $ABCD$ rule obeyed by
Gaussian beam propagation (also the ray propagation in matrix optics)
\cite{Gerrard} in the domain of quantum optics.

\subsection{Single-mode GFO gained via coherent state method}

For the coherent state $\left \vert z\right \rangle $ in quantum optics
\cite{Glauber,Klauder}
\begin{equation}
\left \vert z\right \rangle =\exp \left[  za^{\dagger}-z^{\ast}a\right]
|0\rangle \equiv \left \vert \left(
\begin{array}
[c]{c}%
z\\
z^{\ast}%
\end{array}
\right)  \right \rangle , \label{3.2}%
\end{equation}
which is the eigenstate of annihilation operator $a,$\ $a\left \vert
z\right \rangle =z\left \vert z\right \rangle $, using the IWOP and (\ref{3.1}),
we can put the over-completeness relation of $\left \vert z\right \rangle $ into
normal ordering%
\begin{equation}
\int \frac{d^{2}{z}}{\pi}\left \vert z\right \rangle \left \langle z\right \vert
=\int \frac{d^{2}{z}}{\pi}\colon e^{-\left(  z^{\ast}-a^{\dagger}\right)
\left(  z-a\right)  }\colon=1. \label{3.3}%
\end{equation}
the canonical form of coherent state $\left \vert z\right \rangle $ is expressed
as
\begin{equation}
\left \vert z\right \rangle =\left \vert p,q\right \rangle =\exp \left[  i\left(
pQ-qP\right)  \right]  |0\rangle \equiv \left \vert \left(
\begin{array}
[c]{c}%
q\\
p
\end{array}
\right)  \right \rangle , \label{3.6}%
\end{equation}
where $z=\left(  q+\mathtt{i}p\right)  /\sqrt{2}$. It follows that
$\left \langle p,q\right \vert Q\left \vert p,q\right \rangle =q,$ $\left \langle
p,q\right \vert P\left \vert p,q\right \rangle =p,$ this indicates that the
states $\left \vert p,q\right \rangle $ generate a canonical phase-space
representation for a state $\left \vert \Psi \right \rangle ,$ $\Psi \left(
p,q\right)  =\left \langle p,q\right \vert \left.  \Psi \right \rangle .$ Thus the
coherent state is a good candidate for providing with classical phase-space
description of quantum systems. Remembering that the Fresnel transform's
parameters $\left(  A,B,C,D\right)  $ are elements of a ray transfer matrix
$M$ describing optical systems, $M$ belongs to the unimodular symplectic
group, and the coherent state $\left \vert p,x\right \rangle $ is a good
candidate for providing with classical phase-space description of quantum
systems, we naturally think of that the symplectic transformation $\left(
\begin{array}
[c]{cc}%
A & B\\
C & D
\end{array}
\right)  \left(
\begin{array}
[c]{c}%
q\\
p
\end{array}
\right)  $ in classical phase space may mapping onto a generalized Fresnel
operator in Hilbert space through the coherent state basis. Thus we construct
the following ket-bra projection operator%
\begin{equation}
\iint \limits_{-\infty}^{\infty}dxdp\left \vert \left(
\begin{array}
[c]{cc}%
A & B\\
C & D
\end{array}
\right)  \left(
\begin{array}
[c]{c}%
q\\
p
\end{array}
\right)  \right \rangle \left \langle \left(
\begin{array}
[c]{c}%
q\\
p
\end{array}
\right)  \right \vert \label{5.1}%
\end{equation}
as the GFO. In fact, using notation of $\left \vert z\right \rangle $ (coherent
state)$,$ and introducing complex numbers $s,r,$%
\begin{equation}
s=\frac{1}{2}\left[  A+D-i\left(  B-C\right)  \right]  ,\;r=-\frac{1}%
{2}\left[  A-D+i\left(  B+C\right)  \right]  ,\;|s|^{2}-|r|^{2}=1, \label{5.2}%
\end{equation}
from (\ref{3.6}) we know
\begin{align}
\left \vert \left(
\begin{array}
[c]{cc}%
A & B\\
C & D
\end{array}
\right)  \left(
\begin{array}
[c]{c}%
q\\
p
\end{array}
\right)  \right \rangle  &  =\left \vert \left(
\begin{array}
[c]{cc}%
s & -r\\
-r^{\ast} & s^{\ast}%
\end{array}
\right)  \left(
\begin{array}
[c]{c}%
z\\
z^{\ast}%
\end{array}
\right)  \right \rangle \equiv \left \vert sz-rz^{\ast}\right \rangle \nonumber \\
&  =\exp \left[  -\frac{1}{2}|sz-rz^{\ast}|^{2}+(sz-rz^{\ast})a^{\dagger
}\right]  |0\rangle, \label{5.3}%
\end{align}
$\left(
\begin{array}
[c]{cc}%
s & -r\\
-r^{\ast} & s^{\ast}%
\end{array}
\right)  $ is still a symplectic group element, so (\ref{5.1}) becomes
\cite{fancommun1}%
\begin{equation}
F_{1}\left(  s,r\right)  =\sqrt{s}\int \frac{d^{2}z}{\pi}\left \vert
sz-rz^{\ast}\right \rangle \left \langle z\right \vert ,\; \; \label{5.4}%
\end{equation}
where the factor $\sqrt{s}$ is attached for anticipating the unitarity of the
operator $F_{1}.$ Eq. (\ref{5.1}) tells us that $c$-number transform $\left(
q,p\right)  \rightarrow \left(  Aq+Bp,Cq+Dp\right)  $ in coherent state basis
maps into $F_{1}\left(  s,r\right)  $. Now we prove $F_{1}\left(  s,r\right)
$ is really the FO we want. Using the IWOP technique and Eq. (\ref{5.3}) and
(\ref{3.1}) we can perform the integral%
\begin{align}
F_{1}\left(  s,r\right)   &  =\sqrt{s}\int \frac{d^{2}z}{\pi}\colon \exp \left[
-\left \vert s\right \vert ^{2}\left \vert z\right \vert ^{2}+sza^{\dagger
}+z^{\ast}\left(  a-ra^{\dagger}\right)  +\frac{r^{\ast}s}{2}z^{2}%
+\frac{rs^{\ast}}{2}z^{\ast2}-a^{\dagger}a\right]  \colon \nonumber \\
&  =\frac{1}{\sqrt{s^{\ast}}}\exp \left(  -\frac{r}{2s^{\ast}}a^{\dagger
2}\right)  \colon \exp \left \{  \left(  \frac{1}{s^{\ast}}-1\right)  a^{\dagger
}a\right \}  \colon \exp \left(  \frac{r^{\ast}}{2s^{\ast}}a^{2}\right)
\nonumber \\
&  =\exp \left(  -\frac{r}{2s^{\ast}}a^{\dagger2}\right)  \exp \left \{  \left(
a^{\dagger}a+\frac{1}{2}\right)  \ln \frac{1}{s^{\ast}}\right \}  \exp \left(
\frac{r^{\ast}}{2s^{\ast}}a^{2}\right)  , \label{5.5}%
\end{align}
where we have used the mathematical formula \cite{book}
\begin{align}
&  \int \frac{d^{2}z}{\pi}\exp \{ \zeta \left \vert z\right \vert ^{2}+\xi z+\eta
z^{\ast}+fz^{2}+gz^{\ast2}\} \nonumber \\
&  =\frac{1}{\sqrt{\zeta^{2}-4fg}}\exp \left \{  \frac{-\zeta \xi \eta+\xi
^{2}g+\eta^{2}f}{\zeta^{2}-4fg}\right \}  , \label{5.6}%
\end{align}
with the convergent condition $\operatorname{Re}(\zeta \pm f\pm
g)<0,\operatorname{Re}(\frac{\zeta^{2}-4fg}{\zeta \pm f\pm g})<0.$ It then
follows
\begin{equation}
\left \langle z\right \vert F_{1}\left(  s,r\right)  \left \vert z^{\prime
}\right \rangle =\frac{1}{\sqrt{s^{\ast}}}\exp \left[  -\frac{\left \vert
z\right \vert ^{2}+\left \vert z^{\prime}\right \vert ^{2}}{2}-\frac{rz^{\ast2}%
}{2s^{\ast}}+\frac{r^{\ast}z^{\prime2}}{2s^{\ast}}+\frac{z^{\ast}z^{\prime}%
}{s^{\ast}}\right]  . \label{5.8}%
\end{equation}
Then using
\begin{equation}
\left \langle x_{i}\right \vert \left.  z\right \rangle =\pi^{-1/4}\exp \left(
-\frac{x_{i}^{2}}{2}+\sqrt{2}x_{i}z-\frac{z^{2}}{2}-\frac{\left \vert
z\right \vert ^{2}}{2}\right)  . \label{5.9}%
\end{equation}
and the completeness relation of coherent state as well as (\ref{5.2}) we
obtain the matrix element of $F_{1}\left(  s,r\right)  $ ($\equiv F_{1}\left(
A,B,C\right)  $) in coordinate representation $\left \langle x_{i}\right \vert
$,
\begin{align}
\left \langle x_{2}\right \vert F_{1}\left(  s,r\right)  \left \vert
x_{1}\right \rangle  &  =\int \frac{d^{2}z}{\pi}\left \langle x_{2}\right \vert
\left.  z\right \rangle \left \langle z\right \vert F_{1}\left(  s,r\right)
\int \frac{d^{2}z^{\prime}}{\pi}\left \vert z^{\prime}\right \rangle \left \langle
z^{\prime}\right \vert \left.  x_{1}\right \rangle \nonumber \\
&  =\frac{1}{\sqrt{2\pi iB}}\exp \left[  \frac{i}{2B}\left(  Ax_{1}^{2}%
-2x_{2}x_{1}+Dx_{2}^{2}\right)  \right]  \equiv \mathcal{K}\left(  x_{2}%
,x_{1}\right)  , \label{5.10}%
\end{align}
which is just the kernel of generalized Fresnel transform $\mathcal{K}\left(
x_{2},x_{1}\right)  $ in (\ref{2.12}). The above discussions demonstrate how
to transit classical Fresnel transform to GFO through the coherent state and
the IWOP technique.

Now if we define $g\left(  x_{2}\right)  =\left \langle x_{2}\right \vert
\left.  g\right \rangle $, $f\left(  x_{1}\right)  =\left \langle x_{1}%
\right \vert \left.  f\right \rangle $ and using Eq. (\ref{3.10}), we can
rewrite Eq. (\ref{2.11}) as%
\begin{align}
\left \langle x_{2}\right \vert \left.  g\right \rangle  &  =\int_{-\infty
}^{\infty}dx_{1}\left \langle x_{2}\right \vert F_{1}\left(  A,B,C\right)
\left \vert x_{1}\right \rangle \left \langle x_{1}\right \vert \left.
f\right \rangle \nonumber \\
&  =\left \langle x_{2}\right \vert F_{1}\left(  A,B,C\right)  \left \vert
f\right \rangle , \label{5.11}%
\end{align}
which is just the quantum mechanical version of GFO. Therefore, the
1-dimensional GFT in classical optics corresponds to the 1-mode GFO
$F_{1}\left(  A,B,C\right)  $ operating on state vector $\left \vert
f\right \rangle $ in Hilbert space, i.e. $\left \vert g\right \rangle
=F_{1}\left(  A,B,C\right)  \left \vert f\right \rangle $. One merit of GFO is:
using coordinate-momentum representation transform we can immediately obtain
GFT in \textquotedblleft frequency" domain, i.e.
\begin{align}
&  \left \langle p_{2}\right \vert F\left \vert p_{1}\right \rangle =\int
_{-\infty}^{\infty}dx_{1}dx_{2}\left \langle p_{2}\right \vert \left.
x_{2}\right \rangle \left \langle x_{2}\right \vert F\left \vert x_{1}%
\right \rangle \left \langle x_{1}\right.  \left \vert p_{1}\right \rangle
\nonumber \\
&  =\frac{1}{\sqrt{2\pi iB}}\int_{-\infty}^{\infty}\frac{dx_{1}dx_{2}}{2\pi
}\exp \left[  \frac{iA}{2B}\left(  x_{1}^{2}+\frac{x_{1}}{A}\left(
Bp_{1}-2x_{2}\right)  \right)  +\frac{iD}{2B}x_{2}^{2}-ip_{2}x_{2}\right]
\nonumber \\
&  =\frac{1}{\sqrt{2\pi i\left(  -C\right)  }}\exp \left[  \frac{i}{2\left(
-C\right)  }\left(  Dp_{1}^{2}-2p_{2}p_{1}+Ap_{2}^{2}\right)  \right]  .
\label{5.12}%
\end{align}
Obviously, $F_{1}\left(  A,B,C\right)  $ induces the following transform
\begin{equation}
F_{1}^{-1}\left(  A,B,C\right)  \left(
\begin{array}
[c]{c}%
Q\\
P
\end{array}
\right)  F_{1}\left(  A,B,C\right)  =\left(
\begin{array}
[c]{cc}%
A & B\\
C & D
\end{array}
\right)  \left(
\begin{array}
[c]{c}%
Q\\
P
\end{array}
\right)  . \label{5.13}%
\end{equation}

\subsection{Group Multiplication Rule for Single-mode GFO}

Because two successive optical Fresnel transforms is still a Fresnel
transform, we wonder if the product of two GFO is still a GFO. On the other
hand, we have known that the GFO is the image of the symplectic transform
$z\rightarrow sz-rz^{\ast},$ we expect that the product of two symplectic
transforms maps into the GFO which is just the product of two GFOs. If this is
so, then correspondence between GFT and GFO is perfect. Using (\ref{5.4}),
$\left \langle z\right.  \left \vert z^{\prime}\right \rangle =\exp \left[
-\frac{1}{2}\left(  |z|^{2}+|z^{\prime}|^{2}\right)  +z^{\ast}z^{\prime
}\right]  $ and the IWOP technique we can directly perform the following
integrals%
\begin{align}
F_{1}\left(  s,r\right)  F_{1}\left(  s^{\prime},r^{\prime}\right)   &
=\sqrt{ss^{\prime}}\int \frac{d^{2}zd^{2}z^{\prime}}{\pi^{2}}\left \vert
sz-rz^{\ast}\right \rangle \left \langle z\right.  \left \vert s^{\prime
}z^{\prime}-r^{\prime}z^{\prime \ast}\right \rangle \left \langle z^{\prime
}\right \vert \nonumber \\
&  =\frac{1}{\sqrt{s^{\prime \prime \ast}}}\exp \left[  -\frac{r^{\prime \prime}%
}{2s^{\prime \prime \ast}}a^{\dagger2}\right]  \colon \exp \left \{  \left(
\frac{1}{s^{\prime \prime \ast}}-1\right)  a^{\dagger}a\right \}  \colon
\exp \left[  \frac{r^{\prime \prime \ast}}{2s^{\prime \prime \ast}}a^{2}\right]
\nonumber \\
&  =\sqrt{s^{\prime \prime}}\int \frac{d^{2}z}{\pi}\left \vert s^{\prime
\prime \ast}z-r^{\prime \prime}z^{\ast}\right \rangle \left \langle z\right \vert
=F_{1}\left(  s^{\prime \prime},r^{\prime \prime}\right)  , \label{7.3}%
\end{align}
where we have set%
\begin{equation}
s^{\prime \prime}=ss^{\prime}+rr^{\prime \ast},\;r^{\prime \prime}=r^{\prime
}s+rs^{\prime \ast}, \label{7.2}%
\end{equation}
or
\begin{equation}
M^{\prime \prime}\equiv \left(
\begin{array}
[c]{cc}%
s^{\prime \prime} & -r^{\prime \prime}\\
-r^{\ast \prime \prime} & s^{\ast \prime \prime}%
\end{array}
\right)  =\left(
\begin{array}
[c]{cc}%
s & -r\\
-r^{\ast} & s^{\ast}%
\end{array}
\right)  \left(
\begin{array}
[c]{cc}%
s^{\prime} & -r^{\prime}\\
-r^{\prime \ast} & s^{\prime \ast}%
\end{array}
\right)  =MM^{\prime},\  \left \vert s^{\prime \prime}\right \vert ^{2}-\left \vert
r^{\prime \prime}\right \vert ^{2}=1, \label{7.4}%
\end{equation}
from which we see that it is just the mapping of the above ($A,B,C,D)$
matrices multiplication. Hence $F_{1}\left(  s,r\right)  F_{1}\left(
s^{\prime},r^{\prime}\right)  $ is the loyal representation of the product of
two symplectic group elements shown in (\ref{7.4}). The above discussion
actually reveals an important property of coherent states, though two coherent
state vectors are not orthogonal, but the equation
\begin{equation}
\sqrt{ss^{\prime}}\int \frac{d^{2}zd^{2}z^{\prime}}{\pi^{2}}\left \vert
sz-rz^{\ast}\right \rangle \left \langle z\right.  \left \vert s^{\prime
}z^{\prime}-r^{\prime}z^{\prime \ast}\right \rangle \left \langle z^{\prime
}\right \vert =\sqrt{s^{\prime \prime}}\int \frac{d^{2}z}{\pi}\left \vert
s^{\prime \prime \ast}z-r^{\prime \prime}z^{\ast}\right \rangle \left \langle
z\right \vert \label{7.5}%
\end{equation}
seems as if their overlap $\left \langle z\right.  \left \vert s^{\prime
}z^{\prime}-r^{\prime}z^{\prime \ast}\right \rangle $ was a $\delta$-function.
The coherent state representation for GFOs' product may be visualized very
easily, but it achieves striking importance, because it does not change its
form when treating symplectic transform according to $z\rightarrow
sz-rz^{\ast}$.

As a result of this group multiplication rule of GFO, we immediately obtain
\begin{align}
&  \mathcal{K}^{M^{\prime \prime}}\left(  x_{2},x_{1}\right)  =\left \langle
x_{2}\right \vert F_{1}\left(  s^{\prime \prime},r^{\prime \prime}\right)
\left \vert x_{1}\right \rangle \nonumber \\
&  =\int_{-\infty}^{\infty}dx_{3}\left \langle x_{2}\right \vert F_{1}\left(
s,r\right)  \left \vert x_{3}\right \rangle \left \langle x_{3}\right \vert
F_{1}\left(  s^{\prime},r^{\prime}\right)  \left \vert x_{1}\right \rangle
\nonumber \\
&  =\int_{-\infty}^{\infty}dx_{3}\mathcal{K}^{M}\left(  x_{2},x_{3}\right)
\mathcal{K}^{M^{\prime}}\left(  x_{3},x_{1}\right)  , \label{7.6}%
\end{align}
provided that the parameter matrices $\left(  s^{\prime \prime},r^{\prime
\prime}\right)  $ satisfy (\ref{7.2}). Thus by virtue of the group
multiplication property of GFO we immediately find the successive transform
property of GFTs.

\section{Quantum Optical ABCD Law for optical propagation ---single-mode case}

In classical optics, ray-transfer matrices, $N=\left(
\begin{array}
[c]{cc}%
A & B\\
C & D
\end{array}
\right)  ,$ $AD-BC=1$, have been used to describe the geometrical formation of
images by a centered lens system. For an optical ray (a centered spherical
wavefront) passing through optical instruments there is a famous law, named
ABCD law, governing the relation between input ray $\left(  r_{1},\alpha
_{1}\right)  $ and output ray $\left(  r_{2},\alpha_{2}\right)  ,$ i.e.%

\begin{equation}
\left(
\begin{array}
[c]{c}%
r_{2}\\
\alpha_{2}%
\end{array}
\right)  =N\left(
\begin{array}
[c]{c}%
r_{1}\\
\alpha_{1}%
\end{array}
\right)  , \label{8.1}%
\end{equation}
where $r_{1}$ is the ray height from the optical axis, and $\alpha_{1}$ is
named the optical direction-cosine, $r_{1}/\alpha_{1}\equiv R_{1}$ specifies
the ray's wavefront shape. Eq. (\ref{8.1}) implies
\begin{equation}
R_{2}\equiv \frac{r_{2}}{\alpha_{2}}=\frac{AR_{1}+B}{CR_{1}+D}. \label{8.2}%
\end{equation}
This law is the core of matrix optics, since it tells us how the curvature of
a centered spherical wavefront changes from one reference plane to the next.
Besides, the multiplication rule of matrix optics implies that if the
ray-transfer matrices of the $n$ optical components are $N_{1},N_{2}%
,N_{3},\cdots,N_{n}$, respectively, then the whole system is determined by a
matrix $N=N_{1}N_{2}N_{3}\cdots N_{n}.$

One of the remarkable things of modern optics is the case with which
geometrical ray-transfer methods, constituting the matrix optics, can be
adapted to describe the generation and propagation of Laser beams. In 1965
Kogelnik \cite{Kogelnik} pointed out that propagation of Gaussian beam also
obeys ABCD law via optical diffraction integration, i.e. the input light field
$f\left(  x_{1}\right)  $ and output light field $g\left(  x_{2}\right)  $ are
related to each other by so-called Fresnel integration \cite{Goodman}
$g\left(  x_{2}\right)  =\int_{-\infty}^{\infty}\mathcal{K}\left(
A,B,C;x_{2},x_{1}\right)  f\left(  x_{1}\right)  dx_{1},$ where%
\[
\mathcal{K}\left(  A,B,C;x_{2},x_{1}\right)  =\frac{1}{\sqrt{2\pi iB}}%
\exp \left[  \frac{i}{2B}\left(  Ax_{1}^{2}-2x_{2}x_{1}+Dx_{2}^{2}\right)
\right]  .
\]
The ABCD law for Gaussian beam passing through an optical system is
\cite{ABCD}%
\begin{equation}
q_{2}=\frac{Aq_{1}+B}{Cq_{1}+D}, \label{8.3}%
\end{equation}
where $q_{1}$ $(q_{2})$ represents the complex curvature of the input (output)
Gaussian beam, Eq. (\ref{8.3}) has the similar form as Eq. (\ref{8.2}). An
interesting and important question naturally arises \cite{FANHUOC}: Does ABCD
law also exhibit in quantum optics? Since classical Fresnel transform should
have its quantum optical counterpart?

To see the ABCD law more explicitly, using Eq.(\ref{5.2}) we can re-express
Eq.(\ref{5.5}) as
\begin{align}
F_{1}\left(  A,B,C\right)   &  =\sqrt{\frac{2}{A+D+i\left(  B-C\right)  }%
}\colon \exp \left \{  \frac{A-D+i\left(  B+C\right)  }{2\left[  A+D+i\left(
B-C\right)  \right]  }a^{\dagger2}\right. \nonumber \\
&  \left.  +\left[  \frac{2}{A+D+i\left(  B-C\right)  }-1\right]  a^{\dagger
}a-\frac{A-D-i\left(  B+C\right)  }{2\left[  A+D+i\left(  B-C\right)  \right]
}a^{2}\right \}  \colon, \label{8.4}%
\end{align}
and the multiplication rule for $F_{1}$ is $F\left(  A^{\prime},B^{\prime
},C^{\prime},D^{\prime}\right)  F\left(  A,B,C,D\right)  =F\left(
A^{\prime \prime},B^{\prime \prime},C^{\prime \prime},D^{\prime \prime}\right)  ,$
where%
\begin{equation}
\left(
\begin{array}
[c]{cc}%
A^{\prime \prime} & B^{\prime \prime}\\
C^{\prime \prime} & D^{\prime \prime}%
\end{array}
\right)  =\left(
\begin{array}
[c]{cc}%
A^{\prime} & B^{\prime}\\
C^{\prime} & D^{\prime}%
\end{array}
\right)  \left(
\begin{array}
[c]{cc}%
A & B\\
C & D
\end{array}
\right)  . \label{8.5}%
\end{equation}

Next we directly use the GFO to derive ABCD law in quantum optics. From
Eq.(\ref{8.5}) we see that the GFO generates%

\begin{equation}
F_{1}\left(  A,B,C\right)  \left \vert 0\right \rangle =\sqrt{\frac
{2}{A+iB-i\left(  C+iD\right)  }}\exp \left \{  \frac{A-D+i\left(  B+C\right)
}{2\left[  A+D+i\left(  B-C\right)  \right]  }a^{\dagger2}\right \}  \left \vert
0\right \rangle , \label{8.6}%
\end{equation}
if we identify
\begin{equation}
\frac{A-D+i\left(  B+C\right)  }{A+D+i\left(  B-C\right)  }=\frac{q_{1}%
-i}{q_{1}+i}, \label{8.7}%
\end{equation}
then%
\begin{equation}
F_{1}\left(  A,B,C\right)  \left \vert 0\right \rangle =\sqrt{-\frac{2/\left(
C+iD\right)  }{q_{1}+i}}\exp \left[  \frac{q_{1}-i}{2\left(  q_{1}+i\right)
}a^{\dagger2}\right]  \left \vert 0\right \rangle , \label{8.8}%
\end{equation}
The solution of Eq.(\ref{8.7}) is
\begin{equation}
q_{1}\equiv-\frac{A+iB}{C+iD}. \label{8.9}%
\end{equation}
Let $F_{1}\left(  A,B,C\right)  \left \vert 0\right \rangle $ expressed by
(\ref{8.8}) be an input state for an optical system which is characteristic by
parameters $A^{\prime},B^{\prime},C^{\prime},D^{\prime},$ then the
\textit{quantum optical ABCD law} states that the output state is
\begin{equation}
F_{1}\left(  A^{\prime},B^{\prime},C^{\prime}\right)  F_{1}\left(
A,B,C\right)  \left \vert 0\right \rangle =\sqrt{\frac{-2/\left(  C^{\prime
\prime}+iD^{\prime \prime}\right)  }{q_{2}+i}}\exp \left[  \frac{q_{2}%
-i}{2\left(  q_{2}+i\right)  }a^{\dagger2}\right]  \left \vert 0\right \rangle ,
\label{8.10}%
\end{equation}
which has the similar form as Eq.(\ref{8.8})$,$ where $\left(  C^{\prime
\prime},D^{\prime \prime}\right)  $ is determined by Eq.(\ref{8.5})$,$ and
\begin{equation}
\bar{q}_{2}=\frac{A^{\prime}\bar{q}_{1}+B^{\prime}}{C^{\prime}\bar{q}%
_{1}+D^{\prime}},\text{ \ }\bar{q}_{j}\equiv-q_{j},\text{ \ }\left(
j=1,2\right)  \label{8.11}%
\end{equation}
which resembles Eq.(\ref{8.3}).

Proof:

According to the multiplication rule of two GFOs and Eqs.(\ref{8.4}%
)-(\ref{8.5}) we have%
\begin{align}
&  F_{1}\left(  A^{\prime},B^{\prime},C^{\prime}\right)  F_{1}\left(
A,B,C\right)  \left \vert 0\right \rangle \nonumber \\
&  =\sqrt{\frac{2}{A^{\prime \prime}+D^{\prime \prime}+i\left(  B^{\prime \prime
}-C^{\prime \prime}\right)  }}\exp \left \{  \frac{A^{\prime \prime}%
-D^{\prime \prime}+i\left(  B^{\prime \prime}+C^{\prime \prime}\right)
}{2\left[  A^{\prime \prime}+D^{\prime \prime}+i\left(  B^{\prime \prime
}-C^{\prime \prime}\right)  \right]  }a^{\dagger2}\right \}  \left \vert
0\right \rangle \nonumber \\
&  =\sqrt{\frac{2}{A^{\prime}\left(  A+iB\right)  +B^{\prime}\left(
C+iD\right)  -iC^{\prime}\left(  A+iB\right)  -iD^{\prime}\left(  C+iD\right)
}}\nonumber \\
&  \times \exp \left \{  \frac{A^{\prime}\left(  A+iB\right)  +B^{\prime}\left(
C+iD\right)  +iC^{\prime}\left(  A+iB\right)  +iD^{\prime}\left(  C+iD\right)
}{2\left[  A^{\prime}\left(  A+iB\right)  +B^{\prime}\left(  C+iD\right)
-iC^{\prime}\left(  A+iB\right)  -iD^{\prime}\left(  C+iD\right)  \right]
}a^{\dagger2}\right \}  \left \vert 0\right \rangle \nonumber \\
&  =\sqrt{\frac{-2/\left(  C+iD\right)  }{A^{\prime}q_{1}-B^{\prime}-i\left(
C^{\prime}q_{1}-D^{\prime}\right)  }}\exp \left \{  \frac{A^{\prime}%
q_{1}-B^{\prime}+i\left(  C^{\prime}q_{1}-D^{\prime}\right)  }{2\left[
A^{\prime}q_{1}-B^{\prime}-i\left(  C^{\prime}q_{1}-D^{\prime}\right)
\right]  }a^{\dagger2}\right \}  \left \vert 0\right \rangle . \label{8.12}%
\end{align}
Using Eq.(\ref{8.9}) we see $\frac{2/\left(  C+iD\right)  }{C^{\prime}%
q_{1}-D^{\prime}}=-2/\left(  C^{\prime \prime}+iD^{\prime \prime}\right)  ,$
together using Eq.(\ref{8.11}) we can reach Eq.(\ref{8.10}), thus the law is
proved. Using Eq. (\ref{8.8}) we can re-express Eq.(\ref{8.11}) as
\begin{equation}
q_{2}=-\frac{A^{\prime}(A+iB)+B^{\prime}(C+iD)}{C^{\prime}(A+iB)+D^{\prime
}(C+iD)}=-\frac{A^{\prime \prime}+iB^{\prime \prime}}{C^{\prime \prime
}+iD^{\prime \prime}}, \label{8.13}%
\end{equation}
which is in consistent to Eq.(\ref{8.9}). Eqs. (\ref{8.8})-(\ref{8.13}) are
therefore self-consistent.

As an application of quantum optical ABCD law, we apply it to tackle the
time-evolution of a time-dependent harmonic oscillator whose Hamiltonian is
\begin{equation}
H=\frac{1}{2}e^{-2\gamma t}P^{2}+\frac{1}{2}\omega_{0}^{2}e^{2\gamma t}%
Q^{2},\text{ \  \ }\hbar=1, \label{8.14}%
\end{equation}
where we have set the initial mass $m_{0}=1,$ $\gamma$ denotes damping. Using
$u\left(  t\right)  =e^{\frac{i\gamma}{2}Q^{2}}e^{-\frac{i\gamma t}{2}\left(
QP+PQ\right)  }\ $to perform the transformation%
\begin{align}
u\left(  t\right)  Qu^{-1}\left(  t\right)   &  =e^{-\gamma t}Q,\nonumber \\
u\left(  t\right)  Pu^{-1}\left(  t\right)   &  =e^{\gamma t}P-\gamma
e^{\gamma t}Q, \label{8.15}%
\end{align}
then $i\frac{\partial \left \vert \psi \left(  t\right)  \right \rangle }{\partial
t}=H\left \vert \psi \left(  t\right)  \right \rangle \ $leads to $i\frac
{\partial \left \vert \phi \right \rangle }{\partial t}=\mathcal{H}\left \vert
\phi \right \rangle ,$ $\left \vert \phi \right \rangle =u\left(  t\right)
\left \vert \psi \left(  t\right)  \right \rangle ,$
\begin{equation}
H\rightarrow \mathcal{H}=u\left(  t\right)  Hu^{-1}\left(  t\right)  -iu\left(
t\right)  \frac{\partial u^{-1}\left(  t\right)  }{\partial t}=\frac{1}%
{2}P^{2}+\frac{1}{2}\omega^{2}Q^{2},. \label{8.16}%
\end{equation}
where $\omega^{2}=\omega_{0}^{2}-\gamma^{2}.$ $\mathcal{H}$ does not contain
$t$ explicitly. The dynamic evolution of a mass-varying harmonic oscillator
from the Fock state $\left \vert 0\right \rangle $ at initial time to a squeezed
state at time $t$ is
\begin{equation}
\left \vert \psi \left(  t\right)  \right \rangle _{0}=u^{-1}\left(  t\right)
\left \vert 0\right \rangle =e^{\frac{i\gamma t}{2}\left(  QP+PQ\right)
}e^{-\frac{i\gamma}{2}Q^{2}}\left \vert 0\right \rangle , \label{8.17}%
\end{equation}
if we let $A=D=1,B=0,C=-\gamma;$ and $A^{\prime}=e^{-\gamma t},D^{\prime
}=e^{\gamma t},B^{\prime}=C^{\prime}=0,$ then $q_{1}=\frac{1}{\gamma-i},$
$q_{2}=\frac{e^{-2\gamma t}}{\gamma-i}$, according to Eq.(\ref{8.10}) we
directly obtain%
\begin{equation}
u^{-1}\left(  t\right)  \left \vert 0\right \rangle =\sqrt{\frac{2e^{-\gamma t}%
}{e^{-2\gamma t}+i\gamma+1}}\exp \left[  \frac{e^{-2\gamma t}-1-i\gamma
}{2\left(  e^{-2\gamma t}+1+i\gamma \right)  }a^{\dagger2}\right]  \left \vert
0\right \rangle , \label{8.18}%
\end{equation}
so the time evolution of the damping oscillator embodies the quantum optical
ABCD law.

\section{Optical operator method studied via GFO's decomposition}

Fresnel diffraction is the core of Fourier optics
\cite{Goodman,Collins1,Alieva1,agarwal}, Fresnel transform is frequently used
in optical imaging, optical propagation and optical instrument design. The GFT
represents a class of optical transforms which are of great importance for
their applications to describe various optical systems. It is easily seen that
when we let the transform kernel $\mathcal{K}\left(  x_{2},x_{1}\right)
=\exp \left(  ix_{2}x_{1}\right)  $, the GFT changes into the well-known
Fourier transform, which is adapted to express mathematically the Fraunhofer
diffraction. And if $\mathcal{K}\left(  x_{2},x_{1}\right)  =\exp[i\left(
x_{2}-x_{1}\right)  ^{2}]$, the GFT then describes a Fresnel diffraction. In
studying various optical transformations one also proposed so-called optical
operator method \cite{Nazarathy} which used quantum mechanical operators'
ordered product to express the mechanism of optical systems, such that the ray
transfer through optical instruments and the diffraction can be discussed by
virtue of the commutative relations of operators and the matrix algebra. Two
important questions thus naturally arises: how to directly map the classical
optical transformations to the optical operator method? How to combine the
usual optical transformation operators, such as the square phase operators,
scaling operator, Fourier transform operator and the propagation operator in
free space, into a concise and unified form? In this section we shall solve
these two problems and develop the optical operator method onto a new stage.

\subsection{Four fundamental optical operators derived by decomposing GFO}

The GFO $F_{1}\left(  A,B,C\right)  $ can also be expressed in the form of
quadratic combination of canonical operators $Q$ and $P$ \cite{fanwuncommun},
i.e.,
\begin{equation}
F_{1}\left(  A,B,C\right)  =\exp \left(  \frac{iC}{2A}Q^{2}\right)  \exp \left(
-\frac{i}{2}\left(  QP+PQ\right)  \ln A\right)  \exp \left(  -\frac{iB}%
{2A}P^{2}\right)  , \label{9.1}%
\end{equation}
where we have set $\hbar=1,$ $A\neq0.$ To confirm this, we first calculate
matrix element
\begin{align}
\left \langle x\right \vert F_{1}\left(  A,B,C\right)  \left \vert
p\right \rangle  &  =\exp \left(  \frac{iC}{2A}x^{2}\right)  \left \langle
x\right \vert \exp \left(  -\frac{i}{2}\left(  QP+PQ\right)  \ln A\right)
\left \vert p\right \rangle \exp \left(  -\frac{iB}{2A}p^{2}\right) \nonumber \\
\  &  =\frac{1}{\sqrt{2\pi A}}\exp \left(  \frac{iC}{2A}x^{2}-\frac{iB}%
{2A}p^{2}+\frac{ipx}{A}\right)  , \label{9.2}%
\end{align}
where we have used the squeezing property
\begin{equation}
\exp \left[  -\frac{i}{2}\left(  QP+PQ\right)  \ln A\right]  \left \vert
p\right \rangle =\frac{1}{\sqrt{A}}\left \vert p/A\right \rangle . \label{9.3}%
\end{equation}
It then follows from (\ref{9.2}) and $AD-BC=1$, we have
\begin{equation}
\left \langle x_{2}\right \vert F_{1}\left(  A,B,C\right)  \left \vert
x_{1}\right \rangle =\int_{-\infty}^{\infty}dp\left \langle x_{2}\right \vert
F_{1}\left(  A,B,C\right)  \left \vert p\right \rangle \left \langle p\right \vert
\left.  x_{1}\right \rangle =\mathcal{K}^{M}\left(  x_{2},x_{1}\right)  .
\label{9.4}%
\end{equation}
Thus $F_{1}\left(  A,B,C\right)  $ in (\ref{9.1}) is really the expected GFO.
Next we directly use (\ref{5.1}) and the canonical operator $\left(
Q,P\right)  $ representation (\ref{9.1}) to develop the optical operator method.

By noticing the matrix decompositions \cite{MatrixOptics}%
\begin{equation}
\left(
\begin{array}
[c]{cc}%
A & B\\
C & D
\end{array}
\right)  =\left(
\begin{array}
[c]{cc}%
1 & 0\\
C/A & 1
\end{array}
\right)  \left(
\begin{array}
[c]{cc}%
A & 0\\
0 & A^{-1}%
\end{array}
\right)  \left(
\begin{array}
[c]{cc}%
1 & B/A\\
0 & 1
\end{array}
\right)  , \label{9.5}%
\end{equation}
and comparing (\ref{5.1}) and (\ref{9.1}) as well as using (\ref{8.5}) we know%
\begin{equation}
F_{1}\left(  A,B,C\right)  =F_{1}\left(  1,0,C/A\right)  F_{1}\left(
A,0,0\right)  F_{1}\left(  1,B/A,0\right)  , \label{9.6}%
\end{equation}
where%
\begin{align}
F_{1}\left(  1,0,C/A\right)   &  =\frac{\sqrt{2+iC/A}}{2\sqrt{2}\pi}\int
dxdp\left \vert \left(
\begin{array}
[c]{cc}%
1 & 0\\
C/A & 1
\end{array}
\right)  \left(
\begin{array}
[c]{c}%
x\\
p
\end{array}
\right)  \right \rangle \left \langle \left(
\begin{array}
[c]{c}%
x\\
p
\end{array}
\right)  \right \vert \nonumber \\
&  =\exp \left(  \frac{iC}{2A}Q^{2}\right)  , \label{9.7}%
\end{align}
which is named quadrature phase operator; and
\begin{align}
F_{1}\left(  1,B/A,0\right)   &  =\frac{\sqrt{2-iB/A}}{2\sqrt{2}\pi}\int
dxdp\left \vert \left(
\begin{array}
[c]{cc}%
1 & B/A\\
0 & 1
\end{array}
\right)  \left(
\begin{array}
[c]{c}%
x\\
p
\end{array}
\right)  \right \rangle \left \langle \left(
\begin{array}
[c]{c}%
x\\
p
\end{array}
\right)  \right \vert \nonumber \\
&  =\exp \left(  -\frac{iB}{2A}P^{2}\right)  , \label{9.8}%
\end{align}
which is named Fresnel propagator in free space; as well as%
\begin{align}
F_{1}\left(  A,0,0\right)   &  =\frac{\sqrt{A+A^{-1}}}{2\sqrt{2}\pi}\int
dxdp\left \vert \left(
\begin{array}
[c]{cc}%
A & 0\\
0 & A^{-1}%
\end{array}
\right)  \left(
\begin{array}
[c]{c}%
x\\
p
\end{array}
\right)  \right \rangle \left \langle \left(
\begin{array}
[c]{c}%
x\\
p
\end{array}
\right)  \right \vert \nonumber \\
&  =\exp \left[  -\frac{i}{2}\left(  QP+PQ\right)  \ln A\right]  , \label{9.9}%
\end{align}
which is named scaling operator (squeezed operator \cite{squeezed1,squeezed2}%
). When $A=D=0,B=1,C=-1,$ from (\ref{8.4}) we see
\begin{align}
F_{1}\left(  0,1,-1\right)   &  =\sqrt{-i}\int \frac{dxdp}{2\pi}\left \vert
\left(
\begin{array}
[c]{cc}%
0 & 1\\
-1 & 0
\end{array}
\right)  \left(
\begin{array}
[c]{c}%
x\\
p
\end{array}
\right)  \right \rangle \left \langle \left(
\begin{array}
[c]{c}%
x\\
p
\end{array}
\right)  \right \vert \nonumber \\
&  =\exp \left[  -\left(  a^{\dagger}a+\frac{1}{2}\right)  \ln i\right]
\nonumber \\
&  =\exp \left[  -i\frac{\pi}{2}\left(  a^{\dagger}a+\frac{1}{2}\right)
\right]  , \label{9.10}%
\end{align}
which is named the Fourier operator, since it quantum mechanically transforms
\cite{Fanshu}
\begin{align}
\exp \left[  i\frac{\pi}{2}\left(  a^{\dagger}a+\frac{1}{2}\right)  \right]
Q\exp \left[  -i\frac{\pi}{2}\left(  a^{\dagger}a+\frac{1}{2}\right)  \right]
&  =P,\nonumber \\
\exp \left[  i\frac{\pi}{2}\left(  a^{\dagger}a+\frac{1}{2}\right)  \right]
P\exp \left[  -i\frac{\pi}{2}\left(  a^{\dagger}a+\frac{1}{2}\right)  \right]
&  =-Q. \label{9.11}%
\end{align}

\subsection{Alternate decompositions of GFO}

Note that when $A=0$, the decomposition (\ref{9.1}) is not available, instead,
from%
\begin{equation}
\left(
\begin{array}
[c]{cc}%
A & B\\
C & D
\end{array}
\right)  ^{-1}=\left(
\begin{array}
[c]{cc}%
D & -B\\
-C & A
\end{array}
\right)  , \label{9.12}%
\end{equation}
and (\ref{8.4}), (\ref{8.5}) and (\ref{9.1}) we have%
\begin{equation}
F_{1}^{-1}\left(  A,B,C\right)  =\exp \left(  -\frac{iC}{2D}Q^{2}\right)
\exp \left[  -\frac{i}{2}\left(  QP+PQ\right)  \ln D\right]  \exp \left(
\frac{iB}{2D}P^{2}\right)  , \label{9.13}%
\end{equation}
it then follows
\begin{equation}
F_{1}\left(  A,B,C\right)  =\exp \left(  -\frac{iB}{2D}P^{2}\right)
\exp \left[  \frac{i}{2}\left(  QP+PQ\right)  \ln D\right]  \exp \left(
\frac{iC}{2D}Q^{2}\right)  ,\text{ }D\neq0. \label{9.14}%
\end{equation}
Besides, when we notice
\begin{equation}
\left(
\begin{array}
[c]{cc}%
A & B\\
C & D
\end{array}
\right)  =\left(
\begin{array}
[c]{cc}%
1 & 0\\
D/B & 1
\end{array}
\right)  \left(
\begin{array}
[c]{cc}%
B & 0\\
0 & 1/B
\end{array}
\right)  \left(
\begin{array}
[c]{cc}%
0 & 1\\
-1 & 0
\end{array}
\right)  \left(
\begin{array}
[c]{cc}%
1 & 0\\
A/B & 1
\end{array}
\right)  , \label{9.15}%
\end{equation}
and
\begin{equation}
\left(
\begin{array}
[c]{cc}%
A & B\\
C & D
\end{array}
\right)  =\left(
\begin{array}
[c]{cc}%
1 & A/C\\
0 & 1
\end{array}
\right)  \left(
\begin{array}
[c]{cc}%
-1/C & 0\\
0 & -C
\end{array}
\right)  \left(
\begin{array}
[c]{cc}%
0 & 1\\
-1 & 0
\end{array}
\right)  \left(
\begin{array}
[c]{cc}%
1 & D/C\\
0 & 1
\end{array}
\right)  , \label{9.16}%
\end{equation}
we have another decomposition for $B\neq0,$
\begin{align}
F_{1}\left(  A,B,C\right)   &  =\exp \left(  \frac{iD}{2B}Q^{2}\right)
\exp \left(  -\frac{i}{2}\left(  QP+PQ\right)  \ln B\right) \nonumber \\
&  \times \exp \left[  -\frac{i\pi}{2}\left(  a^{\dagger}a+\frac{1}{2}\right)
\right]  \exp \left(  \frac{iA}{2B}Q^{2}\right)  ,\text{ } \label{9.17}%
\end{align}
and for $C\neq0$
\begin{align}
F_{1}\left(  A,B,C\right)   &  =\exp \left(  -\frac{iA}{2C}P^{2}\right)
\exp \left[  -\frac{i}{2}\left(  QP+PQ\right)  \ln \left(  \frac{-1}{C}\right)
\right] \nonumber \\
&  \times \exp \left[  -\frac{i\pi}{2}\left(  a^{\dagger}a+\frac{1}{2}\right)
\right]  \exp \left(  -\frac{iD}{2C}P^{2}\right)  . \label{9.18}%
\end{align}

\subsection{Some optical operator identities}

For a special optical systems with the parameter $A=0,$ $C=-B^{-1},$%
\begin{equation}
\left(
\begin{array}
[c]{cc}%
0 & B\\
-B^{-1} & D
\end{array}
\right)  =\left(
\begin{array}
[c]{cc}%
1 & 0\\
D/B & 1
\end{array}
\right)  \left(
\begin{array}
[c]{cc}%
B & 0\\
0 & B^{-1}%
\end{array}
\right)  \left(
\begin{array}
[c]{cc}%
0 & 1\\
-1 & 0
\end{array}
\right)  , \label{9.20}%
\end{equation}
we have%
\begin{align}
&  \exp \left(  -\frac{iB}{2D}P^{2}\right)  \exp \left(  \frac{i}{2}\left(
QP+PQ\right)  \ln D\right)  \exp \left(  \frac{-i}{2DB}Q^{2}\right) \nonumber \\
&  =\exp \left(  \frac{-iD}{2B}Q^{2}\right)  \exp \left(  \frac{i}{2}\left(
QP+PQ\right)  \ln B\right)  \exp \left[  -\frac{i\pi}{2}\left(  a^{\dagger
}a+\frac{1}{2}\right)  \right]  . \label{9.21}%
\end{align}
In particular, when $A=D=0,$ $C=-B^{-1},$ from%
\begin{equation}
\left(
\begin{array}
[c]{cc}%
0 & B\\
-\frac{1}{B} & 0
\end{array}
\right)  =\left(
\begin{array}
[c]{cc}%
B & 0\\
0 & \frac{1}{B}%
\end{array}
\right)  \left(
\begin{array}
[c]{cc}%
0 & 1\\
-1 & 0
\end{array}
\right)  , \label{9.22}%
\end{equation}
we have%
\begin{align}
&  \exp \left[  -\frac{B^{2}-1}{2\left(  B^{2}+1\right)  }a^{\dagger2}\right]
\exp \left[  \left(  a^{\dagger}a+\frac{1}{2}\right)  \ln \left(  \frac
{-2Bi}{B^{2}+1}\right)  \right]  \exp \left[  -\frac{B^{2}-1}{2\left(
B^{2}+1\right)  }a^{2}\right] \nonumber \\
&  =\exp \left[  -\frac{i}{2}\left(  QP+PQ\right)  \ln B\right]  \exp \left[
-i\frac{\pi}{2}\left(  a^{\dagger}a+\frac{1}{2}\right)  \right]  .
\label{9.23}%
\end{align}
Using the following relations%

\begin{equation}
\left(
\begin{array}
[c]{cc}%
A & B\\
C & D
\end{array}
\right)  =\left(
\begin{array}
[c]{cc}%
1 & \left(  A-1\right)  /C\\
0 & 1
\end{array}
\right)  \left(
\begin{array}
[c]{cc}%
1 & 0\\
C & 1
\end{array}
\right)  \left(
\begin{array}
[c]{cc}%
1 & \left(  D-1\right)  /C\\
0 & 1
\end{array}
\right)  , \label{9.24}%
\end{equation}
it then follows that
\begin{equation}
F_{1}\left(  A,B,C\right)  =\exp \left(  -\frac{i\left(  A-1\right)  }{2C}%
P^{2}\right)  \exp \left(  \frac{iC}{2}Q^{2}\right)  \exp \left(  -\frac
{i\left(  D-1\right)  }{2C}P^{2}\right)  ,\text{ } \label{9.25}%
\end{equation}
while from%
\begin{equation}
\left(
\begin{array}
[c]{cc}%
A & B\\
C & D
\end{array}
\right)  =\left(
\begin{array}
[c]{cc}%
1 & 0\\
\left(  D-1\right)  /B & 1
\end{array}
\right)  \left(
\begin{array}
[c]{cc}%
1 & B\\
0 & 1
\end{array}
\right)  \left(
\begin{array}
[c]{cc}%
1 & 0\\
\left(  A-1\right)  /B & 1
\end{array}
\right)  \label{9.26}%
\end{equation}
we obtain%
\begin{equation}
F_{1}\left(  A,B,C\right)  =\exp \left(  \frac{i\left(  D-1\right)  }{2B}%
Q^{2}\right)  \exp \left(  -\frac{iB}{2}P^{2}\right)  \exp \left(
\frac{i\left(  A-1\right)  }{2B}Q^{2}\right)  ,\text{ } \label{9.27}%
\end{equation}
so we have
\begin{align}
&  \exp \left(  \frac{i\left(  D-1\right)  }{2B}Q^{2}\right)  \exp \left(
-\frac{iB}{2}P^{2}\right)  \exp \left(  \frac{i\left(  A-1\right)  }{2B}%
Q^{2}\right) \nonumber \\
&  =\exp \left(  -\frac{i\left(  A-1\right)  }{2C}P^{2}\right)  \exp \left(
\frac{iC}{2}Q^{2}\right)  \exp \left(  -\frac{i\left(  D-1\right)  }{2C}%
P^{2}\right)  . \label{9.28}%
\end{align}

In this section, based on a one-to-one correspondence between classical
Fresnel transform in phase space and quantum unitary transform in state-vector
space and the IWOP technique as well as the coherent state representation we
have found a way to directly map the classical optical transformations to the
optical operator method. We have combined the usual optical transformation
operators, such as the square phase operators, scaling operator, Fourier
transform operator and the propagation operator in free space, into a concise
and unified form. The various decompositions of Fresnel operator into the
exponential canonical operators are also obtained.

\section{Quantum tomography and probability distribution for the Fresnel
quadrature phase}

In quantum optics theory all possible linear combinations of quadratures $Q$
and $P$\ of the oscillator field mode $a$ and $a^{\dagger}$ can be measured by
the homodyne measurement just by varying the phase of the local oscillator.
The average of the random outcomes of the measurement, at a given local
oscillator phase, is connected with the marginal distribution of Wigner
function (WF), thus the homodyne measurement of light field permits the
reconstruction of the WF of a quantum system by varying the phase shift
between two oscillators. In Ref. \cite{r1} Vogel and Risken pointed out that
the probability distribution for the rotated quadrature phase $Q_{\theta
}\equiv \lbrack a^{\dagger}\exp(i\theta)+a\exp(-i\theta)]/\sqrt{2},$ $\left[
a,a^{\dagger}\right]  =1,$which depends on only one $\theta$ angle, can be
expressed in terms of WF, and that the reverse is also true (named as
Vogel-Risken relation), i.e., one can obtain the Wigner distribution by
tomographic inversion of a set of measured probability distributions,
$P_{\theta}\left(  q_{\theta}\right)  ,$ of the quadrature amplitude. Once the
distribution $P_{\theta}\left(  q_{\theta}\right)  $ are obtained, one can use
the inverse Radon transformation familiar in tomographic imaging to obtain the
Wigner distribution and density matrix. The Radon transform of the WF is
closely related to the expectation values or densities formed with the
eigenstates to the rotated canonical observables. The field of problems of the
reconstruction of the density operator from such data is called quantum
tomography. (Optical tomographic imaging techniques derive two-dimensional
data from a three-dimensional object to obtain a slice image of the internal
structure and thus have the ability to peer inside the object noninvasively,
the slice image is equated with tomogram.) The theoretical development in
quantum tomography in the last decade has progressed in the direction of
determining more physical relevant parameters of the density from tomographic
data\cite{r1,r2,r3,r4,r5}.

\subsection{Relation between Fresnel transform and Radon transform of WF}

In \cite{r6,r7} the Radon transform of WF which depends on two continuous
parameters is introduced, this has the advantage in conveniently associating
quantum tomography theory with squeezed coherent state theory. In this
subsection we want to derive relations between the Fresnel transform and the
Radon transform of WF in quantum optics in tomography theory.

By extending the rotated quadrature phase $Q_{\theta}$ to the Fresnel
quadrature phase
\begin{equation}
Q_{F}\equiv \left(  s^{\ast}a+ra^{\dagger}+sa^{\dagger}+r^{\ast}a\right)
/\sqrt{2}=F_{1}QF_{1}^{\dagger}, \label{10.1}%
\end{equation}
where $s$ and $r$ are related to ABCD through(\ref{5.2}),
\begin{equation}
s=\frac{1}{2}\left[  A+D-i\left(  B-C\right)  \right]  ,\;r=-\frac{1}%
{2}\left[  A-D+i\left(  B+C\right)  \right]  ,\;|s|^{2}-|r|^{2}=1,
\end{equation}
we shall prove that the $(D,B)$ related Radon transform of Wigner operator
$\Delta \left(  q,p\right)  $ is just the pure state density operator
$\left \vert q\right \rangle _{s,rs,r}\left \langle q\right \vert $ (named as the
tomographic density operator) formed with the eigenstates belonging to the
quadrature $Q_{F}$, ( $\left \vert q\right \rangle _{s,r}=F_{1}\left \vert
q\right \rangle ,$ $Q$ is the coordinate operator),%
\begin{equation}
F_{1}\left \vert q\right \rangle \left \langle q\right \vert F_{1}^{\dagger
}=\left \vert q\right \rangle _{s,rs,r}\left \langle q\right \vert =\int_{-\infty
}^{\infty}dq^{\prime}dp^{\prime}\delta \left[  q-\left(  Dq^{\prime}%
-Bp^{\prime}\right)  \right]  \Delta \left(  q^{\prime},p^{\prime}\right)  ,
\label{10.2}%
\end{equation}%
\begin{equation}
D=\frac{1}{2}\left(  s+s^{\ast}+r+r^{\ast}\right)  ,\ B=\frac{1}{2i}\left(
s^{\ast}-s+r^{\ast}-r\right)  , \label{10.3}%
\end{equation}
Since $F$ corresponds to classical Fresnel transform in optical diffraction
theory, so Eq. (\ref{10.2}) indicates that the probability distribution for
the Fresnel quadrature phase is the Radon transform of WF \cite{fanhuoc2}.

Proof:

Firstly, from (\ref{5.5}) we see
\begin{equation}
F_{1}\left(  s,r\right)  aF_{1}^{\dag}\left(  s,r\right)  =s^{\ast
}a+ra^{\dagger}, \label{10.4}%
\end{equation}
so from $Q=\frac{a+a^{\dagger}}{\sqrt{2}},P=i\frac{a^{\dagger}-a}{\sqrt{2}},$
indeed we have%
\begin{equation}
F_{1}QF_{1}^{\dagger}=F_{1}\frac{a+a^{\dagger}}{\sqrt{2}}F_{1}^{\dagger
}=\left(  s^{\ast}a+ra^{\dagger}+sa^{\dagger}+r^{\ast}a\right)  /\sqrt
{2}=Q_{F}. \label{10.5}%
\end{equation}
Secondly, we can derive the explicit form of $\left \vert q\right \rangle
_{s,r}.$ Starting from $s^{\ast}+r^{\ast}=D+iB,$\  \ $s^{\ast}-r^{\ast}=A-iC,$
we set up the eigenvector equation%
\begin{equation}
Q_{F}\left \vert q\right \rangle _{s,r}=\left(  DQ-BP\right)  \left \vert
q\right \rangle _{s,r}=q\left \vert q\right \rangle _{s,r},\text{ } \label{10.6}%
\end{equation}
it follows%
\begin{equation}
\left \vert q\right \rangle _{s,r}=F_{1}\left(  s,r\right)  \left \vert
q\right \rangle . \label{10.7}%
\end{equation}
\ In the coordinate and momentum representations we have%
\begin{align}
\left \langle q^{\prime}\right \vert Q_{F}\left \vert q\right \rangle _{s,r}  &
=\left(  Dq^{\prime}+iB\frac{d}{dq^{\prime}}\right)  \left \langle q^{\prime
}\right \vert \left.  q\right \rangle _{s,r}=q\left \langle q^{\prime}\right \vert
\left.  q\right \rangle _{s,r}.\label{10.9}\\
\left \langle p\right \vert Q_{F}\left \vert q\right \rangle _{s,r}  &  =\left(
iD\frac{d}{dp}-Bp\right)  \left \langle p\right \vert \left.  q\right \rangle
_{s,r}=q\left \langle p\right \vert \left.  q\right \rangle _{s,r}. \label{10.10}%
\end{align}
The normalizable solutions to (\ref{10.9}) and (\ref{10.10}) are%
\begin{align}
\left \langle q^{\prime}\right \vert \left.  q\right \rangle _{s,r}  &  =c\left(
q\right)  \exp \left[  \frac{iq^{\prime}\left(  Dq^{\prime}-2q\right)  }%
{2B}\right]  ,\label{10.11}\\
\left \langle p\right \vert \left.  q\right \rangle _{s,r}  &  =d\left(
q\right)  \exp \left[  \frac{ip\left(  -Bp-2q\right)  }{2D}\right]  .
\label{10.12}%
\end{align}
Using the Fock representation of $\left \vert q\right \rangle $ and $\left \vert
p\right \rangle $ in Eqs.(\ref{3.7}) and (\ref{3.8}), we obtain%
\begin{align}
\left \vert q\right \rangle _{s,r}  &  =\int_{-\infty}^{\infty}dq^{\prime
}\left \vert q^{\prime}\right \rangle \left \langle x^{\prime}\right \vert \left.
q\right \rangle _{s,r}\nonumber \\
&  =\pi^{-1/4}c\left(  q\right)  \sqrt{\frac{2B\pi}{B-iD}}\exp \left[
-\frac{q^{2}}{2B\left(  B-iD\right)  }+\frac{\sqrt{2}a^{\dagger}q}{D+iB}%
-\frac{D-iB}{D+iB}\frac{a^{\dagger2}}{2}\right]  \left \vert 0\right \rangle ,
\label{10.13}%
\end{align}
and%
\begin{align}
\left \vert q\right \rangle _{s,r}  &  =\int_{-\infty}^{\infty}dp\left \vert
p\right \rangle \left \langle p\right \vert \left.  q\right \rangle _{s,r}%
\nonumber \\
&  =d\left(  q\right)  \pi^{-1/4}\sqrt{\frac{2\pi D}{D+iB}}\exp \left[
-\frac{q^{2}}{2D\left(  D+iB\right)  }+\allowbreak \frac{\sqrt{2}a^{\dagger}%
q}{D+iB}-\frac{D-iB}{D+iB}\frac{a^{\dagger2}}{2}\right]  \left \vert
0\right \rangle . \label{10.14}%
\end{align}
Comparing Eq.(\ref{10.13}) with (\ref{10.14}) we see%
\begin{equation}
\frac{c\left(  q\right)  }{d\left(  q\right)  }=\sqrt{\allowbreak \frac{D}{iB}%
}\exp \left[  \frac{\allowbreak iA}{2B}q^{2}-\frac{iCq^{2}}{2D}\right]  .
\label{10.15}%
\end{equation}
On the other hand, according to the orthogonalization of $\left \vert
q\right \rangle _{s,r}$, $_{s,r}\left \langle q^{\prime}\right.  \left \vert
q^{\prime \prime}\right \rangle _{s,r}=\delta \left(  q^{\prime}-q^{\prime \prime
}\right)  ,$ we have
\begin{equation}
\left \vert c\left(  q\right)  \right \vert ^{2}=\frac{1}{2\pi B},\text{
\ }\left \vert d\left(  q\right)  \right \vert ^{2}=\frac{1}{2\pi D}.
\label{10.16}%
\end{equation}
Thus combining Eq.(\ref{10.15}) and (\ref{10.16}) we deduce%

\begin{equation}
c\left(  q\right)  =\frac{1}{\sqrt{2\pi iB}}\exp \left[  \frac{\allowbreak
iA}{2B}q^{2}\right]  ,\text{ }d\left(  q\right)  =\frac{1}{\sqrt{2\pi D}}%
\exp \left[  \frac{iCq^{2}}{2D}\right]  , \label{10.17}%
\end{equation}
and
\begin{equation}
\left \vert q\right \rangle _{s,r}=\frac{\pi^{-1/4}}{\sqrt{D+iB}}\exp \left \{
-\frac{A-iC}{D+iB}\frac{q^{2}}{2}+\frac{\sqrt{2}q}{D+iB}a^{\dagger}%
-\frac{D-iB}{D+iB}\frac{a^{\dagger2}}{2}\right \}  \left \vert 0\right \rangle ,
\label{10.18}%
\end{equation}
or%
\begin{equation}
\left \vert q\right \rangle _{s,r}\equiv \frac{\pi^{-1/4}}{\sqrt{s^{\ast
}+\allowbreak r^{\ast}}}\exp \left \{  -\frac{s^{\ast}-r^{\ast}}{s^{\ast
}+r^{\ast}}\frac{q^{2}}{2}+\frac{\sqrt{2}q}{s^{\ast}+r^{\ast}}a^{\dagger
}-\frac{s+r}{s^{\ast}+r^{\ast}}\frac{a^{\dagger2}}{2}\right \}  \left \vert
0\right \rangle . \label{10.19}%
\end{equation}
It is easily seen that that $\left \vert q\right \rangle _{s,r}$ make up a
complete set (so $\left \vert q\right \rangle _{s,r}$ can be named as the
tomography representation),%
\begin{equation}
\int_{-\infty}^{\infty}dq\left \vert q\right \rangle _{s,r}{}_{s,r}\left \langle
q\right \vert =1. \label{10.20}%
\end{equation}
Then according to the Weyl quantization scheme \cite{Weyl}%
\begin{equation}
H\left(  Q,P\right)  =\int_{-\infty}^{\infty}dpdq\Delta \left(  q,p\right)
h\left(  q,p\right)  , \label{10.21}%
\end{equation}
where $h\left(  q,p\right)  $ is the Weyl correspondence of $H\left(
Q,P\right)  ,$
\begin{equation}
h\left(  q,p\right)  =2\pi T_{r}\left[  H\left(  Q,P\right)  \Delta \left(
q,p\right)  \right]  , \label{10.22}%
\end{equation}
$\Delta \left(  q,p\right)  $ is the Wigner operator \cite{r13,r14},%
\begin{equation}
\Delta \left(  q,p\right)  =\frac{1}{2\pi}\int_{-\infty}^{\infty}%
due^{ipu}\left \vert q+\frac{u}{2}\right \rangle \left \langle q-\frac{u}%
{2}\right \vert , \label{10.23}%
\end{equation}
and using (\ref{10.22}), (\ref{10.23}) and (\ref{10.13}) we know that the
classical Weyl correspondence (Weyl image) of the projection operator
$\left \vert q\right \rangle _{s,rs,r}\left \langle q\right \vert $ is
\begin{align}
&  2\pi Tr\left[  \Delta \left(  q^{\prime},p^{\prime}\right)  \left \vert
q\right \rangle _{s,rs,r}\left \langle q\right \vert \right] \nonumber \\
&  =\left.  _{s,r}\left \langle q\right \vert \right.  \int_{-\infty}^{\infty
}due^{ip^{\prime}u}\left \vert q^{\prime}+\frac{u}{2}\right \rangle \left \langle
q^{\prime}-\frac{u}{2}\right \vert \left.  q\right \rangle _{s,r}\nonumber \\
&  =\frac{1}{2\pi B}\int_{-\infty}^{\infty}du\exp \left[  ip^{\prime}u+\frac
{i}{B}u\left(  q-Dq^{\prime}\right)  \right] \nonumber \\
&  =\delta \left[  q-\left(  Dq^{\prime}-Bp^{\prime}\right)  \right]  ,
\label{10.24}%
\end{align}
which means%
\begin{equation}
\left \vert q\right \rangle _{s,rs,r}\left \langle q\right \vert =\int_{-\infty
}^{\infty}dq^{\prime}dp^{\prime}\delta \left[  q-\left(  Dq^{\prime}%
-Bp^{\prime}\right)  \right]  \Delta \left(  q^{\prime},p^{\prime}\right)  .
\label{10.25}%
\end{equation}
Combining Eqs. (\ref{10.4})-(\ref{10.7}) together we complete the proof.
Therefore, the probability distribution for the Fresnel quadrature phase is
the Radon transform of WF%
\begin{equation}
|\left \langle q\right \vert F_{1}^{\dagger}\left \vert \psi \right \rangle
|^{2}=|_{s,r}\left \langle q\right \vert \left.  \psi \right \rangle |^{2}%
=\int_{-\infty}^{\infty}dq^{\prime}dp^{\prime}\delta \left[  q-\left(
Dq^{\prime}-Bp^{\prime}\right)  \right]  \left \langle \psi \right \vert
\Delta \left(  q^{\prime},p^{\prime}\right)  \left \vert \psi \right \rangle ,
\label{10.26}%
\end{equation}
so we name $\left \vert q\right \rangle _{s,rs,r}\left \langle q\right \vert $
\textbf{the tomographic density}. Moreover, the tomogram of quantum state
$\left \vert \psi \right \rangle $ is just the squared modulus of the wave
function $_{s,r}\left \langle q\right \vert \left.  \psi \right \rangle ,$ this
new relation between quantum tomography and optical Fresnel transform may
provide experimentalists to figure out new approach for generating tomography.

The introduction of $\left \vert q\right \rangle _{s,r}$ also bring convenience
to obtain the inverse of Radon transformation, using (\ref{10.20}) we have%
\begin{equation}
e^{-igQ_{F}}=\int_{-\infty}^{\infty}dq\left \vert q\right \rangle _{s,r}{}%
_{s,r}\left \langle q\right \vert e^{-igq}=\int_{-\infty}^{\infty}%
dqdp\Delta \left(  q,p\right)  e^{-ig\left(  Dq-Bp\right)  }.
\end{equation}
Considering its right hand-side as a Fourier transformation, its reciprocal
transform is%
\begin{align}
\Delta \left(  q,p\right)   &  =\frac{1}{4\pi^{2}}\int_{-\infty}^{\infty
}dq^{\prime}\int_{-\infty}^{\infty}dg^{\prime}|g^{\prime}|\int_{0}^{\pi
}d\varphi \left \vert q^{\prime}\right \rangle _{s,r}{}_{s,r}\left \langle
q^{\prime}\right \vert \nonumber \\
&  \times \exp \left[  -ig^{\prime}\left(  \frac{q^{\prime}}{\sqrt{D^{2}+B^{2}}%
}-q\cos \varphi-p\sin \varphi \right)  \right]  ,
\end{align}
where $g^{\prime}=g\sqrt{D^{2}+B^{2}},$ $\cos \varphi=\frac{D}{\sqrt
{D^{2}+B^{2}}},$ $\sin \varphi=\frac{-B}{\sqrt{D^{2}+B^{2}}}.$So once the
distribution $|_{s,r}\left \langle q\right \vert \left.  \psi \right \rangle
|^{2}$ are obtained, one can use the inverse Radon transformation familiar in
tomographic imaging to obtain the Wigner distribution.

By analogy, we can conclude that the $(A,C)$ related Radon transform of
$\Delta \left(  q,p\right)  $ is just the pure state density operator
$\left \vert p\right \rangle _{s,rs,r}\left \langle p\right \vert $ formed with
the eigenstates belonging to the conjugate quadrature of $Q_{F},$%
\begin{align}
F_{1}\left \vert p\right \rangle \left \langle p\right \vert F_{1}^{\dagger}  &
=\left \vert p\right \rangle _{s,rs,r}\left \langle p\right \vert =\int_{-\infty
}^{\infty}dq^{\prime}dp^{\prime}\delta \left[  p-\left(  Aq^{\prime}%
-Cp^{\prime}\right)  \right]  \Delta \left(  q^{\prime},p^{\prime}\right)
,\nonumber \\
A  &  =\frac{1}{2}\left(  s^{\ast}-r^{\ast}+s-r\right)  ,\text{ \ }C=\frac
{1}{2i}\left(  s-r-s^{\ast}+r^{\ast}\right)  . \label{10.27}%
\end{align}
Similarly, we find that for the momentum density,
\begin{equation}
F_{1}\left \vert p\right \rangle \left \langle p\right \vert F_{1}^{\dagger
}=\left \vert p\right \rangle _{s,rs,r}\left \langle p\right \vert =\int_{-\infty
}^{\infty}dq^{\prime}dp^{\prime}\delta \left[  p-\left(  Ap^{\prime}%
-Cq^{\prime}\right)  \right]  \Delta \left(  q^{\prime},p^{\prime}\right)  ,
\label{10.28}%
\end{equation}
where
\begin{equation}
F_{1}\left \vert p\right \rangle =\left \vert p\right \rangle _{s,r}=\frac
{\pi^{-1/4}}{\sqrt{A-iC}}\exp \left \{  -\frac{D+iB}{A-iC}\frac{p^{2}}{2}%
+\frac{\sqrt{2}ip}{A-iC}a^{\dagger}+\frac{A+iC}{A-iC}\frac{a^{\dagger2}}%
{2}\right \}  \left \vert 0\right \rangle . \label{10.29}%
\end{equation}
As an application of the relation (\ref{10.2}), recalling that the
$F_{1}(r,s)$ makes up a faithful representation of the symplectic group
\cite{r10}, it then follows from (\ref{10.2}) that
\begin{align}
&  F_{1}^{\prime}(r^{\prime},s^{\prime})F_{1}(r,s)\left \vert q\right \rangle
\left \langle q\right \vert F_{1}^{\dagger}(r,s)F_{1}^{\prime \dagger}(r^{\prime
},s^{\prime})=\left \vert q\right \rangle _{s^{\prime \prime},r^{\prime \prime
}\text{ }s^{\prime \prime},r^{\prime \prime}}\left \langle q\right \vert
\nonumber \\
&  =\int \int_{-\infty}^{\infty}dq^{\prime}dp^{\prime}\delta \left[  q-\left(
\left(  B^{\prime}C+DD^{\prime}\right)  q^{\prime}-\left(  AB^{\prime
}+BD^{\prime}\right)  p^{\prime}\right)  \right]  \Delta \left(  q^{\prime
},p^{\prime}\right)  , \label{10.30}%
\end{align}
In this way a complicated Radon transform of tomography can be viewed as the
sequential operation of two Fresnel transforms. This confirms that the
continuous Radon transformation corresponds to the symplectic group
transformation \cite{r6,r7}, this is an advantage of introducing the Fresnel
operator. The group property of Fresnel operators help us to analyze
complicated Radon transforms in terms of some sequential Fresnel
transformations. The new relation may provide experimentalists to figure out
new approach for realizing tomography.

\subsection{Another new theorem to calculating the tomogram}

In this subsection, we introduce a new theorem, i.e., the tomogram of a
density operator $\rho$ is equal to the marginal integration of the classical
Weyl correspondence function of $F^{\dagger}\rho F,$ where $F$ is the Fresnel operator.

Multiplying both sides of Eq. (\ref{10.25}) by a density matrix $\rho$ and
then performing the trace, noting the Wigner function $W(p,q)=\mathtt{Tr}%
\left[  \rho \Delta(p,q)\right]  ,$ one can see%
\begin{align}
&  Tr\left[
{\displaystyle \iint \nolimits_{-\infty}^{\infty}}
dq%
\acute{}%
dp%
\acute{}%
\delta \left[  q-\left(  Dq%
\acute{}%
-Bp%
\acute{}%
\right)  \right]  \Delta \left(  q%
\acute{}%
,p%
\acute{}%
\right)  \rho \right] \nonumber \\
&  =Tr\left(  \left \vert q\right \rangle _{s,rs,r}\left \langle q\right \vert
\rho \right)  =_{s,r}\left \langle q\right \vert \rho \left \vert q\right \rangle
_{s,r}=\left \langle q\right \vert F^{\dagger}\rho F\left \vert q\right \rangle
\nonumber \\
&  =%
{\displaystyle \iint \nolimits_{-\infty}^{\infty}}
dq%
\acute{}%
dp%
\acute{}%
\delta \left[  q-\left(  Dq%
\acute{}%
-Bp%
\acute{}%
\right)  \right]  W(p,q). \label{j1}%
\end{align}
The right hand side of Eq. (\ref{j1}) is commonly defined as the tomogram of
quantum states in $(B,D)$ direction, so in our view the calculation of
tomogram in $(B,D)$ direction is ascribed to calculating
\begin{equation}
\left \langle q\right \vert F^{\dagger}\rho F\left \vert q\right \rangle \equiv
\Xi. \label{j14}%
\end{equation}
This is a concise and neat formula. Similarly, the tomogram in $(A,C)$
direction is ascribed to $\left \langle p\right \vert F^{\dagger}\rho
F\left \vert p\right \rangle $.

According to the Weyl correspondence rule
\begin{equation}
H\left(  X,P\right)  =\iint_{-\infty}^{\infty}dpdx\mathfrak{h}(p,x)\Delta
(p,x), \label{15}%
\end{equation}
and the Weyl ordering form of $\Delta(p,q)$
\begin{equation}
\Delta(p,q)=%
\genfrac{}{}{0pt}{}{:}{:}%
\delta \left(  q-Q\right)  \delta \left(  p-P\right)
\genfrac{}{}{0pt}{}{:}{:}%
, \label{16}%
\end{equation}
where the symbol$%
\genfrac{}{}{0pt}{}{:}{:}%
\
\genfrac{}{}{0pt}{}{:}{:}%
$ denotes Weyl ordering, the classical correspondence of a Weyl ordered
operator $%
\genfrac{}{}{0pt}{}{:}{:}%
\mathfrak{h}(Q,P)%
\genfrac{}{}{0pt}{}{:}{:}%
$ is obtained just by replacing $Q\rightarrow q,P\rightarrow p$ in $h,$ i.e.,
\begin{equation}%
\genfrac{}{}{0pt}{}{:}{:}%
\mathfrak{h}(Q,P)%
\genfrac{}{}{0pt}{}{:}{:}%
=\iint_{-\infty}^{\infty}dpdq\mathfrak{h}(p,q)\Delta(p,q), \label{j17}%
\end{equation}
Let the classical Weyl correspondence of $F^{\dagger}\rho F$ be $h(p,q)$%
\[
F^{\dagger}\rho F=\iint_{-\infty}^{\infty}dpdqh(p,q)\Delta(p,q),
\]
then using (\ref{j14}) and (\ref{10.23}) we have%

\begin{align}
\Xi &  =\left \langle q\right \vert F^{\dagger}\rho F\left \vert q\right \rangle
\nonumber \\
&  =\left \langle q\right \vert
{\displaystyle \iint}
dpdq%
\acute{}%
h(p,q%
\acute{}%
)\Delta \left(  p,q%
\acute{}%
\right)  \left \vert q\right \rangle \nonumber \\
&  =%
{\displaystyle \iint}
dpdq%
\acute{}%
h(p,q%
\acute{}%
)\int_{-\infty}^{+\infty}\frac{dv}{2\pi}e^{ipv}\left \langle x\right.  \left.
q%
\acute{}%
+\frac{v}{2}\right \rangle \left \langle q%
\acute{}%
-\frac{v}{2}\right.  \left.  q\right \rangle \nonumber \\
&  =%
{\displaystyle \iint}
dpdq%
\acute{}%
h(p,q%
\acute{}%
)\int_{-\infty}^{+\infty}\frac{dv}{2\pi}e^{ipv}\delta \left(  q%
\acute{}%
-q+\frac{v}{2}\right)  \delta \left(  q%
\acute{}%
-q-\frac{v}{2}\right) \nonumber \\
&  =\frac{1}{\pi}%
{\displaystyle \iint}
dpdq%
\acute{}%
h(p,q%
\acute{}%
)e^{i2p\left(  q%
\acute{}%
-q\right)  }\delta \left(  2q%
\acute{}%
-2q\right)  =\int_{-\infty}^{\infty}\frac{dp}{2\pi}h(p,q). \label{j18}%
\end{align}
Thus we reach a theorem:

The tomogram of a density operator $\rho$ is equal to the marginal integration
of the classical Weyl correspondence $h(p,q)$ of $F^{\dagger}\rho F,$ where
$F$ is the Fresnel operator, expressed by%
\begin{equation}
\mathtt{Tr}\left[  \rho \left \vert q\right \rangle _{s,rs,r}\left \langle
q\right \vert \right]  =\int_{-\infty}^{\infty}\frac{dp}{2\pi}h(p,q),
\label{19}%
\end{equation}
or%
\begin{equation}
\mathtt{Tr}\left[  \rho \left \vert p\right \rangle _{s,rs,r}\left \langle
p\right \vert \right]  =\int_{-\infty}^{\infty}\frac{dx}{2\pi}h(p,q).
\label{20}%
\end{equation}
In this way the relationship between tomogram of a density operator $\rho$ and
the Fresnel transformed $\rho^{\prime}$s classical Weyl function is established.

\section{Two-mode GFO and Its Application}

For two-dimensional optical Fresnel transforms (see (\ref{2.5})) in the $x-y$
plane one may naturally think that the 2-mode GFO is just the direct product
of two independent 1-mode GFOs but with the same $(A,B,C,D)$ matrix. However,
here we present another 2-mode Fresnel operator which can not only lead to the
usual 2-dimensional optical Fresnel transforms in some appropriate quantum
mechanical representations, but also provide us with some new classical
transformations (we name them entangled Fresnel transformations).

\subsection{Two-mode GFO gained via coherent state representation}

Similar in spirit to the single-mode case, we introduce the two-mode GFO
$F_{2}\left(  r,s\right)  $ through the following 2-mode coherent state
representation \cite{fancommun1}
\begin{equation}
F_{2}\left(  r,s\right)  =s\int \frac{d^{2}z_{1}d^{2}z_{2}}{\pi^{2}}\left \vert
sz_{1}+rz_{2}^{\ast},rz_{1}^{\ast}+sz_{2}\right \rangle \left \langle
z_{1},z_{2}\right \vert , \label{11.1}%
\end{equation}
which indicates that $F_{2}\left(  r,s\right)  $ is a mapping of classical
sympletic transform $\left(  z_{1},z_{2}\right)  \rightarrow \left(
sz_{1}+rz_{2}^{\ast},rz_{1}^{\ast}+sz_{2}\right)  $ in phase space.
Concretely, the ket in (\ref{11.1}) is
\begin{equation}
\left \vert sz_{1}+rz_{2}^{\ast},rz_{1}^{\ast}+sz_{2}\right \rangle
\equiv \left \vert sz_{1}+rz_{2}^{\ast}\right \rangle _{1}\otimes \left \vert
rz_{1}^{\ast}+sz_{2}\right \rangle _{2},\text{ }ss^{\ast}-rr^{\ast}=1,
\label{11.2}%
\end{equation}
$s$ and $r$ are complex and satisfy the unimodularity condition. Using the
IWOP technique we perform the integral in (\ref{11.1}) and obtain%
\begin{align}
F_{2}\left(  r,s\right)   &  =s\int \frac{1}{\pi^{2}}d^{2}z_{1}d^{2}z_{2}%
\colon \exp[-|s|^{2}\left(  |z_{1}|^{2}+|z_{2}|^{2}\right)  -r^{\ast}%
sz_{1}z_{2}-rs^{\ast}z_{1}^{\ast}z_{2}^{\ast}\nonumber \\
&  +\left(  sz_{1}+rz_{2}^{\ast}\right)  a_{1}^{\dagger}+\left(  rz_{1}^{\ast
}+sz_{2}\right)  a_{2}^{\dagger}+z_{1}^{\ast}a_{1}+z_{2}^{\ast}a_{2}%
-a_{1}^{\dagger}a_{1}-a_{2}^{\dagger}a_{2}]\colon \nonumber \\
&  =\frac{1}{s^{\ast}}\exp \left(  \frac{r}{s^{\ast}}a_{1}^{\dagger}%
a_{2}^{\dagger}\right)  \colon \exp \left[  \left(  \frac{1}{s^{\ast}}-1\right)
\left(  a_{1}^{\dagger}a_{1}+a_{2}^{\dagger}a_{2}\right)  \right]  \colon
\exp \left(  -\frac{r^{\ast}}{s^{\ast}}a_{1}a_{2}\right) \nonumber \\
&  =\exp \left(  \frac{r}{s^{\ast}}a_{1}^{\dagger}a_{2}^{\dagger}\right)
\exp[\left(  a_{1}^{\dagger}a_{1}+a_{2}^{\dagger}a_{2}+1\right)  \ln \left(
s^{\ast}\right)  ^{-1}]\exp \left(  -\frac{r^{\ast}}{s^{\ast}}a_{1}%
a_{2}\right)  . \label{11.3}%
\end{align}
Thus $F_{2}\left(  r,s\right)  $ induces the transform
\begin{equation}
F_{2}\left(  r,s\right)  a_{1}F_{2}^{-1}\left(  r,s\right)  =s^{\ast}%
a_{1}-ra_{2}^{\dagger},F_{2}\left(  r,s\right)  a_{2}F_{2}^{-1}\left(
r,s\right)  =s^{\ast}a_{2}-ra_{1}^{\dagger}. \label{11.4}%
\end{equation}
and $F_{2}$ is actually a general 2-mode squeezing operator. Recall that
(\ref{3.15}) implies the intrinsic relation between the EPR entangled state
and the two-mode squeezed\ state, which has physical implementation, i.e. in
the output of a parametric down-conversion the idler-mode and the signal-mode
constitute a two-mode squeezed state, meanwhile are entangled with each other
in frequency domain, we naturally select the entangled state representation to
relate $F_{2}\left(  r,s\right)  $ to two-dimensional GFT. Letting $\left \vert
g\right \rangle =F_{2}\left(  r,s\right)  \left \vert f\right \rangle ,$ and then
projecting $\left \vert \psi \right \rangle $ onto the entangled state
$\left \langle \eta^{\prime}\right \vert $ defined by (\ref{3.11}) and using the
completeness relation (\ref{3.13}) of $\left \vert \eta \right \rangle $, we
obtain
\begin{align}
g\left(  \eta^{\prime}\right)   &  \equiv \left \langle \eta^{\prime}\right \vert
\left.  g\right \rangle =\left \langle \eta^{\prime}\right \vert F_{2}\left(
r,s\right)  \left \vert f\right \rangle \nonumber \\
&  =\int \frac{d^{2}\eta}{\pi}\left \langle \eta^{\prime}\right \vert
F_{2}\left(  r,s\right)  \left \vert \eta \right \rangle \left \langle
\eta \right.  \left \vert f\right \rangle \equiv \int d^{2}\eta \mathcal{K}%
_{2}^{\left(  r,s\right)  }\left(  \eta^{\prime},\eta \right)  f\left(
\eta \right)  . \label{11.5}%
\end{align}
Then using the overcompleteness relation of the coherent state, and%
\begin{align*}
\left \langle z_{1}^{\prime},z_{2}^{\prime}\right \vert F_{2}\left(  r,s\right)
\left \vert z_{1},z_{2}\right \rangle  &  =\frac{1}{s^{\ast}}\exp \left \{
-\frac{1}{2}(\left \vert z_{1}\right \vert ^{2}+\left \vert z_{2}\right \vert
^{2}+\left \vert z_{1}^{\prime}\right \vert ^{2}+\left \vert z_{2}^{\prime
}\right \vert ^{2})\right. \\
&  \left.  +\frac{r}{s^{\ast}}z_{1}^{\prime \ast}z_{2}^{\prime \ast}%
-\frac{r^{\ast}}{s^{\ast}}z_{1}z_{2}+\frac{1}{s^{\ast}}\left(  z_{1}%
^{\prime \ast}z_{1}+z_{2}^{\prime \ast}z_{2}\right)  \right \}  ,
\end{align*}
we can calculate the integral kernel%
\begin{align}
\mathcal{K}_{2}^{\left(  r,s\right)  }\left(  \eta^{\prime},\eta \right)   &
=\frac{1}{\pi}\left \langle \eta^{\prime}\right \vert F_{2}\left(  r,s\right)
\left \vert \eta \right \rangle \nonumber \\
&  =\int \frac{d^{2}z_{1}d^{2}z_{2}d^{2}z_{1}^{\prime}d^{2}z_{2}^{\prime}}%
{\pi^{5}}\left \langle \eta^{\prime}\right \vert \left.  z_{1}^{\prime}%
,z_{2}^{\prime}\right \rangle \left \langle z_{1}^{\prime},z_{2}^{\prime
}\right \vert F_{2}\left(  r,s\right)  \left \vert z_{1},z_{2}\right \rangle
\left \langle z_{1},z_{2}\right.  \left \vert \eta \right \rangle \nonumber \\
&  =\frac{1}{s^{\ast}}\int \frac{d^{2}z_{1}d^{2}z_{2}d^{2}z_{1}^{\prime}%
d^{2}z_{2}^{\prime}}{\pi^{5}}\exp \left[  -\left(  \left \vert z_{1}\right \vert
^{2}+\left \vert z_{2}\right \vert ^{2}+\left \vert z_{1}^{\prime}\right \vert
^{2}+\left \vert z_{2}^{\prime}\right \vert ^{2}\right)  -\frac{1}{2}\left(
\left \vert \eta^{\prime}\right \vert ^{2}+\left \vert \eta \right \vert
^{2}\right)  \right] \nonumber \\
&  \times \exp \left[  -\frac{r^{\ast}}{s^{\ast}}z_{1}z_{2}+z_{1}^{\ast}%
z_{2}^{\ast}+\eta z_{1}^{\ast}+\frac{1}{s^{\ast}}\left(  z_{1}^{\prime \ast
}z_{1}+z_{2}^{\prime \ast}z_{2}\right)  +\frac{r}{s^{\ast}}z_{1}^{\prime \ast
}z_{2}^{\prime \ast}+z_{1}^{\prime}z_{2}^{\prime}+\eta^{\prime \ast}%
z_{1}^{\prime}-\eta^{\prime}z_{2}^{\prime}-\eta^{\ast}z_{2}^{\ast}\right]
\nonumber \\
&  =\frac{1}{\left(  -r-s+r^{\ast}+s^{\ast}\right)  \pi}\exp \left[
\frac{\left(  -s+r^{\ast}\right)  \left \vert \eta \right \vert ^{2}-\left(
r+s\right)  \left \vert \eta^{\prime}\right \vert ^{2}+\eta \eta^{\prime \ast
}+\eta^{\ast}\eta^{\prime}}{-r-s+r^{\ast}+s^{\ast}}-\frac{1}{2}\left(
\left \vert \eta^{\prime}\right \vert ^{2}+\left \vert \eta \right \vert
^{2}\right)  \right]  . \label{11.6}%
\end{align}
Using the relation between $s,r$ and $\left(  A,B,C,D\right)  $ in
Eq.(\ref{5.2}) Eq. (\ref{11.6}) becomes%
\begin{equation}
\mathcal{K}_{2}^{\left(  r,s\right)  }\left(  \eta^{\prime},\eta \right)
=\frac{1}{2iB\pi}\exp \left[  \frac{i}{2B}\left(  A\left \vert \eta \right \vert
^{2}-\left(  \eta \eta^{\prime \ast}+\eta^{\ast}\eta^{\prime}\right)
+D\left \vert \eta^{\prime}\right \vert ^{2}\right)  \right]  \equiv
\mathcal{K}_{2}^{M}\left(  \eta^{\prime},\eta \right)  , \label{11.7}%
\end{equation}
where the superscript $M$ only means the parameters of $K_{2}^{M}$ are
$\left[  A,B;C,D\right]  $, and the subscript $2$ means the two-dimensional
kernel. Eq. (\ref{11.7}) has the similar form as (\ref{2.12}) except for its
complex form. Taking $\eta_{1}=x_{1},$ $\eta_{2}=x_{2}$ and $\eta_{1}^{\prime
}=x_{1}^{\prime}$, $\eta_{2}^{\prime}=x_{2}^{\prime}$, we have%
\begin{equation}
\mathcal{K}_{2}^{M}\left(  \eta^{\prime},\eta \right)  =\mathcal{K}_{2}%
^{M}\left(  x_{1}^{\prime},x_{2}^{\prime};x_{1},x_{2}\right)  =\mathcal{K}%
_{1}^{M}\left(  x_{1},x_{1}^{\prime}\right)  \otimes \mathcal{K}_{1}^{M}\left(
x_{2},x_{2}^{\prime}\right)  . \label{11.8}%
\end{equation}
This shows that $F_{2}\left(  r,s\right)  $ is really the counterpart of the
2-dimensional GFT.

If taking the matrix element of $F_{2}\left(  r,s\right)  $ in the $\left \vert
\xi \right \rangle $ representation which is conjugate to $\left \vert
\eta \right \rangle $, we obtain the 2-dimensional GFT in its `frequency
domain', i.e.,
\begin{align}
\left \langle \xi^{\prime}\right \vert F_{2}\left(  r,s\right)  \left \vert
\xi \right \rangle  &  =\int \frac{d^{2}\eta^{\prime}d^{2}\eta}{\pi^{2}%
}\left \langle \xi^{\prime}\right \vert \left.  \eta^{\prime}\right \rangle
\left \langle \sigma \right \vert F_{2}\left(  r,s\right)  \left \vert
\eta \right \rangle \left \langle \eta \right \vert \left.  \xi \right \rangle
\nonumber \\
&  =\frac{1}{8iB\pi}\int \frac{d^{2}\eta^{\prime}d^{2}\eta}{\pi^{2}}%
\mathcal{K}_{2}^{\left(  r,s\right)  }\left(  \eta^{\prime},\eta \right)
\exp \left(  \frac{\xi^{\prime \ast}\eta^{\prime}-\xi^{\prime}\eta^{\prime \ast
}+\xi \eta^{\ast}-\xi^{\ast}\eta}{2}\right) \nonumber \\
&  =\frac{1}{2i\left(  -C\right)  \pi}\exp \left[  \frac{i}{2\left(  -C\right)
}\left(  D\left \vert \xi \right \vert ^{2}+A\left \vert \xi^{\prime}\right \vert
^{2}-\xi^{\prime \ast}\xi-\xi^{\prime}\xi^{\ast}\right)  \right]
\equiv \mathcal{K}_{2}^{N}\left(  \xi^{\prime},\xi \right)  , \label{11.9}%
\end{align}
where the superscript $N$ means that this transform kernel corresponds to the
parameter matrix $N=\left[  D,-C,-B,A\right]  $.

The two-mode GFO also abides by the group multiplication rule. Using the IWOP
technique and (\ref{11.1}) we obtain%
\begin{align}
&  F_{2}\left(  r,s\right)  F_{2}\left(  r^{\prime},s^{\prime}\right)
\nonumber \\
&  =ss^{\prime}\int \frac{d^{2}z_{1}d^{2}z_{2}d^{2}z_{1}^{\prime}d^{2}%
z_{2}^{\prime}}{\pi^{4}}\colon \exp \{-|s|^{2}\left(  |z_{1}|^{2}+|z_{2}%
|^{2}\right)  -r^{\ast}sz_{1}z_{2}\nonumber \\
&  -rs^{\ast}z_{1}^{\ast}z_{2}^{\ast}-\frac{1}{2}[|z_{1}^{\prime}|^{2}%
+|z_{2}^{\prime}|^{2}+|s^{\prime}z_{1}^{\prime}+r^{\prime}z_{2}^{\prime \ast
}|^{2}+|r^{\prime}z_{1}^{\prime \ast}+s^{\prime}z_{2}^{\prime}|^{2}]\nonumber \\
&  +\left(  sz_{1}+rz_{2}^{\ast}\right)  a_{1}^{\dagger}+\left(  rz_{1}^{\ast
}+sz_{2}\right)  a_{2}^{\dagger}+z_{1}^{\prime \ast}a_{1}+z_{2}^{\prime \ast
}a_{2}\nonumber \\
&  +z_{1}^{\ast}\left(  s^{\prime}z_{1}^{\prime}+r^{\prime}z_{2}^{\prime \ast
}\right)  +z_{2}^{\ast}\left(  r^{\prime}z_{1}^{\prime \ast}+s^{\prime}%
z_{2}^{\prime}\right)  -a_{1}^{\dagger}a_{1}-a_{2}^{\dagger}a_{2}\}
\colon \nonumber \\
&  =\frac{1}{s^{\prime \prime \ast}}\exp \left(  \frac{r^{\prime \prime}%
}{2s^{\prime \prime \ast}}a_{1}^{\dagger}a_{2}^{\dagger}\right)  \colon
\exp \left \{  \left(  \frac{1}{s^{\prime \prime \ast}}-1\right)  \left(
a_{1}^{\dagger}a_{1}+a_{2}^{\dagger}a_{2}\right)  \right \}  \colon \exp \left(
-\frac{r^{\prime \prime \ast}}{2s^{\prime \prime \ast}}a_{1}a_{2}\right)
\nonumber \\
&  =F_{2}\left(  r^{\prime \prime},s^{\prime \prime}\right)  , \label{11.10}%
\end{align}
where $\left(  r^{\prime \prime},s^{\prime \prime}\right)  $ are given by
Eq.(\ref{7.2}) or (\ref{7.4}). Therefore, (\ref{11.10}) is a loyal
representation of the multiplication rule for ray transfer matrices in the
sense of \textit{Matrix Optics}.$\ $

\subsection{Quantum Optical ABCD Law for two-mode GFO}

Next we extended quantum optical ABCD Law to two-mode case. Operating with
$F_{2}(r,s)$ on two-mode number state $\left \vert m,n\right \rangle $ and using
the overlap between coherent state and number state, i.e.
\begin{equation}
\left \langle z_{1},z_{2}\right.  \left \vert m,n\right \rangle =\frac
{z_{1}^{\ast m}z_{2}^{\ast n}}{\sqrt{m!n!}}\exp \left[  -\frac{1}{2}\left(
\left \vert z_{1}\right \vert ^{2}+\left \vert z_{2}\right \vert ^{2}\right)
\right]  , \label{11.11}%
\end{equation}
and the integral formula \cite{Fanshu}%
\begin{equation}
H_{m,n}\left(  \xi,\eta \right)  =(-1)^{n}e^{\xi \eta}\int \frac{d^{2}z}{\pi
}z^{n}z^{\ast m}e^{-\left \vert z\right \vert ^{2}+\xi z-\eta z^{\ast}},
\label{11.12}%
\end{equation}
we can calculate%
\begin{align}
&  F_{2}(r,s)\left \vert m,n\right \rangle \nonumber \\
&  =s\int \frac{d^{2}z_{1}d^{2}z_{2}}{\pi^{2}}\left \vert sz_{1}+rz_{2}^{\ast
},rz_{1}^{\ast}+sz_{2}\right \rangle \left \langle z_{1},z_{2}\right.
\left \vert m,n\right \rangle \nonumber \\
&  =\frac{s}{\sqrt{m!n!}}\int \frac{d^{2}z_{1}d^{2}z_{2}}{\pi^{2}}z_{1}^{\ast
m}z_{2}^{\ast n}\exp \left[  -\left \vert s\right \vert ^{2}\left(  \left \vert
z_{1}\right \vert ^{2}+\left \vert z_{2}\right \vert ^{2}\right)  \right]
\nonumber \\
&  \times \exp \left[  -sr^{\ast}z_{1}z_{2}-rs^{\ast}z_{1}^{\ast}z_{2}^{\ast
}+\left(  sz_{1}+rz_{2}^{\ast}\right)  a_{1}^{\dagger}+\left(  rz_{1}^{\ast
}+sz_{2}\right)  a_{2}^{\dagger}\right]  \left \vert 00\right \rangle
\nonumber \\
&  =\frac{s}{\left \vert s\right \vert ^{2m+2}\sqrt{m!n!}}\int \frac{d^{2}z_{2}%
}{\pi}z_{2}^{\ast n}\left(  sa_{1}^{\dagger}-sr^{\ast}z_{2}\right)  ^{m}%
\exp \left(  -\left \vert z_{2}\right \vert ^{2}+\frac{1}{s^{\ast}}a_{2}%
^{\dagger}z_{2}+\frac{ra_{1}^{\dagger}a_{2}^{\dagger}}{s^{\ast}}\right)
\left \vert 00\right \rangle \nonumber \\
&  =\frac{r^{\ast n}}{s^{\ast n+1}\sqrt{m!n!}}H_{m,n}\left[  \frac
{a_{1}^{\dagger}}{s^{\ast}},\frac{a_{2}^{\dagger}}{r^{\ast}}\right]
\exp \left(  \frac{ra_{1}^{\dagger}a_{2}^{\dagger}}{s^{\ast}}\right)
\left \vert 00\right \rangle , \label{11.13}%
\end{align}
where $H_{m,n}\left(  \epsilon,\varepsilon \right)  $ is the two variables
Hermite polynomial \cite{r6,r7}, shown in (\ref{4.20}) and (\ref{4.21}).

Using Eqs.(\ref{5.2}) and (\ref{8.9}), we recast Eq.(\ref{11.13}) into%
\begin{align}
F_{2}(r,s)\left \vert m,n\right \rangle  &  =\frac{-2/\left(  C+iD\right)
}{\left(  q_{1}+i\right)  \sqrt{m!n!}}\left(  -\frac{q_{1}^{\ast}+i}{q_{1}%
+i}\frac{C-iD}{C+iD}\right)  ^{n}\nonumber \\
&  \times H_{m,n}\left[  -\frac{2a_{1}^{\dagger}/\left(  C+iD\right)  }%
{q_{1}+i},\frac{2a_{2}^{\dagger}/\left(  C-iD\right)  }{q_{1}^{\ast}%
+i}\right]  \exp \left(  -\frac{q_{1}-i}{q_{1}+i}a_{1}^{\dagger}a_{2}^{\dagger
}\right)  \left \vert 00\right \rangle . \label{11.14}%
\end{align}
Noticing the multiplication rule of $F_{2}\left(  r,s\right)  $ in
Eq.(\ref{11.10}), which is equivalent to
\begin{equation}
F_{2}\left(  A^{\prime},B^{\prime},C^{\prime}\right)  F_{2}\left(
A,B,C\right)  =F_{2}\left(  A^{\prime \prime},B^{\prime \prime},C^{\prime \prime
}\right)  , \label{11.15}%
\end{equation}
where $\left(  A^{\prime},B^{\prime},C^{\prime}\right)  ,\left(  A,B,C\right)
$ and $\left(  A^{\prime \prime},B^{\prime \prime},C^{\prime \prime}\right)  $
are related to each other by Eq.(\ref{8.5}).

Next we directly use the GFO to derive ABCD rule in quantum optics for
Gaussian beam in two-mode case. According to Eq.(\ref{11.13}) and Eq.
(\ref{11.15}) we obtain%
\begin{align}
&  F_{2}\left(  A^{\prime},B^{\prime},C^{\prime}\right)  F_{2}\left(
A,B,C\right)  \left \vert m,n\right \rangle \nonumber \\
&  =\frac{r^{\prime \prime \ast n}}{s^{\prime \prime \ast n+1}\sqrt{m!n!}}%
H_{m,n}\left[  \frac{a_{1}^{\dagger}}{s^{\prime \prime \ast}},\frac
{a_{2}^{\dagger}}{r^{\prime \prime \ast}}\right]  \exp \left[  \frac
{r^{\prime \prime}a_{1}^{\dagger}a_{2}^{\dagger}}{s^{\prime \prime \ast}}\right]
\left \vert 00\right \rangle , \label{11.16}%
\end{align}
Similar to the way of deriving Eq. (\ref{11.14}), we can simplify Eq.
(\ref{11.16}) as%
\begin{align}
&  F_{2}\left(  A^{\prime},B^{\prime},C^{\prime}\right)  F_{2}\left(
A,B,C\right)  \left \vert 00\right \rangle \nonumber \\
&  =\frac{-2/\left(  C^{\prime \prime}+iD^{\prime \prime}\right)  }{\left(
q_{2}+i\right)  \sqrt{m!n!}}\left(  -\frac{q_{2}^{\ast}+i}{q_{2}+i}%
\frac{C^{\prime \prime}-iD^{\prime \prime}}{C^{\prime \prime}+iD^{\prime \prime}%
}\right)  ^{n}\nonumber \\
&  \times H_{m,n}\left[  -\frac{2a_{1}^{\dagger}/\left(  C^{\prime \prime
}+iD^{\prime \prime}\right)  }{q_{2}+i},\frac{2a_{2}^{\dagger}/\left(
C^{\prime \prime}-iD^{\prime \prime}\right)  }{q_{2}^{\ast}+i}\right]
\exp \left[  -\frac{q_{2}-i}{q_{2}+i}a_{1}^{\dagger}a_{2}^{\dagger}\right]
\left \vert 00\right \rangle , \label{11.17}%
\end{align}
where the relation between $q_{2}$ and $q_{1}$ are determined by
Eq.(\ref{8.11}) which resembles Eq.(\ref{8.3}), this is just the new ABCD law
for two-mode case in quantum optics.

\subsection{Optical operators derived by decomposing GFO}

\subsubsection{GFO as quadratic combinations of canonical operators}

In order to obtain the quadratic combinations of canonical operators, let
first derive an operator identity. Note $Q_{i}=(a_{i}+a_{i}^{\dagger}%
)/\sqrt{2},$ $P_{i}=(a_{i}-a_{i}^{\dagger})/(\sqrt{2}\mathtt{i}),$ and
Eq.(\ref{3.12}),(\ref{3.13}) we can prove the operator identity%
\begin{align}
e^{\frac{\lambda}{2}\left[  \left(  Q_{1}-Q_{2}\right)  ^{2}+\left(
P_{1}+P_{2}\right)  ^{2}\right]  }  &  =\int \frac{d^{2}\eta}{\pi}%
e^{\frac{\lambda}{2}\left[  \left(  Q_{1}-Q_{2}\right)  ^{2}+\left(
P_{1}+P_{2}\right)  ^{2}\right]  }\left \vert \eta \right \rangle \left \langle
\eta \right \vert \nonumber \\
&  =\frac{1}{1-\lambda}\colon \exp \left[  \frac{2\lambda}{1-\lambda}%
K_{+}\right]  \colon, \label{11.18}%
\end{align}
where we have set
\begin{equation}
K_{+}\equiv \frac{1}{4}[\left(  Q_{1}-Q_{2}\right)  ^{2}+\left(  P_{1}%
+P_{2}\right)  ^{2}]. \label{K1}%
\end{equation}
When $B=0$, $A=1,$ $C\rightarrow C/A,$ $D=1,$ and using Eq.(\ref{11.3}) we see
that
\begin{align}
F_{2}\left(  1,0,C/A\right)   &  =\frac{2}{2-iC/A}\colon \exp \left \{
\frac{iC/A}{2-iC/A}\left(  a_{1}^{\dagger}a_{1}+a_{2}^{\dagger}a_{2}%
-a_{1}^{\dagger}a_{2}^{\dagger}-a_{1}a_{2}\right)  \right \}  \colon \nonumber \\
&  =\exp \left \{  \frac{iC}{A}K_{+}\right \}  , \label{11.19}%
\end{align}
which is corresponding to the square phase operator in single-mode case. In a
similar way, using (\ref{3.18}) and (\ref{3.19}) we can derive another
operator identity
\begin{align}
e^{\frac{\lambda}{2}\left[  \left(  Q_{1}+Q_{2}\right)  ^{2}+\left(
P_{1}-P_{2}\right)  ^{2}\right]  }  &  =\int \frac{d^{2}\xi}{\pi}%
e^{\frac{\lambda}{2}\left[  \left(  Q_{1}+Q_{2}\right)  ^{2}+\left(
P_{1}-P_{2}\right)  ^{2}\right]  }\left \vert \xi \right \rangle \left \langle
\xi \right \vert \nonumber \\
&  =\frac{1}{1-\lambda}\colon \exp \left[  \frac{2\lambda}{1-\lambda}%
K_{-}\right]  \colon, \label{11.20}%
\end{align}
where
\begin{equation}
K_{-}=\frac{1}{4}[\left(  Q_{1}+Q_{2}\right)  ^{2}+\left(  P_{1}-P_{2}\right)
^{2}]. \label{K2}%
\end{equation}
It then follows from Eqs.(\ref{11.3}) and (\ref{11.20})
\begin{align}
F_{2}\left(  1,B/A,0\right)   &  =\frac{2}{2+iB/A}\colon \exp \left \{
-\frac{iB/A}{2+iB/A}\left(  a_{1}^{\dagger}a_{2}^{\dagger}+a_{2}^{\dagger
}a_{2}+a_{1}^{\dagger}a_{1}+a_{1}a_{2}\right)  \right \}  \colon \nonumber \\
&  =\frac{2}{2+iB/A}\colon \exp \left \{  -\frac{2iB/A}{2+iB/A}K_{-}\right \}
\colon \nonumber \\
&  =\exp \left \{  -\frac{iB}{A}K_{-}\right \}  , \label{11.21}%
\end{align}
which is corresponding to Fresnel propagator in free space (single-mode case).
In particular, when $B=C=0,$ and $D=A^{-1},$ Eq. (\ref{11.3}) becomes
\begin{equation}
F_{2}\left(  A,0,0\right)  =\operatorname{sech}\lambda \colon \exp \left[
-a_{1}^{\dagger}a_{2}^{\dagger}\tanh \lambda+\left(  \operatorname{sech}%
\lambda-1\right)  \left(  a_{1}^{\dagger}a_{1}+a_{2}^{\dagger}a_{2}\right)
+a_{1}a_{2}\tanh \lambda \right]  \colon, \label{11.22}%
\end{equation}
where$\frac{A-A^{-1}}{A+A^{-1}}=\tanh \lambda,$ $A=e^{\lambda}.$ Eq.
(\ref{11.22}) is just the two-mode squeezing operator,
\begin{align}
F_{2}\left(  A,0,0\right)   &  =\exp \left[  i\left(  Q_{1}P_{2}+Q_{2}%
P_{1}\right)  \ln A\right]  \equiv \exp \left[  -2K_{0}\ln A\right]
,\label{11.22b}\\
K_{0}  &  \equiv-\frac{\mathtt{i}}{2}\left(  Q_{1}P_{2}+Q_{2}P_{1}\right)  ,
\label{K0}%
\end{align}
which actually squeezes the entangled state $\left \vert \xi \right \rangle $
(its conjugate state is $\left \vert \eta \right \rangle $),%
\begin{equation}
F_{2}\left(  A,0,0\right)  \left \vert \xi \right \rangle =\int \frac{d^{2}%
\xi^{\prime}}{\pi A}\left \vert \xi^{\prime}/A\right \rangle \left \langle
\xi^{\prime}\right \vert \left.  \xi \right \rangle =\frac{1}{A}\left \vert
\xi/A\right \rangle . \label{11.23}%
\end{equation}
Using the decomposition (\ref{9.5}) of the matrix and combining equations
(\ref{11.19}), (\ref{11.21}) and (\ref{11.22b}) together, we see that
\begin{align}
F_{2}\left(  A,B,C\right)   &  =F_{2}\left(  1,0,C/A\right)  F_{2}\left(
A,0,0\right)  F_{2}\left(  1,B/A,0\right) \nonumber \\
&  =\exp \left \{  \frac{iC}{A}K_{+}\right \}  \exp \left \{  -2K_{0}\ln A\right \}
\exp \left \{  -\frac{iB}{A}K_{-}\right \}  . \label{11.24}%
\end{align}
This is the two-mode quadratic canonical operator representation of
$F_{2}\left(  A,B,C\right)  $.

To prove Eq.(\ref{11.24}), using (\ref{11.23}) and (\ref{3.20}) we see%
\begin{align}
\left \langle \eta \right \vert F_{2}\left(  A,B,C\right)  \left \vert
\xi \right \rangle  &  =\exp \left(  \frac{iC}{2A}\left \vert \eta \right \vert
^{2}-\frac{iB}{2A}\left \vert \xi \right \vert ^{2}\right)  \left \langle
\eta \right \vert \int \frac{d^{2}\xi^{\prime}}{A\pi}\left \vert \xi^{\prime
}/A\right \rangle \left \langle \xi^{\prime}\right \vert \left.  \xi \right \rangle
\nonumber \\
\  &  =\frac{1}{A}\exp \left(  \frac{iC}{2A}\left \vert \eta \right \vert
^{2}-\frac{iB}{2A}\left \vert \xi \right \vert ^{2}\right)  \left \langle
\eta \right \vert \left.  \xi/A\right \rangle \nonumber \\
\  &  =\frac{1}{2A}\exp \left(  \frac{iC}{2A}\left \vert \eta \right \vert
^{2}-\frac{iB}{2A}\left \vert \xi \right \vert ^{2}\right)  \exp \left[  \frac
{1}{2A}\left(  \eta^{\ast}\xi-\eta \xi^{\ast}\right)  \right]  . \label{11.25}%
\end{align}
It then follows%
\begin{align}
\left \langle \eta^{\prime}\right \vert F_{2}\left(  A,B,C\right)  \left \vert
\eta \right \rangle  &  =\int_{\infty}^{\infty}\frac{d^{2}\xi}{\pi}\left \langle
\eta^{\prime}\right \vert F_{2}\left \vert \xi \right \rangle \left \langle
\xi \right \vert \left.  \eta \right \rangle \nonumber \\
&  =\frac{1}{2iB}\exp \left[  \frac{i}{2B}\left(  A\left \vert \eta \right \vert
^{2}-i\left(  \eta \eta^{\prime \ast}+\eta^{\ast}\eta^{\prime}\right)
+D\left \vert \eta^{\prime}\right \vert ^{2}\right)  \right] \nonumber \\
&  \equiv \mathcal{K}_{2}^{M}\left(  \eta^{\prime},\eta \right)  , \label{11.26}%
\end{align}
which is just the transform kernel of a 2-dimensional GFT and the definition
given in (\ref{11.24}) is true.

Note that the quadratic combinations in Eqs.(\ref{K1}), (\ref{K2}) and
(\ref{K0}) of the four canonical operators $\left(  Q_{1},Q_{2};P_{1}%
,P_{2}\right)  $ obey the commutative relations $\left[  K_{+},K_{-}\right]
=2K_{0},$ $\left[  K_{0},K_{\pm}\right]  =\pm K_{\pm},$\ so $F_{2}\left(
A,B,C\right)  $ involves a SU(2) Lie algebra structure (this structure is also
compiled by $Q^{2}/2$, $P^{2}/2$ and $-i\left(  QP+PQ\right)  /2$ that have
been used in constructing $F_{1}\left(  A,B,C\right)  $).

\subsubsection{Alternate decompositions of GFO and new optical operator
identities}

When $A=D=0,B=1,C=-1,$ from Eq.(\ref{11.3}) we see
\begin{align}
F_{2}\left(  0,1,-1\right)   &  =\exp \left[  -\left(  a_{1}^{\dagger}%
a_{1}+a_{2}^{\dagger}a_{2}+1\right)  \ln i\right] \nonumber \\
&  =\exp \left[  -\mathtt{i}\frac{\pi}{2}\left(  a_{1}^{\dagger}a_{1}%
+a_{2}^{\dagger}a_{2}+1\right)  \right]  \equiv \mathcal{F}, \label{11.27}%
\end{align}
which can also be named the Fourier operator, since it induces the quantum
mechanically transforms \cite{Fanshu}%
\begin{equation}
\mathcal{F}^{\dagger}Q_{i}\mathcal{F}=P_{i},\text{ }\mathcal{F}^{\dagger}%
P_{i}\mathcal{F}^{\dagger}=-Q_{i}. \label{11.28}%
\end{equation}
it then follows that
\begin{equation}
\mathcal{F}^{\dagger}K_{+}\mathcal{F}=K_{-}. \label{11.29}%
\end{equation}

On the other hand, in order to obtain the decomposition of $F_{2}\left(
A,B,C\right)  $ for $A=0,$ similar to deriving Eq.(\ref{9.14}), we have
\begin{equation}
F_{2}\left(  A,B,C\right)  =\exp \left[  -\frac{iB}{D}K_{-}\right]  \exp \left[
2K_{0}\ln D\right]  \exp \left[  \frac{iC}{D}K_{+}\right]  ,\text{ for }D\neq0.
\label{11.30}%
\end{equation}
While for $B\neq0$ or $C\neq0$, using Eqs.(\ref{9.15}) and (\ref{9.16}) we
have another decomposition of $F_{2}\left(  A,B,C\right)  $, i.e.,%
\begin{equation}
F_{2}\left(  A,B,C\right)  =\exp \left[  \frac{iD}{B}K_{+}\right]  \exp \left[
-2K_{0}\ln B\right]  \mathcal{F}\exp \left[  \frac{iA}{B}K_{+}\right]  ,\text{
}B\neq0, \label{11.31}%
\end{equation}
and
\begin{equation}
F_{2}\left(  A,B,C\right)  =\exp \left[  -\frac{iA}{C}K_{-}\right]  \exp \left[
-2K_{0}\ln \frac{-1}{C}\right]  \mathcal{F}\exp \left[  -\frac{iD}{C}%
K_{-}\right]  ,C\neq0. \label{11.32}%
\end{equation}

In addition, noticing Eqs.(\ref{9.26}) and (\ref{9.24}), we can rewrite
Eqs.(\ref{11.31}) and (\ref{11.32}) as follows
\begin{equation}
F_{2}\left(  A,B,C\right)  =\exp \left[  \frac{i}{B}\left(  D-1\right)
K_{+}\right]  \exp \left[  -iBK_{-}\right]  \exp \left[  \frac{i}{B}\left(
A-1\right)  K_{+}\right]  ,B\neq0, \label{11.33}%
\end{equation}
and
\begin{equation}
F_{2}\left(  A,B,C\right)  =\exp \left[  \frac{-i}{C}\left(  A-1\right)
K_{-}\right]  \exp \left[  iCK_{+}\right]  \exp \left[  \frac{-i}{C}\left(
D-1\right)  K_{-}\right]  ,C\neq0, \label{11.34}%
\end{equation}
respectively.

Next, according to some optical systems used frequently in physical optics, we
derive some new entangled optical operator identities. For a special optical
system with the parameter $A=0,$ $C=-B^{-1},$ (\ref{9.20}) which corresponds
to the Fourier transform system, we have%
\begin{equation}
\exp \left[  -\frac{iB}{D}K_{-}\right]  \exp \left[  2K_{0}\ln D\right]
\exp \left[  -\frac{i}{BD}K_{+}\right]  =\exp \left[  \frac{iD}{B}K_{+}\right]
\exp \left[  -2K_{0}\ln B\right]  \mathcal{F}. \label{11.35}%
\end{equation}
In particular, when $A=D=0,$ $C=-B^{-1},$ Eq.(\ref{9.22}) corresponding to the
ideal spectrum analyzer, we have
\begin{equation}
\exp \left[  -iBK_{-}\right]  \exp \left[  -\frac{i}{B}K_{+}\right]  \exp \left[
-iBK_{-}\right]  =\exp \left[  -2K_{0}\ln B\right]  \mathcal{F}. \label{11.36}%
\end{equation}
When $B=0,$ $D=A^{-1},$%
\[
\left(
\begin{array}
[c]{cc}%
A & 0\\
C & A^{-1}%
\end{array}
\right)  =\left(
\begin{array}
[c]{cc}%
1 & 0\\
C/A & 1
\end{array}
\right)  \left(
\begin{array}
[c]{cc}%
A & 0\\
0 & A^{-1}%
\end{array}
\right)  ,
\]
which corresponds to the form of image system, another operator identity is
given by%
\begin{equation}
\exp \left[  -2K_{0}\ln A\right]  \exp \left[  iACK_{+}\right]  =\exp \left[
\frac{iC}{A}K_{+}\right]  \exp \left[  -2K_{0}\ln A\right]  . \label{11.37}%
\end{equation}
When $C=0,$ $A=D^{-1},$%
\[
\left(
\begin{array}
[c]{cc}%
D^{-1} & B\\
0 & D
\end{array}
\right)  =\left(
\begin{array}
[c]{cc}%
1 & B/D\\
0 & 1
\end{array}
\right)  \left(
\begin{array}
[c]{cc}%
D^{-1} & 0\\
0 & D
\end{array}
\right)  ,
\]
which corresponds to the far foci system,
\begin{equation}
\exp \left[  \frac{iD}{B}K_{+}\right]  \exp \left[  -2K_{0}\ln B\right]
\mathcal{F}\exp \left[  \frac{i}{BD}K_{+}\right]  =\exp \left[  -\frac{iB}%
{D}K_{-}\right]  \exp \left[  2K_{0}\ln D\right]  . \label{11.38}%
\end{equation}
When $D=0,$ $C=-B^{-1},$ corresponding to the Fresnel transform system,
\[
\left(
\begin{array}
[c]{cc}%
A & B\\
-B^{-1} & 0
\end{array}
\right)  =\left(
\begin{array}
[c]{cc}%
B & 0\\
0 & B^{-1}%
\end{array}
\right)  \left(
\begin{array}
[c]{cc}%
0 & 1\\
-1 & 0
\end{array}
\right)  \left(
\begin{array}
[c]{cc}%
1 & 0\\
A/B & 1
\end{array}
\right)  ,
\]
we have
\begin{equation}
\exp \left[  -\frac{i}{AB}K_{+}\right]  \exp \left[  -2K_{0}\ln A\right]
\exp \left[  -\frac{iB}{A}K_{-}\right]  =\exp \left[  -2K_{0}\ln B\right]
\mathcal{F}\exp \left[  \frac{iA}{B}K_{+}\right]  . \label{11.39}%
\end{equation}

The GFO can unify those optical operators in two-mode case. Various
decompositions of the GFO into the exponential canonical operators,
corresponding to the decomposition of ray transfer matrix $\left[
A,B,C,D\right]  ,$ are also derived. In our derivation, the entangled state
representation is of useness in our research.

\subsection{Quantum tomography and probability distribution for the Fresnel
quadrature phase---two-mode entangled case}

In section 8 we have found that under the Fresnel transformation the pure
position density $\left \vert q\right \rangle \left \langle q\right \vert
$\ becomes the tomographic density $\left \vert q\right \rangle _{s,rs,r}%
\left \langle q\right \vert $, which is just the Radon transform of the Wigner
operator $\Delta \left(  q,p\right)  .$ In this section we want to generalize
the above conclusion to two-mode entangled case. Here we shall prove
\begin{equation}
F_{2}\left \vert \eta \right \rangle \left \langle \eta \right \vert F_{2}^{\dagger
}=\left \vert \eta \right \rangle _{s,rs,r}\left \langle \eta \right \vert =\pi \int
d^{2}\gamma d^{2}\sigma \delta \left(  \eta_{2}-D\sigma_{2}+B\gamma_{1}\right)
\delta \left(  \eta_{1}-D\sigma_{1}-B\gamma_{2}\right)  \Delta \left(
\sigma,\gamma \right)  , \label{9}%
\end{equation}
i.e., we show that $\left \vert \eta \right \rangle _{s,rs,r}\left \langle
\eta \right \vert $ is just the Radon transform of the entangled Wigner operator
$\Delta \left(  \sigma,\gamma \right)  .$

Similar in spirit to the single-mode case, operating $F_{2}\left(  r,s\right)
$ on entangled state representation $\left \vert \eta \right \rangle $ we see%

\begin{align}
F_{2}\left(  r,s\right)  \left \vert \eta \right \rangle  &  =\frac{1}{s^{\ast}%
}\int \frac{d^{2}z_{1}d^{2}z_{2}}{\pi^{2}}\exp \left[  \frac{r}{s^{\ast}}%
a_{1}^{\dagger}a_{2}^{\dagger}+\left(  \frac{1}{s^{\ast}}-1\right)  \left(
a_{1}^{\dagger}z_{1}+a_{2}^{\dagger}z_{2}\right)  -\frac{r^{\ast}}{s^{\ast}%
}z_{1}z_{2}\right]  \left \vert z_{1},z_{2}\right \rangle \left \langle
z_{1},z_{2}\right \vert \left.  \eta \right \rangle \nonumber \\
&  =\frac{1}{s^{\ast}}\int \frac{d^{2}z_{1}d^{2}z_{2}}{\pi^{2}}\exp \left[
-\left \vert z_{1}\right \vert ^{2}+\frac{1}{s^{\ast}}\left(  a_{1}^{\dagger
}-r^{\ast}z_{2}\right)  z_{1}+\left(  \eta+z_{2}^{\ast}\right)  z_{1}^{\ast
}\right] \nonumber \\
&  \times \exp \left[  -\frac{1}{2}\left \vert \eta \right \vert ^{2}-\left \vert
z_{2}\right \vert ^{2}+\frac{1}{s^{\ast}}z_{2}a_{2}^{\dagger}-\eta^{\ast}%
z_{2}^{\ast}+\frac{r}{s^{\ast}}a_{1}^{\dagger}a_{2}^{\dagger}\right]
\left \vert 00\right \rangle \nonumber \\
&  =\frac{1}{s^{\ast}}\int \frac{d^{2}z_{2}}{\pi}\exp \left[  -\frac{s^{\ast
}+r^{\ast}}{s^{\ast}}\left \vert z_{2}\right \vert ^{2}+\frac{1}{s^{\ast}%
}\left(  a_{2}^{\dagger}-\eta r^{\ast}\right)  z_{2}+\frac{1}{s^{\ast}}\left(
a_{1}^{\dagger}-s^{\ast}\eta^{\ast}\right)  z_{2}^{\ast}\right] \nonumber \\
&  \times \exp \left[  +\frac{\eta}{s^{\ast}}a_{1}^{\dagger}+\frac{r}{s^{\ast}%
}a_{1}^{\dagger}a_{2}^{\dagger}-\frac{1}{2}\left \vert \eta \right \vert
^{2}\right]  \left \vert 00\right \rangle \nonumber \\
&  =\frac{1}{s^{\ast}+r^{\ast}}\exp \left \{  -\allowbreak \frac{s^{\ast}%
-r^{\ast}}{2\left(  s^{\ast}+r^{\ast}\right)  }\left \vert \eta \right \vert
^{2}+\allowbreak \frac{\eta a_{1}^{\dagger}}{s^{\ast}+r^{\ast}}\allowbreak
-\allowbreak \frac{\eta^{\ast}a_{2}^{\dagger}}{s^{\ast}+r^{\ast}}+\frac
{s+r}{s^{\ast}+r^{\ast}}\allowbreak a_{1}^{\dagger}a_{2}^{\dagger}\right \}
\left \vert 00\right \rangle \equiv \left \vert \eta \right \rangle _{s,r},
\label{22}%
\end{align}
or
\begin{equation}
\left \vert \eta \right \rangle _{s,r}=\frac{1}{\allowbreak D+iB}\exp \left \{
-\frac{\allowbreak A-iC}{2\left(  \allowbreak D+iB\right)  }\left \vert
\eta \right \vert ^{2}+\frac{\eta a_{1}^{\dagger}}{\allowbreak D+iB}-\frac
{\eta^{\ast}a_{2}^{\dagger}}{\allowbreak D+iB}+\frac{\allowbreak
D-iB}{\allowbreak D+iB}a_{1}^{\dagger}a_{2}^{\dagger}\right \}  \left \vert
00\right \rangle , \label{23}%
\end{equation}
where we have used the integration formula%
\begin{equation}
\int \frac{d^{2}z}{\pi}\exp \left(  \zeta \left \vert z\right \vert ^{2}+\xi z+\eta
z^{\ast}\right)  =-\frac{1}{\zeta}e^{-\frac{\xi \eta}{\zeta}},\text{Re}\left(
\zeta \right)  <0. \label{24}%
\end{equation}
Noticing the completeness relation and the orthogonality of $\left \vert
\eta \right \rangle $ we immediately derive
\begin{equation}
\int \frac{d^{2}\eta}{\pi}\left \vert \eta \right \rangle _{s,rs,r}\left \langle
\eta \right \vert =1,\text{ }_{s,r}\left \langle \eta \right \vert \left.
\eta^{\prime}\right \rangle _{s,r}=\pi \delta \left(  \eta-\eta^{\prime}\right)
\delta \left(  \eta^{\ast}-\eta^{\prime \ast}\right)  , \label{25}%
\end{equation}
a generalized entangled state representation $\left \vert \eta \right \rangle
_{s,r}$ with the completeness relation (\ref{25}). From (\ref{23}) we can see
that%
\begin{align}
a_{1}\left \vert \eta \right \rangle _{s,r}  &  =\left(  \frac{\eta}{\allowbreak
D+iB}+\frac{\allowbreak D-iB}{\allowbreak D+iB}a_{2}^{\dagger}\right)
\left \vert \eta \right \rangle _{s,r},\label{26}\\
a_{2}\left \vert \eta \right \rangle _{s,r}  &  =\left(  -\frac{\eta^{\ast}%
}{\allowbreak D+iB}+\frac{\allowbreak D-iB}{\allowbreak D+iB}a_{1}^{\dagger
}\right)  \left \vert \eta \right \rangle _{s,r}, \label{27}%
\end{align}
so we have the eigen-equations for $\left \vert \eta \right \rangle _{s,r}$ as
follows%
\begin{align}
\left[  D\left(  Q_{1}-Q_{2}\right)  -B\left(  P_{1}-P_{2}\right)  \right]
\left \vert \eta \right \rangle _{s,r}  &  =\sqrt{2}\eta_{1}\left \vert
\eta \right \rangle _{s,r},\text{ }\label{28}\\
\left[  B\left(  Q_{1}+Q_{2}\right)  +D\left(  P_{1}+P_{2}\right)  \right]
\left \vert \eta \right \rangle _{s,r}  &  =\sqrt{2}\eta_{2}\left \vert
\eta \right \rangle _{s,r}, \label{29}%
\end{align}
We can also check Eqs.(\ref{26})-(\ref{29}) by another way.

\subsubsection{$\left \vert \eta \right \rangle _{s,r\text{ }s,r}\left \langle
\eta \right \vert $ as Radon transform of the entangled Wigner operator}

For two-mode correlated system, we have introduced the Wigner operator in
(\ref{3.27}). According to the Wely correspondence rule \cite{Weyl}%
\begin{equation}
H\left(  a_{1}^{\dagger},a_{2}^{\dagger};a_{1},a_{2}\right)  =\int d^{2}\gamma
d^{2}\sigma h\left(  \sigma,\gamma \right)  \Delta \left(  \sigma,\gamma \right)
, \label{31}%
\end{equation}
where $h\left(  \sigma,\gamma \right)  $ is the Weyl correspondence of
$H\left(  a_{1}^{\dagger},a_{2}^{\dagger};a_{1},a_{2}\right)  ,$ and
\begin{equation}
h\left(  \sigma,\gamma \right)  =4\pi^{2}\mathtt{Tr}\left[  H\left(
a_{1}^{\dagger},a_{2}^{\dagger};a_{1},a_{2}\right)  \Delta \left(
\sigma,\gamma \right)  \right]  , \label{32}%
\end{equation}
the classical Weyl correspondence of the projection operator $\left \vert
\eta \right \rangle _{r,sr,s}\left \langle \eta \right \vert $ can be calculated
as
\begin{align}
&  4\pi^{2}\mathtt{Tr}\left[  \left \vert \eta \right \rangle _{r,sr,s}%
\left \langle \eta \right \vert \Delta \left(  \sigma,\gamma \right)  \right]
\nonumber \\
&  =4\pi^{2}\int \frac{d^{2}\eta^{\prime}}{\pi^{3}}\left.  _{r,s}\left \langle
\eta \right \vert \left.  \sigma-\eta^{\prime}\right \rangle \left \langle
\sigma+\eta^{\prime}\right \vert \left.  \eta \right \rangle _{r,s}\right.
\exp(\eta^{\prime}\gamma^{\ast}-\eta^{\prime \ast}\gamma)\nonumber \\
&  =4\pi^{2}\int \frac{d^{2}\eta^{\prime}}{\pi^{3}}\left \langle \eta \right \vert
F_{2}^{\dagger}\left \vert \sigma-\eta^{\prime}\right \rangle \left \langle
\sigma+\eta^{\prime}\right \vert F_{2}\left \vert \eta \right \rangle \exp
(\eta^{\prime}\gamma^{\ast}-\eta^{\prime \ast}\gamma). \label{33}%
\end{align}
Then using Eq.(\ref{11.7}) we have%
\begin{equation}
4\pi^{2}\mathtt{Tr}\left[  \left \vert \eta \right \rangle _{s,rs,r}\left \langle
\eta \right \vert \Delta \left(  \sigma,\gamma \right)  \right]  =\pi \delta \left(
\eta_{2}-D\sigma_{2}+B\gamma_{1}\right)  \delta \left(  \eta_{1}-D\sigma
_{1}-B\gamma_{2}\right)  , \label{34}%
\end{equation}
which means the following Weyl correspondence
\begin{equation}
\left \vert \eta \right \rangle _{s,rs,r}\left \langle \eta \right \vert =\pi \int
d^{2}\gamma d^{2}\sigma \delta \left(  \eta_{2}-D\sigma_{2}+B\gamma_{1}\right)
\delta \left(  \eta_{1}-D\sigma_{1}-B\gamma_{2}\right)  \Delta \left(
\sigma,\gamma \right)  , \label{35}%
\end{equation}
so the projector operator $\left \vert \eta \right \rangle _{s,rs,r}\left \langle
\eta \right \vert $ is just the Radon transformation of $\Delta \left(
\sigma,\gamma \right)  $, $D$ and $B$ are the Radon transformation parameter.
Combining Eqs. (\ref{22})-(\ref{35}) together we complete the proof (\ref{9}).
Therefore, the quantum tomography in two-mode entangled case is expressed as
\begin{equation}
|_{s,r}\left \langle \eta \right \vert \left.  \psi \right \rangle |^{2}%
=|\left \langle \eta \right \vert F^{\dagger}\left \vert \psi \right \rangle
|^{2}=\pi \int d^{2}\gamma d^{2}\sigma \delta \left(  \eta_{2}-D\sigma
_{2}+B\gamma_{1}\right)  \delta \left(  \eta_{1}-D\sigma_{1}-B\gamma
_{2}\right)  \left \langle \psi \right \vert \Delta \left(  \sigma,\gamma \right)
\left \vert \psi \right \rangle . \label{36}%
\end{equation}
where $\left \langle \psi \right \vert \Delta \left(  \sigma,\gamma \right)
\left \vert \psi \right \rangle $ is the Wigner function. So the probability
distribution for the Fresnel quadrature phase is the tomography (Radon
transform of the two-mode Wigner function).{\small \ }This new relation
between quantum tomography and optical Fresnel transform may provide
experimentalists to figure out new approach for generating tomography.

Next we turn to the \textquotedblleft frequency\textquotedblright \ domain,
that is to say, we shall prove that the $(A,C)$ related Radon transform of
entangled Wigner operator $\Delta \left(  \sigma,\gamma \right)  $ is just the
pure state density operator $\left \vert \xi \right \rangle _{s,rs,r}\left \langle
\xi \right \vert ,$ i.e.,
\begin{equation}
F_{2}\left \vert \xi \right \rangle \left \langle \xi \right \vert F_{2}^{\dagger
}=\left \vert \xi \right \rangle _{s,rs,r}\left \langle \xi \right \vert =\pi
\int \delta \left(  \xi_{1}-A\sigma_{1}-C\gamma_{2}\right)  \delta \left(
\xi_{2}-A\sigma_{2}+C\gamma_{1}\right)  \Delta \left(  \sigma,\gamma \right)
d^{2}\sigma d^{2}\gamma, \label{37}%
\end{equation}
where $\left \vert \xi \right \rangle $ is the conjugated entangled state to
$\left \vert \eta \right \rangle $.

By analogy with the above procedure, we obtain the 2-dimensional Fresnel
transformation in its `frequency domain', i.e.,
\begin{align}
\mathcal{K}_{2}^{N}\left(  \xi^{\prime},\xi \right)   &  \equiv \frac{1}{\pi
}\left \langle \xi^{\prime}\right \vert F_{2}\left(  r,s\right)  \left \vert
\xi \right \rangle \nonumber \\
&  =\int \frac{d^{2}\eta d^{2}\sigma}{\pi^{2}}\left \langle \xi^{\prime
}\right \vert \left.  \eta^{\prime}\right \rangle \left \langle \eta^{\prime
}\right \vert F_{2}\left(  r,s\right)  \left \vert \eta \right \rangle
\left \langle \eta \right \vert \left.  \xi \right \rangle \nonumber \\
&  =\frac{1}{8iB\pi}\int \frac{d^{2}\sigma d^{2}\eta}{\pi^{2}}\exp \left(
\frac{\xi^{\prime \ast}\eta^{\prime}-\xi^{\prime}\eta^{\prime \ast}+\xi
\eta^{\ast}-\xi^{\ast}\eta}{2}\right)  \mathcal{K}_{2}^{\left(  \mathtt{r}%
,s\right)  }\left(  \sigma,\eta \right) \nonumber \\
&  =\frac{1}{2i\left(  -C\right)  \pi}\exp \left[  \frac{i}{2\left(  -C\right)
}\left(  D\left \vert \xi \right \vert ^{2}+A\left \vert \xi^{\prime}\right \vert
^{2}-\xi^{\prime \ast}\xi-\xi^{\prime}\xi^{\ast}\right)  \right]  , \label{39}%
\end{align}
where the superscript $N$ means that this transform kernel corresponds to the
parameter matrix $N=\left[  D,-C,-B,A\right]  $. Thus the 2D Fresnel
transformation in its `frequency domain' is given by%
\begin{equation}
\Psi \left(  \xi^{\prime}\right)  =\int \mathcal{K}_{2}^{N}\left(  \xi^{\prime
},\xi \right)  \Phi \left(  \xi \right)  d^{2}\xi. \label{40}%
\end{equation}
Operating $F_{2}\left(  r,s\right)  $ on $\left \vert \xi \right \rangle $ we
also have%
\begin{equation}
\left \vert \xi \right \rangle _{s,r}=\frac{1}{\allowbreak \allowbreak A-iC}%
\exp \left \{  -\frac{D+iB}{2\left(  \allowbreak A-iC\right)  }\left \vert
\eta \right \vert ^{2}+\frac{\xi a_{1}^{\dagger}}{A-iC}+\frac{\xi^{\ast}%
a_{2}^{\dagger}}{\allowbreak A-iC}-\frac{\allowbreak A+iC}{\allowbreak
A-iC}a_{1}^{\dagger}a_{2}^{\dagger}\right \}  \left \vert 00\right \rangle ,
\label{41}%
\end{equation}
or
\begin{equation}
\left \vert \xi \right \rangle _{s,r}=\frac{1}{s^{\ast}-r^{\ast}}\exp \left \{
-\allowbreak \frac{s^{\ast}+r^{\ast}}{2\left(  s^{\ast}-r^{\ast}\right)
}\left \vert \xi \right \vert ^{2}+\allowbreak \frac{\xi a_{1}^{\dagger}}{s^{\ast
}-r^{\ast}}\allowbreak+\allowbreak \frac{\xi^{\ast}a_{2}^{\dagger}}{s^{\ast
}-r^{\ast}}-\frac{s-r}{s^{\ast}-r^{\ast}}\allowbreak a_{1}^{\dagger}%
a_{2}^{\dagger}\right \}  \left \vert 00\right \rangle . \label{42}%
\end{equation}
Noticing that the entangled Wigner operator in $\left \langle \xi \right \vert $
representation is expressed as
\begin{equation}
\Delta \left(  \sigma,\gamma \right)  =\int \frac{d^{2}\xi}{\pi^{3}}\left \vert
\gamma+\xi \right \rangle \left \langle \gamma-\xi \right \vert \exp(\xi^{\ast
}\sigma-\sigma^{\ast}\xi), \label{43}%
\end{equation}
and using the classical correspondence of $\left \vert \xi \right \rangle
_{s,rs,r}\left \langle \xi \right \vert $ which is calculated by
\begin{align}
h(\sigma,\gamma)  &  =4\pi^{2}\mathtt{Tr}\left[  \left \vert \xi \right \rangle
_{s,r\text{ }s,r}\left \langle \xi \right \vert \Delta \left(  \sigma
,\gamma \right)  \right] \nonumber \\
\  &  =4\int \frac{d^{2}\xi}{\pi}\left \langle \gamma-\xi \right \vert
F_{2}\left \vert \xi \right \rangle \left \langle \xi \right \vert F_{2}^{\dag
}|\gamma+\xi \rangle \exp(\xi^{\ast}\sigma-\sigma^{\ast}\xi)\nonumber \\
&  =\pi \delta \left(  \xi_{1}-A\sigma_{1}-C\gamma_{2}\right)  \delta \left(
\xi_{2}-A\sigma_{2}+C\gamma_{1}\right)  , \label{44}%
\end{align}
we obtain
\begin{equation}
\left \vert \xi \right \rangle _{s,r\text{ }s,r}\left \langle \xi \right \vert
=\pi \int \delta \left(  \xi_{1}-A\sigma_{1}-C\gamma_{2}\right)  \delta \left(
\xi_{2}-A\sigma_{2}+C\gamma_{1}\right)  \Delta \left(  \sigma,\gamma \right)
d^{2}\sigma d^{2}\gamma, \label{45}%
\end{equation}
so the projector operator $\left \vert \xi \right \rangle _{s,r\text{ }%
s,r}\left \langle \xi \right \vert $ is another Radon transformation of the
two-mode Wigner operator, with $A$ and $C$ being the Radon transformation
parameter (`frequency' domain). Therefore, the quantum tomography in
$_{s,r}\left \langle \xi \right \vert $ representation is expressed as the Radon
transformation of the Wigner function%
\begin{equation}
|\left \langle \xi \right \vert F^{\dagger}\left \vert \psi \right \rangle
|^{2}=|_{s,r}\left \langle \xi \right \vert \left.  \psi \right \rangle |^{2}%
=\pi \int d^{2}\gamma d^{2}\sigma \delta \left(  \xi_{1}-A\sigma_{1}-C\gamma
_{2}\right)  \delta \left(  \xi_{2}-A\sigma_{2}+C\gamma_{1}\right)
\left \langle \psi \right \vert \Delta \left(  \sigma,\gamma \right)  \left \vert
\psi \right \rangle , \label{46}%
\end{equation}
and $_{s,r}\left \langle \xi \right \vert =\left \langle \xi \right \vert
F^{\dagger}.$

\subsubsection{Inverse Radon transformation}

Now we consider the inverse Radon transformation. For instance, using
Eq.(\ref{35}) we see the Fourier transformation of $\left \vert \eta
\right \rangle _{s,rs,r}\left \langle \eta \right \vert $ is%
\begin{align}
&  \int d^{2}\eta \left \vert \eta \right \rangle _{s,rs,r}\left \langle
\eta \right \vert \exp(-i\zeta_{1}\eta_{1}-i\zeta_{2}\eta_{2})\nonumber \\
&  =\pi \int d^{2}\gamma d^{2}\sigma \Delta \left(  \sigma,\gamma \right)
\exp \left[  -i\zeta_{1}\left(  D\sigma_{1}+B\gamma_{2}\right)  -i\zeta
_{2}\left(  D\sigma_{2}-B\gamma_{1}\right)  \right]  , \label{47}%
\end{align}
the right-hand side of (\ref{47}) can be regarded as a special Fourier
transformation of $\Delta \left(  \sigma,\gamma \right)  $, so by making its
inverse Fourier transformation, we get%
\begin{align}
\Delta \left(  \sigma,\gamma \right)   &  =\frac{1}{(2\pi)^{4}}\int_{-\infty
}^{\infty}dr_{1}\left \vert r_{1}\right \vert \int_{-\infty}^{\infty}%
dr_{2}\left \vert r_{2}\right \vert \int_{0}^{\pi}d\theta_{1}d\theta
_{2}\nonumber \\
&  \times \int_{-\infty}^{\infty}\frac{d^{2}\eta}{\pi}\left \vert \eta
\right \rangle _{s,rs,r}\left \langle \eta \right \vert K\left(  r_{1}%
,r_{2},\theta_{1},\theta_{2}\right)  , \label{48}%
\end{align}
where $\cos \theta_{1}=\cos \theta_{2}=\frac{D}{\sqrt{B^{2}+D^{2}}},r_{1}%
=\zeta_{1}\sqrt{B^{2}+D^{2}},r_{2}=\zeta_{2}\sqrt{B^{2}+D^{2}}$ and
\begin{align}
K\left(  r_{1},r_{2},\theta_{1},\theta_{2}\right)   &  \equiv \exp \left[
-ir_{1}\left(  \frac{\eta_{1}}{\sqrt{B^{2}+D^{2}}}-\sigma_{1}\cos \theta
_{1}-\gamma_{2}\sin \theta_{1}\right)  \right] \nonumber \\
&  \times \exp \left[  -ir_{2}\left(  \frac{\eta_{2}}{\sqrt{B^{2}+D^{2}}}%
-\sigma_{2}\cos \theta_{2}+\gamma_{1}\sin \theta_{2}\right)  \right]  .
\label{49}%
\end{align}
Eq.(\ref{48}) is just the inverse Radon transformation of entangled Wigner
operator in the entangled state representation. This is different from the two
independent Radon transformations' direct product of the two independent
single-mode Wigner operators, because in (\ref{23}) the $\left \vert
\eta \right \rangle _{s,r}$ is an entangled state. Therefore the Wigner function
of quantum state $\left \vert \psi \right \rangle $ can be reconstructed from the
tomographic inversion of a set of measured probability distributions
$\left \vert _{s,r}\left \langle \eta \right.  \left \vert \psi \right \rangle
\right \vert ^{2}$, i.e.,%
\begin{align}
W_{\psi}  &  =\frac{1}{(2\pi)^{4}}\int_{-\infty}^{\infty}dr_{1}\left \vert
r_{1}\right \vert \int_{-\infty}^{\infty}dr_{2}\left \vert r_{2}\right \vert
\int_{0}^{\pi}d\theta_{1}d\theta_{2}\nonumber \\
&  \times \int_{-\infty}^{\infty}\frac{d^{2}\eta}{\pi}\left \vert _{s,r}%
\left \langle \eta \right.  \left \vert \psi \right \rangle \right \vert
^{2}K\left(  r_{1},r_{2},\theta_{1},\theta_{2}\right)  . \label{50}%
\end{align}

Thus, based on the previous section, we have further extended the relation
connecting optical Fresnel transformation with quantum tomography to the
entangled case{\small .} The tomography representation $_{s,r}\left \langle
\eta \right \vert =\left \langle \eta \right \vert F_{2}^{\dagger}$ is set up,
based on which the tomogram of quantum state $\left \vert \psi \right \rangle $
is just the squared modulus of the wave function $_{s,r}\left \langle
\eta \right \vert \left.  \psi \right \rangle .$ i.e. the probability distribution
for the Fresnel quadrature phase is the tomogram (Radon transform of the
Wigner function).

\section{Fractional Fourier Transformation (FrFT) for 1-D case}

The fractional Fourier transform (FrFT) has been shown to be a very useful
tool in Fourier optics and information optics. The concept of FrFT was firstly
introduced mathematically in 1980 by Namias \cite{Namias} as a mathematical
tool for solving theoretical physical problems \cite{f0}, but did not brought
enough attention until Mendlovic and Ozaktas \cite{Mendlovic,Ozaktas} defined
the $\alpha$-th FrFT physically, based on propagation in quadratic
graded-index media (GRIN media with medium parameters $n(r)=n_{1}-n_{2}%
r^{2}/2$). Since then a lot of works have been done on its properties, optical
implementations and applications \cite{f1,f2,f3,f4}.

\subsection{Quantum version of FrFT}

The FrFT of $\theta$-order is defined in a manner, i.e.,
\begin{equation}
\mathcal{F}_{\theta}\left[  f\left(  x\right)  \right]  =\sqrt{\frac
{e^{i\left(  \frac{\pi}{2}-\theta \right)  }}{2\pi \sin \theta}}\int_{-\infty
}^{\infty}\exp \left \{  -i\frac{x^{2}+y^{2}}{2\tan \theta}+\frac{ixx^{\prime}%
}{\sin \theta}\right \}  f\left(  x\right)  dx, \label{12.1}%
\end{equation}
where the exponential function is an integral kernel. In order to find the
quantum correspondence of FrFT, multiplying the function $\exp \left \{
-i\frac{x^{2}+y^{2}}{2\tan \theta}+\frac{ixx^{\prime}}{\sin \theta}\right \}  $
by the ket $\int dy\left \vert y\right \rangle $ and bra $\int dx\left \langle
x\right \vert $ from left and right, respectively, where $\left \vert
y\right \rangle $ and $\left \vert x\right \rangle $ are coordinate eigenvectors,
$X\left \vert x\right \rangle =x\left \vert x\right \rangle $, and then using
(\ref{3.7}) and the IWOP technique to perform the integration, we obtain%
\begin{align}
&  \int_{-\infty}^{\infty}dxdy\left \vert y\right \rangle \exp \left \{
-i\frac{x^{2}+y^{2}}{2\tan \theta}+\frac{ixy}{\sin \theta}\right \}  \left \langle
x\right \vert \nonumber \\
&  =\sqrt{-2\pi i\sin \theta e^{i\theta}}\colon \exp \left \{  \left(  e^{i\theta
}-1\right)  a^{\dagger}a\right \}  \colon \nonumber \\
&  =\sqrt{-2\pi i\sin \theta e^{i\theta}}\exp \left \{  i\theta a^{\dagger
}a\right \}  , \label{12.2}%
\end{align}
where we have used the operator identity in the last step of Eq.(\ref{12.2})%

\begin{equation}
\exp \left \{  fa^{\dagger}a\right \}  =\colon \exp \left \{  \left(  e^{f}%
-1\right)  a^{\dagger}a\right \}  \colon. \label{12.3}%
\end{equation}
From the orthogonal relation $\left \langle x^{\prime}\right.  \left \vert
x\right \rangle =\delta \left(  x-x^{\prime}\right)  ,$ we know that
Eq.(\ref{12.2}) indicates
\begin{equation}
\sqrt{\frac{e^{i\left(  \frac{\pi}{2}-\theta \right)  }}{2\pi \sin \theta}}%
\exp \left \{  -i\frac{x^{2}+y^{2}}{2\tan \theta}+\frac{ixy}{\sin \theta}\right \}
=\left \langle y\right \vert e^{i\theta a^{\dagger}a}\left \vert x\right \rangle ,
\label{12.4}%
\end{equation}
which implies that the integral kernel in Eq.(\ref{12.1}) is just the matrix
element of operator $\exp \left \{  i\theta a^{\dagger}a\right \}  $\ in
coordinate state ($\exp \left \{  i\theta a^{\dagger}a\right \}  $ called as
Fractional Fourier Operator \cite{r15}. Therefore, if we consider $f\left(
x\right)  $ as $\left \langle x\right.  \left \vert f\right \rangle $, the wave
function of quantum state $\left \vert f\right \rangle $ in the coordinate
representation, from Eqs. (\ref{12.1}) and (\ref{12.4}) it then follows
\begin{equation}
\mathcal{F}_{\theta}\left[  f\left(  x\right)  \right]  =\int_{-\infty
}^{\infty}dx\left \langle y\right \vert e^{i\theta a^{\dagger}a}\left \vert
x\right \rangle f\left(  x\right)  =\left \langle y\right \vert e^{i\theta
a^{\dagger}a}\left \vert f\right \rangle \equiv g\left(  y\right)  ,
\label{12.5}%
\end{equation}
which suggests
\begin{equation}
\left \vert g\right \rangle =e^{i\theta a^{\dagger}a}\left \vert f\right \rangle .
\label{12.6}%
\end{equation}
From Eqs.(\ref{12.5})\ and (\ref{12.1}) one can see that the FrFT in
Eq.(\ref{12.1}) corresponds actually to the rotating operator $\left(
e^{i\theta a^{\dagger}a}\right)  $ transform in Eq.(\ref{12.5}) between two
quantum states, which is just the quantum version of FrFT.

In fact, using quantum version of FrFT, one can directly derive various
properties of the FrFTs. An important feature of the FrFT is that they are
composed according to $\mathcal{F}_{\theta^{\prime}}\mathcal{F}_{\theta
}=\mathcal{F}_{\theta^{\prime}+\theta}$ (the additivity property). Without
losing generality, we examine%
\begin{equation}
\mathcal{F}_{\theta+\theta^{\prime}}\left[  f\left(  x\right)  \right]
\equiv \int_{-\infty}^{\infty}\frac{d^{2}\eta}{\pi}\left \langle y\right \vert
e^{i\left(  \theta+\theta^{\prime}\right)  a^{\dagger}a}\left \vert
x\right \rangle f\left(  x\right)  . \label{12.7}%
\end{equation}
According to the completeness relation of coordinate eigenvector,
$\int_{-\infty}^{\infty}dx^{\prime}\left \vert x^{\prime}\right \rangle
\left \langle x^{\prime}\right \vert =1,$ Eq.(\ref{12.7}) yields%
\begin{align}
\mathcal{F}_{\theta+\theta^{\prime}}\left[  f\left(  x\right)  \right]   &
=\int_{-\infty}^{\infty}dx\left \langle y\right \vert e^{i\theta a^{\dagger}%
a}e^{i\theta^{\prime}a^{\dagger}a}\left \vert x\right \rangle f\left(  x\right)
\nonumber \\
&  =\int_{-\infty}^{\infty}dx^{\prime}\left \langle y\right \vert e^{i\theta
a^{\dagger}a}\left \vert x^{\prime}\right \rangle \int_{-\infty}^{\infty
}dx\left \langle x^{\prime}\right \vert e^{i\theta^{\prime}a^{\dagger}%
a}\left \vert x\right \rangle f\left(  x\right) \nonumber \\
&  =\int_{-\infty}^{\infty}dx^{\prime}\left \langle y\right \vert e^{i\theta
a^{\dagger}a}\left \vert x^{\prime}\right \rangle \mathcal{F}_{\theta}\left[
f\left(  x\right)  \right]  =\mathcal{F}_{\theta}\mathcal{F}_{\theta^{\prime}%
}\left[  f\left(  x\right)  \right]  , \label{12.8}%
\end{align}
which is just the additivity of FrFT.

In particular, when $\left \vert f\right \rangle $ is the number state,
$\left \vert f\right \rangle =\left \vert n\right \rangle =\frac{a^{\dagger n}%
}{\sqrt{n!}}\left \vert 0\right \rangle $, its wavefunction in coordinate
representation is
\begin{equation}
f\left(  x\right)  =\left \langle x\right.  \left \vert n\right \rangle =\frac
{1}{\sqrt{2^{n}n!\sqrt{\pi}}}e^{-x^{2}/2}H_{n}(x), \label{12.9}%
\end{equation}
the FrFT of $\left \langle x\right.  \left \vert n\right \rangle $ is
\begin{equation}
\mathcal{F}_{\theta}\left[  \left \langle x\right.  \left \vert n\right \rangle
\right]  =\left \langle y\right \vert e^{i\theta a^{\dagger}a}\left \vert
n\right \rangle =e^{in\theta}\left \langle y\right.  \left \vert n\right \rangle ,
\label{12.10}%
\end{equation}
or
\begin{equation}
\mathcal{F}_{\theta}\left[  e^{-x^{2}/2}H_{n}(x)\right]  =e^{in\theta
}e^{-y^{2}/2}H_{n}(y), \label{12.11}%
\end{equation}
which indicates that the eigenfunction is Hermite-Gaussian function with the
corresponding eigenvalue being $e^{in\theta}$.

\subsection{On the Scaled FrFT Operator}

In studying various optical transformations the optical operator method is
proposed \cite{r16} as mapping of ray-transfer ABCD matrix, such that the ray
transfer through optical instruments and the diffraction can be discussed by
virtue of the commutative relations of operators and the matrix algebra. The
square phase operators, scaling operator, Fourier transform operator and the
propagation operator in free space have been proposed in the literature, two
important questions thus naturally arise: 1. what is the scaled FrFT (SFrFT)
operator which corresponds to the SFrFT's integration kernel \cite{Namias}%
\begin{equation}
\frac{1}{\sqrt{2\pi if_{e}\sin \phi}}\exp \left \{  \frac{i\left(  x^{2}%
+x^{\prime2}\right)  }{2f_{e}\tan \phi}-\frac{ix^{\prime}x}{f_{e}\sin \phi
}\right \}  , \label{12.12}%
\end{equation}
where $f_{e}$ is standard focal length (or a scaled parameter); 2. If this
operator is found, can it be further decomposed into simpler operators and
what are their physical meaning? Since SFrFT has wide application in optical
information detection and can be implemented even by using a thick lens
\cite{r17},\ so our questions are worth of paying attention \cite{FHCPL}.
\begin{figure}[ptb]
\label{Fig2}
\centering \includegraphics[width=1.5cm]{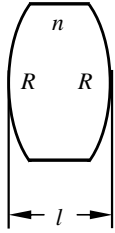}\caption{{\protect \small A
thick lens as a kind of fractional Fourier transform device. }}%
\end{figure}

Let start with a thick lens (shown in Fig.1) which represents a transfer
matrix \cite{r17}
\begin{equation}
\left(
\begin{array}
[c]{cc}%
\mathcal{A} & \mathcal{B}\\
\mathcal{C} & \mathcal{D}%
\end{array}
\right)  =\left(
\begin{array}
[c]{cc}%
1-\frac{\left(  1-1/n\right)  l}{R_{1}} & \frac{l}{n}\\
-[\left(  n-1\right)  \frac{R_{1}+R_{2}}{R_{1}R_{2}}-\frac{l\left(
n-1\right)  ^{2}}{nR_{1}R_{2}}] & 1-\frac{\left(  1-1/n\right)  l}{R_{2}}%
\end{array}
\right)  , \label{12.13}%
\end{equation}
where $n$ is the reflective index; $l$ is the thickness of thick lens; $R_{1}$
and $R_{2}$ denotes the curvature radius of the two surfaces of the lens,
respectively. When we choose $R_{1}=R_{2}=R,$ then Eq.(\ref{12.13}) reduces to%
\begin{equation}
\left(
\begin{array}
[c]{cc}%
\mathcal{A} & \mathcal{B}\\
\mathcal{C} & \mathcal{D}%
\end{array}
\right)  =\left(
\begin{array}
[c]{cc}%
1-\frac{\left(  1-1/n\right)  l}{R} & \frac{l}{n}\\
-[\left(  n-1\right)  \frac{2}{R}-\frac{l\left(  n-1\right)  ^{2}}{nR^{2}}] &
1-\frac{\left(  1-1/n\right)  l}{R}%
\end{array}
\right)  . \label{12.14}%
\end{equation}
By defining $1-\frac{\left(  1-1/n\right)  l}{R}=\cos \phi,$ $\frac{l}{n}%
=f_{e}\sin \phi,$ and $\frac{l}{R}=\frac{n\left(  1-\cos \phi \right)  }{n-1},$
$l=nf_{e}\sin \phi,$ we can recast (\ref{12.14}) into the simple form%
\begin{equation}
\left(
\begin{array}
[c]{cc}%
\mathcal{A} & \mathcal{B}\\
\mathcal{C} & \mathcal{D}%
\end{array}
\right)  =\left(
\begin{array}
[c]{cc}%
\cos \phi & f_{e}\sin \phi \\
-\sin \phi/f_{e} & \cos \phi
\end{array}
\right)  ,\text{ }\det \left(
\begin{array}
[c]{cc}%
\mathcal{A} & \mathcal{B}\\
\mathcal{C} & \mathcal{D}%
\end{array}
\right)  =1. \label{12.15}%
\end{equation}
According to (\ref{8.4}) we immediately know that the operator of SFrFT is%
\begin{align}
F_{1}\left(  \mathcal{A},\mathcal{B},\mathcal{C}\right)   &  =\exp \left \{
\frac{i\left(  f_{e}-1/f_{e}\right)  \tan \phi}{2V}a^{\dagger2}\right \}
\nonumber \\
&  \times \exp \left \{  \left(  a^{\dagger}a+\frac{1}{2}\right)  \ln \left(
\frac{2\sec \phi}{V}\right)  \right \} \nonumber \\
&  \times \exp \left \{  \frac{i\left(  f_{e}-1/f_{e}\right)  \tan \phi}{2V}%
a^{2}\right \}  ,\text{ \ }\label{12.16}\\
\text{ }V  &  =\left[  2+i\left(  f_{e}+1/f_{e}\right)  \tan \phi \right]
.\nonumber
\end{align}
Noting that the matrix $\left(
\begin{array}
[c]{cc}%
\mathcal{A} & \mathcal{B}\\
\mathcal{C} & \mathcal{D}%
\end{array}
\right)  $ can be decomposed into%
\begin{equation}
\left(
\begin{array}
[c]{cc}%
\mathcal{A} & \mathcal{B}\\
\mathcal{C} & \mathcal{D}%
\end{array}
\right)  =\left(
\begin{array}
[c]{cc}%
1 & 0\\
-\frac{1}{f_{e}}\tan \phi & 1
\end{array}
\right)  \left(
\begin{array}
[c]{cc}%
\cos \phi & 0\\
0 & \cos \phi
\end{array}
\right)  \left(
\begin{array}
[c]{cc}%
1 & f_{e}\tan \phi \\
0 & 1
\end{array}
\right)  , \label{12.17}%
\end{equation}
according to the previous section we have%
\begin{align}
F_{1}\left(  \mathcal{A},\mathcal{B},\mathcal{C}\right)   &  =F\left(
1,0,-\frac{1}{f_{e}}\tan \phi \right)  F\left(  \cos \phi,0,0\right)  F\left(
1,f_{e}\tan \phi,0\right) \nonumber \\
&  =\exp \left(  \frac{\tan \phi}{2if_{e}}Q^{2}\right)  \exp \left \{  -\frac
{i}{2}\left(  QP+PQ\right)  \ln \cos \phi \right \}  \exp \left(  \frac{f_{e}%
\tan \phi}{2i}P^{2}\right)  , \label{12.18}%
\end{align}
where $Q=\left(  a+a^{\dagger}\right)  /\sqrt{2},$ $P=\left(  a-a^{\dagger
}\right)  /\left(  \sqrt{2}i\right)  $ and $\exp \left(  -\frac{i\tan \phi
}{2f_{e}}Q^{2}\right)  ,$ $\exp \left \{  -\frac{i}{2}\left(  QP+PQ\right)
\ln \cos \phi \right \}  $ and $\exp \left(  -\frac{if_{e}\tan \phi}{2}P^{2}\right)
$ are the quadrature phase operator, the squeezing operator and the free
propagation operator, respectively. On the other hand, from $\left(
\begin{array}
[c]{cc}%
\mathcal{A} & \mathcal{B}\\
\mathcal{C} & \mathcal{D}%
\end{array}
\right)  ^{-1}=\left(
\begin{array}
[c]{cc}%
\mathcal{D} & \mathcal{-B}\\
\mathcal{-C} & \mathcal{A}%
\end{array}
\right)  ,$ and (\ref{12.18}) we see%
\begin{equation}
F_{1}^{-1}\left(  \mathcal{A},\mathcal{B},\mathcal{C}\right)  =\exp \left(
-\frac{i\mathcal{C}}{2\mathcal{D}}Q^{2}\right)  \exp \left(  -\frac{i}%
{2}\left(  QP+PQ\right)  \ln \mathcal{D}\right)  \exp \left(  \frac
{i\mathcal{B}}{2\mathcal{D}}P^{2}\right)  , \label{12.19}%
\end{equation}
it then follows
\begin{equation}
F_{1}\left(  \mathcal{A},\mathcal{B},\mathcal{C}\right)  =\exp \left(
\frac{f_{e}\tan \phi}{2i}P^{2}\right)  \exp \left(  \frac{i}{2}\left(
QP+PQ\right)  \ln \cos \phi \right)  \exp \left(  \frac{\tan \phi}{2if_{e}}%
Q^{2}\right)  . \label{12.20}%
\end{equation}
Using the canonical operator form (\ref{12.18}) or (\ref{12.20}) of
$F_{1}\left(  \mathcal{A},\mathcal{B},\mathcal{C}\right)  $ we can deduce its
matrix element in the coordinate states $\left \vert x\right \rangle $ (its
conjugate state is $\left \vert p\right \rangle $)
\begin{equation}
\left \langle x^{\prime}\right \vert F_{1}\left(  \mathcal{A},\mathcal{B}%
,\mathcal{C}\right)  \left \vert x\right \rangle =\frac{1}{\sqrt{2\pi if_{e}%
\sin \phi}}\exp \left \{  \left(  \frac{i\left(  x^{2}+x^{\prime2}\right)
}{2f_{e}\tan \phi}-\frac{ix^{\prime}x}{f_{e}\sin \phi}\right)  \right \}  ,
\label{12.21}%
\end{equation}
which is just the kernel of SFrFT, thus we name $F_{1}\left(  \mathcal{A}%
,\mathcal{B},\mathcal{C}\right)  $ SFrFT operator. Noticing that the
$Q^{2}/2,$ $P^{2}/2$ and $\frac{i}{4}\left(  QP+PQ\right)  $ construct a close
SU(2) Lie algebra, we can put Eq.(\ref{12.20}) into a more compact form, i.e.,%
\begin{equation}
F_{1}\left(  \mathcal{A},\mathcal{B},\mathcal{C}\right)  =\exp \left \{
-i\frac{\phi f_{e}}{2}\left(  P^{2}+\frac{Q^{2}}{f_{e}^{2}}\right)  \right \}
, \label{12.22}%
\end{equation}
Eqs. (\ref{12.18}), (\ref{12.20}) and (\ref{12.22}) are different forms of the
same operator of SFrFT. Especially, when $f_{e}=1,$ $F_{1}\left(
\mathcal{A},\mathcal{B},\mathcal{C}\right)  \rightarrow \exp \left \{  -i\phi
a^{\dagger}a\right \}  ,$ which is the usual FrFT operator.

Using (\ref{12.21}) the SFrFT of $f\left(  x\right)  =\left \langle
x\right \vert \left.  f\right \rangle ,$ denoted as $\mathcal{F}_{f_{e}}^{\phi
}\left[  f\left(  x\right)  \right]  ,$ can be expressed as an matrix element
in quantum optics context,
\begin{equation}
\mathcal{F}_{f_{e}}^{\phi}\left[  f\left(  x\right)  \right]  =\int
dx\left \langle x^{\prime}\right \vert F_{1}\left(  \mathcal{A},\mathcal{B}%
,\mathcal{C}\right)  \left \vert x\right \rangle \left \langle x\right \vert
\left.  f\right \rangle =\left \langle x^{\prime}\right \vert F_{1}\left(
\mathcal{A},\mathcal{B},\mathcal{C}\right)  \left \vert f\right \rangle .
\label{12.23}%
\end{equation}
The above discussions are useful since any unimodular matrix $\left(
\begin{array}
[c]{cc}%
A & B\\
C & D
\end{array}
\right)  $ can be decomposed into \cite{r18}%
\begin{equation}
\left(
\begin{array}
[c]{cc}%
A & B\\
C & D
\end{array}
\right)  =\left(
\begin{array}
[c]{cc}%
1 & 0\\
-\mathcal{P} & 1
\end{array}
\right)  \left(
\begin{array}
[c]{cc}%
m & 0\\
0 & m^{-1}%
\end{array}
\right)  \left(
\begin{array}
[c]{cc}%
\cos \phi & f_{e}\sin \phi \\
-\sin \phi/f_{e} & \cos \phi
\end{array}
\right)  , \label{12.24}%
\end{equation}
where the parameters $m,\mathcal{P},\phi$ are all real,
\begin{equation}
m^{2}=A^{2}+\frac{B^{2}}{f_{e}^{2}},\text{ }\tan \phi=\frac{B}{Af_{e}},\text{
}\mathcal{P}=-\frac{AC+DB/f_{e}^{2}}{A^{2}+\frac{B^{2}}{f_{e}^{2}}}.
\label{12.25}%
\end{equation}
Correspondingly, the operator of FrFT is given by%
\begin{align}
F_{1}\left(  A,B,C\right)   &  =F_{1}\left(  1,0,-\mathcal{P}\right)
F_{1}\left(  m,0,0\right)  F_{1}\left(  \mathcal{A},\mathcal{B},\mathcal{C}%
\right) \nonumber \\
&  =\exp \left(  -\frac{i}{2}\mathcal{P}Q^{2}\right)  \exp \left(  -\frac{i}%
{2}\left(  QP+PQ\right)  \ln m\right)  \exp \left \{  \frac{\phi f_{e}}%
{2i}\left(  P^{2}+\frac{Q^{2}}{f_{e}^{2}}\right)  \right \}  , \label{12.26}%
\end{align}
where $F_{1}\left(  1,0,-\mathcal{P}\right)  =\exp \left[  -\frac{i}%
{2}\mathcal{P}Q^{2}\right]  $ is the quadratic phase operator. Thus the
general Fresnel transform can always be expressed by SFrFT as follows
\begin{align}
g\left(  x^{\prime}\right)   &  =\int dy\left \langle x^{\prime}\right \vert
F_{1}\left(  1,0,-\mathcal{P}\right)  F_{1}\left(  m,0,0\right)  \left \vert
y\right \rangle \int dx\left \langle y\right \vert F_{1}\left(  \mathcal{A}%
,\mathcal{B},\mathcal{C}\right)  \left \vert x\right \rangle f\left(  x\right)
\nonumber \\
&  =\sqrt{m}\int dx^{\prime \prime}dy\left \langle x^{\prime}\right \vert
\exp \left(  -\frac{i}{2}\mathcal{P}X^{2}\right)  \left \vert mx^{\prime \prime
}\right \rangle \left \langle x^{\prime \prime}\right.  \left \vert y\right \rangle
\int dx\left \langle y\right \vert F_{1}\left(  \mathcal{A},\mathcal{B}%
,\mathcal{C}\right)  \left \vert x\right \rangle f\left(  x\right) \nonumber \\
&  =\exp \left(  -\frac{i}{2}\mathcal{P}x^{\prime2}\right)  \int \frac{dy}%
{\sqrt{m}}\delta \left(  \frac{x^{\prime}}{m}-y\right)  \int dx\left \langle
y\right \vert F_{1}\left(  \mathcal{A},\mathcal{B},\mathcal{C}\right)
\left \vert x\right \rangle f\left(  x\right) \nonumber \\
&  =\frac{1}{\sqrt{m}}\exp \left(  -\frac{i}{2}\mathcal{P}x^{\prime2}\right)
\int dx\left \langle \frac{x^{\prime}}{m}\right \vert F_{1}\left(
\mathcal{A},\mathcal{B},\mathcal{C}\right)  \left \vert x\right \rangle f\left(
x\right) \nonumber \\
&  =\frac{1}{\sqrt{m}}\exp \left(  -\frac{i}{2}\mathcal{P}x^{\prime2}\right)
\mathcal{F}_{f_{e}}^{\phi}\left[  f\right]  \left(  \frac{x^{\prime}}%
{m}\right)  . \label{12.27}%
\end{align}
i.e., the output $g\left(  x^{\prime}\right)  $ is the SFrFT of the input
$f\left(  x\right)  $ plus a quadratic phase term $\exp \left(  -\frac{i}%
{2}\mathcal{P}x^{\prime2}\right)  .$

\subsection{An integration transformation from Chirplet to FrFT kernel}

In the history of developing optics we have known that each optical setup
corresponds to an optical transformation, for example, thick lens as a
fractional Fourier transformer. In turn, once a new integration transform is
found, its experimental implementation is expected. In this subsection we
report a new integration transformation which can convert chirplet function to
FrFT kernel \cite{FHJMO}, as this new transformation is invertible and obeys
Parseval theorem, we expect it be realized by experimentalists.

The new transform we propose here is
\begin{equation}
\iint_{-\infty}^{\infty}\frac{dpdq}{\pi}e^{2i\left(  p-x\right)  \left(
q-y\right)  }h(p,q)\equiv f\left(  x,y\right)  , \label{12.28}%
\end{equation}
which differs from the usual two-fold Fourier transformation $\iint_{-\infty
}^{\infty}\frac{dxdy}{4\pi^{2}}e^{ipx+iqy}f(x,y).$ In particular, when
$h(p,q)=1,$ Eq. (\ref{12.28}) reduces to
\begin{equation}
\iint_{-\infty}^{\infty}\frac{dpdq}{\pi}e^{2i\left(  p-x\right)  \left(
q-y\right)  }=\int_{-\infty}^{\infty}dq\delta \left(  q-y\right)
e^{-2xi\left(  q-y\right)  }=1, \label{12.29}%
\end{equation}
so $e^{2i\left(  p-x\right)  \left(  q-y\right)  }$ can be considered a basis
funtion in $p-q$ phase space, or Eq. (\ref{12.28}) can be looked as an
expansion of $f\left(  x,y\right)  $ with the expansion coefficient being
$h(p,q).$ We can prove that the reciprocal transformation of (\ref{12.28}) is%
\begin{equation}
\iint_{-\infty}^{\infty}\frac{dxdy}{\pi}e^{-2i(p-x)(q-y)}f(x,y)=h(p,q).
\label{12.30}%
\end{equation}
In fact, substituting (\ref{12.28}) into the left-hand side of (\ref{12.30})
yields%
\begin{align}
&  \iint_{-\infty}^{\infty}\frac{dp^{\prime}dq^{\prime}}{\pi}h(p^{\prime
},q^{\prime})\iint \frac{dxdy}{\pi}e^{2i\left[  \left(  p^{\prime}-x\right)
\left(  q^{\prime}-y\right)  -\left(  p-x\right)  \left(  q-y\right)  \right]
}\nonumber \\
&  =\iint_{-\infty}^{\infty}dp^{\prime}dq^{\prime}h(p^{\prime},q^{\prime
})e^{2i\left(  p^{\prime}q^{\prime}-pq\right)  }\nonumber \\
&  \times \delta \left(  p-p^{\prime}\right)  \delta \left(  q-q^{\prime}\right)
\left.  =h(p,q)\right.  . \label{12.31}%
\end{align}
This transformation's Parseval-like theorem is
\begin{align}
&  \iint_{-\infty}^{\infty}\frac{dpdq}{\pi}|h(p,q)|^{2}\nonumber \\
&  =\iint \frac{dxdy}{\pi}|f\left(  x,y\right)  |^{2}\iint \frac{dx^{\prime
}dy^{\prime}}{\pi}e^{2i\left(  x^{\prime}y^{\prime}-xy\right)  }\nonumber \\
&  \times \iint_{-\infty}^{\infty}\frac{dpdq}{\pi}e^{2i\left[  \left(
-y^{\prime}p-x^{\prime}q\right)  +\left(  py+xq\right)  \right]  }\nonumber \\
&  =\iint \frac{dxdy}{\pi}|f\left(  x,y\right)  |^{2}\iint dx^{\prime
}dy^{\prime}e^{2i\left(  x^{\prime}y^{\prime}-xy\right)  }\nonumber \\
&  \times \delta \left(  x-x^{\prime}\right)  \delta \left(  p-p^{\prime}\right)
\nonumber \\
&  =\iint \frac{dxdy}{\pi}|f\left(  x,y\right)  |^{2}. \label{12.32}%
\end{align}
Now we apply Eq. (\ref{12.28}) to phase space transformation in quantum
optics. Recall that a signal $\psi \left(  q\right)  $'s Wigner transform
\cite{r5,r13,r14,r19} is
\begin{equation}
\psi \left(  q\right)  \rightarrow \int \frac{du}{2\pi}e^{ipu}\psi^{\ast}\left(
q+\frac{u}{2}\right)  \psi \left(  q-\frac{u}{2}\right)  . \label{12.33}%
\end{equation}
Using Dirac's symbol \cite{r20} to write $\psi \left(  q\right)  =\left \langle
q\right \vert \left.  \psi \right \rangle ,$ $\left \vert q\right \rangle $ is the
eigenvector of coordinate $Q$, the Wigner operator emerges from (\ref{12.33}),%
\begin{equation}
\frac{1}{2\pi}\int_{-\infty}^{\infty}due^{-ipu}\left \vert q-\frac{u}%
{2}\right \rangle \left \langle q+\frac{u}{2}\right \vert =\Delta \left(
p,q\right)  ,\text{ }\hbar=1. \label{12.34}%
\end{equation}
If $h\left(  q,p\right)  $ is quantized as the operator $\hat{H}\left(
P,Q\right)  $ through the Weyl-Wigner correspondence \cite{Weyl}%
\begin{equation}
H\left(  P,Q\right)  =\iint_{-\infty}^{\infty}dpdq\Delta \left(  p,q\right)
h\left(  q,p\right)  , \label{12.35}%
\end{equation}
then%
\begin{equation}
h\left(  q,p\right)  =\int_{-\infty}^{\infty}due^{-ipu}\left \langle q+\frac
{u}{2}\right \vert \hat{H}\left(  Q,P\right)  \left \vert q-\frac{u}%
{2}\right \rangle , \label{12.36}%
\end{equation}
this in the literature is named the Weyl transform, $h\left(  q,p\right)  $ is
the Weyl classical correspondence of the operator $\hat{H}\left(  Q,P\right)
$. Substituting (\ref{12.36}) into (\ref{12.28}) we have
\begin{align}
&  \iint_{-\infty}^{\infty}\frac{dpdq}{\pi}e^{2i\left(  p-x\right)  \left(
q-y\right)  }h(p,q)\nonumber \\
&  =\iint_{-\infty}^{\infty}\frac{dpdq}{\pi}e^{2i\left(  p-x\right)  \left(
q-y\right)  }\int_{-\infty}^{\infty}due^{-ipu}\nonumber \\
&  \times \left \langle q+\frac{u}{2}\right \vert \hat{H}\left(  Q,P\right)
\left \vert q-\frac{u}{2}\right \rangle \nonumber \\
&  =\int_{-\infty}^{\infty}dq\int_{-\infty}^{\infty}du\left \langle q+\frac
{u}{2}\right \vert \hat{H}\left(  Q,P\right)  \left \vert q-\frac{u}%
{2}\right \rangle \nonumber \\
&  \times \delta \left(  q-y-\frac{u}{2}\right)  e^{-2ix\left(  q-y\right)
}\nonumber \\
&  =\int_{-\infty}^{\infty}due^{-ixu}\left \langle y+u\right \vert \hat
{H}\left(  Q,P\right)  \left \vert y\right \rangle . \label{12.37}%
\end{align}
Using $\left \langle y+u\right \vert =\left \langle u\right \vert e^{iPy}$ and
$(\sqrt{2\pi})^{-1}e^{-ixu}=\left \langle p_{=x}\right \vert \left.
u\right \rangle ,$ where $\left \langle p\right \vert $ is the momentum
eigenvector, and%
\begin{align}
\int_{-\infty}^{\infty}due^{-ixu}\left \langle y+u\right \vert  &
=\int_{-\infty}^{\infty}due^{-ixu}\left \langle u\right \vert e^{iPy}\nonumber \\
&  =\sqrt{2\pi}\int_{-\infty}^{\infty}du\left \langle p_{=x}\right \vert \left.
u\right \rangle \left \langle u\right \vert e^{iPy}\nonumber \\
&  =\sqrt{2\pi}\left \langle p_{=x}\right \vert e^{ixy}, \label{12.38}%
\end{align}
then Eq. (\ref{12.37}) becomes%
\begin{equation}
\iint_{-\infty}^{\infty}\frac{dpdq}{\pi}e^{2i\left(  p-x\right)  \left(
q-y\right)  }h(p,q)=\sqrt{2\pi}\left \langle p_{=x}\right \vert \hat{H}\left(
Q,P\right)  \left \vert y\right \rangle e^{ixy}, \label{12.39}%
\end{equation}
thus through the new integration transformation a new relationship between a
phase space function $h(p,q)$ and its Weyl-Wigner correspondence operator
$\hat{H}\left(  Q,P\right)  $ is revealed. The inverse of (\ref{12.39}),
according to (\ref{12.30}), is%
\begin{equation}
\iint_{-\infty}^{\infty}\frac{dxdy}{\sqrt{\pi/2}}e^{-2i\left(  p-x\right)
\left(  q-y\right)  }\left \langle p_{=x}\right \vert \hat{H}\left(  Q,P\right)
\left \vert y\right \rangle e^{ixy}=h(p,q). \label{12.40}%
\end{equation}
For example, when $\hat{H}\left(  Q,P\right)  =e^{f(P^{2}+Q^{2}-1)/2},$ its
classical correspondence is%
\begin{equation}
e^{f\left(  P^{2}+Q^{2}-1\right)  /2}\rightarrow h(p,q)=\frac{2}{e^{f}+1}%
\exp \left \{  2\frac{e^{f}-1}{e^{f}+1}\left(  p^{2}+q^{2}\right)  \right \}  .
\label{12.41}%
\end{equation}
Substituting (\ref{12.41}) into (\ref{12.39}) we have%
\begin{align}
&  \frac{2}{e^{f}+1}\iint_{-\infty}^{\infty}\frac{dpdq}{\pi}e^{2i\left(
p-x\right)  \left(  q-y\right)  }\exp \left \{  2\frac{e^{f}-1}{e^{f}+1}\left(
p^{2}+q^{2}\right)  \right \} \nonumber \\
&  =\sqrt{2\pi}\left \langle p_{=x}\right \vert e^{f\left(  P^{2}+Q^{2}%
-1\right)  /2}\left \vert y\right \rangle e^{ixy}. \label{12.42}%
\end{align}
Using the Gaussian integration formula
\begin{align}
&  \iint_{-\infty}^{\infty}\frac{dpdq}{\pi}e^{2i\left(  p-x\right)  \left(
q-y\right)  }e^{-\lambda \left(  p^{2}+q^{2}\right)  }\nonumber \\
&  =\frac{1}{\sqrt{\lambda^{2}+1}}\exp \left \{  \frac{-\lambda \left(
x^{2}+y^{2}\right)  }{\lambda^{2}+1}+\frac{2i\lambda^{2}}{\lambda^{2}%
+1}xy\right \}  , \label{12.43}%
\end{align}
in particular, when $\lambda=-i\tan \left(  \frac{\pi}{4}-\frac{\alpha}%
{2}\right)  ,$ with $\frac{-\lambda}{\lambda^{2}+1}=\frac{i}{2\tan \alpha},$
$\frac{2\lambda^{2}}{\lambda^{2}+1}=1-\frac{1}{\sin \alpha},$ Eq. (\ref{12.43})
becomes$\allowbreak$%
\begin{align}
&  \frac{2}{ie^{-i\alpha}+1}\iint_{-\infty}^{\infty}\frac{dpdq}{\pi
}e^{2i\left(  p-x\right)  \left(  q-y\right)  }\nonumber \\
&  \times \exp \left \{  i\left(  p^{2}+q^{2}\right)  \tan(\frac{\pi}{4}%
-\frac{\alpha}{2})\right \} \nonumber \\
&  =\frac{1}{\sqrt{ie^{-i\alpha}\sin \alpha}}\exp \left \{  \frac{i\left(
x^{2}+y^{2}\right)  }{2\tan \alpha}-\frac{ixy}{\sin \alpha}\right \}  e^{ixy},
\label{12.44}%
\end{align}
where $\exp \{i\tan \left(  \frac{\pi}{4}-\frac{\alpha}{2}\right)  \left(
p^{2}+q^{2}\right)  \}$ represents an infinite long chirplet function.
Comparing (\ref{12.44}) with (\ref{12.42}) we see $ie^{-i\alpha}=e^{f},$
$f=i\left(  \frac{\pi}{2}-\alpha \right)  ,$ it then follows%
\begin{align}
&  \left \langle p_{=x}\right \vert e^{i(\frac{\pi}{2}-\alpha)\left(
P^{2}+Q^{2}-1\right)  /2}\left \vert y\right \rangle \nonumber \\
&  =\frac{1}{\sqrt{2\pi ie^{-i\alpha}\sin \alpha}}\exp \left \{  \frac{i\left(
x^{2}+y^{2}\right)  }{2\tan \alpha}-\frac{ixy}{\sin \alpha}\right \}  ,
\label{12.45}%
\end{align}
where the right-hand side of (\ref{12.45}) is just the FrFT kernel. Therefore
the new integration transformation (\ref{12.28}) can convert spherical wave to
FrFT kernel. We expect this transformation could be implemented by experimentalists.

Moreover, this transformation can also serve for solving some operator
ordering problems. We notice
\begin{align}
&  \frac{1}{\pi}\exp[2i\left(  p-x\right)  \left(  q-y\right)  ]\nonumber \\
&  =\int_{-\infty}^{\infty}\frac{dv}{2\pi}\delta \left(  q-y-\frac{v}%
{2}\right)  \exp \left \{  i\left(  p-x\right)  v\right \}  , \label{12.46}%
\end{align}
so the transformation (\ref{12.28}) is equivalent to%
\begin{align}
h(p,q)  &  \rightarrow \iint_{-\infty}^{\infty}\frac{dpdq}{\pi}e^{2i\left(
p-x\right)  \left(  q-y\right)  }h(p,q)\nonumber \\
&  =\iint_{-\infty}^{\infty}dpdq\int_{-\infty}^{\infty}\frac{dv}{2\pi}%
\delta \left(  q-y-\frac{v}{2}\right)  e^{i\left(  p-x\right)  v}%
h(p,q)\nonumber \\
&  =\iint_{-\infty}^{\infty}\frac{dpdq}{2\pi}h(p+x,y+\frac{q}{2})e^{ipq}.
\label{12.47}%
\end{align}
For example, using (\ref{12.34}) and (\ref{12.46}) we have
\begin{align}
\Delta(p,q)  &  \rightarrow \iint_{-\infty}^{\infty}\frac{dpdq}{2\pi}%
\Delta(p+x,y+\frac{q}{2})e^{ipq}\nonumber \\
&  =\iint_{-\infty}^{\infty}\frac{dpdq}{4\pi^{2}}\int_{-\infty}^{\infty
}due^{-i\left(  p+x\right)  u}\nonumber \\
&  \times \left \vert y+\frac{q}{2}-\frac{u}{2}\right \rangle \left \langle
y+\frac{q}{2}+\frac{u}{2}\right \vert e^{ipq}\nonumber \\
&  =\int_{-\infty}^{\infty}\frac{dq}{2\pi}\int_{-\infty}^{\infty}%
due^{-ixu}\delta \left(  q-u\right) \nonumber \\
&  \times \left \vert y+\frac{q}{2}-\frac{u}{2}\right \rangle \left \langle
y+\frac{q}{2}+\frac{u}{2}\right \vert \nonumber \\
&  =\int_{-\infty}^{\infty}\frac{du}{2\pi}e^{-ixu}\left \vert y\right \rangle
\left \langle y+u\right \vert =\left \vert y\right \rangle \left \langle
y\right \vert \int_{-\infty}^{\infty}\frac{du}{2\pi}e^{iu\left(  P-u\right)
}\nonumber \\
&  =\delta \left(  y-Q\right)  \delta \left(  x-P\right)  , \label{12.48}%
\end{align}
so%
\begin{equation}
\frac{1}{\pi}\iint \mathtt{d}p^{\prime}\mathtt{d}q^{\prime}\Delta \left(
q^{\prime},p^{\prime}\right)  e^{2\mathtt{i}\left(  p-p^{\prime}\right)
\left(  q-q^{\prime}\right)  }=\delta \left(  q-Q\right)  \delta \left(
p-P\right)  , \label{12.49}%
\end{equation}
thus this new transformation can convert the Wigner operator to $\delta \left(
q-Q\right)  \delta \left(  p-P\right)  .$ Similarly, we have
\[
\frac{1}{\pi}\iint \mathtt{d}p^{\prime}\mathtt{d}q^{\prime}\Delta \left(
q^{\prime},p^{\prime}\right)  e^{-2\mathtt{i}\left(  p-p^{\prime}\right)
\left(  q-q^{\prime}\right)  }=\delta \left(  p-P\right)  \delta \left(
q-Q\right)  .
\]
Then for the Wigner function of a density operator $\rho$, $W_{\psi
}(p,q)\equiv \mathtt{Tr}\left[  \rho \Delta(p,q)\right]  ,$ we have%
\begin{align}
&  \iint_{-\infty}^{\infty}\frac{dp^{\prime}dq^{\prime}}{\pi}\mathtt{Tr}%
\left[  \rho \Delta(p^{\prime},q^{\prime})\right]  e^{2\mathtt{i}\left(
p-p^{\prime}\right)  \left(  q-q^{\prime}\right)  }\nonumber \\
&  =\mathtt{Tr}\left[  \rho \delta \left(  q-Q\right)  \delta \left(  p-P\right)
\right] \nonumber \\
&  =\int \frac{dudv}{4\pi^{2}}\mathtt{Tr}\left[  \rho e^{i\left(  q-Q\right)
u}e^{i\left(  p-P\right)  v}\right]  , \label{12.50}%
\end{align}
we may define $\mathtt{Tr}\left[  \rho e^{i\left(  q-Q\right)  u}e^{i\left(
p-P\right)  v}\right]  $ as the $Q-P$ characteristic function. Similarly,
\begin{align}
&  \iint_{-\infty}^{\infty}\frac{dp^{\prime}dq^{\prime}}{\pi}\mathtt{Tr}%
\left[  \rho \Delta(p^{\prime},q^{\prime})\right]  e^{-2\mathtt{i}\left(
p-p^{\prime}\right)  \left(  q-q^{\prime}\right)  }\nonumber \\
&  =\mathtt{Tr}\left[  \rho \delta \left(  p-P\right)  \delta \left(  q-Q\right)
\right] \nonumber \\
&  =\int \frac{dudv}{4\pi^{2}}\mathtt{Tr}\left[  \rho e^{i\left(  p-P\right)
v}e^{i\left(  q-Q\right)  u}\right]  \label{12.51}%
\end{align}
we name $\mathtt{Tr}\left[  \rho e^{i\left(  p-P\right)  v}e^{i\left(
q-Q\right)  u}\right]  $ as the $P-Q$ characteristic function.

\section{Complex Fractional Fourier Transformation}

In this section, we extend 1-D FrFT to the complex fractional Fourier
transformation (CFrFT).

\subsection{Quantum version of CFrFT}

According to Ref. \cite{r21}, based on the entangled state $\left \vert
\eta \right \rangle $ in two-mode Fock space and its orthonormal property, we
can take the matrix element of $\exp \left[  -i\alpha \left(  a_{1}^{\dagger
}a_{1}+a_{2}^{\dagger}a_{2}\right)  \right]  $ in the entangled state
$\left \vert \eta \right \rangle ,$
\begin{equation}
\mathcal{K}^{F}\left(  \eta^{\prime},\eta \right)  =\left \langle \eta^{\prime
}\right \vert \exp \left[  -i\alpha \left(  a_{1}^{\dagger}a_{1}+a_{2}^{\dagger
}a_{2}\right)  \right]  \left \vert \eta \right \rangle , \label{13.1}%
\end{equation}
as the integral transform kernel of CFrFT,
\begin{equation}
\mathcal{F}_{\alpha}\left[  f\right]  \left(  \eta \right)  =\int \frac
{d^{2}\eta}{\pi}\mathcal{K}^{F}\left(  \eta^{\prime},\eta \right)  f\left(
\eta \right)  . \label{13.2}%
\end{equation}
Using the normally ordered expansion of $e^{\lambda a_{1}^{\dagger}a_{1}%
}=\colon \exp \left[  \left(  e^{\lambda}-1\right)  a_{1}^{\dagger}a_{1}\right]
\colon$ and the completeness relation of the coherent state representation,
$\left \vert z_{i}\right \rangle =\exp \left \{  -\frac{1}{2}\left \vert
z_{i}\right \vert ^{2}+z_{i}a_{i}^{\dagger}\right \}  \left \vert 0\right \rangle
_{i},$ we calculate that $\mathcal{K}^{F}\left(  \eta^{\prime},\eta \right)  $
is
\begin{align}
\mathcal{K}^{F}\left(  \eta^{\prime},\eta \right)   &  =\left \langle
\eta^{\prime}\right \vert \frac{d^{2}z_{1}^{\prime}d^{2}z_{2}^{\prime}}{\pi
^{2}}\left \vert z_{1}^{\prime},z_{2}^{\prime}\right \rangle \left \langle
z_{1}^{\prime},z_{2}^{\prime}\right \vert \colon \exp \left[  \left(
e^{-i\alpha}-1\right)  \left(  a_{1}^{\dagger}a_{1}+a_{2}^{\dagger}%
a_{2}\right)  \right]  \colon \nonumber \\
&  \times \int \frac{d^{2}z_{1}d^{2}z_{2}}{\pi^{2}}\left \vert z_{1}%
,z_{2}\right \rangle \left \langle z_{1},z_{2}\right \vert \left.  \eta
\right \rangle \nonumber \\
&  =\frac{e^{i(\alpha-\frac{\pi}{2})}}{2\sin \alpha}\exp \left[  \frac
{i(\left \vert \eta^{\prime}\right \vert ^{2}+\left \vert \eta \right \vert ^{2}%
)}{2\tan \alpha}-\frac{i\left(  \eta^{\prime \ast}\eta+\eta^{\ast}\eta^{\prime
}\right)  }{2\sin \alpha}\right]  , \label{13.3}%
\end{align}
which is just the integral kernel of the CFrFT in \cite{r22}. Thus we see that
the matrix element of $\exp \left[  -i\alpha \left(  a_{1}^{\dagger}a_{1}%
+a_{2}^{\dagger}a_{2}\right)  \right]  $ between two entangled state
representations $\left \vert \eta \right \rangle $ and $\left \vert \eta^{\prime
}\right \rangle $ corresponds to CFrFT$.$ This is a new route from quantum
optical transform to classical CFrFT transform. Let $\eta=x_{2}+iy_{2},$
$\eta^{\prime}=x_{1}+iy_{1},$ (\ref{13.2}) becomes%
\begin{align}
\mathcal{F}_{\alpha}\left[  f\right]  \left(  x_{2},y_{2}\right)   &
=\frac{e^{i(\alpha-\frac{\pi}{2})}}{2\sin \alpha}\exp \left[  \frac{i\left(
x_{2}^{2}+y_{2}^{2}\right)  }{2\tan \alpha}\right] \nonumber \\
&  \times \int \frac{dx_{1}dy_{1}}{\pi}\exp \left[  \frac{i\left(  x_{1}%
^{2}+y_{1}^{2}\right)  }{2\tan \alpha}-i\frac{\left(  x_{1}x_{2}+y_{1}%
y_{2}\right)  }{\sin \alpha}\right]  f\left(  x_{1},y_{1}\right)  .
\label{13.4}%
\end{align}
In fact, letting $f\left(  \eta \right)  =\left \langle \eta \right \vert \left.
f\right \rangle ,$ then using Eqs.(\ref{3.13}) and (\ref{13.3}) we have
\begin{align}
&  \left \langle \eta^{\prime}\right \vert \exp \left[  -i\alpha \left(
a_{1}^{\dagger}a_{1}+a_{2}^{\dagger}a_{2}\right)  \right]  \left \vert
f\right \rangle \nonumber \\
&  =\int \frac{d^{2}\eta}{\pi}\left \langle \eta^{\prime}\right \vert \exp \left[
-i\alpha \left(  a_{1}^{\dagger}a_{1}+a_{2}^{\dagger}a_{2}\right)  \right]
\left \vert \eta \right \rangle \left \langle \eta \right \vert \left.
f\right \rangle =\int \frac{d^{2}\eta}{\pi}\mathcal{K}^{F}\left(  \eta^{\prime
},\eta \right)  f\left(  \eta \right) \nonumber \\
&  =\frac{e^{i(\alpha-\frac{\pi}{2})}}{2\sin \alpha}\int \frac{d^{2}\eta}{\pi
}\exp \left[  \frac{i(\left \vert \eta^{\prime}\right \vert ^{2}+\left \vert
\eta \right \vert ^{2})}{2\tan \alpha}-\frac{i\left(  \eta^{\prime \ast}\eta
+\eta^{\ast}\eta^{\prime}\right)  }{2\sin \alpha}\right]  f\left(  \eta \right)
. \label{13.5}%
\end{align}
Thus the quantum mechanical version of CFrFT is given by%
\begin{equation}
\mathcal{F}_{\alpha}\left[  f\right]  \left(  \eta^{\prime}\right)
\equiv \left \langle \eta^{\prime}\right \vert \exp \left[  -i\alpha \left(
a_{1}^{\dagger}a_{1}+a_{2}^{\dagger}a_{2}\right)  \right]  \left \vert
f\right \rangle . \label{13.6}%
\end{equation}
The standard complex Fourier transform is $\mathcal{F}_{\pi/2}.$
$\mathcal{F}_{0}$ is the identity operator.

\subsection{Additivity property and eigenmodes of CFrFT}

We will show later that this CFrFT can help us to reveal some new property
which has been overlooked in the formulation of the direct product of two real
FrFTs \cite{r23}. The definition (\ref{13.16}) is of course required to
satisfy the basic postulate that $\mathcal{F}_{\alpha}\mathcal{F}_{\beta
}\left[  f\right]  \left(  \eta^{\prime}\right)  =\mathcal{F}_{\alpha+\beta
}\left[  f\left(  \eta \right)  \right]  $ (the additivity property). For this
purpose, using Eq.(\ref{13.16}) and Eq.(\ref{3.13}) we see%
\begin{align}
\mathcal{F}_{\alpha+\beta}\left[  f\left(  \eta \right)  \right]   &
\equiv \left \langle \eta^{\prime}\right \vert e^{-i(\alpha+\beta)\left(
a_{1}^{\dagger}a_{1}+a_{2}^{\dagger}a_{2}\right)  }\left \vert f\right \rangle
\nonumber \\
&  =\int_{-\infty}^{\infty}\frac{d^{2}\eta^{\prime \prime}}{\pi}\left \langle
\eta^{\prime}\right \vert e^{-i\alpha \left(  a_{1}^{\dagger}a_{1}%
+a_{2}^{\dagger}a_{2}\right)  }\left \vert \eta^{\prime \prime}\right \rangle
\nonumber \\
&  \times \int_{-\infty}^{\infty}\frac{d^{2}\eta}{\pi}\left \langle \eta
^{\prime \prime}\right \vert e^{-i\beta \left(  a_{1}^{\dagger}a_{1}%
+a_{2}^{\dagger}a_{2}\right)  }\left \vert \eta \right \rangle f\left(
\eta \right) \nonumber \\
&  =\int_{-\infty}^{\infty}\frac{d^{2}\eta^{\prime \prime}}{\pi}\left \langle
\eta^{\prime}\right \vert e^{-i\alpha \left(  a_{1}^{\dagger}a_{1}%
+a_{2}^{\dagger}a_{2}\right)  }\left \vert \eta^{\prime \prime}\right \rangle
\mathcal{F}_{\beta}\left[  f\left(  \eta \right)  \right] \nonumber \\
&  =\mathcal{F}_{\alpha}\mathcal{F}_{\beta}\left[  f\left(  \eta \right)
\right]  . \label{13.7}%
\end{align}
\ This derivation is clear and concise by employing the $\left \vert
\eta \right \rangle $ representation and quantum mechanical version of CFrFT.

On the other hand, the formula (\ref{13.6}) can help us to derive CFrFT of
some wave functions easily. For example, when $\left \vert f\right \rangle $ is
a two-mode number state $\left \vert m,n\right \rangle =a_{1}^{\dag m}%
a_{2}^{\dag n}/\sqrt{m!n!}\left \vert 00\right \rangle $, then the CFrFT of the
wave function $\left \langle \eta \right \vert \left.  m,n\right \rangle $ is
\begin{align}
\mathcal{F}_{\alpha}\left[  \left \langle \eta \right \vert \left.
m,n\right \rangle \right]   &  =\left \langle \eta^{\prime}\right \vert
e^{i(\alpha+\beta)\left(  a_{1}^{\dagger}a_{1}+a_{2}^{\dagger}a_{2}\right)
}\left \vert m,n\right \rangle \nonumber \\
&  =e^{i(\alpha+\beta)\left(  m+n\right)  }\left \langle \eta^{\prime
}\right \vert \left.  m,n\right \rangle . \label{13.8}%
\end{align}
To calculate $\left \langle \eta^{\prime}\right \vert \left.  m,n\right \rangle
$, let us recall the definition of two-variable Hermite polynomial
$H_{m,n}\left(  \xi,\xi^{\ast}\right)  $ (\ref{4.20,4.21}), we can expand
$\left \langle \eta^{\prime}\right \vert $ as%
\begin{equation}
\left \langle \eta^{\prime}\right \vert =\left \langle 00\right \vert \sum
_{m,n=0}^{\infty}i^{m+n}\frac{a_{1}^{m}a_{2}^{n}}{m!n!}H_{m,n}\left(
-i\eta^{\prime \ast},i\eta^{\prime}\right)  e^{-\left \vert \eta^{\prime
}\right \vert ^{2}/2}, \label{13.9}%
\end{equation}
thus%
\begin{equation}
\left \langle \eta^{\prime}\right \vert \left.  m,n\right \rangle =\frac{i^{m+n}%
}{\sqrt{m!n!}}H_{m,n}\left(  -i\eta^{\prime \ast},i\eta^{\prime}\right)
e^{-\left \vert \eta^{\prime}\right \vert ^{2}/2}. \label{13.10}%
\end{equation}
As a result of (\ref{13.10}) we see that equation (\ref{13.8}) becomes%
\begin{equation}
\mathcal{F}_{\alpha}\left[  H_{m,n}\left(  -i\eta^{\ast},i\eta \right)
e^{-\left \vert \eta \right \vert ^{2}/2}\right]  =e^{i(\alpha+\beta)\left(
m+n\right)  }H_{m,n}\left(  -i\eta^{\prime \ast},i\eta^{\prime}\right)
e^{-\left \vert \eta^{\prime}\right \vert ^{2}/2}. \label{13.11}%
\end{equation}
If we consider the operation $\mathcal{F}_{\alpha}$ as an operator, one can
say that the eigenfunction of $\mathcal{F}_{\alpha}$ (the eigenmodes of CFrFT)
is the two-variable Hermite polynomials $H_{m,n}$ with the eigenvalue being
$e^{i(\alpha+\beta)\left(  m+n\right)  }$. This is a new property of CFrFT.
Since the function space spanned by $H_{m,n}\left(  \eta,\eta^{\ast}\right)  $
is complete,%
\begin{equation}
\int \frac{d^{2}\eta}{\pi}e^{-|\eta|^{2}}H_{m,n}\left(  \eta,\eta^{\ast
}\right)  \left[  H_{m,n}\left(  \eta,\eta^{\ast}\right)  \right]  ^{\ast
}=\sqrt{m!n!m^{\prime}!n^{\prime}!}\delta_{m,m^{\prime}}\delta_{n,n^{\prime}},
\label{13.12}%
\end{equation}
and
\begin{equation}
\sum_{m,n=0}^{\infty}\frac{1}{m!n!}H_{m,n}\left(  \eta,\eta^{\ast}\right)
\left[  H_{m,n}\left(  \eta^{\prime},\eta^{\prime \ast}\right)  \right]
^{\ast}e^{-\left \vert \eta \right \vert ^{2}}=\pi \delta \left(  \eta-\eta
^{\prime}\right)  \delta \left(  \eta^{\ast}-\eta^{\prime \ast}\right)  ,
\label{13.13}%
\end{equation}
one can confirms that the eigenmodes of CFrFT form an orthogonal and complete
basis set \cite{r24a}. Note that the two variable Hermite polynomial
$H_{m,n}\left(  \eta,\eta^{\ast}\right)  $ is not the direct product of two
independent ordinary Hermite polynomials, so CFrFT differs from the direct
product of two FrFTs.

\subsection{From Chirplet to CFrFT kernel}

In this subsection, by developing Eq. (\ref{12.28}) to more general case which
can be further related to the transformation between two mutually conjugate
entangled state representations $\left \vert \xi \right \rangle $ and $\left \vert
\eta \right \rangle $, we shall propose a new integration transformation in
$\xi-\eta$ phase space (see Eq. (\ref{13.14}) below) and its inverse
transformation. We find that Eq. (\ref{13.14}) also possesses some
well-behaved transformation properties and can be used to obtain the CFrFT
kernel from a chirplet \cite{r24}.

\subsubsection{New complex integration transformation}

Corresponding to the structure of phase space spanned by $\left \vert
\xi \right \rangle $ and $\left \vert \eta \right \rangle $ and enlightened by Eq.
(\ref{12.28}), we propose a new complex integration transformation in
$\xi-\eta$ phase space
\begin{equation}
\int \frac{d^{2}\xi d^{2}\eta}{\pi^{2}}e^{\left(  \xi-\mu \right)  \left(
\eta^{\ast}-\nu^{\ast}\right)  -\left(  \eta-\nu \right)  \left(  \xi^{\ast
}-\mu^{\ast}\right)  }\mathcal{F}(\eta,\xi)\equiv D\left(  \nu,\mu \right)  .
\label{13.14}%
\end{equation}
When $\mathcal{F}(\xi,\eta)=1,$ (\ref{13.14}) becomes%
\begin{align}
&  \int \frac{d^{2}\xi d^{2}\eta}{\pi^{2}}e^{\left(  \xi-\mu \right)  \left(
\eta^{\ast}-\nu^{\ast}\right)  -\left(  \eta-\nu \right)  \left(  \xi^{\ast
}-\mu^{\ast}\right)  }\nonumber \\
&  =\int d^{2}\xi \delta \left(  \xi-\mu \right)  \delta \left(  \xi^{\ast}%
-\mu^{\ast}\right)  e^{\nu \left(  \xi^{\ast}-\mu^{\ast}\right)  -\nu^{\ast
}\left(  \xi-\mu \right)  }=1, \label{13.15}%
\end{align}
so $e^{\left(  \xi-\mu \right)  \left(  \eta^{\ast}-\nu^{\ast}\right)  -\left(
\eta-\nu \right)  \left(  \xi^{\ast}-\mu^{\ast}\right)  }$ can be considered a
basis function in $\xi-\eta$ phase space, or Eq. (\ref{13.14}) can be looked
as an expansion of $D\left(  \nu,\mu \right)  $ in terms of $e^{\left(  \xi
-\mu \right)  \left(  \eta^{\ast}-\nu^{\ast}\right)  -\left(  \eta-\nu \right)
\left(  \xi^{\ast}-\mu^{\ast}\right)  },$ with the expansion coefficient being
$\mathcal{F}(\eta,\xi).$

We can prove that the inverse transform of (\ref{13.14}) is
\begin{equation}
\int \frac{d^{2}\mu d^{2}\nu}{\pi^{2}}e^{\left(  \xi^{\ast}-\mu^{\ast}\right)
\left(  \eta-\nu \right)  -\left(  \eta^{\ast}-\nu^{\ast}\right)  \left(
\xi-\mu \right)  }D\left(  \nu,\mu \right)  \equiv \mathcal{F}(\eta,\xi).
\label{13.16}%
\end{equation}
In fact, substituting (\ref{13.14}) into the left-hand side of (\ref{13.16})
yields%
\begin{align}
&  \int \frac{d^{2}\xi^{\prime}d^{2}\eta^{\prime}}{\pi^{2}}\mathcal{F}%
(\eta^{\prime},\xi^{\prime})\int \frac{d^{2}\mu d^{2}\nu}{\pi^{2}}\nonumber \\
&  \times e^{\left(  \xi^{\prime}-\mu \right)  \left(  \eta^{\prime \ast}%
-\nu^{\ast}\right)  -\left(  \eta^{\prime}-\nu \right)  \left(  \xi^{\prime
\ast}-\mu^{\ast}\right)  +\left(  \xi^{\ast}-\mu^{\ast}\right)  \left(
\eta-\nu \right)  -\left(  \eta^{\ast}-\nu^{\ast}\right)  \left(  \xi
-\mu \right)  }\nonumber \\
&  =\int \frac{d^{2}\xi^{\prime}d^{2}\eta^{\prime}}{\pi^{2}}\mathcal{F}%
(\eta^{\prime},\xi^{\prime})e^{\left(  \xi^{\prime}\eta^{\prime \ast}%
-\eta^{\prime}\xi^{\prime \ast}+\xi^{\ast}\eta-\eta^{\ast}\xi \right)
}\nonumber \\
&  \times \iint \frac{d^{2}\mu d^{2}\nu}{\pi^{2}}e^{\left(  \eta^{\ast}%
-\eta^{\prime \ast}\right)  \mu+\left(  \eta^{\prime}-\eta \right)  \mu^{\ast}%
}e^{\left(  \xi^{\prime \ast}-\xi^{\ast}\right)  \nu+\left(  \xi-\xi^{\prime
}\right)  \nu^{\ast}}\nonumber \\
&  =\int d^{2}\xi^{\prime}d^{2}\eta^{\prime}\mathcal{F}(\eta^{\prime}%
,\xi^{\prime})e^{\left(  \xi^{\prime}\eta^{\prime \ast}-\eta^{\prime}%
\xi^{\prime \ast}+\xi^{\ast}\eta-\eta^{\ast}\xi \right)  }\nonumber \\
&  \times \delta^{\left(  2\right)  }\left(  \eta^{\prime}-\eta \right)
\delta^{\left(  2\right)  }\delta \left(  \xi-\xi^{\prime}\right)  \left.
=\mathcal{F}(\eta,\xi)\right.  . \label{13.17}%
\end{align}
This Parseval-like theorem for this transformation can also be demonstrated,
\begin{align}
&  \int \frac{d^{2}\xi d^{2}\eta}{\pi^{2}}|\mathcal{F}(\eta,\xi)|^{2}%
\nonumber \\
&  =\int \frac{d^{2}\mu d^{2}\nu}{\pi^{2}}|D\left(  \nu,\mu \right)  |^{2}%
\iint \frac{d^{2}\mu^{\prime}d^{2}\nu^{\prime}}{\pi^{2}}\nonumber \\
&  \times \exp \left[  \left(  \mu^{\ast}\nu-\nu^{\ast}\mu \right)  +\left(
\mu^{\prime}\nu^{\prime \ast}-\nu^{\prime}\mu^{\prime \ast}\right)  \right]
\nonumber \\
&  \times \int \frac{d^{2}\xi d^{2}\eta}{\pi^{2}}\exp \left[  \left(  \mu
^{\prime \ast}-\mu^{\ast}\right)  \eta+\left(  \mu-\mu^{\prime}\right)
\eta^{\ast}\right] \nonumber \\
&  \times \exp \left[  (\nu^{\ast}-\nu^{\prime \ast})\xi+(\nu^{\prime}-\nu
)\xi^{\ast}\right] \nonumber \\
&  =\int \frac{d^{2}\mu d^{2}\nu}{\pi^{2}}|D\left(  \nu,\mu \right)  |^{2}\iint
d^{2}\mu^{\prime}d^{2}\nu^{\prime}\nonumber \\
&  \times \exp \left[  \left(  \mu^{\ast}\nu-\nu^{\ast}\mu \right)  +\left(
\mu^{\prime}\nu^{\prime \ast}-\nu^{\prime}\mu^{\prime \ast}\right)  \right]
\nonumber \\
&  \times \delta^{\left(  2\right)  }\left(  \mu-\mu^{\prime}\right)
\delta^{\left(  2\right)  }\delta(\nu^{\prime}-\nu)\nonumber \\
&  =\int \frac{d^{2}\mu d^{2}\nu}{\pi^{2}}|D\left(  \nu,\mu \right)  |^{2}.
\label{13.18}%
\end{align}

\subsubsection{Complex integration transformation and complex Weyl
transformation}

In Ref. \cite{r25} for correlated two-body systems, we have successfully
established the so-called entangled Wigner operator, expressed in the
entangled state $\left \langle \eta \right \vert $ representation as
(\ref{3.27}),
\begin{equation}
\Delta \left(  \sigma,\gamma \right)  \rightarrow \Delta(\eta,\xi)=\int
\frac{d^{2}\sigma}{\pi^{3}}\left \vert \eta-\sigma \right \rangle \left \langle
\eta+\sigma \right \vert e^{\sigma \xi^{\ast}-\sigma^{\ast}\xi}, \label{13.19}%
\end{equation}
the advantage of introducing $\Delta(\eta,\xi)$ can be seen in Ref.
\cite{r26}. The corresponding Wigner function for a density matrix $\rho$ is%
\begin{equation}
W_{\rho}(\eta,\xi)=\int \frac{d^{2}\eta}{\pi^{3}}\left \langle \eta
+\sigma \right \vert \rho \left \vert \eta-\sigma \right \rangle e^{\sigma \xi^{\ast
}-\sigma^{\ast}\xi}. \label{13.20}%
\end{equation}
If $F(\eta,\xi)$ is quantized as the operator $F\left(  Q_{1},Q_{2}%
,P_{1},P_{2}\right)  $ through the Weyl-Wigner correspondence%
\begin{equation}
F\left(  Q_{1},Q_{2},P_{1},P_{2}\right)  =\int d^{2}\eta d^{2}\xi
\mathcal{F}(\eta,\xi)\Delta(\eta,\xi), \label{13.21}%
\end{equation}
then using (\ref{13.20}) we see%
\begin{align}
\mathcal{F}(\eta,\xi)  &  =4\pi^{2}\mathtt{Tr}\left[  F\left(  Q_{1}%
,Q_{2},P_{1},P_{2}\right)  \Delta(\eta,\xi)\right] \nonumber \\
&  =4\int \frac{d^{2}\sigma}{\pi}e^{\sigma \xi^{\ast}-\sigma^{\ast}\xi
}\left \langle \eta+\sigma \right \vert F\left(  Q_{1},Q_{2},P_{1},P_{2}\right)
\left \vert \eta-\sigma \right \rangle , \label{13.22}%
\end{align}
which is named as the complex Weyl transform, and $\mathcal{F}(\eta,\xi)$ is
the Weyl classical correspondence of $F\left(  Q_{1},Q_{2},P_{1},P_{2}\right)
$. Substituting (\ref{13.22}) into (\ref{13.14}) we get%
\begin{align}
&  \iint \frac{d^{2}\xi d^{2}\eta}{\pi^{2}}e^{\left(  \xi-\mu \right)  \left(
\eta^{\ast}-\nu^{\ast}\right)  -\left(  \eta-\nu \right)  \left(  \xi^{\ast
}-\mu^{\ast}\right)  }\mathcal{F}(\eta,\xi)\nonumber \\
&  =\iint \frac{d^{2}\xi d^{2}\eta}{\pi^{2}}e^{\left(  \xi-\mu \right)  \left(
\eta^{\ast}-\nu^{\ast}\right)  -\left(  \eta-\nu \right)  \left(  \xi^{\ast
}-\mu^{\ast}\right)  }\nonumber \\
&  \times4\int \frac{d^{2}\sigma}{\pi}e^{\sigma \xi^{\ast}-\sigma^{\ast}\xi
}\left \langle \eta+\sigma \right \vert F\left(  Q_{1},Q_{2},P_{1},P_{2}\right)
\left \vert \eta-\sigma \right \rangle \nonumber \\
&  =4\int \frac{d^{2}\sigma d^{2}\eta}{\pi}e^{-\mu \left(  \eta^{\ast}-\nu
^{\ast}\right)  +\mu^{\ast}\left(  \eta-\nu \right)  }\delta \left(  \eta^{\ast
}-\nu^{\ast}-\sigma^{\ast}\right)  \delta \left(  \eta-\nu-\sigma \right)
\nonumber \\
&  \times \left \langle \eta+\sigma \right \vert F\left(  Q_{1},Q_{2},P_{1}%
,P_{2}\right)  \left \vert \eta-\sigma \right \rangle \nonumber \\
&  =4\int \frac{d^{2}\sigma}{\pi}e^{\mu^{\ast}\sigma-\mu \sigma^{\ast}%
}\left \langle \nu+2\sigma \right \vert F\left(  Q_{1},Q_{2},P_{1},P_{2}\right)
\left \vert \nu \right \rangle . \label{13.23}%
\end{align}
Using (\ref{3.12}), we have \
\begin{align}
\left \langle \nu+2\sigma \right \vert  &  =\left \langle 2\sigma \right \vert
\exp \left \{  \frac{i}{\sqrt{2}}\left[  \nu_{1}\left(  P_{1}-P_{2}\right)
-\nu_{2}\left(  Q_{1}+Q_{2}\right)  \right]  \right \}  ,\label{13.24}\\
\nu &  =\nu_{1}+i\nu_{2}.\nonumber
\end{align}
As a result of (\ref{13.24}) and $\frac{1}{2}e^{\mu^{\ast}\sigma-\mu
\sigma^{\ast}}=\left \langle \xi_{=\mu}\right \vert \left.  2\sigma \right \rangle
,$ we see%

\begin{align}
&  4\int d^{2}\sigma e^{\mu^{\ast}\sigma-\mu \sigma^{\ast}}\left \langle
\nu+2\sigma \right \vert \nonumber \\
&  =8\int d^{2}\sigma \left \langle \xi_{=\mu}\right \vert \left.  2\sigma
\right \rangle \left \langle 2\sigma \right \vert \exp \{ \frac{i}{\sqrt{2}}\left[
\nu_{1}\left(  P_{1}-P_{2}\right)  -\nu_{2}\left(  Q_{1}+Q_{2}\right)
\right]  \} \nonumber \\
&  =2\pi \left \langle \xi_{=\mu}\right \vert \exp \{ \frac{i}{\sqrt{2}}\left[
\nu_{1}\left(  P_{1}-P_{2}\right)  -\nu_{2}\left(  Q_{1}+Q_{2}\right)
\right]  \} \nonumber \\
&  =2\pi \left \langle \xi_{=\mu}\right \vert e^{i\left(  \mu_{2}\nu_{1}-\mu
_{1}\nu_{2}\right)  }. \label{13.25}%
\end{align}
Using (\ref{13.25}), we convert Eq. (\ref{13.23}) as%
\begin{align}
&  \iint \frac{d^{2}\xi d^{2}\eta}{\pi^{2}}e^{\left(  \xi-\mu \right)  \left(
\eta^{\ast}-\nu^{\ast}\right)  -\left(  \eta-\nu \right)  \left(  \xi^{\ast
}-\mu^{\ast}\right)  }F(\eta,\xi)\nonumber \\
&  =2\pi \left \langle \xi_{=\mu}\right \vert F\left(  Q_{1},Q_{2},P_{1}%
,P_{2}\right)  \left \vert \nu \right \rangle e^{i\left(  \nu_{1}\mu_{2}-\nu
_{2}\mu_{1}\right)  }. \label{13.26}%
\end{align}
The inverse of (\ref{13.26}), according to (\ref{13.16}), is%
\begin{align}
\mathcal{F}(\eta,\xi)  &  =\iint \frac{2d^{2}\mu d^{2}\nu}{\pi}e^{\left(
\xi^{\ast}-\mu^{\ast}\right)  \left(  \eta-\nu \right)  -\left(  \eta^{\ast
}-\nu^{\ast}\right)  \left(  \xi-\mu \right)  }\nonumber \\
&  \times \left \langle \xi_{=\mu}\right \vert F\left(  Q_{1},Q_{2},P_{1}%
,P_{2}\right)  \left \vert \nu \right \rangle e^{i\left(  \nu_{1}\mu_{2}-\nu
_{2}\mu_{1}\right)  }. \label{13.27}%
\end{align}
Thus through the new integration transformation, a new relationship between a
phase space function $\mathcal{F}(\eta,\xi)$ and its Weyl-Wigner
correspondence operator $F\left(  Q_{1},Q_{2},P_{1},P_{2}\right)  $ is revealed.

For example, from the following Weyl-Wigner correspondence%
\begin{equation}
\frac{4}{\left(  e^{f}+1\right)  ^{2}}\exp \left[  \frac{e^{f}-1}{e^{f}%
+1}(\left \vert \eta \right \vert ^{2}+\left \vert \xi \right \vert ^{2})\right]
\rightarrow \exp \{f[K_{+}+K_{-}-1]\}, \label{13.28}%
\end{equation}
($K_{+}$ and $K_{-}$ are defined in Eqs.(\ref{K1}) and (\ref{K2})) and
(\ref{13.27}) we have%
\begin{align}
&  \frac{4}{\left(  e^{f}+1\right)  ^{2}}\iint \frac{d^{2}\xi d^{2}\eta}%
{\pi^{2}}e^{\left(  \xi-\mu \right)  \left(  \eta^{\ast}-\nu^{\ast}\right)
-\left(  \eta-\nu \right)  \left(  \xi^{\ast}-\mu^{\ast}\right)  }\nonumber \\
&  \times \exp \left[  \frac{e^{f}-1}{e^{f}+1}(\left \vert \eta \right \vert
^{2}+\left \vert \xi \right \vert ^{2})\right] \nonumber \\
&  =2\pi \left \langle \xi_{=\mu}\right \vert F\left(  Q_{1},Q_{2},P_{1}%
,P_{2}\right)  \left \vert \nu \right \rangle e^{i\left(  \nu_{1}\mu_{2}-\nu
_{2}\mu_{1}\right)  }. \label{13.29}%
\end{align}
Using the Gaussian integration formula%
\begin{align}
&  \iint \frac{d^{2}\xi d^{2}\eta}{\pi^{2}}e^{\left(  \xi-\mu \right)  \left(
\eta^{\ast}-\nu^{\ast}\right)  -\left(  \eta-\nu \right)  \left(  \xi^{\ast
}-\mu^{\ast}\right)  }e^{-\lambda(\left \vert \eta \right \vert ^{2}+\left \vert
\xi \right \vert ^{2})}\nonumber \\
&  =\frac{1}{1+\lambda^{2}}\exp \left[  -\frac{\lambda(\left \vert
\mu \right \vert ^{2}+\left \vert \nu \right \vert ^{2})}{1+\lambda^{2}}%
+\frac{\lambda^{2}\left(  \mu \nu^{\ast}-\mu^{\ast}\nu \right)  }{1+\lambda^{2}%
}\right]  , \label{13.30}%
\end{align}
in particular, when $\lambda=-i\tan \left(  \frac{\pi}{4}-\frac{\alpha}%
{2}\right)  ,$ with $\frac{-\lambda}{\lambda^{2}+1}=\frac{i}{2\tan \alpha},$
$\frac{\lambda^{2}}{\lambda^{2}+1}=\frac{1}{2}-\frac{1}{2\sin \alpha},$Eq.
(\ref{13.30}) becomes$\allowbreak$%
\begin{align}
&  \iint \frac{d^{2}\xi d^{2}\eta}{\pi^{2}}e^{\left(  \xi-\mu \right)  \left(
\eta^{\ast}-\nu^{\ast}\right)  -\left(  \eta-\nu \right)  \left(  \xi^{\ast
}-\mu^{\ast}\right)  }\exp \left[  i\tan(\frac{\pi}{4}-\frac{\alpha}%
{2})(\left \vert \eta \right \vert ^{2}+\left \vert \xi \right \vert ^{2})\right]
\nonumber \\
&  =\frac{e^{i\alpha}}{i\sin \alpha}\exp \left[  \frac{i(\left \vert
\mu \right \vert ^{2}+\left \vert \nu \right \vert ^{2})}{2\tan \alpha}-\frac{\mu
\nu^{\ast}-\mu^{\ast}\nu}{2\sin \alpha}+i\mu_{2}\nu_{1}-i\mu_{1}\nu_{2}\right]
, \label{13.31}%
\end{align}
where $\exp[i\tan \left(  \frac{\pi}{4}-\frac{\alpha}{2}\right)  (\left \vert
\eta \right \vert ^{2}+\left \vert \xi \right \vert ^{2})]$ represents an infinite
long chirplet function. By taking $f=i(\frac{\pi}{2}-\alpha)$ in
(\ref{13.29}), such that $ie^{-i\alpha}=e^{f},$ and comparing with
(\ref{13.31}) we obtain%
\begin{align}
&  \left \langle \xi_{=\mu}\right \vert F\left(  Q_{1},Q_{2},P_{1},P_{2}\right)
\left \vert \nu \right \rangle \nonumber \\
&  =\frac{-ie^{i\alpha}}{2\pi \sin \alpha}\exp \left[  \frac{i(\left \vert
\mu \right \vert ^{2}+\left \vert \nu \right \vert ^{2})}{2\tan \alpha}-\frac{\mu
\nu^{\ast}-\mu^{\ast}\nu}{2\sin \alpha}\right]  , \label{13.32}%
\end{align}
where the right-hand side of (\ref{13.32}) is just the CFrFT kernel whose
properties can be seen in Ref. \cite{r26}. (One may compare the forms
(\ref{13.3}) and (\ref{13.32}) to see their slight difference. For the
relation between them we refer to Ref.\cite{r24,r26}). Dragoman has shown that
the kernel of the CFrFT can be classically produced with rotated astigmatic
optical systems that mimic the quantum entanglement. Therefore the new
integration transformation (\ref{13.14}) can convert spherical wave to CFrFT
kernel. We expect this transformation could be implemented by experimentalists.

\subsection{Squeezing for the generalized scaled FrFT}

In some practical applications it is necessary to introduce input and output
scale parameters \cite{17,18} into FrFT, i.e., scaled FrFT. The reason lies in
that two facts: (1) the scaled FrFT may be more useful and convenient for
optical information processing due to the scale parameters (free parameters)
introduced into FrFT; (2) it can be reduced to the conventional FrFT under a
given condition. In this subsection, by establishing the relation between the
optical scaled FrFT and quantum mechanical squeezing-rotating operator
transform in one-mode case, we employ the IWOP technique and the bipartite
entangled state representation of two-mode squeezing operator to extend the
scaled FrFT to more general cases, such as scaled complex FrFT and entangled
scaled FrFT. The properties of scaled FrFTs can be seen more clearly from the
viewpoint of representation transform in quantum mechanics.

\subsubsection{Quantum correspondence of the scaled FrFT}

The scaled FrFT \cite{r17} of $\alpha$-order is defined in a manner such that
the usual FrFT is its special case, i.e.,
\begin{equation}
\mathcal{F}_{\alpha}\left[  f\left(  x\right)  \right]  =\sqrt{\frac
{e^{i\left(  \frac{\pi}{2}-\alpha \right)  }}{2\pi \mu \nu \sin \alpha}}%
\int_{-\infty}^{\infty}\exp \left \{  -i\frac{x^{2}/\mu^{2}+y^{2}/\nu^{2}}%
{2\tan \alpha}+\frac{ixx^{\prime}}{\mu \nu \sin \alpha}\right \}  f\left(
x\right)  dx, \label{13.33}%
\end{equation}
where the exponential function is an integral kernel. In a similar way to
deriving the quantum correspondence of FrFT in (\ref{12.4}), and using the
natural repression of single-mode squeezing operator $S_{1}$ in coordinate
representation \cite{r27},
\begin{equation}
S_{1}\left(  \mu \right)  =\frac{1}{\sqrt{\mu}}\int_{-\infty}^{\infty
}dx\left \vert \frac{x}{\mu}\right \rangle \left \langle x\right \vert ,
\label{13.34}%
\end{equation}
we have
\begin{align}
&  \exp \left \{  -i\frac{x^{2}/\mu^{2}+y^{2}/\nu^{2}}{2\tan \alpha}%
+\frac{ixx^{\prime}}{\mu \nu \sin \alpha}\right \} \nonumber \\
&  =\sqrt{-2\pi i\mu \nu e^{i\alpha}\sin \alpha}\left \langle y\right \vert
S_{1}^{\dagger}\left(  \nu \right)  \exp \left \{  i\alpha a^{\dagger}a\right \}
S_{1}\left(  \mu \right)  \left \vert x\right \rangle , \label{13.35}%
\end{align}
which implies that the integral kernel in Eq.(\ref{13.33}) is just the matrix
element of operator $S_{1}^{\dagger}\left(  \nu \right)  \exp \left \{  i\theta
a^{\dagger}a\right \}  S_{1}\left(  \mu \right)  $\ in coordinate states. From
Eq.(\ref{13.35}) it then follows that
\begin{align}
\mathcal{F}_{\alpha}\left[  f\left(  x\right)  \right]   &  =\int_{-\infty
}^{\infty}dx\left \langle y\right \vert S_{1}^{\dagger}\left(  \nu \right)
e^{i\alpha a^{\dagger}a}S_{1}\left(  \mu \right)  \left \vert x\right \rangle
f\left(  x\right) \nonumber \\
&  =\left \langle y\right \vert S_{1}^{\dagger}\left(  \nu \right)  e^{i\alpha
a^{\dagger}a}S_{1}\left(  \mu \right)  \left \vert f\right \rangle \equiv
g\left(  y\right)  , \label{13.36}%
\end{align}
which suggests
\begin{equation}
\left \vert g\right \rangle =S_{1}^{\dagger}\left(  \nu \right)  e^{i\alpha
a^{\dagger}a}S_{1}\left(  \mu \right)  \left \vert f\right \rangle .
\label{13.37}%
\end{equation}
From Eqs.(\ref{13.36})\ and (\ref{13.33}) one can see that the scaled FrFT in
Eq.(\ref{13.36}) corresponds actually to the squeezing-rotating operator
$\left(  S_{1}^{\dagger}\left(  \nu \right)  e^{i\theta a^{\dagger}a}%
S_{1}\left(  \mu \right)  \right)  $ transform in Eq.(\ref{13.36}) between two
quantum states.

\subsubsection{The Scaled CFrFT}

On the basis of quantum mechanical version of one-mode scaled FrFT, we
generalize it to two-mode case, i.e., we can introduce the integral
\begin{align}
\mathcal{F}_{\alpha}^{C}\left[  f\left(  \eta \right)  \right]   &
\equiv \left \langle \eta^{\prime}\right \vert S_{2}^{\dagger}\left(  \nu \right)
e^{i\alpha \left(  a_{1}^{\dagger}a_{1}+a_{2}^{\dagger}a_{2}\right)  }%
S_{2}\left(  \mu \right)  \left \vert f\right \rangle \nonumber \\
&  =\int_{-\infty}^{\infty}\frac{d^{2}\eta}{\pi}\left \langle \eta^{\prime
}\right \vert S_{2}^{\dagger}\left(  \nu \right)  e^{i\alpha \left(
a_{1}^{\dagger}a_{1}+a_{2}^{\dagger}a_{2}\right)  }S_{2}\left(  \mu \right)
\left \vert \eta \right \rangle f\left(  \eta \right)  , \label{13.38}%
\end{align}
where $f\left(  \eta \right)  =$ $\left \langle \eta \right.  \left \vert
f\right \rangle $. Using the natural expression of the two-mode squeezing
operator $S_{2}$ (\ref{3.16}), and noticing that $\left \langle \eta^{\prime
}\right \vert e^{i\theta \left(  a_{1}^{\dagger}a_{1}+a_{2}^{\dagger}%
a_{2}\right)  }\left \vert \eta \right \rangle $ is just the integral kernel of
CFrFT (\ref{13.3}), we can reform (\ref{13.38}) as
\begin{align}
\mathcal{F}_{\alpha}^{C}\left[  f\left(  \eta \right)  \right]   &
=\frac{e^{i\left(  \frac{\pi}{2}-\alpha \right)  }}{2\mu \nu \sin \alpha}\int
\frac{d^{2}\eta}{\pi}f\left(  \eta \right) \nonumber \\
&  \times \exp \left \{  -\frac{i(\left \vert \eta^{\prime}\right \vert ^{2}%
/\nu^{2}+\allowbreak \left \vert \eta \right \vert ^{2}/\mu^{2})}{2\tan \alpha
}+\frac{i\left(  \eta^{\prime}{}^{\ast}\allowbreak \eta+\eta^{\ast}%
\allowbreak \eta^{\prime}\right)  }{2\mu \nu \sin \alpha}\right \}  . \label{13.39}%
\end{align}
It is obvious that Eq.(\ref{13.39}) is just a generalized CFrFT with squeezing
parameters, we name it the scaled CFrFT.\ Thus we link a two-mode
squeezing-rotating operator transform to the scaled CFrFT of complex functions.

\subsubsection{Entangled scaled FrFT}

On the other hand, recall that the entangled state $\left \vert \eta
\right \rangle $ can be Schmidt-decomposed as \cite{r28}
\begin{equation}
\left \vert \eta \right \rangle =e^{-i\eta_{1}\eta_{2}}\int_{-\infty}^{\infty
}dx\left \vert x\right \rangle _{1}\otimes \left \vert x-\sqrt{2}\eta
_{1}\right \rangle _{2}e^{i\sqrt{2}x\eta_{2}}, \label{13.40}%
\end{equation}
we see that
\begin{align}
\left \langle x_{1}^{\prime},x_{2}^{\prime}\right \vert \left.  \eta^{\prime
}\right \rangle  &  =e^{-i\eta_{1}^{\prime}\eta_{2}^{\prime}}\delta \left(
\sqrt{2}\eta_{1}^{\prime}+x_{2}^{\prime}-x_{1}^{\prime}\right)  e^{i\sqrt
{2}x_{1}^{\prime}\eta_{2}^{\prime}},\nonumber \\
\left \langle \eta \right \vert \left.  x_{1},x_{2}\right \rangle  &
=e^{i\eta_{1}\eta_{2}}\delta \left(  \sqrt{2}\eta_{1}+x_{2}-x_{1}\right)
e^{-i\sqrt{2}x_{1}\eta_{2}}. \label{13.41}%
\end{align}
Using Eq.(\ref{3.13}) we have
\begin{align}
K\left(  x_{1}^{\prime},x_{2}^{\prime},x_{1},x_{2}\right)   &  \equiv
\left \langle x_{1}^{\prime},x_{2}^{\prime}\right \vert S_{2}^{\dagger}\left(
\nu \right)  e^{i\alpha \left(  a_{1}^{\dagger}a_{1}+a_{2}^{\dagger}%
a_{2}\right)  }S_{2}\left(  \mu \right)  \left \vert x_{1},x_{2}\right \rangle
.\nonumber \\
&  =\int \frac{d^{2}\eta d^{2}\eta^{\prime}}{\pi^{2}}\left \langle x_{1}%
^{\prime},x_{2}^{\prime}\right \vert \left.  \eta^{\prime}\right \rangle
\left \langle \eta^{\prime}\right \vert S_{2}^{\dagger}\left(  \nu \right)
e^{i\alpha \left(  a_{1}^{\dagger}a_{1}+a_{2}^{\dagger}a_{2}\right)  }%
S_{2}\left(  \mu \right)  \left \vert \eta \right \rangle \left \langle
\eta \right.  \left \vert x_{1},x_{2}\right \rangle , \label{13.42}%
\end{align}
where $\left \vert x_{1},x_{2}\right \rangle =\left \vert x_{1}\right \rangle
\otimes \left \vert x_{2}\right \rangle .$ On substituting Eqs. (\ref{13.40}) and
(\ref{13.41}) into Eq.(\ref{13.42}), we can derive%
\begin{align}
K\left(  x_{1}^{\prime},x_{2}^{\prime},x_{1},x_{2}\right)   &  =\left \{
\sqrt{\frac{e^{i\left(  \frac{\pi}{2}-\alpha \right)  }}{2\pi \sin \alpha}}%
\exp \left[  -i\frac{\lambda_{\nu}^{\prime2}+\lambda_{\mu}^{2}}{2\tan \alpha
}+\frac{i\lambda_{\mu}\lambda_{\nu}^{\prime}}{\sin \alpha}\right]  \right \}
\nonumber \\
&  \times \left \{  \sqrt{\frac{e^{i\left(  \frac{\pi}{2}-\alpha \right)  }}%
{2\pi \sin \alpha}}\exp \left[  -i\frac{\kappa_{\nu}^{\prime2}+\kappa_{\mu}^{2}%
}{2\tan \alpha}+\frac{i\kappa_{\mu}\kappa_{\nu}^{\prime}}{\sin \alpha}\right]
\right \}  , \label{13.43}%
\end{align}
where $\lambda_{\mu}=\frac{x_{1}-x_{2}}{\sqrt{2}\mu},$ $\lambda_{\nu}^{\prime
}=\frac{x_{1}^{\prime}-x_{2}^{\prime}}{\sqrt{2}\nu};\kappa_{\mu}=\frac
{\mu \left(  x_{2}+x_{1}\right)  }{\sqrt{2}},\kappa_{\nu}^{\prime}=\frac
{\nu \left(  x_{1}^{\prime}+x_{2}^{\prime}\right)  }{\sqrt{2}}.$

From Eq.(\ref{13.43}) one can see that a new 2-dimensional (2D) scaled FrFT
can be composed of one 1D scaled FFT in its space domain and the other in its
"frequency" domain, while the transform variables being the combination of two
coordinates as shown in Eq.(\ref{13.44}), so Eq.(\ref{13.43}) is quite
different from the direct product two 1D scaled FrFTs that are both in `space
domain` are indicated in Eq.(\ref{13.36}). Note that the new 2D scaled FFT is
still characterized by only 3-parameter. Therefore, for any function $f\left(
x_{1},x_{2}\right)  =\left \langle x_{1},x_{2}\right.  \left \vert
f\right \rangle $ we can define an entangled scaled FrFT, i.e.,
\begin{align}
\mathcal{F}_{\alpha}^{E}\left[  f\left(  x_{1},x_{2}\right)  \right]   &
=\int_{-\infty}^{\infty}K\left(  x_{1}^{\prime},x_{2}^{\prime},x_{1}%
,x_{2}\right)  f\left(  x_{1},x_{2}\right)  dx_{1}dx_{2}\nonumber \\
&  =\left \langle x_{1}^{\prime},x_{2}^{\prime}\right \vert S_{2}^{\dagger
}\left(  \nu \right)  e^{i\alpha \left(  a_{1}^{\dagger}a_{1}+a_{2}^{\dagger
}a_{2}\right)  }S_{2}\left(  \mu \right)  \left \vert f\right \rangle .
\label{13.45}%
\end{align}

Next we examine the properties of these scaled FrFTs in the quantum optics
context. Without losing generality, for the additivity property, we consider
the scaled CFrFT,
\begin{equation}
\mathcal{F}_{\alpha+\beta}^{C}\left[  f\left(  \eta \right)  \right]
\equiv \int \frac{d^{2}\eta}{\pi}\left \langle \eta^{\prime}\right \vert
S_{2}^{\dagger}\left(  \nu \right)  e^{i(\alpha+\beta)\left(  a_{1}^{\dagger
}a_{1}+a_{2}^{\dagger}a_{2}\right)  }S_{2}\left(  \mu \right)  \left \vert
\eta \right \rangle f\left(  \eta \right)  . \label{13.46}%
\end{equation}
Inserting the completeness relation of $\left \vert \eta \right \rangle $ into
Eq.(\ref{13.46}) yields%
\begin{align}
\mathcal{F}_{\alpha+\beta}^{C}\left[  f\left(  \eta \right)  \right]   &
=\int \frac{d^{2}\eta}{\pi}\left \langle \eta^{\prime}\right \vert S_{2}%
^{\dagger}\left(  \nu \right)  e^{i\alpha \left(  a_{1}^{\dagger}a_{1}%
+a_{2}^{\dagger}a_{2}\right)  }S_{2}\left(  \tau \right)  S_{2}^{\dagger
}\left(  \tau \right)  e^{i\beta \left(  a_{1}^{\dagger}a_{1}+a_{2}^{\dagger
}a_{2}\right)  }S_{2}\left(  \mu \right)  \left \vert \eta \right \rangle f\left(
\eta \right) \nonumber \\
&  =\int \frac{d^{2}\eta^{\prime \prime}}{\pi}\left \langle \eta^{\prime
}\right \vert S_{2}^{\dagger}\left(  \nu \right)  e^{i\alpha \left(
a_{1}^{\dagger}a_{1}+a_{2}^{\dagger}a_{2}\right)  }S_{2}\left(  \mu_{=\tau
}^{\prime}\right)  \left \vert \eta^{\prime \prime}\right \rangle \nonumber \\
&  \times \int \frac{d^{2}\eta}{\pi}\left \langle \eta^{\prime \prime}\right \vert
S_{2}^{\dagger}\left(  \nu_{=\tau}^{\prime}\right)  e^{i\beta \left(
a_{1}^{\dagger}a_{1}+a_{2}^{\dagger}a_{2}\right)  }S_{2}\left(  \mu \right)
\left \vert \eta \right \rangle f\left(  \eta \right) \nonumber \\
&  =\int \frac{d^{2}\eta^{\prime \prime}}{\pi}\left \langle \eta^{\prime
}\right \vert S_{2}^{\dagger}\left(  \nu \right)  e^{i\alpha \left(
a_{1}^{\dagger}a_{1}+a_{2}^{\dagger}a_{2}\right)  }S_{2}\left(  \tau \right)
\left \vert \eta_{=\tau}^{\prime \prime}\right \rangle \mathcal{F}_{\beta}\left[
f\left(  \eta \right)  \right]  =\mathcal{F}_{\alpha}\mathcal{F}_{\beta}\left[
f\left(  \eta \right)  \right]  , \label{13.47}%
\end{align}
which is just the additivity property. It should be pointed out that the
condition of additive operator for the scaled FrFTs is that the parameter
$\nu^{\prime}$ of the prior cascade should be equal to the parameter
$\mu^{\prime}$ of the next one, i.e., $\mu^{\prime}=\nu^{\prime}.$ For other
scaled FrFTs, the properties can also be discussed in the similar way
(according to their quantum versions).

To this end, we should emphasize that different scaled FrFTs correspond to
different quantum mechanical squeezing operators or representations. That is
to say, it is possible that some other scaled FrFT can be presented by using
different quantum mechanical squeezing operators or representations.

\section{Adaption of Collins diffraction formula and CFrFT}

The connection between the Fresnel diffraction in free space and the FrFT had
been bridged by Pellat-Finet \cite{r29} who found that FrFTs are adapted to
the mathematical expression of Fresnel diffraction, just as the standard
Fourier transform is adapted to Fraunhofer diffraction. In previous sections,
a new formulation of the CFrFT and the Collins diffraction formula are
respectively derived in the context of representation transform of quantum
optics. In this section we inquire if the adaption problem of Collins
diffraction formula to the CFrFT can also be tackled in the context of quantum
optics. We shall treat this topic with the use of two-mode (3 parameters)
squeezing operator and in the entangled state representation of continuous
variables, in so doing the quantum mechanical version of associated theory of
classical diffraction and classical CFrFT is obtained, which connects
classical optics and quantum optics in this aspect.

\begin{figure}[ptb]
\label{Fig1}
\centering \includegraphics[width=8cm]{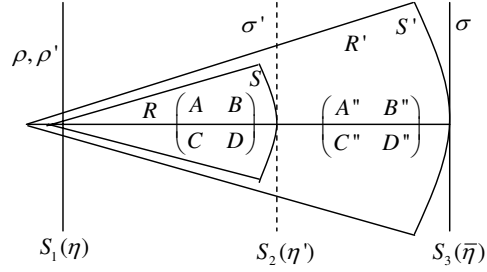}\caption{{\protect \small The
Fresnel diffraction through ABCD optical system.}}%
\end{figure}

For Gaussian beam, the $ABCD$ rule is equally derived via optical diffraction
integral theory---the Collins integral formula. As shown in Fig.2, if
$f\left(  \eta \right)  $ represents the input field amplitude at point $\eta$
on $S_{1}$, and $g\left(  \eta^{\prime}\right)  $ denotes the diffraction
field amplitude at point $\eta^{\prime}$ on $S_{2},$ then Collins formula in
complex form takes the form (\ref{11.5,11.7}). Next we shall examine adaption
of the Collins formula to the CFrFT by virtue of the entangled state
representation in quantum optics \cite{r23}.

\subsection{Adaption of the Collins formula to CFrFT}

Using the completeness relation of $\left \vert \eta \right \rangle ,$ we can
further put Eq.(\ref{11.5}) into%
\begin{equation}
g\left(  \eta^{\prime}\right)  =\left \langle \eta^{\prime}\right \vert
U_{2}\left(  r,s\right)  \left \vert f\right \rangle =\left \langle \eta^{\prime
}\right \vert U_{2}\left(  r,s\right)  \mu_{1}^{2}\int \frac{d^{2}\eta}{\pi
}\left \vert \mu_{1}\eta \right \rangle \left \langle \mu_{1}\eta \right.
\left \vert f\right \rangle , \label{14.1}%
\end{equation}
and taking $\eta^{\prime}=\sqrt{\frac{B}{D}}\frac{\sigma}{K},$ $\mu_{1}%
=\sqrt{\frac{B}{A}}/L$ as well as writing
\begin{equation}
g\left(  \eta^{\prime}\right)  \rightarrow \left \langle \sqrt{\frac{B}{D}}%
\frac{\sigma}{K}\right \vert \left.  g\right \rangle \equiv G\left(
\sigma \right)  ,\text{ }f\left(  \mu_{1}\eta \right)  \equiv F\left(
\eta \right)  , \label{14.2}%
\end{equation}
where $K$ and $L$ are two constants to be determined later, then according to
Eqs. (\ref{11.5}) and (\ref{14.1}) we have%
\begin{align}
G\left(  \sigma \right)   &  =\mu_{1}^{2}\int \frac{d^{2}\eta}{\pi}\left \langle
\eta^{\prime}\right \vert U_{2}\left(  r,s\right)  \left \vert \mu_{1}%
\eta \right \rangle F\left(  \eta \right) \nonumber \\
&  =\frac{1}{2iAL^{2}}\exp \left[  \frac{i\left \vert \sigma \right \vert ^{2}%
}{2K^{2}}\right]  \int \frac{d^{2}\eta}{\pi}\exp \left \{  \frac{i\left \vert
\eta \right \vert ^{2}}{2L^{2}}-\frac{i\left(  \sigma^{\ast}\eta+\sigma
\eta^{\ast}\right)  }{2LK\sqrt{AD}}\right \}  F\left(  \eta \right)  .
\label{14.3}%
\end{align}
Comparing Eq.(\ref{14.3}) with Eq.(\ref{11.7}) leads us to choose%
\begin{equation}
L^{2}=\tan \alpha,\ K=\sqrt{\sin2\alpha/\left(  2AD\right)  }. \label{14.4}%
\end{equation}
Then Eq.(\ref{14.3}) becomes%
\begin{align}
G\left(  \sigma \right)   &  =\frac{\cos \alpha}{i2A\sin \alpha}\exp \left[
i\frac{AD-\cos^{2}\alpha}{\sin2\alpha}\left \vert \sigma \right \vert ^{2}\right]
\nonumber \\
&  \times \int \frac{d^{2}\eta}{\pi}\exp \left \{  \frac{i\left(  \left \vert
\eta \right \vert ^{2}+\left \vert \sigma \right \vert ^{2}\right)  }{2\tan \alpha
}-\frac{i\left(  \sigma^{\ast}\eta+\sigma \eta^{\ast}\right)  }{2\sin \alpha
}\right \}  F\left(  \eta \right) \nonumber \\
&  =\frac{\cos \alpha}{A}e^{-i\alpha}\exp \left[  i\frac{AD-\cos^{2}\alpha}%
{\sin2\alpha}\left \vert \sigma \right \vert ^{2}\right]  \mathcal{F}_{\alpha
}\left[  F\right]  \left(  \sigma \right)  , \label{14.5}%
\end{align}
so Eq. (\ref{14.5}) is a standard CFrFT up to a quadratic phase term
$\exp \left[  i\frac{AD-\cos^{2}\alpha}{\sin2\alpha}\left \vert \sigma
\right \vert ^{2}\right]  $, according to Eq.(\ref{14.4}) and $\sqrt{\frac
{B}{D}}\frac{\sigma}{K}=\eta^{\prime}$, it can also be written as%
\begin{equation}
\exp \left[  i\frac{AD-\cos^{2}\alpha}{\sin2\alpha}\left \vert \sigma \right \vert
^{2}\right]  =\exp \left[  \frac{i}{R}\left \vert \eta^{\prime}\right \vert
^{2}\right]  , \label{14.6}%
\end{equation}
which represents a quadratic approximation to a sphere wave diverging from a
luminous point at distance%
\begin{equation}
R=\frac{2AB}{AD-\cos^{2}\alpha} \label{14.7}%
\end{equation}
from $S_{2}.$ Let $S$ be the sphere tangent to $S_{2}$ with radius $R$ (see
Fig.2). A point on $S$ is located by its\ projection on $S_{2}$, this means
that coordinates on $S_{2}$ can also be used as coordinates on $S$. Therefore,
the quadratic phase term can be compensated if the output field is observed on
$S$ but $S_{2}$. Then, after considering the phase compensation, the field
transforms from $S_{1}$ to $S$ is
\begin{equation}
G_{S}\left(  \sigma \right)  =\frac{\cos \alpha}{A}e^{-i\alpha}\mathcal{F}%
_{\alpha}\left[  F\right]  \left(  \sigma \right)  , \label{14.8}%
\end{equation}
In this way, the field amplitude on $S$ is the perfect $\alpha-th$ FFT-C\ of
the field amplitude on $S_{1}$.

\subsection{Adaption of the additivity property of CFrFT to the Collins
formula for two successive Fresnel diffractions}

The most important property of FrFT is that $\mathcal{F}_{\alpha}$ obeys the
additivity rule, i.e., two successive FrFT of order $\alpha$\ and $\beta$
makes up the FFT of order $\alpha+\beta$. For the CFrFT, its additivity
property is proven in Eq.(\ref{13.7}). For Collins diffraction from $S_{1}$ to
$S^{\prime}$ (see Fig.1), the additivity means that the diffraction pattern
observed on $S^{\prime}$($\bar{\eta}$) \ (the sphere tangent to $S_{3}$ with
radius $R^{\prime}$) and associated with $\mathcal{F}_{\alpha+\beta}$ should
be the result of a first diffraction phenomenon (associated with
$\mathcal{F}_{\alpha})$ on $S$ (with $\eta^{\prime}$), followed by a second
diffraction phenomenon (associated with $\mathcal{F}_{\beta})$ from $S$ to
$S^{\prime}$. This is a necessary consequence of the Huygens principle. Next
we prove that such is indeed the case.

Firstly, let us consider the field transform from $S_{1}$ (with $\eta$) to
$S^{\prime}$ (see Fig.2) described by the ray transfer matrix [$A^{\prime
},B^{\prime},C^{\prime},D^{\prime}$]. Similar to deriving Eq.(\ref{14.7}),
after the squeezing transform and the phase compensation,%
\begin{equation}
R^{\prime}=\frac{2A^{\prime}B^{\prime}}{A^{\prime}D^{\prime}-\cos^{2}%
\alpha^{\prime}}\rightarrow \exp \left(  \frac{i}{R^{\prime}}\left \vert
\bar{\eta}\right \vert ^{2}\right)  , \label{14.9}%
\end{equation}
thus we can obtain the expression of CFrFT for Collins diffraction from
$S_{1}$ to $S^{\prime}$ (not $S_{3}$)$,$%
\begin{equation}
G_{S^{\prime}}\left(  \sigma^{\prime}\right)  =\frac{\cos \alpha^{\prime}%
}{A^{\prime}}e^{-i\alpha^{\prime}}\mathcal{F}_{\alpha^{\prime}}\left[
f\left(  \mu_{1}^{\prime}\eta \right)  \right]  \left(  \sigma^{\prime}\right)
, \label{14.10}%
\end{equation}
where $\bar{\eta}=\sqrt{\frac{B^{\prime}}{D^{\prime}}}\frac{\sigma^{\prime}%
}{K^{\prime}}\ $and $\mu_{1}^{\prime}=\sqrt{\frac{B^{\prime}}{A^{\prime}}%
}/L^{\prime},$
\begin{equation}
L^{\prime2}=\tan \alpha^{\prime},\ K^{\prime}=\sqrt{\sin2\alpha^{\prime
}/\left(  2A^{\prime}D^{\prime}\right)  }. \label{14.11}%
\end{equation}
Eq.(\ref{14.10}) is the same in form as Eq.(\ref{14.8}) but with primed
variables. Using Eqs.(\ref{14.9}) and (\ref{11.5}) one can prove that the
transform from $S_{1}$ to $S^{\prime}$ is (see Eqs. (\ref{11.5}),
(\ref{14.5})-(\ref{14.7}))$\ $%
\begin{equation}
g_{S^{\prime}}\left(  \bar{\eta}\right)  =\exp \left(  -\frac{i}{R^{\prime}%
}\left \vert \bar{\eta}\right \vert ^{2}\right)  \left \langle \bar{\eta
}\right \vert U_{2}\left(  r^{\prime},s^{\prime}\right)  \left \vert
f\right \rangle \equiv G_{S^{\prime}}\left(  \sigma^{\prime}\right)  .
\label{14.12}%
\end{equation}
In Eq.(\ref{14.12}) we have taken the phase compensation term (\ref{14.9})
into account.

Secondly, let us consider the second diffraction from $S$ to $S_{3}$
determined by the ray transfer matrix [$A^{\prime \prime},B^{\prime \prime
},C^{\prime \prime},D^{\prime \prime}$]. For this purpose, using the group
multiplication rule of $F_{2}\left(  r,s\right)  $, we can decompose the
diffraction from $S_{1}$\ to $S^{\prime}$ into two parts: one is described as
the matrix $[A,B,C,D]$ from plane $S_{1}$ (with $\eta$) to $S_{2}\left(
S\right)  $ (with $\eta^{\prime}$), the other is $[A^{\prime \prime}%
,B^{\prime \prime},C^{\prime \prime},D^{\prime \prime}]$ from plane $S_{2}$ to
$S_{3}\left(  S^{\prime}\right)  $ (with $\bar{\eta}$), then the total matrix
from $S_{1}$ to $S_{3}$ is%
\begin{equation}
\left(
\begin{array}
[c]{cc}%
A^{\prime} & B^{\prime}\\
C^{\prime} & D^{\prime}%
\end{array}
\right)  =\left(
\begin{array}
[c]{cc}%
A^{\prime \prime} & B^{\prime \prime}\\
C^{\prime \prime} & D^{\prime \prime}%
\end{array}
\right)  \left(
\begin{array}
[c]{cc}%
A & B\\
C & D
\end{array}
\right)  . \label{14.13}%
\end{equation}
Using Eq.(\ref{3.13}) and the group multiplication rule of $F_{2}\left(
r^{\prime},s^{\prime}\right)  $, we can further put Eq.(\ref{14.12}) into
another form
\begin{align}
G_{S^{\prime}}\left(  \sigma^{\prime}\right)   &  =\exp \left[  -\frac
{i}{R^{\prime}}\left \vert \bar{\eta}\right \vert ^{2}\right]  \left \langle
\bar{\eta}\right \vert U_{2}\left(  r^{\prime \prime},s^{\prime \prime}\right)
U_{2}\left(  r,s\right)  \left \vert f\right \rangle \nonumber \\
&  =\exp \left[  -\frac{i}{R^{\prime}}\left \vert \bar{\eta}\right \vert
^{2}\right]  \left \langle \bar{\eta}\right \vert U_{2}\left(  r^{\prime \prime
},s^{\prime \prime}\right)  \mu_{2}^{\prime2}\int \frac{d^{2}\sigma}{\pi
}\left \vert \mu_{2}^{\prime}\sigma \right \rangle \left \langle \mu_{2}^{\prime
}\sigma \right \vert U_{2}\left(  r,s\right)  \mu_{1}^{\prime2}\int \frac
{d^{2}\eta}{\pi}\left \vert \mu_{1}^{\prime}\eta \right \rangle \left \langle
\mu_{1}^{\prime}\eta \right.  \left \vert f\right \rangle \nonumber \\
&  =\mu_{2}^{\prime2}\int \frac{d^{2}\sigma}{\pi}\exp \left(  -\frac{i\left \vert
\bar{\eta}\right \vert ^{2}}{R^{\prime}}\right)  \left \langle \bar{\eta
}\right \vert U_{2}\left(  r^{\prime \prime},s^{\prime \prime}\right)  \left \vert
\mu_{2}^{\prime}\sigma \right \rangle \left[  \mu_{1}^{\prime2}\int \frac
{d^{2}\eta}{\pi}\left \langle \mu_{2}^{\prime}\sigma \right \vert U_{2}\left(
r,s\right)  \left \vert \mu_{1}^{\prime}\eta \right \rangle F\left(  \eta \right)
\right] \nonumber \\
&  =\frac{B}{DK^{2}}\int \frac{d^{2}\sigma}{\pi}\exp \left(  -\frac{i\left \vert
\bar{\eta}\right \vert ^{2}}{R^{\prime}}\right)  \left \langle \bar{\eta
}\right \vert U_{2}\left(  r^{\prime \prime},s^{\prime \prime}\right)  \left \vert
\mu_{2}^{\prime}\sigma \right \rangle G\left(  \sigma \right)  , \label{14.14}%
\end{align}
where $\mu_{2}^{\prime}=\sqrt{\frac{B}{D}}\frac{1}{K}$ and we have made a
reasonable assumption that $\mu_{1}^{\prime}=\mu_{1}$(so $f\left(  \mu
_{1}^{\prime}\eta \right)  =F\left(  \eta \right)  $), which means that there
are same scaled variants for the input field amplitudes on $S_{1}$ of the
diffractions from $S_{1}$ to $S$ and from $S_{1}$ to $S^{\prime}$.

In order to examine the second diffraction domain from $S$ to $S^{\prime}$
(not $S_{3}$), we need to translate the output field amplitude $G\left(
\sigma \right)  $ observed on plane $S_{2}$ to the field amplitude observed on
sphere plane $S,$ i.e., putting $G\left(  \sigma \right)  $ into $G_{S}\left(
\sigma \right)  $ (see Eq.(\ref{14.8})) by taking the phase compensation (see
Eq.(\ref{14.3})) into account. Thus the field transform from $S$ to
$S^{\prime}$ is ($G_{S}\left(  \sigma \right)  \rightarrow G_{S^{\prime}%
}\left(  \sigma^{\prime}\right)  $)%
\begin{align}
G_{S^{\prime}}\left(  \sigma^{\prime}\right)   &  =\frac{B}{DK^{2}}\int
\frac{d^{2}\sigma}{\pi}\exp \left(  \frac{i}{R}\frac{B\left \vert \sigma
\right \vert ^{2}}{DK^{2}}-\frac{i\left \vert \bar{\eta}\right \vert ^{2}%
}{R^{\prime}}\right)  \left \langle \bar{\eta}\right \vert U_{2}\left(
r^{\prime \prime},s^{\prime \prime}\right)  \left \vert \mu_{2}^{\prime}%
\sigma \right \rangle G_{S}\left(  \sigma \right) \nonumber \\
&  =\frac{B}{DK^{2}}\frac{1}{2iB^{\prime \prime}}\int \frac{d^{2}\sigma}{\pi
}\exp \left \{  \frac{iB^{\prime}}{K^{\prime2}D^{\prime}}\left(  \frac
{D^{\prime \prime}}{2B^{\prime \prime}}-\frac{1}{R^{\prime}}\right)  \left \vert
\sigma^{\prime}\right \vert ^{2}\right. \nonumber \\
&  \left.  +\frac{iB\left \vert \sigma \right \vert ^{2}}{DK^{2}}\left(
\frac{A^{\prime \prime}}{2B^{\prime \prime}}+\frac{1}{R}\right)  -\frac{i\left(
\sigma \sigma^{\prime \ast}+\sigma^{\prime}\sigma^{\ast}\right)  }%
{2B^{\prime \prime}K^{\prime}K\sqrt{\frac{DD^{\prime}}{BB^{\prime}}}}\right \}
G_{S}\left(  \sigma \right)  . \label{14.15}%
\end{align}
Comparing Eq.(\ref{14.15}) with Eq.(\ref{11.7}) leads us to choose%
\begin{equation}
\sin \beta=B^{\prime \prime}K^{\prime}K\sqrt{\frac{DD^{\prime}}{BB^{\prime}}%
}=\frac{B^{\prime \prime}}{2}\sqrt{\frac{\sin2\alpha^{\prime}\sin2\alpha
}{A^{\prime}ABB^{\prime}}}, \label{14.16}%
\end{equation}
and noticing that $\mu_{1}^{\prime}=\mu_{1}$ yields $\frac{B^{\prime}%
}{A^{\prime}\tan \alpha^{\prime}}=\frac{B}{A\tan \alpha},$ thus we have
\begin{equation}
\text{ }A^{\prime}=\frac{B^{\prime \prime}}{B}\frac{\cos \alpha^{\prime}%
\sin \alpha}{\sin \beta},\text{ }A=\frac{B^{\prime \prime}}{B^{\prime}}\frac
{\sin \alpha^{\prime}\cos \alpha}{\sin \beta}. \label{14.17}%
\end{equation}
Combining Eqs.(\ref{14.1}) and (\ref{14.17}) it then follows (letting
$\alpha^{\prime}=\alpha+\beta$)%
\begin{equation}
\frac{B}{DK^{2}}\frac{1}{2iB^{\prime \prime}}=\frac{AB}{\sin2\alpha}\frac
{1}{iB^{\prime \prime}}=\frac{A}{iA^{\prime}\cos \alpha}\frac{\cos \alpha
^{\prime}}{2\sin \beta}, \label{14.18}%
\end{equation}
and
\begin{align}
\frac{B^{\prime}}{K^{\prime2}D^{\prime}}\left(  \frac{D^{\prime \prime}%
}{2B^{\prime \prime}}-\frac{1}{R^{\prime}}\right)   &  =\frac{1}{2}\cot
\beta,\nonumber \\
\frac{B}{DK^{2}}\left(  \frac{A^{\prime \prime}}{2B^{\prime \prime}}+\frac{1}%
{R}\right)   &  =\frac{1}{2}\cot \beta, \label{14.19}%
\end{align}
where we have used Eqs.(\ref{14.7}), (\ref{14.9}) and (\ref{14.13}).

Substitution of Eqs.(\ref{14.8}), (\ref{14.16}), (\ref{14.18}) and
(\ref{14.19}) into Eq.(\ref{14.15}) yields%
\begin{align}
G_{S^{\prime}}\left(  \sigma^{\prime}\right)   &  =\frac{A}{iA^{\prime}%
\cos \alpha}\frac{\cos \alpha^{\prime}}{2\sin \beta}\int \frac{d^{2}\sigma}{\pi
}\exp \left \{  \frac{i\left(  \left \vert \sigma^{\prime}\right \vert
^{2}+\left \vert \sigma \right \vert ^{2}\right)  }{2\tan \beta}-\frac{i\left(
\sigma \sigma^{\prime \ast}+\sigma^{\prime}\sigma^{\ast}\right)  }{2\sin \beta
}\right \}  G_{S}\left(  \sigma \right) \nonumber \\
&  =\frac{\cos \alpha^{\prime}e^{-i\alpha}}{iA^{\prime}}\frac{1}{2\sin \beta
}\int \frac{d^{2}\sigma}{\pi}\exp \left \{  \frac{i\left(  \left \vert
\sigma^{\prime}\right \vert ^{2}+\left \vert \sigma \right \vert ^{2}\right)
}{2\tan \beta}-\frac{i\left(  \sigma \sigma^{\prime \ast}+\sigma^{\prime}%
\sigma^{\ast}\right)  }{2\sin \beta}\right \}  \mathcal{F}_{\alpha}\left[
F\right]  \left(  \sigma \right) \nonumber \\
&  =\frac{\cos \alpha^{\prime}}{A^{\prime}}e^{-i\left(  \alpha+\beta \right)
}\mathcal{F}_{\beta}\mathcal{F}_{\alpha}\left[  F\right]  \left(
\sigma^{\prime}\right)  . \label{14.20}%
\end{align}
The first equation of Eq.(\ref{14.20}) indicates that it is just a CFrFT of
$G_{S}\left(  \sigma \right)  $ from $S$ to $S^{\prime}.$ Comparing
Eq.(\ref{14.20}) with Eq.(\ref{14.10}), we see
\begin{equation}
\mathcal{F}_{\beta}\mathcal{F}_{\alpha}\left[  F\right]  \left(
\sigma^{\prime}\right)  =\mathcal{F}_{\alpha+\beta}\left[  F\right]  \left(
\sigma^{\prime}\right)  . \label{14.21}%
\end{equation}
Thus we complete the study of adaption of CFrFT to the mathematical
representation of Collins diffraction formula in quantum optics context.

\section{The Fractional Radon transform}

Optical tomographic imaging techniques derive two-dimensional data from a
three-dimensional object to obtain a slice image of the internal structure and
thus have the ability to peer inside the object noninvasively. The
mathematical method which complete this task is the Radon transformation.
Similarly, one can use the inverse Radon transformation to obtain the Wigner
distribution by tomographic inversion of a set of measured probability
distributions of the quadrature amplitude \cite{vogel,Smithey}. Based on the
Radon transform \cite{Radon} and the FrFT we can introduce the conception of
fractional Radon transformation (FRT) which combines both of them in a
reasonable way. We notice the well-known fact that the usual Radon transform
of a function $f\left(  \vec{r}\right)  $ can be proceeded in two successive
steps, the first step is an $n-$dimensional ordinary Fourier transform, i.e.
performing a usual FT of $f\left(  \vec{r}\right)  $ in $n$-dimensional
$\vec{k}$ space,
\begin{equation}
F\left(  \vec{k}\right)  =F\left(  t\hat{e}\right)  =\int f\left(  \vec
{r}\right)  e^{-2\pi i\vec{k}\cdot \vec{r}}d\vec{r}, \label{18.1}%
\end{equation}
where $\vec{k}=t\hat{e},$ $\hat{e}$ is a unit vector, $t$ is a real number.
Its inverse is
\begin{equation}
f\left(  \vec{r}\right)  =\int F\left(  \vec{k}\right)  e^{2\pi i\vec{k}%
\cdot \vec{r}}d\vec{k}. \label{18.2}%
\end{equation}
Letting $s=t\lambda$ and rewriting (\ref{18.1}) as
\begin{equation}
F\left(  t\hat{e}\right)  =\int_{-\infty}^{\infty}ds\int d\vec{r}f\left(
\vec{r}\right)  e^{-2\pi is}\delta \left(  s-\vec{k}\cdot \vec{r}\right)
=\int_{-\infty}^{\infty}d\lambda e^{-2\pi it\lambda}\int f\left(  \vec
{r}\right)  \delta \left(  \lambda-\hat{e}\cdot \vec{r}\right)  d\vec{r},
\label{18.3}%
\end{equation}
one can see that the integration over $d\vec{r}$ has been defined as a Radon
transform of $f\left(  \vec{r}\right)  $, denoted as
\begin{equation}
\int f\left(  \vec{r}\right)  \delta \left(  \lambda-\hat{e}\cdot \vec
{r}\right)  d\vec{r}=f_{R}\left(  \lambda,\hat{e}\right)  . \label{18.4}%
\end{equation}
So $F\left(  t\hat{e}\right)  $ can be considered as a $1-$dimensional Fourier
transform of $f_{R}\left(  \lambda,\hat{e}\right)  ,$
\begin{equation}
F\left(  t\hat{e}\right)  =\int_{-\infty}^{\infty}d\lambda e^{-2\pi it\lambda
}f_{R}\left(  \lambda,\hat{e}\right)  . \label{18.5}%
\end{equation}
Its inverse transform is%
\begin{equation}
f_{R}\left(  \lambda,\hat{e}\right)  =\int_{-\infty}^{\infty}F\left(  t\hat
{e}\right)  e^{2\pi it\lambda}dt, \label{18.6}%
\end{equation}
this ordinary $1-$dimensional Fourier transform is considered as the second
step. Combining result of (\ref{18.1}) and (\ref{18.6}) we have%
\begin{equation}
f_{R}\left(  \lambda,\hat{e}\right)  =\int_{-\infty}^{\infty}\int f\left(
\vec{r}\right)  e^{-2\pi it\hat{e}\cdot \vec{r}}e^{2\pi it\lambda}d\vec{r}dt.
\label{18.7}%
\end{equation}
i. e. two usual FTs make up a Radon transform, The inverse of (\ref{18.7}) is
\begin{equation}
\int_{-\infty}^{\infty}\int f_{R}\left(  \lambda,\hat{e}\right)  e^{2\pi
i\vec{k}\cdot \vec{r}}e^{-2\pi it\lambda}d\vec{k}d\lambda=f\left(  \vec
{r}\right)  . \label{18.8}%
\end{equation}
By analogy with these procedures we can make two successively FRFTs to realize
the new fractional Radon transformation \cite{FRT} . The $n$-dimensional FrFT
of $f\left(  \vec{r}\right)  $ is defined as
\begin{equation}
\mathfrak{F}_{\alpha,\vec{k}}\left[  f\right]  =\left(  C_{\alpha}\right)
^{n}\int \exp \left(  \frac{i\left(  \vec{r}^{2}+\vec{k}^{2}\right)  }%
{2\tan \alpha}-\frac{i\vec{k}\cdot \vec{r}}{\sin \alpha}\right)  f\left(  \vec
{r}\right)  d\vec{r}\equiv F_{\alpha}\left(  t\hat{e}\right)  ,\text{ \ }%
\vec{k}=t\hat{e}. \label{18.9}%
\end{equation}
where $\alpha$ is named as the order of FrFT, $C_{\alpha}=\left[
\frac{e^{i\alpha}}{2\pi i\sin \alpha}\right]  ^{1/2}.$

Firstly, we perform an $1-$dimensional inverse fractional Fourier transform
for $F_{\alpha}\left(  t\hat{e}\right)  \ $in $t$-space,%
\begin{align}
f_{R,\alpha}\left(  \lambda,\hat{e}\right)   &  =\left[  C_{\alpha}\right]
^{1-n}\mathfrak{F}_{-\alpha,t}\left[  F_{\alpha}\left(  t\hat{e}\right)
\right] \nonumber \\
&  =\left[  C_{\alpha}\right]  ^{1-n}C_{-\alpha}\int_{-\infty}^{\infty}%
\exp \left(  -\frac{i\left(  \lambda^{2}+t^{2}\right)  }{2\tan \alpha}%
+\frac{i\lambda t}{\sin \alpha}\right)  F_{\alpha}\left(  t\hat{e}\right)  dt,
\label{18.11}%
\end{align}
$\left[  C_{\alpha}\right]  ^{1-n}$ was introduced for later's convenience.
Then substituting (\ref{18.9}) into (\ref{18.11}) we have%
\begin{align}
f_{R,\alpha}\left(  \lambda,\hat{e}\right)   &  =\left[  C_{\alpha}\right]
^{1-n}C_{-\alpha}\left(  C_{\alpha}\right)  ^{n}\int \exp \left(  -\frac
{i\left(  t^{2}+\lambda^{2}\right)  }{2\tan \alpha}+\frac{i\lambda t}%
{\sin \alpha}+\frac{i\left(  \vec{r}^{2}+t^{2}\right)  }{2\tan \alpha}%
-\frac{it\hat{e}\cdot \vec{r}}{\sin \alpha}\right)  f\left(  \vec{r}\right)
d\vec{r}dt\nonumber \\
&  =\int \exp \left(  \frac{i\left(  \vec{r}^{2}-\lambda^{2}\right)  }%
{2\tan \alpha}\right)  \delta \left(  \lambda-\hat{e}\cdot \vec{r}\right)
f\left(  \vec{r}\right)  d\vec{r}, \label{18.12}%
\end{align}
which completes the $n$-dimensional fractional Radon transformation.
Especially, when $\alpha=\pi/2,$ (\ref{18.12}) reduces to the usual Radon
transform (\ref{18.4}). Now we examine if the additive property of FrFT is
consistent with (\ref{18.12}). According to the additive property of FrFT
$\mathfrak{F}_{\alpha}\mathfrak{F}_{\alpha}=\mathfrak{F}_{\alpha+\beta},$ and
(\ref{18.9}) we see%
\begin{align}
F_{\alpha+\beta}\left(  \vec{k}=t\hat{e}\right)   &  =\mathfrak{F}_{\beta
,\vec{k}}\mathfrak{F}_{\alpha,\vec{\xi}}\left[  f\right] \nonumber \\
&  =\left(  C_{\beta}\right)  ^{n}\left(  C_{\alpha}\right)  ^{n}\int \int
\exp \left(  \frac{i\left(  \vec{r}^{2}+\vec{\xi}^{2}\right)  }{2\tan \alpha
}+\frac{i\left(  \vec{\xi}^{2}+\vec{k}^{2}\right)  }{2\tan \beta}\right)
\nonumber \\
&  \times \exp \left(  -\frac{i\vec{\xi}\cdot \vec{r}}{\sin \alpha}-\frac{i\vec
{k}\cdot \vec{\xi}}{\sin \beta}\right)  f\left(  \vec{r}\right)  d\vec{r}%
d\vec{\xi}\nonumber \\
&  =\mathfrak{F}_{\beta+\alpha,\vec{k}}\left[  f\right]  . \label{18.13}%
\end{align}
The corresponding one-dimensional inverse FrFT should be%
\begin{align}
&  \left[  \frac{e^{i\left(  \alpha+\beta \right)  }}{2\pi i\sin \left(
\alpha+\beta \right)  }\right]  ^{\left(  1-n\right)  /2}C_{-\beta}C_{-\alpha
}\int_{-\infty}^{\infty}\exp \left(  -\frac{i\left(  \lambda^{2}+\mu
^{2}\right)  }{2\tan \beta}-\frac{i\left(  t^{2}+\mu^{2}\right)  }{2\tan \alpha
}\right) \nonumber \\
&  \times \exp \left(  +\frac{i\lambda \mu}{\sin \beta}+\frac{i\mu t}{\sin \alpha
}\right)  F_{\alpha+\beta}\left(  t\hat{e}\right)  dtd\mu \nonumber \\
&  =\left[  \frac{e^{i\left(  \alpha+\beta \right)  }}{2\pi i\sin \left(
\alpha+\beta \right)  }\right]  ^{\left(  1-n\right)  /2}C_{-\left(
\alpha+\beta \right)  }\int_{-\infty}^{\infty}F_{\alpha+\beta}\left(  t\hat
{e}\right)  \exp \left(  -\frac{i\left(  t^{2}+\lambda^{2}\right)  }%
{2\tan \left(  \alpha+\beta \right)  }+\frac{i\lambda t}{\sin \left(
\alpha+\beta \right)  }\right)  dt\nonumber \\
&  =\int \exp \left(  \frac{i\left(  \vec{r}^{2}-\lambda^{2}\right)  }%
{2\tan \left(  \alpha+\beta \right)  }\right)  \delta \left(  \lambda-\hat
{e}\cdot \vec{r}\right)  f\left(  \vec{r}\right)  d\vec{r}=f_{R,\alpha+\beta
}\left(  \lambda,\hat{e}\right)  , \label{18.14}%
\end{align}
\  \ which coincides with (\ref{18.12}). From (\ref{18.12}) and (\ref{18.14}),
we can confirm that the transform kernel of $\alpha$th FrFT is
\begin{equation}
\exp \left(  \frac{i\left(  \vec{r}^{2}-\lambda^{2}\right)  }{2\tan \alpha
}\right)  \delta \left(  \lambda-\hat{e}\cdot \vec{r}\right)  . \label{18.15}%
\end{equation}
For example, one can calculate the fractional Radon transform of the $n-$mode
Wigner operator to obtain some new quantum mechanical representations. Finally
we give the inversion of the fractional Radon transformation, From
(\ref{18.12}) we have
\begin{equation}
\frac{1}{\left(  2\pi \sin^{2}\alpha \right)  ^{n/2}}\int \int f_{R,\alpha
}\left(  \lambda,\hat{e}\right)  \exp \left(  \frac{i\left(  \lambda^{2}%
-\vec{r}^{2}\right)  }{2\tan \alpha}-\frac{i\lambda t}{\sin \alpha}+\frac
{it\hat{e}\cdot \vec{r}}{\sin \alpha}\right)  d\vec{k}d\lambda=f\left(  \vec
{r}\right)  , \label{18.16}%
\end{equation}
which is an extension of (\ref{18.8}).

In summary, based on the Radon transform and fractional Fourier transform we
have naturally introduced the $n$-dimensional FRFT, in Ref. Zalevsky and
Mendlovic \cite{zeev} also defined 2-dimensional FRFT, but in different
approach. We have identified the transform kernel for FrFT. The generalization
to complex fractional Radon transformation is also possible \cite{jiang}.

\section{Wavelet transformation and the IWOP\ technique}

In recent years wavelet transforms \cite{wavelet,wavelet5} have been developed
which can overcome some shortcomings of the classical Fourier analysis and
therefore has been widely used in Fourier optics and information science since
1980s. Here we present a quantum optical version of classical wavelet
transform (WT) by virtue of the IWOP\ technique.

\subsection{Quantum optical version of classical WTs}

A wavelet has its energy concentrated in time to give a tool for the analysis
of transient, nonstationary, or time-varying phenomena. (It is a wavelet
because it is localized and it resembles a wave because it oscillates.)
Mathematically, wavelets are defined by starting with a function $\psi$ of the
real variable $x$, named a mother wavelet which is required to decrease
rapidly to zero as $|x|$ tends to infinity,
\begin{equation}
\int_{-\infty}^{\infty}\psi \left(  x\right)  dx=0, \label{15.1}%
\end{equation}
A more general requirement for a mother\ wavelet is to demanded $\psi \left(
x\right)  $ to have vanishing moments $\int_{-\infty}^{\infty}x^{l}\psi \left(
x\right)  dx=0,$ $l=0,1,2...,L.$ (A greater degree of smoothness than
continuity also leads to vanishing moments for the mother wavelet). The theory
of wavelets is concerned with the representation of a function in terms of a
two-parameter family of dilates and translates of a fixed function. The mother
wavelet $\psi$ generates the other wavelets of the family $\psi_{\left(
\mu,s\right)  }$, ($\mu$ is scaling parameter, $s$ is a translation parameter,
$s\in \mathrm{R),}$ the dilated-translated function is defined as
\begin{equation}
\psi_{\left(  \mu,s\right)  }\left(  x\right)  =\frac{1}{\sqrt{\left \vert
\mu \right \vert }}\psi \left(  \frac{x-s}{\mu}\right)  , \label{15.2}%
\end{equation}
while the wavelet integral transform of a signal function $f\left(  x\right)
\in L^{2}\left(  \mathrm{R}\right)  $ by $\psi$ is defined by
\begin{equation}
W_{\psi}f\left(  \mu,s\right)  =\frac{1}{\sqrt{\left \vert \mu \right \vert }%
}\int_{-\infty}^{\infty}f\left(  x\right)  \psi^{\ast}\left(  \frac{x-s}{\mu
}\right)  dx. \label{15.3}%
\end{equation}
We can express (\ref{15.3}) as
\begin{equation}
W_{\psi}f\left(  \mu,s\right)  =\left \langle \psi \right \vert U\left(
\mu,s\right)  \left \vert f\right \rangle . \label{15.4}%
\end{equation}
where $\left \langle \psi \right \vert $ is the state vector corresponding to the
given mother wavelet, $\left \vert f\right \rangle $ is the state to be
transformed, and
\begin{equation}
U\left(  \mu,s\right)  \equiv \frac{1}{\sqrt{\left \vert \mu \right \vert }}%
\int_{-\infty}^{\infty}\left \vert \frac{x-s}{\mu}\right \rangle \left \langle
x\right \vert dx \label{15.5}%
\end{equation}
is the squeezing-translating operator \cite{wavelet,wavelet1,wavelet2},
$\left \langle x\right \vert $ is the eigenvector of coordinate operator. In
order to combine the wavelet transform with quantum states transform more
tightly, using the IWOP technique we can directly perform the integral in
(\ref{15.5}) ($Q=(a+a^{\dag})/\sqrt{2},\mu>0$)%
\begin{align}
U\left(  \mu,s\right)   &  =\frac{1}{\sqrt{\pi \mu}}\int_{-\infty}^{\infty
}dx\colon \exp \left[  -\frac{\mu^{2}+1}{2\mu^{2}}x^{2}+\frac{xs}{\mu^{2}}%
+\sqrt{2}\frac{x-s}{\mu}a^{\dagger}+\sqrt{2}xa-\frac{s^{2}}{2\mu^{2}}%
-Q^{2}\right]  \colon \nonumber \\
&  =\sqrt{\frac{2\mu}{1+\mu^{2}}}\colon \exp \left[  \frac{1}{2\left(  1+\mu
^{2}\right)  }\left(  \frac{s}{\mu}+\sqrt{2}a^{\dagger}+\sqrt{2}\mu a\right)
^{2}-\sqrt{2}\frac{s}{\mu}a^{\dagger}-\frac{s^{2}}{2\mu^{2}}-Q^{2}\right]
\colon. \label{15.6}%
\end{align}
This is the explicitly normal product form. Let $\mu=e^{\lambda}$,
$\operatorname{sech}\lambda=\frac{2\mu}{1+\mu^{2}},$ $\tanh \lambda=\frac
{\mu^{2}-1}{\mu^{2}+1},$ using the operator identity $e^{ga^{\dagger}a}%
=\colon \exp \left[  \left(  e^{g}-1\right)  a^{\dagger}a\right]  \colon,$ Eq.
(\ref{15.6}) becomes%
\begin{align}
U\left(  \mu,s\right)   &  =\exp \left[  \frac{-s^{2}}{2\left(  1+\mu
^{2}\right)  }-\frac{a^{\dagger2}}{2}\tanh \lambda-\frac{a^{\dagger}s}{\sqrt
{2}}\operatorname{sech}\lambda \right] \nonumber \\
&  \times \exp \left[  \left(  a^{\dagger}a+\frac{1}{2}\right)  \ln
\operatorname{sech}\lambda \right] \nonumber \\
&  \times \exp \left[  \frac{a^{2}}{2}\tanh \lambda+\frac{sa}{\sqrt{2}%
}\operatorname{sech}\lambda \right]  . \label{15.7}%
\end{align}
In particular, when $s=0$, it reduces to the well-known squeezing operator,
\begin{equation}
U\left(  \mu,0\right)  =\frac{1}{\sqrt{\mu}}\int_{-\infty}^{\infty}\left \vert
\frac{x}{\mu}\right \rangle \left \langle x\right \vert dx=\exp[\frac{\lambda}%
{2}\left(  a^{2}-a^{\dagger2}\right)  . \label{15.8}%
\end{equation}
For a review of the squeezed state theory we refer to \cite{squeezed1}.

\subsection{The condition of\ mother wavelet in the context of quantum optics}

Now we analyze the condition (\ref{15.1}) for mother wavelet from the point of
view of quantum optics. Due to
\begin{equation}
\int_{-\infty}^{\infty}\left \vert x\right \rangle dx=\left \vert
p=0\right \rangle , \label{15.9}%
\end{equation}
where $\left \vert p\right \rangle $ is the momentum eigenstate, we can recast
the condition into quantum mechanics as
\begin{equation}
\int_{-\infty}^{\infty}\psi \left(  x\right)  dx=0\rightarrow \left \langle
p=0\right \vert \left.  \psi \right \rangle =0, \label{15.10}%
\end{equation}
which indicates that the probability of a measurement of $\left \vert
\psi \right \rangle $ by the projection operator $\left \vert p\right \rangle
\left \langle p\right \vert $ with value $p=0$ is zero. Without loss of
generality, we suppose
\begin{equation}
\left \vert \psi \right \rangle _{M}=G\left(  a^{\dagger}\right)  \left \vert
0\right \rangle =\sum_{n=0}^{\infty}g_{n}a^{\dagger n}\left \vert 0\right \rangle
, \label{15.11}%
\end{equation}
where $g_{n}$ are such chosen as to letting $\left \vert \psi \right \rangle $
obeying the condition (\ref{15.1}). Using the coherent states'
overcompleteness relation we have%
\begin{align}
\left \langle p=0\right \vert \left.  \psi \right \rangle  &  =\left \langle
p=0\right \vert \int \frac{d^{2}z}{\pi}\left \vert z\right \rangle \left \langle
z\right \vert \sum \limits_{n}g_{n}a^{\dagger n}\left \vert 0\right \rangle
\nonumber \\
&  =\sum \limits_{n}g_{n}\int \frac{d^{2}z}{\pi}e^{-|z|^{2}}z^{\ast n}%
\sum \limits_{m}\frac{\left(  \frac{z^{2}}{2}\right)  ^{m}}{m!}\nonumber \\
&  =\sum \limits_{m}\sum \limits_{n}\frac{1}{m!2^{m}}g_{n}\delta_{n,2m}%
n!=\sum \limits_{n}g_{2n}=0. \label{15.12}%
\end{align}
Eq.(\ref{15.12}) provides a general formalism to find the qualified wavelets.
For example, assuming $g_{2n}=0$ for $n>3$, so the coefficients of the
survived terms should satisfy
\begin{equation}
g_{0}+g_{2}+3g_{4}+15g_{6}=0, \label{15.64}%
\end{equation}
and $\left \vert \psi \right \rangle $ becomes
\begin{equation}
\left \vert \psi \right \rangle =\left(  g_{0}+g_{2}a^{\dagger2}+g_{4}%
a^{\dagger4}+g_{6}a^{\dagger6}\right)  \left \vert 0\right \rangle .
\label{15.65}%
\end{equation}
Projecting it onto the coordinate representation, we get the qualified
wavelets%
\begin{align}
\psi \left(  x\right)   &  =\pi^{-1/4}e^{-x^{2}/2}\left[  g_{0}+g_{2}\left(
2x^{2}-1\right)  +g_{4}\left(  4x^{4}-12x^{2}+3\right)  \right. \nonumber \\
&  \left.  +g_{6}\left(  8x^{6}-60x^{4}+90x^{2}-15\right)  \right]  ,
\label{15.66}%
\end{align}
where we have used $\left \langle x\right.  \left \vert n\right \rangle =\left(
2^{n}n!\sqrt{\pi}\right)  ^{-1/2}H_{n}\left(  x\right)  e^{-x^{2}/2}$, and
$H_{n}\left(  x\right)  $ is the Hermite polynomials. Now we take some examples.

Case 1: in (\ref{15.12}) by taking $g_{0}=\frac{1}{2},$ $g_{2}=-\frac{1}{2},$
$g_{2n}=0$ (otherwise)$,$ we have
\begin{equation}
\left \vert \psi \right \rangle _{M}=\frac{1}{2}\left(  1-a^{\dagger2}\right)
\left \vert 0\right \rangle , \label{15.13}%
\end{equation}
it then follows
\begin{equation}
\psi_{M}\left(  x\right)  \equiv \frac{1}{2}\left \langle x\right \vert \left(
1-a^{\dagger2}\right)  \left \vert 0\right \rangle =\frac{1}{2}\left \langle
x\right \vert \left(  \left \vert 0\right \rangle -\sqrt{2}\left \vert
2\right \rangle \right)  =\pi^{-1/4}e^{-x^{2}/2}\left(  1-x^{2}\right)  ,
\label{15.14}%
\end{equation}
which is just the Maxican hat wavelet, satisfying the condition $\int
_{-\infty}^{\infty}e^{-x^{2}/2}\left(  1-x^{2}\right)  dx=0.$Hence $\frac
{1}{2}\left(  1-a^{\dagger2}\right)  \left \vert 0\right \rangle $ is the state
vector corresponding to the Maxican hat mother wavelet (see Fig. 3). Once the
state vector $\left \langle \psi \right \vert $ corresponding to mother wavelet
is known, for any state $\left \vert f\right \rangle $ the matrix element
$_{M}\left \langle \psi \right \vert U\left(  \mu,s\right)  \left \vert
f\right \rangle $ is just the wavelet transform of $f(x)$ with respect to
$\left \langle \psi \right \vert .$

\begin{figure}[ptb]
\label{Fig3}
\centering \includegraphics[width=6cm]{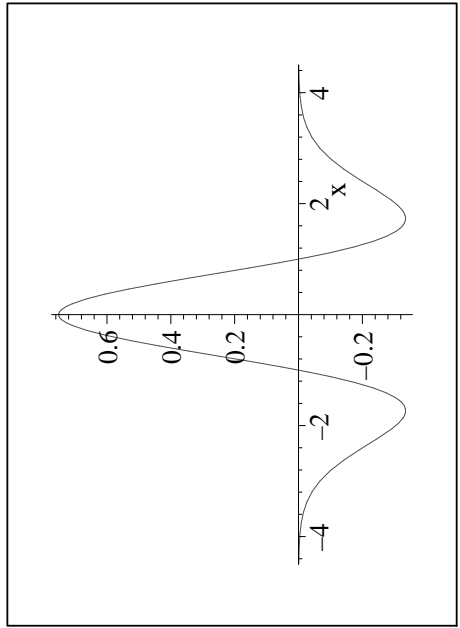}\caption{{\protect \small Traditional
Mexican hat wavelet.}}%
\end{figure}

\begin{figure}[ptb]
\label{Fig4}
\centering \includegraphics[width=6cm]{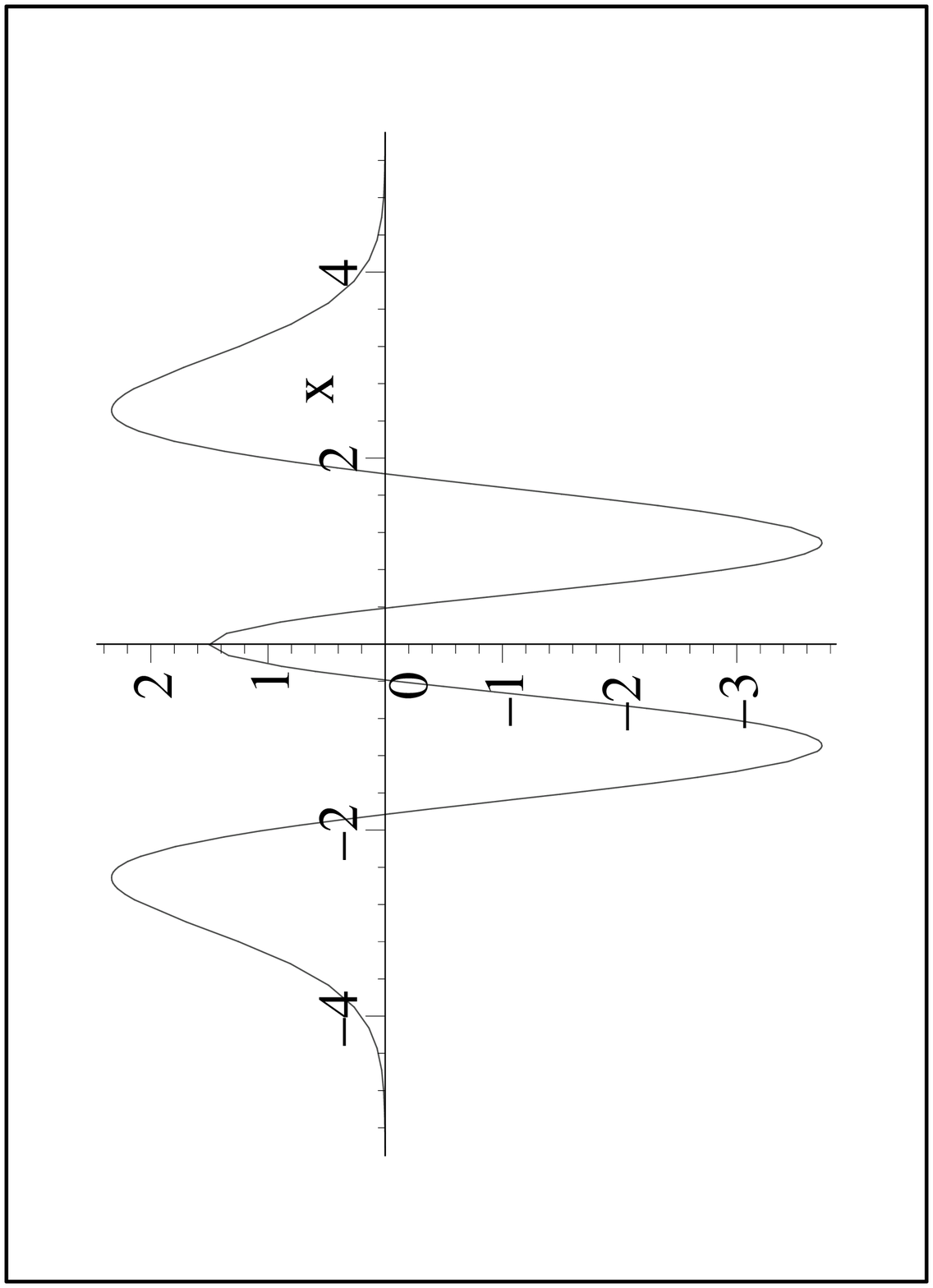}\caption{{\protect \small Generalized
Mexican hat wavelet }$\psi_{2}\left(  x\right)  ${\protect \small when }%
$g_{0}=-2${\protect \small , }$g_{2}=-1${\protect \small , }$g_{4}%
=1${\protect \small and }$g_{6}=0${\protect \small .}}%
\end{figure}

\begin{figure}[ptb]
\label{Fig5}
\centering \includegraphics[width=6cm]{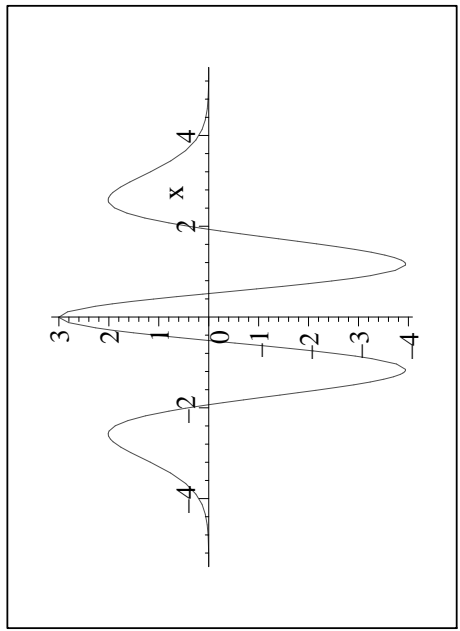}\caption{{\protect \small Generalized
Mexican hat wavelet when }$g_{0}=-1${\protect \small , }$g_{2}=-2$%
{\protect \small , }$g_{4}=1${\protect \small and }$g_{6}=0${\protect \small .}}%
\end{figure}

Case 2: when $g_{0}=-2$, $g_{2}=-1$, $g_{4}=1$ and $g_{6}=0$, from
(\ref{15.66}) we obtain (see Fig. 4)%
\begin{equation}
\psi_{2}\left(  x\right)  =2\pi^{-1/4}e^{-x^{2}/2}\left(  2x^{4}%
-7x^{2}+1\right)  , \label{15.67}%
\end{equation}
which also satisfies $\int_{-\infty}^{\infty}\psi_{2}\left(  x\right)  dx=0$.
Note that when $g_{0}=-1$, $g_{2}=-2$, $g_{4}=1$ and $g_{6}=0$, we obtain a
slightly different wavelet (see Fig. 5). Therefore, as long as the parameters
$g_{2n}$ conforms to condition (\ref{15.64}), we can adjust their values to
control the shape of the wavelets.

Case 3: when $g_{0}=1$, $g_{2}=2$, $g_{4}=4$ and $g_{6}=-1$, we get (see Fig.
6)%
\begin{equation}
\psi_{3}\left(  x\right)  =\pi^{-1/4}e^{-x^{2}/2}\left(  -8x^{6}%
+76x^{4}-134x^{2}+26\right)  , \label{15.68}%
\end{equation}
and $\int_{-\infty}^{\infty}\psi_{3}\left(  x\right)  dx=0$. From these
figures we observe that the number of the nodes of the curves at the $x$-axis
is equal to the highest power of the wavelet functions.\begin{figure}[ptb]
\label{Fig6}
\centering \includegraphics[width=6cm]{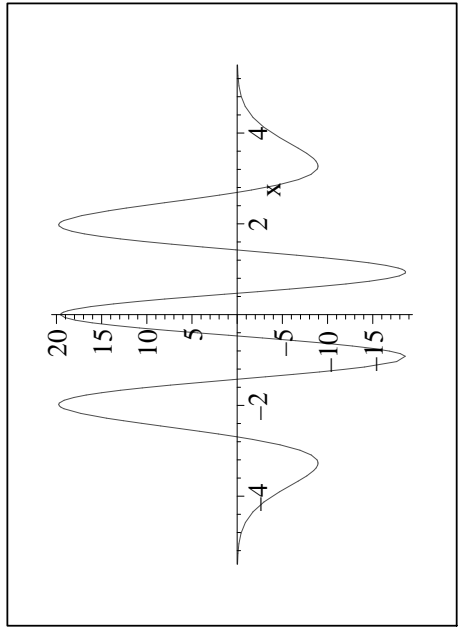}\caption{{\protect \small Generalized
Mexican hat wavelet }$\psi_{3}\left(  x\right)  ${\protect \small when }%
$g_{0}=1${\protect \small , }$g_{2}=2${\protect \small , }$g_{4}=4$%
{\protect \small and }$g_{6}=-1${\protect \small .}}%
\end{figure}

To further reveal the properties of the newly found wavelets, we compare the
wavelet transform computed with the well-known Mexican hat wavelet $\psi
_{1}\left(  x\right)  $ and that with our new wavelet $\psi_{2}\left(
x\right)  $. Concretely, we map a simple cosine signal $\cos \pi x$ by
performing the wavelet transforms with $\psi_{i}\left[  T\left(  x-X\right)
\right]  $, $i=1,2$, into a two-dimensional space $\left(  X,T\right)  $,
where $X$ denotes the location of a wavelet and $a$ its size. The resulting
wavelet transforms by $\psi_{1}\left(  x\right)  $ (=$\psi_{M}\left(
x\right)  $) and $\psi_{2}\left(  x\right)  $ are%
\begin{align}
\Omega_{1}\left(  X,T\right)   &  =\frac{2}{\sqrt{3}}\int_{-\infty}^{\infty
}dx\psi_{1}\left[  T\left(  x-X\right)  \right]  \cos \pi x,\label{15.69}\\
\Omega_{2}\left(  X,T\right)   &  =\frac{1}{\sqrt{30}}\int_{-\infty}^{\infty
}dx\psi_{2}\left[  T\left(  x-X\right)  \right]  \cos \pi x, \label{15.70}%
\end{align}
where $2/\sqrt{3}$ and $1/\sqrt{30}$ are the normalization factors for
$\psi_{1}$ and $\psi_{2}$ respectively, the wavelet integral $\Omega
_{i}\left(  X,T\right)  $ are also called wavelet coefficient which measures
the variation of cos$\pi x$ in a neighborhood of $X,$ whose size is
proportional to $1/T$. The contour line representation of $\Omega_{1}\left(
X,T\right)  $ and $\Omega_{2}\left(  X,T\right)  $ are depicted in Fig. 7 and
Fig. 8, respectively, where the transverse axis is $X$-axis (time axis), while
the longitudinal axis ($T$-axis) is the frequency axis. \begin{figure}[ptb]
\label{Fig7}
\centering \includegraphics[width=8cm]{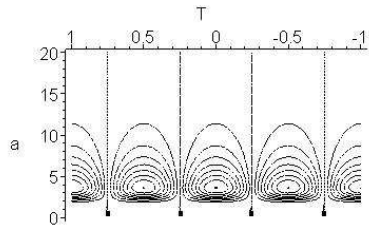}\caption{{\protect \small Contour
line representation of }$\Omega_{1}\left(  X,a\right)  .$}%
\end{figure}

\begin{figure}[ptb]
\label{Fig8}
\centering \includegraphics[width=8cm]{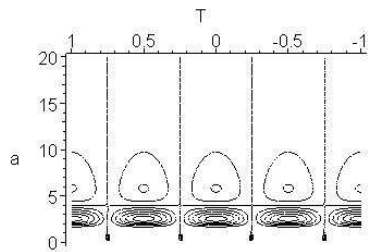}\caption{{\protect \small Contour
line representation of }$\Omega_{2}\left(  X,a\right)  .$}%
\end{figure}

It is remarkable that although two overall shapes of the two contour lines
look similar, there exist two notable differences between these two figures :
1) Along $T$-axis $\Omega_{1}\left(  X,T\right)  $ has one maximum, while
$\Omega_{2}\left(  X,T\right)  $ has one main maximum and one subsidiary
maximum (\textquotedblleft two islands"), so when $\psi_{2}$ scales its size
people have one more chance to identify the frequency information of the
cosine wave than using $\psi_{M}$. Interesting enough, the \textquotedblleft
two islands" of $\Omega_{2}\left(  X,T\right)  $ in Fig. 8 can be imagined as
if they were produced while the figure of $\Omega_{1}\left(  X,T\right)  $
deforms into two sub-structures along $a$-axis. 2) Near the maximum of
$\Omega_{2}\left(  X,T\right)  $ the density of the contour lines along
$a$-axis is higher than that of $\Omega_{1}\left(  X,T\right)  $, which
indicates that the new wavelet $\psi_{2}$ is more sensitive in detecting
frequency information of the signal at this point. Therefore, $\psi_{2}\left(
x\right)  $ may be superior to $\psi_{M}\left(  x\right)  $ in analyzing some
signals. Finally, we mention that there exist some remarkable qualitative
similarities between the mother wavelets presented in Figs. 3 through 6 and
some of the amplitude envelopes of higher order laser spatial modes and
spatial supermodes of phase locked diode laser arrays \cite{las1,las2,las3},
which are due to spatial coherence.

\subsection{Quantum mechanical version of Parseval theorem for WT}

In this subsection, we shall prove that the Parseval theorem of 1D WT
\cite{wavelet1,wavelet2,r30}:%
\begin{equation}
\int_{-\infty}^{\infty}\frac{d\mu}{\mu^{2}}\int_{-\infty}^{\infty}dsW_{\psi
}f_{1}\left(  \mu,s\right)  W_{\psi}^{\ast}f_{2}\left(  \mu,s\right)
=2C_{\psi}\int f_{1}\left(  x\right)  f_{2}^{\ast}\left(  x\right)  dx,
\label{15.16}%
\end{equation}
where $\psi \left(  x\right)  $ is a mother wavelet whose Fourier transform is
$\psi \left(  p\right)  ,$ $C_{\psi}=2\pi \int_{0}^{\infty}\frac{\left \vert
\psi \left(  p\right)  \right \vert ^{2}}{p}dp<\infty$. In the context of
quantum mechanics, according to Eq.(\ref{15.4}) we see that the quantum
mechanical version of Parseval theorem should be
\begin{equation}
\int_{-\infty}^{\infty}\frac{d\mu}{\mu^{2}}\int_{-\infty}^{\infty
}ds\left \langle \psi \right \vert U\left(  \mu,s\right)  \left \vert
f_{1}\right \rangle \left \langle f_{2}\right \vert U^{\dagger}\left(
\mu,s\right)  \left \vert \psi \right \rangle =2C_{\psi}\left \langle
f_{2}\right.  \left \vert f_{1}\right \rangle . \label{15.17}%
\end{equation}
and since $\psi \left(  x\right)  =\left \langle x\right \vert \left.
\psi \right \rangle $, so $\psi \left(  p\right)  $ involved in $C_{\psi}$ is
$\left \langle p\right \vert \left.  \psi \right \rangle ,$ $\left \langle
p\right \vert $ is the momentum eigenvector%
\begin{equation}
\psi \left(  p\right)  =\left \langle p\right \vert \left.  \psi \right \rangle
=\frac{1}{\sqrt{2\pi}}\int_{-\infty}^{\infty}dx\psi \left(  x\right)  e^{-ipx}.
\label{15.18}%
\end{equation}
Eq.(\ref{15.17}) indicates that once the state vector $\left \langle
\psi \right \vert $ corresponding to mother wavelet is known, for any two states
$\left \vert f_{1}\right \rangle $ and $\left \vert f_{2}\right \rangle $, their
overlap up to the factor $C_{\psi}$ (determined by Eq.(\ref{15.31})) is just
their corresponding overlap of WTs in the ($\mu,s)$ parametric space.

\textit{Proof of Equation}\textbf{ (\ref{15.17})}: In order to show
Eq.(\textbf{\ref{15.17}}), we calculate
\begin{align}
U^{\dagger}\left(  \mu,s\right)  \left \vert p\right \rangle  &  =\frac{1}%
{\sqrt{\left \vert \mu \right \vert }}\int_{-\infty}^{\infty}dx\left \vert
x\right \rangle \left \langle \frac{x-s}{\mu}\right \vert \left.  p\right \rangle
\nonumber \\
&  =\frac{e^{-i\frac{ps}{\mu}}}{\sqrt{2\pi \left \vert \mu \right \vert }}%
\int_{-\infty}^{\infty}dx\left \vert x\right \rangle e^{i\frac{p}{\mu}%
x}\nonumber \\
&  =\frac{1}{\sqrt{\left \vert \mu \right \vert }}e^{-i\frac{ps}{\mu}}\left \vert
\frac{p}{\mu}\right \rangle , \label{15.19}%
\end{align}
which leads to
\begin{equation}
\int_{-\infty}^{\infty}dsU^{\dagger}\left(  \mu,s\right)  \left \vert
p^{\prime}\right \rangle \left \langle p\right \vert U\left(  \mu,s\right)
=2\pi \delta \left(  p-p^{\prime}\right)  \left \vert \frac{p^{\prime}}{\mu
}\right \rangle \left \langle \frac{p}{\mu}\right \vert , \label{15.20}%
\end{equation}
where we have used the formula%
\begin{equation}
\int_{-\infty}^{\infty}\frac{dx}{2\pi}e^{ix\left(  p-p^{\prime}\right)
}=\delta \left(  p-p^{\prime}\right)  . \label{15.21}%
\end{equation}
Inserting the completeness relation $\int_{-\infty}^{\infty}dp\left \vert
p\right \rangle \left \langle p\right \vert =1$ into the left side of
Eq.(\ref{15.16}) and then using Eq.(\ref{15.20}) we have%
\begin{align}
\text{L.H.S of Eq.(\ref{15.16})}  &  =\int_{-\infty}^{\infty}\frac{d\mu}%
{\mu^{2}}\int_{-\infty}^{\infty}dsdpdp^{\prime}\psi^{\ast}\left(  p\right)
\psi \left(  p^{\prime}\right)  \left \langle f_{2}\right \vert U^{\dagger
}\left(  \mu,s\right)  \left \vert p^{\prime}\right \rangle \left \langle
p\right \vert U\left(  \mu,s\right)  \left \vert f_{1}\right \rangle \nonumber \\
&  =2\pi \int_{-\infty}^{\infty}\frac{d\mu}{\mu^{2}}\int_{-\infty}^{\infty
}dp\psi^{\ast}\left(  p\right)  \psi \left(  p\right)  \left \langle
f_{2}\right.  \left \vert \frac{p}{\mu}\right \rangle \left \langle \frac{p}{\mu
}\right \vert \left.  f_{1}\right \rangle \nonumber \\
&  \equiv I_{1}+I_{2}, \label{15.22}%
\end{align}
where
\begin{align}
I_{1}  &  =2\pi \int_{0}^{\infty}\frac{d\mu}{\mu^{2}}\int_{-\infty}^{\infty
}dp\psi^{\ast}\left(  p\right)  \psi \left(  p\right)  \left \langle
f_{2}\right.  \left \vert \frac{p}{\mu}\right \rangle \left \langle \frac{p}{\mu
}\right \vert \left.  f_{1}\right \rangle \nonumber \\
&  =2\pi \int_{-\infty}^{\infty}dp\left[  \int_{0}^{\infty}\left \vert
\psi \left(  \mu p\right)  \right \vert ^{2}\frac{d\mu}{\mu}\right]
\left \langle f_{2}\right.  \left \vert p\right \rangle \left \langle p\right \vert
\left.  f_{1}\right \rangle , \label{15.23}%
\end{align}
and
\begin{align}
I_{2}  &  =2\pi \int_{-\infty}^{0}\frac{d\mu}{\mu^{2}}\int_{-\infty}^{\infty
}dp\left \vert \psi \left(  p\right)  \right \vert ^{2}\left \langle f_{2}\right.
\left \vert \frac{p}{\mu}\right \rangle \left \langle \frac{p}{\mu}\right \vert
\left.  f_{1}\right \rangle \nonumber \\
&  =2\pi \int_{-\infty}^{\infty}dp\left[  \int_{0}^{\infty}\left \vert
\psi \left(  -\mu p\right)  \right \vert ^{2}\frac{d\mu}{\mu}\right]
\left \langle f_{2}\right.  \left \vert p\right \rangle \left \langle p\right \vert
\left.  f_{1}\right \rangle . \label{15.24}%
\end{align}
Further, we can put Eqs.(\ref{15.23}) and (\ref{15.24}) into the following
forms,%
\begin{align}
I_{1}  &  =2\pi \int_{0}^{\infty}\frac{d\mu}{\mu^{2}}\int_{-\infty}^{\infty
}dp\psi^{\ast}\left(  p\right)  \psi \left(  p\right)  \left \langle
f_{2}\right.  \left \vert \frac{p}{\mu}\right \rangle \left \langle \frac{p}{\mu
}\right \vert \left.  f_{1}\right \rangle \nonumber \\
&  =C_{\psi}\int_{0}^{\infty}dp\left \langle f_{2}\right.  \left \vert
p\right \rangle \left \langle p\right \vert \left.  f_{1}\right \rangle +2\pi
\int_{-\infty}^{0}dp\left[  \int_{0}^{\infty}\left \vert \psi \left(  \mu
p\right)  \right \vert ^{2}\frac{d\mu p}{\mu p}\right]  \left \langle
f_{2}\right.  \left \vert p\right \rangle \left \langle p\right \vert \left.
f_{1}\right \rangle \nonumber \\
&  =C_{\psi}\int_{0}^{\infty}dp\left \langle f_{2}\right.  \left \vert
p\right \rangle \left \langle p\right \vert \left.  f_{1}\right \rangle +2\pi
\int_{-\infty}^{0}dp\left[  \int_{0}^{-\infty}\left \vert \psi \left(
p^{\prime}\right)  \right \vert ^{2}\frac{dp^{\prime}}{p^{\prime}}\right]
\left \langle f_{2}\right.  \left \vert p\right \rangle \left \langle p\right \vert
\left.  f_{1}\right \rangle \nonumber \\
&  =C_{\psi}\int_{0}^{\infty}dp\left \langle f_{2}\right.  \left \vert
p\right \rangle \left \langle p\right \vert \left.  f_{1}\right \rangle +C_{\psi
}^{\prime}\int_{-\infty}^{0}dp\left \langle f_{2}\right.  \left \vert
p\right \rangle \left \langle p\right \vert \left.  f_{1}\right \rangle
\label{15.25}%
\end{align}
and%
\begin{equation}
I_{2}=C_{\psi}^{\prime}\int_{0}^{\infty}dp\left \langle f_{2}\right.
\left \vert p\right \rangle \left \langle p\right \vert \left.  f_{1}\right \rangle
+C_{\psi}\int_{-\infty}^{0}dp\left \langle f_{2}\right.  \left \vert
p\right \rangle \left \langle p\right \vert \left.  f_{1}\right \rangle ,
\label{15.26}%
\end{equation}
where
\begin{align}
C_{\psi}  &  =2\pi \int_{0}^{\infty}\left \vert \psi \left(  \mu p\right)
\right \vert ^{2}\frac{d\mu}{\mu}=2\pi \int_{0}^{\infty}\left \vert \psi \left(
p\right)  \right \vert ^{2}\frac{dp}{p},\nonumber \\
C_{\psi}^{\prime}  &  =2\pi \int_{0}^{-\infty}\left \vert \psi \left(  p^{\prime
}\right)  \right \vert ^{2}\frac{dp^{\prime}}{p^{\prime}}=2\pi \int_{0}^{\infty
}\left \vert \psi \left(  -p\right)  \right \vert ^{2}\frac{dp}{p}, \label{15.27}%
\end{align}
thus when the definite integration satisfies the admissible condition, i.e.,
\begin{equation}
\int_{0}^{\infty}\left \vert \psi \left(  p\right)  \right \vert ^{2}\frac{dp}%
{p}=\int_{0}^{\infty}\left \vert \psi \left(  -p\right)  \right \vert ^{2}%
\frac{dp}{p}, \label{15.28}%
\end{equation}
which leads to
\begin{equation}
2\pi \int_{-\infty}^{\infty}\left \vert \psi \left(  p\right)  \right \vert
^{2}\frac{dp}{\left \vert p\right \vert }=2C_{\psi}. \label{15.29}%
\end{equation}
Eq. (\ref{15.22}) can be transformed to%
\begin{equation}
\text{L.H.S of Eq.(\ref{15.16})}=2C_{\psi}\int_{-\infty}^{\infty
}dp\left \langle f_{2}\right.  \left \vert p\right \rangle \left \langle
p\right \vert \left.  f_{1}\right \rangle =\text{R.H.S of Eq.(\ref{15.16}).}
\label{15.30}%
\end{equation}
where%
\begin{equation}
C_{\psi}\equiv2\pi \int_{0}^{\infty}\frac{\left \vert \psi \left(  p\right)
\right \vert ^{2}}{p}dp<\infty,\text{ } \label{15.31}%
\end{equation}
thus the theorem is proved. Especially, when $\left \vert f_{1}\right \rangle =$
$\left \vert f_{2}\right \rangle ,$ Eq.(\ref{15.17}) becomes%
\begin{equation}
\int_{-\infty}^{\infty}\frac{d\mu}{\mu^{2}}\int_{-\infty}^{\infty
}ds|\left \langle \psi \right \vert U\left(  \mu,s\right)  \left \vert
f_{1}\right \rangle |^{2}=2C_{\psi}\left \langle f_{1}\right.  \left \vert
f_{1}\right \rangle , \label{15.32}%
\end{equation}
which is named isometry of energy.

\subsection{Inversion formula of WT}

Now we can directly derive the inversion formula of WT, i.e. we take
$\left \langle f_{2}\right \vert =\left \langle x\right \vert $ in Eq.(\ref{15.17}%
), then using Eq.(\ref{15.4}) we see that Eq.(\ref{15.17}) reduces to
\begin{equation}
\int_{-\infty}^{\infty}\frac{d\mu}{\mu^{2}}\int_{-\infty}^{\infty}dsW_{\psi
}f_{1}\left(  \mu,s\right)  \left \langle x\right \vert U^{\dagger}\left(
\mu,s\right)  \left \vert \psi \right \rangle =2C_{\psi}\left \langle x\right.
\left \vert f_{1}\right \rangle . \label{15.33}%
\end{equation}
Due to Eq.(\ref{15.5}) we have%
\begin{equation}
\left \langle x\right \vert U^{\dagger}\left(  \mu,s\right)  =\frac{1}%
{\sqrt{\left \vert \mu \right \vert }}\left \langle x\right \vert \int_{-\infty
}^{\infty}dx^{\prime}\left \vert x^{\prime}\right \rangle \left \langle
\frac{x^{\prime}-s}{\mu}\right \vert =\frac{1}{\sqrt{\left \vert \mu \right \vert
}}\left \langle \frac{x-s}{\mu}\right \vert . \label{15.34}%
\end{equation}
It then follows%
\begin{equation}
\int_{-\infty}^{\infty}\frac{d\mu}{\mu^{2}}\int_{-\infty}^{\infty}dsW_{\psi
}f_{1}\left(  \mu,s\right)  \frac{1}{\sqrt{\left \vert \mu \right \vert }%
}\left \langle \frac{x-s}{\mu}\right \vert \left.  \psi \right \rangle =2C_{\psi
}\left \langle x\right.  \left \vert f_{1}\right \rangle , \label{15.35}%
\end{equation}
which means%
\begin{equation}
f_{1}\left(  x\right)  =\frac{1}{2C_{\psi}}\int_{-\infty}^{\infty}\frac{d\mu
}{\mu^{2}\sqrt{\left \vert \mu \right \vert }}\int_{-\infty}^{\infty}%
ds\psi \left(  \frac{x-s}{\mu}\right)  W_{\psi}f_{1}\left(  \mu,s\right)  ,
\label{15.36}%
\end{equation}
this is the inversion formula of WT.

\subsection{New orthogonal property of mother wavelet in parameter space}

Form the Parserval theorem (\ref{15.16}) of WT in quantum mechanics we can
derive some new property of mother wavelet \cite{r31}. Taking $\left \vert
f_{1}\right \rangle =\left \vert x\right \rangle $, $\left \vert f_{2}%
\right \rangle =\left \vert x^{\prime}\right \rangle $ in (\ref{15.16}) one can
see that%
\begin{equation}
\int_{-\infty}^{\infty}\frac{d\mu}{\mu^{2}\left \vert \mu \right \vert }%
\int_{-\infty}^{\infty}ds\psi \left(  \frac{x-s}{\mu}\right)  \psi^{\ast
}\left(  \frac{x^{\prime}-s}{\mu}\right)  =2C_{\psi}\delta \left(  x-x^{\prime
}\right)  , \label{15.37}%
\end{equation}
which is a new orthogonal property of mother wavelet in parameter space
spanned by $\left(  \mu,s\right)  $. In a similar way, we take $\left \vert
f_{1}\right \rangle =\left \vert f_{2}\right \rangle =\left \vert n\right \rangle
,$ a number state, since $\left \langle n\right.  \left \vert n\right \rangle
=1,$ then we have
\begin{equation}
\int_{-\infty}^{\infty}\frac{d\mu}{\mu^{2}}\int_{-\infty}^{\infty
}ds|\left \langle \psi \right \vert U\left(  \mu,s\right)  \left \vert
n\right \rangle |^{2}=2C_{\psi}, \label{15.38}%
\end{equation}
or take $\left \vert f_{1}\right \rangle =\left \vert f_{2}\right \rangle
=\left \vert z\right \rangle ,$ $\left \vert z\right \rangle =\exp \left(
-\left \vert z\right \vert ^{2}/2+za^{\dagger}\right)  \left \vert 0\right \rangle
$ is the coherent state, then
\begin{equation}
\int_{-\infty}^{\infty}\frac{d\mu}{\mu^{2}}\int_{-\infty}^{\infty
}ds|\left \langle \psi \right \vert U\left(  \mu,s\right)  \left \vert
z\right \rangle |^{2}=2C_{\psi}. \label{15.39}%
\end{equation}
This indicates that $C_{\psi}$ is $\left \vert f_{1}\right \rangle
$-independent, which coincides with the expression in (\ref{15.31}). Next, we
consider a special example. When the mother wavelet is the Mexican hat
(\ref{15.14}), we have%
\begin{equation}
\psi_{M}\left(  p\right)  \equiv \left \langle p\right.  \left \vert \psi
_{M}\right \rangle =\frac{1}{2}\left(  \left \langle p\right.  \left \vert
0\right \rangle -\sqrt{2}\left \langle p\right.  \left \vert 2\right \rangle
\right)  =\pi^{-1/4}\allowbreak p^{2}e^{-\frac{1}{2}p^{2}}. \label{15.40}%
\end{equation}
where%
\begin{equation}
\left \langle p\right.  \left \vert n\right \rangle =\frac{\left(  -i\right)
^{n}}{\sqrt{2^{n}n!\sqrt{\pi}}}e^{-p^{2}/2}H_{n}\left(  p\right)  .
\label{15.41}%
\end{equation}
Here $H_{n}\left(  p\right)  $ is the single-variable Hermit polynomial
\cite{r32}. Substituting Eq.(\ref{15.40}) into Eq.(\ref{15.31}) we have%
\begin{equation}
C_{\psi}\equiv2\pi \int_{0}^{\infty}\frac{\left \vert \psi_{M}\left(  p\right)
\right \vert ^{2}}{p}dp=\sqrt{\pi}. \label{15.42}%
\end{equation}
Thus, for the Mexican hat wavelet (\ref{15.14}), we see
\begin{equation}
\int_{-\infty}^{\infty}\frac{d\mu}{\mu^{2}\left \vert \mu \right \vert }%
\int_{-\infty}^{\infty}ds\psi_{M}\left(  \frac{x-s}{\mu}\right)  \psi
_{M}^{\ast}\left(  \frac{x^{\prime}-s}{\mu}\right)  =2\sqrt{\pi}\delta \left(
x-x^{\prime}\right)  . \label{15.43}%
\end{equation}
Eq.(\ref{15.43}) can be checked as follows. Using Eq.(\ref{15.14}) and
noticing that $\psi_{M}\left(  x\right)  =\psi_{M}\left(  -x\right)  $, we can
put the left hand side of Eq.(\ref{15.43}) into%
\begin{align}
&  \mathtt{L.H.S.of}\text{ }(\text{\ref{15.43}})\nonumber \\
&  =2\int_{0}^{\infty}du\int_{-\infty}^{\infty}ds\psi_{M}\left(  ux-s\right)
\psi_{M}^{\ast}\left(  ux^{\prime}-s\right) \nonumber \\
&  =\left \{
\begin{array}
[c]{cc}%
0, & x\neq x^{\prime}\\
\frac{3}{2}\int_{0}^{\infty}du\rightarrow \infty, & x=x^{\prime}%
\end{array}
\right.  =\mathtt{R.H.S.of}\text{ }(\ref{15.43}). \label{15.44}%
\end{align}
where we have used the integration formula$\allowbreak$s
\begin{equation}
\int_{-\infty}^{\infty}\left(  1-s^{2}\right)  ^{2}\exp \left(  -s^{2}\right)
ds=\allowbreak \frac{3}{4}\sqrt{\pi}, \label{15.45}%
\end{equation}
and%
\begin{align}
&  \int_{-\infty}^{\infty}\left(  1-s^{2}\right)  \left[  1-\left(
s-b\right)  ^{2}\right]  e^{-s^{2}/2-\left(  b-s\right)  ^{2}/2}ds\nonumber \\
&  =\frac{\sqrt{\pi}}{16}e^{-\frac{b^{2}}{4}}\left[  12+b^{2}\left(
b^{2}-12\right)  \right]  . \label{15.46}%
\end{align}

Next, we examine if the Morlet wavelet obey the formalism (\ref{15.37}). The
the Morlet wavelet is defined as \cite{r33,r34,r35}
\begin{equation}
\psi_{mor}\left(  x\right)  =\pi^{-1/4}\left(  e^{ifx}-e^{-f^{2}/2}\right)
e^{-x^{2}/2}. \label{15.47}%
\end{equation}
Substituting (\ref{15.47}) into the left hand side of (\ref{15.37}) yields%
\begin{align}
I  &  \equiv \int_{-\infty}^{\infty}\frac{d\mu}{\mu^{2}\left \vert
\mu \right \vert }\int_{-\infty}^{\infty}ds\psi_{mor}\left(  \frac{x-s}{\mu
}\right)  \psi_{mor}^{\ast}\left(  \frac{x^{\prime}-s}{\mu}\right) \nonumber \\
&  =\left \{
\begin{array}
[c]{cc}%
0, & x\neq x^{\prime}\\
2\left(  1+e^{-f^{2}}-2e^{-3f^{2}/4}\right)  \int_{0}^{\infty}\frac{d\mu}%
{\mu^{2}}\rightarrow \infty, & x=x^{\prime}%
\end{array}
\right.  . \label{15.48}%
\end{align}
Thus the Morlet wavelet satisfies Eq.(\ref{15.37}).

\subsection{WT and Wigner-Husimi Distribution Function}

Phase space technique has been proved very useful in various branches of
physics. Distribution functions in phase space have been a major topic in
studying quantum mechanics and quantum statistics. Among various phase space
distributions the Wigner function $F_{w}\left(  q,p\right)  $ \cite{r13,r14}
is the most popularly used, since its two marginal distributions lead to
measuring probability density in coordinate space and momentum space,
respectively. But the Wigner distribution function itself is not a probability
distribution due to being both positive and negative. In spite of its some
attractive formal properties, it needs to be improved. To overcome this
inconvenience, the Husimi distribution function $F_{h}\left(  q^{\prime
},p^{\prime}\right)  $ is introduced \cite{r36}, which is defined in a manner
that guarantees it to be nonnegative. Its definition is smoothing out the
Wigner function by averaging over a \textquotedblleft coarse graining"
function,
\begin{equation}
F_{h}\left(  q,p,\kappa \right)  =\int \int_{-\infty}^{\infty}dq^{\prime
}dp^{\prime}F_{w}\left(  q^{\prime},p^{\prime}\right)  \exp \left[
-\kappa \left(  q^{\prime}-q\right)  ^{2}-\frac{\left(  p^{\prime}-p\right)
^{2}}{\kappa}\right]  , \label{15.49}%
\end{equation}
where $\kappa>0$ is the Gaussian spatial width parameter, which is free to be
chosen and which determines the relative resolution in $p$-space versus $q$-space.

In the following, we shall employ the optical wavelet transformation to study
the Husimi distribution function, this is to say, we shall show that the
Husimi distribution function of a quantum state $\left \vert \psi \right \rangle
$ can be obtained by making a WT of the Gaussian function $e^{-x^{2}/2},$
i.e.,
\begin{equation}
\left \langle \psi \right \vert \Delta_{h}\left(  q,p,\kappa \right)  \left \vert
\psi \right \rangle =\frac{e^{-\frac{p^{2}}{\kappa}}}{\sqrt{\pi \kappa}%
}\left \vert \int_{-\infty}^{\infty}dx\psi^{\ast}\left(  \frac{x-s}{\mu
}\right)  e^{-x^{2}/2}\right \vert ^{2}, \label{15.50}%
\end{equation}
where
\begin{equation}
s=\frac{-1}{\sqrt{\kappa}}\left(  \kappa q+ip\right)  ,\text{ }\mu
=\sqrt{\kappa}, \label{15.51}%
\end{equation}
and $\left \langle \psi \right \vert \Delta_{h}\left(  q,p\right)  \left \vert
\psi \right \rangle $ is the Husimi distribution function as well as $\Delta
_{h}\left(  q,p,\kappa \right)  $ is the Husimi operator,
\begin{equation}
\Delta_{h}\left(  q,p,\kappa \right)  =\frac{2\sqrt{\kappa}}{1+\kappa}%
\colon \exp \left \{  \frac{-\kappa \left(  q-Q\right)  ^{2}}{1+\kappa}%
-\frac{\left(  p-P\right)  ^{2}}{1+\kappa}\right \}  \colon, \label{15.52}%
\end{equation}
here $\colon \colon$ denotes normal ordering; $Q$ and $P$ are the coordinate
and the momentum operator.

\textbf{Proof of Eq.(\ref{15.50}). }According to Eqs.(\ref{15.3}) and
(\ref{15.4}), when $\left \vert f\right \rangle $ is the vacuum state
$\left \vert 0\right \rangle $, $e^{-x^{2}/2}=\pi^{1/4}\left \langle x\right.
\left \vert 0\right \rangle $, we see that
\begin{equation}
\pi^{-1/4}\int_{-\infty}^{\infty}\frac{dx}{\sqrt{\mu}}\psi^{\ast}\left(
\frac{x-s}{\mu}\right)  e^{-x^{2}/2}dx=\left \langle \psi \right \vert U\left(
\mu,s\right)  \left \vert 0\right \rangle . \label{15.53}%
\end{equation}
From Eq.(\ref{15.7}) it then follows that
\begin{equation}
U\left(  \mu,s\right)  \left \vert 0\right \rangle =\operatorname{sech}%
^{1/2}\lambda \exp \left[  \frac{-s^{2}}{2\left(  1+\mu^{2}\right)  }%
-\frac{a^{\dagger}s}{\sqrt{2}}\operatorname{sech}\lambda-\frac{a^{\dagger2}%
}{2}\tanh \lambda \right]  \left \vert 0\right \rangle . \label{15.54}%
\end{equation}
Substituting Eq.(\ref{15.51}) and $\tanh \lambda=\frac{\kappa-1}{\kappa+1},$
$\cosh \lambda=\frac{1+\kappa}{2\sqrt{\kappa}}$ into Eq.(\ref{15.54}) yields%
\begin{align}
&  e^{-\frac{p^{2}}{2\kappa}+\frac{ipq}{\kappa+1}}U\left(  \mu=\sqrt{\kappa
},s=-\sqrt{\kappa}q-ip/\sqrt{\kappa}\right)  \left \vert 0\right \rangle
\nonumber \\
&  =\left(  \frac{2\sqrt{\kappa}}{1+\kappa}\right)  ^{1/2}\exp \left \{
\frac{-\kappa q^{2}}{2\left(  1+\kappa \right)  }-\frac{p^{2}}{2\left(
1+\kappa \right)  }\right. \nonumber \\
&  \left.  +\frac{\sqrt{2}a^{\dagger}}{1+\kappa}\left(  \kappa q+ip\right)
+\frac{1-\kappa}{2\left(  1+\kappa \right)  }a^{\dagger2}\right \}  \left \vert
0\right \rangle \left.  \equiv \left \vert p,q\right \rangle _{\kappa}\right.  ,
\label{14.55}%
\end{align}
then the WT of Eq.(\ref{15.53}) can be further expressed as%
\begin{equation}
e^{-\frac{p^{2}}{2\kappa}+\frac{ipq}{\kappa+1}}\int_{-\infty}^{\infty}%
\frac{dx}{\left(  \kappa \pi \right)  ^{1/4}}\psi^{\ast}\left(  \frac{x-s}{\mu
}\right)  e^{-x^{2}/2}=\left \langle \psi \right.  \left \vert p,q\right \rangle
_{\kappa}. \label{15.56}%
\end{equation}

Using normally ordered form of the vacuum state projector $\left \vert
0\right \rangle \left \langle 0\right \vert =\colon e^{-a^{\dagger}a}\colon,$and
the IWOP method as well as Eq.(\ref{15.55}) we have%
\begin{equation}
\left \vert p,q\right \rangle _{\kappa \kappa}\left \langle p,q\right \vert
=\frac{2\sqrt{\kappa}}{1+\kappa}\colon \exp \left[  \frac{-\kappa \left(
q-Q\right)  ^{2}}{1+\kappa}-\frac{\left(  p-P\right)  ^{2}}{1+\kappa}\right]
\colon=\Delta_{h}\left(  q,p,\kappa \right)  . \label{15.57}%
\end{equation}
Now we explain why $\Delta_{h}\left(  q,p,\kappa \right)  $ is the Husimi
operator. Using the formula for converting an operator $A$ into its Weyl
ordering form \cite{r37}%
\begin{align}
A  &  =2\int \frac{d^{2}\beta}{\pi}\left \langle -\beta \right \vert A\left \vert
\beta \right \rangle
\genfrac{}{}{0pt}{}{:}{:}%
\exp \{2\left(  \beta^{\ast}a-a^{\dagger}\beta+a^{\dagger}a\right)  \}%
\genfrac{}{}{0pt}{}{:}{:}%
,\label{15.58}\\
d^{2}\beta &  =d\beta_{1}d\beta_{2},\text{ }\beta=\beta_{1}+i\beta
_{2},\nonumber
\end{align}
where the symbol $%
\genfrac{}{}{0pt}{}{:}{:}%
\genfrac{}{}{0pt}{}{:}{:}%
$ denotes the Weyl ordering, $\left \vert \beta \right \rangle $ is the usual
coherent state, substituting\ Eq.(\ref{15.57}) into Eq.(\ref{15.58}) and
performing the integration by virtue of the technique of integration within a
Weyl ordered product of operators, we obtain%
\begin{equation}
\left \vert p,q\right \rangle _{\kappa \kappa}\left \langle p,q\right \vert =2%
\genfrac{}{}{0pt}{}{:}{:}%
\exp \left[  -\kappa \left(  q-Q\right)  ^{2}-\frac{\left(  p-P\right)  ^{2}%
}{\kappa}\right]
\genfrac{}{}{0pt}{}{:}{:}%
. \label{15.59}%
\end{equation}
This is the Weyl ordering form of $\left \vert p,q\right \rangle _{\kappa \kappa
}\left \langle p,q\right \vert .$ Then according to Weyl quantization scheme
\cite{Weyl} we know the classical corresponding function of a Weyl ordered
operator is obtained by just replacing $Q\rightarrow q^{\prime},P\rightarrow
p^{\prime},$
\begin{equation}%
\genfrac{}{}{0pt}{}{:}{:}%
\exp \left[  -\kappa \left(  q-Q\right)  ^{2}-\frac{\left(  p-P\right)  ^{2}%
}{\kappa}\right]
\genfrac{}{}{0pt}{}{:}{:}%
\rightarrow \exp \left[  -\kappa \left(  q-q^{\prime}\right)  ^{2}-\frac{\left(
p-p^{\prime}\right)  ^{2}}{\kappa}\right]  , \label{15.60}%
\end{equation}
and in this case the Weyl rule is expressed as
\begin{align}
\left \vert p,q\right \rangle _{\kappa \kappa}\left \langle p,q\right \vert  &
=2\int dq^{\prime}dp^{\prime}%
\genfrac{}{}{0pt}{}{:}{:}%
\delta \left(  q^{\prime}-Q\right)  \delta \left(  p^{\prime}-P\right)
\genfrac{}{}{0pt}{}{:}{:}%
\exp \left[  -\kappa \left(  q-q^{\prime}\right)  ^{2}-\frac{\left(
p-p^{\prime}\right)  ^{2}}{\kappa}\right] \nonumber \\
&  =2\int dq^{\prime}dp^{\prime}\Delta_{w}\left(  q^{\prime},p^{\prime
}\right)  \exp \left[  -\kappa \left(  q^{\prime}-q\right)  ^{2}-\frac{\left(
p^{\prime}-p\right)  ^{2}}{\kappa}\right]  , \label{15.61}%
\end{align}
where at the last step we used the Weyl ordering form of the Wigner operator
$\Delta_{w}\left(  q,p\right)  $ \cite{r38}
\begin{equation}
\Delta_{w}\left(  q,p\right)  =%
\genfrac{}{}{0pt}{}{:}{:}%
\delta \left(  q-Q\right)  \delta \left(  p-P\right)
\genfrac{}{}{0pt}{}{:}{:}%
. \label{15.62}%
\end{equation}
In reference to Eq.(\ref{15.49}) in which the relation between the Husimi
function and the WF is shown, we know that the right-hand side of Eq.
(\ref{15.61}) should be just the Husimi operator, i.e.
\begin{align}
\left \vert p,q\right \rangle _{\kappa \kappa}\left \langle p,q\right \vert  &
=2\int dq^{\prime}dp^{\prime}\Delta_{w}\left(  q^{\prime},p^{\prime}\right)
\exp \left[  -\kappa \left(  q^{\prime}-q\right)  ^{2}-\frac{\left(  p^{\prime
}-p\right)  ^{2}}{\kappa}\right] \nonumber \\
&  =\Delta_{h}\left(  q,p,\kappa \right)  , \label{15.63}%
\end{align}
thus Eq. (\ref{15.50}) is proved by combining Eqs.(\ref{15.63}) and
(\ref{15.56}).

Thus the optical WT can be used to study the Husimi distribution function in
quantum optics phase space theory \cite{r39a}.

\section{Complex Wavelet transformation in entangled state representations}

We now turn to 2-dimensional complex wavelet transform (CWT) \cite{r39}.

\subsection{CWT and the condition of Mother Wavelet}

Since wavelet family involves squeezing transform, we recall that the two-mode
squeezing operator has a natural representation in the entangled state
representation (ESR), $\exp[\lambda \left(  a_{1}^{\dagger}a_{2}^{\dagger
}-a_{1}a_{2}\right)  ]=\frac{1}{\mu}\int_{-\infty}^{\infty}\left \vert
\frac{\eta}{\mu}\right \rangle \left \langle \eta \right \vert dx,$ $\mu
=e^{\lambda}$, thus we are naturally led to studying 2-dimensional CWT in ESR.
Using ESR we can derive some new results more conveniently than using the
direct-product of two single-particle coordinate eigenstates. To be concrete,
we impose the condition on qualified mother wavelets also in $\left \vert
\eta \right \rangle $ representation,
\begin{equation}
\int_{-\infty}^{\infty}\frac{d^{2}\eta}{2\pi}\psi \left(  \eta \right)  =0,
\label{16.1}%
\end{equation}
where $\psi \left(  \eta \right)  =$ $\left \langle \eta \right \vert \left.
\psi \right \rangle .$ Thus we see
\begin{equation}
\int_{-\infty}^{\infty}\frac{d^{2}\eta}{2\pi}\left \vert \eta \right \rangle
=\exp \{-a_{1}^{\dagger}a_{2}^{\dagger}\} \left \vert 00\right \rangle
=\left \vert \xi=0\right \rangle , \label{16.2}%
\end{equation}
and the condition (\ref{16.1}) becomes
\begin{equation}
\left \langle \xi=0\right \vert \left.  \psi \right \rangle =0. \label{16.3}%
\end{equation}
Without loss of generality, assuming
\begin{equation}
\left \vert \psi \right \rangle =\sum_{n,m=0}^{\infty}K_{n,m}a_{1}^{\dagger
n}a_{2}^{\dagger m}\left \vert 00\right \rangle , \label{16.4}%
\end{equation}
then using the two-mode coherent $\left \vert z_{1}z_{2}\right \rangle $ state
we can write (\ref{16.3}) as
\begin{align}
\left \langle \xi=0\right \vert \left.  \psi \right \rangle  &  =\left \langle
\xi=0\right \vert \int \frac{d^{2}z_{1}d^{2}z_{2}}{\pi^{2}}\left \vert z_{1}%
z_{2}\right \rangle \left \langle z_{1}z_{2}\right \vert \sum_{n,m=0}^{\infty
}K_{n,m}a_{1}^{\dagger n}a_{2}^{\dagger m}\left \vert 00\right \rangle
\nonumber \\
\  &  =\sum_{n,m=0}^{\infty}K_{n,m}\int \frac{d^{2}z_{1}d^{2}z_{2}}{\pi^{2}%
}z_{1}^{\ast n}z_{2}^{\ast m}\exp \left[  -|z_{1}|^{2}-|z_{2}|^{2}-z_{1}%
z_{2}\right] \nonumber \\
\  &  =\sum_{n,m=0}^{\infty}K_{n,m}\int \frac{d^{2}z_{2}}{\pi}\exp \left[
-|z_{2}|^{2}\right]  z_{2}^{n}z_{2}^{\ast m}\left(  -1\right)  ^{n}\nonumber \\
&  =\sum_{n=0}^{\infty}n!K_{n,n}\left(  -1\right)  ^{n}=0, \label{16.5}%
\end{align}
this is the constraint on the coefficient $K_{n,n}$ in (\ref{16.5}), i.e., the
admissibility condition for $\left \vert \psi \right \rangle $. Thus Eq.
(\ref{16.4}) is in the form:%
\begin{equation}
\left \vert \psi \right \rangle =\sum_{n=0}^{\infty}n!K_{n,n}\left \vert
n,n\right \rangle . \label{16.6}%
\end{equation}
To derive the qualified mother wavelet $\psi \left(  \eta \right)  =$
$\left \langle \eta \right \vert \left.  \psi \right \rangle $ from $\left \vert
\psi \right \rangle $, noticing Eq.(\ref{13.9}) and (\ref{16.6}) we have%
\begin{align}
\psi \left(  \eta \right)   &  =e^{-\left \vert \eta \right \vert ^{2}/2}\sum
_{n=0}^{\infty}K_{n,n}H_{n,n}\left(  \eta^{\ast},\eta \right)  \left(
-1\right)  ^{n}\nonumber \\
&  =e^{-\left \vert \eta \right \vert ^{2}/2}\sum_{n=0}^{\infty}n!K_{n,n}%
L_{n}\left(  \left \vert \eta \right \vert ^{2}\right)  , \label{16.7}%
\end{align}
where $L_{n}\left(  x\right)  $ is the\ Laguerre polynomial. In this case, we
may name the wavelet in Eq. (\ref{16.7}) as the Laguerre--Gaussian mother
wavelets, analogous to the name of Laguerre--Gaussian modes in optical
propagation. For example:

(1) When taking $K_{0,0}=\frac{1}{2},$ $K_{1,1}=\frac{1}{2},$ $K_{n,n}=0$ for
$n\geqslant2,$ so we see
\begin{equation}
\left \vert \psi \right \rangle _{1}=\frac{1}{2}\left(  1+a_{1}^{\dagger}%
a_{2}^{\dagger}\right)  \left \vert 00\right \rangle , \label{16.8}%
\end{equation}
which differs from the direct-product state $\left(  1-a_{1}^{\dagger
2}\right)  \left \vert 0\right \rangle _{1}\otimes \left(  1-a_{2}^{\dagger
2}\right)  \left \vert 0\right \rangle _{2}$. It then follows from Eq.
(\ref{16.7}) that
\begin{equation}
\psi_{1}\left(  \eta \right)  \equiv \frac{1}{2}\left \langle \eta \right \vert
\left(  \left \vert 00\right \rangle +\left \vert 11\right \rangle \right)
=e^{-\frac{1}{2}\left \vert \eta \right \vert ^{2}}\{1-\frac{1}{2}\left \vert
\eta \right \vert ^{2}\}, \label{16.9}%
\end{equation}
which differs from $e^{-\left(  x^{2}+y^{2}\right)  /2}(1-x^{2})\left(
1-y^{2}\right)  $, the direct product of two 1D Mexican hat wavelets (see also
the difference between Figs. 9 and 10). \begin{figure}[ptb]
\centering \includegraphics[width=3.1in, height=3in,
angle=0]{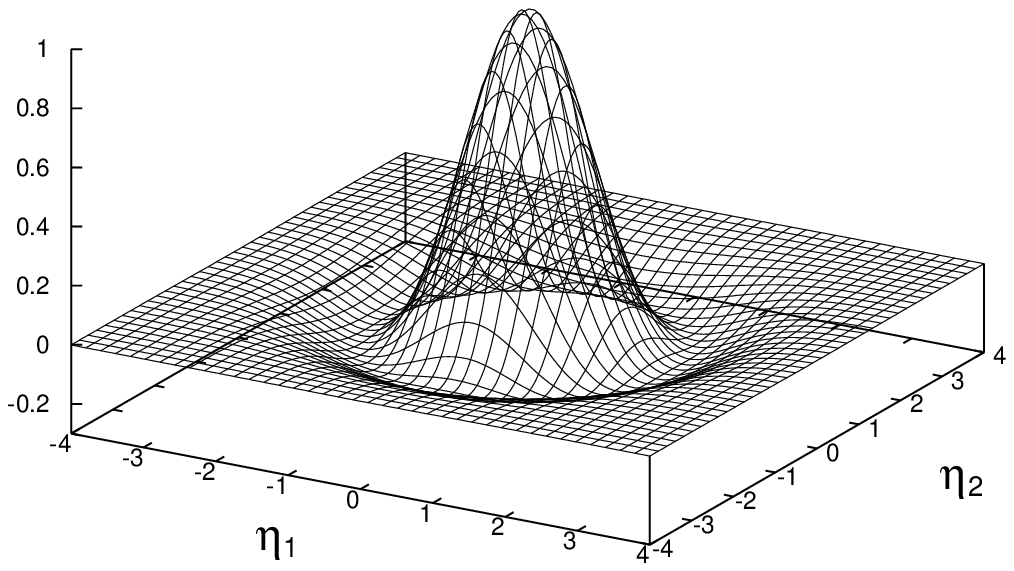}\caption{{The Laguree-Gaussian mother wavelet }$\psi
_{1}\left(  \eta \right)  ${.}}%
\label{Fig9}%
\end{figure}

\begin{figure}[ptb]
\centering \includegraphics[width=3.1in, height=3in,
angle=0]{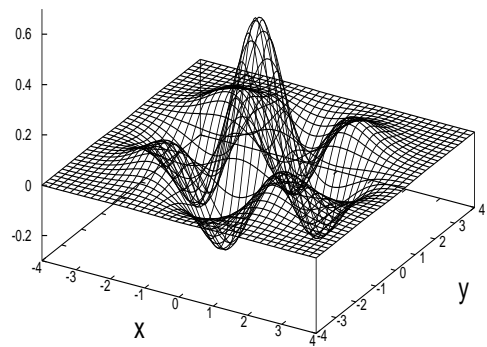}\caption{{2D Mexican hat mother wavelet (Hermite
Gaussian mother wavelet).}}%
\label{Fig10}%
\end{figure}

(2) when $K_{0,0}=1$, $K_{1,1}=3$, $K_{2,2}=1$, $K_{n,n}=0$ for $n\geqslant3,$
we have (see Fig. 11)%
\begin{equation}
\psi_{2}\left(  \eta \right)  \equiv \left(  6-7\left \vert \eta \right \vert
^{2}+\left \vert \eta \right \vert ^{4}\right)  e^{-\frac{1}{2}\left \vert
\eta \right \vert ^{2}}. \label{16.10}%
\end{equation}
\begin{figure}[ptb]
\centering \includegraphics[width=3.1in, height=3in,
angle=0]{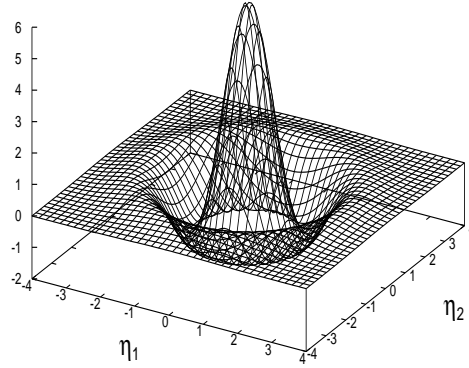}\caption{Laguerre-Gaussian {mother wavelet }$\psi
_{2}\left(  \eta \right)  .$}%
\label{Fig11}%
\end{figure}

(3) when $K_{0,0}=1$, $K_{1,1}=1$, $K_{2,2}=3$, $K_{3,3}=3$, $K_{n,n}=0$ for
$n\geqslant4,$ the mother wavelet $\psi_{3}\left(  \eta \right)  $ (see Fig.
12) reads
\begin{equation}
\psi_{3}\left(  \eta \right)  =\left(  14-31\left \vert \eta \right \vert
^{2}+12\left \vert \eta \right \vert ^{4}-\left \vert \eta \right \vert ^{6}\right)
e^{-\frac{1}{2}\left \vert \eta \right \vert ^{2}}. \label{16.11}%
\end{equation}
\begin{figure}[ptb]
\centering \includegraphics[width=3.1in, height=3in,
angle=0]{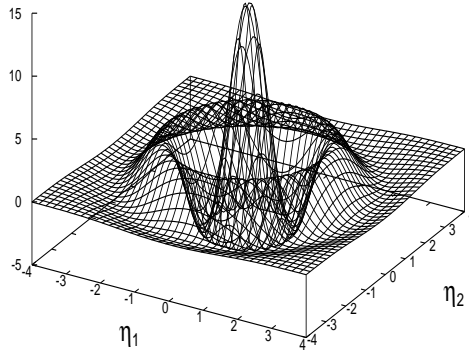}\caption{Laguerre-Gaussian {mother wavelet }$\psi
_{3}\left(  \eta \right)  .$}%
\end{figure}From the figures we can see that as long as the coefficients
$K_{n,n}$ satisfy condition (\ref{16.7}), we can construct arbitrary complex
mother wavelet by adding or reducing the number of coefficients, or by
adjusting the value of them. And since only $K_{n,m}$ ($m=n$) survive in all
the coefficients, the mother wavelets obtained are all circularly symmetric on
the complex plane.

Moreover, the CWT of a signal function $F\left(  \eta \right)  $ by $\Psi$ is
defined by
\begin{equation}
W_{\psi}F\left(  \mu,\kappa \right)  =\frac{1}{\mu}\int \frac{d^{2}\eta}{\pi
}F\left(  \eta \right)  \psi^{\ast}\left(  \frac{\eta-\kappa}{\mu}\right)  .
\label{16.12}%
\end{equation}
Using the $\left \langle \eta \right \vert $ representation we can treat it from
the quantum mechanically,
\begin{equation}
W_{\psi}F\left(  \mu,\kappa \right)  =\frac{1}{\mu}\int \frac{d^{2}\eta}{\pi
}\left \langle \psi \right \vert \left.  \frac{\eta-\kappa}{\mu}\right \rangle
\left \langle \eta \right \vert \left.  F\right \rangle =\left \langle
\Psi \right \vert U_{2}\left(  \mu,\kappa \right)  \left \vert F\right \rangle ,
\label{16.13}%
\end{equation}
where
\begin{equation}
U_{2}\left(  \mu,\kappa \right)  \equiv \frac{1}{\mu}\int \frac{d^{2}\eta}{\pi
}\left \vert \frac{\eta-\kappa}{\mu}\right \rangle \left \langle \eta \right \vert
,\; \mu=e^{\lambda}, \label{16.14}%
\end{equation}
is the two-mode squeezing-displacing operator. Using the IWOP technique we can
calculate its normally ordered form,%
\begin{align}
U_{2}\left(  \mu,\kappa \right)   &  =\operatorname{sech}\lambda \colon \exp \{
\left(  a_{1}^{\dagger}a_{2}^{\dagger}-a_{1}a_{2}\right)  \tanh \lambda+\left(
\operatorname{sech}\lambda-1\right)  \left(  a_{1}^{\dagger}a_{1}%
+a_{2}^{\dagger}a_{2}\right) \nonumber \\
&  +\frac{1}{2}\left(  \sigma^{\ast}a_{2}^{\dagger}-\sigma a_{1}^{\dagger
}\right)  \operatorname{sech}\lambda+\frac{1}{1+\mu^{2}}\left(  \kappa^{\ast
}a_{1}-\kappa a_{2}-\frac{1}{2}\left \vert \kappa \right \vert ^{2}\right)  \}
\colon. \label{16.15}%
\end{align}
When $\kappa=0,$ it reduces to the usual normally ordered two-mode squeezing
operator. Once the state vector $_{M}\left \langle \Psi \right \vert $
corresponding to mother wavelet is known, for any state $\left \vert
F\right \rangle $ the matrix element $_{M}\left \langle \Psi \right \vert
U_{2}\left(  \mu,\kappa \right)  \left \vert F\right \rangle $ is just the
wavelet transform of $F(\eta)$ with respect to $_{M}\left \langle
\Psi \right \vert .$ Therefore, various quantum optical field states can then be
analyzed by their wavelet transforms.

\subsection{Parseval Theorem in CWT}

In order to complete the CWT theory, we must ask if the corresponding Parseval
theorem exists \cite{r40}. This is important since the inversion formula of
CWT may appear as a lemma of this theorem. Noting that CWT involves two-mode
squeezing transform, so the corresponding Parseval theorem differs from that
of the direct-product of two 1D wavelet transforms, too.

Next let us prove the Parseval theorem for CWT,%
\begin{equation}
\int_{0}^{\infty}\frac{d\mu}{\mu^{3}}\int \frac{d^{2}\kappa}{\pi}W_{\psi}%
g_{1}\left(  \mu,\kappa \right)  W_{\psi}^{\ast}g_{2}\left(  \mu,\kappa \right)
=C_{\psi}^{\prime}\int \frac{d^{2}\eta}{\pi}g_{2}^{\ast}\left(  \eta \right)
g_{1}\left(  \eta \right)  , \label{16.16}%
\end{equation}
where $\kappa=\kappa_{1}+i\kappa_{2},$ and
\begin{equation}
C_{\psi}^{\prime}=4\int_{0}^{\infty}\frac{d\left \vert \xi \right \vert
}{\left \vert \xi \right \vert }\left \vert \psi \left(  \xi \right)  \right \vert
^{2}. \label{16.17}%
\end{equation}
$\psi \left(  \xi \right)  $ is the complex Fourier transform of $\psi \left(
\eta \right)  ,\psi \left(  \xi \right)  =\left \langle \xi \right \vert \left.
\psi \right \rangle =\int_{-\infty}^{\infty}\frac{d^{2}\eta}{\pi}\left \langle
\xi \right \vert \left.  \eta \right \rangle \left \langle \eta \right \vert \left.
\psi \right \rangle $. According to (\ref{16.13}) and (\ref{16.14}) the quantum
mechanical version of Parseval theorem should be%
\begin{equation}
\int_{0}^{\infty}\frac{d\mu}{\mu^{3}}\int \frac{d^{2}\kappa}{\pi}\left \langle
\psi \right \vert U_{2}\left(  \mu,\kappa \right)  \left \vert g_{1}\right \rangle
\left \langle g_{2}\right \vert U_{2}^{\dagger}\left(  \mu,\kappa \right)
\left \vert \psi \right \rangle =C_{\psi}^{\prime}\left \langle g_{2}\right.
\left \vert g_{1}\right \rangle . \label{16.18}%
\end{equation}
Eq.(\ref{16.18}) indicates that once the state vector $\left \langle
\psi \right \vert $ corresponding to mother wavelet is known, for any two states
$\left \vert g_{1}\right \rangle $ and $\left \vert g_{2}\right \rangle $, their
overlap up to the factor $C_{\psi}$ (determined by (\ref{16.17})) is just
their corresponding overlap of CWTs in the ($\mu,\kappa$) parametric space.

Next we prove Eq.(\textbf{\ref{16.16}) or (\ref{16.18}). }In the same
procedure as\textbf{ }the proof of Eq.(\ref{15.17})\textbf{. }We start with
calculating $U_{2}^{\dagger}\left(  \mu,\kappa \right)  \left \vert
\xi \right \rangle .$ Using (\ref{3.20}) and (\ref{16.14}), we have%
\begin{equation}
U_{2}^{\dagger}\left(  \mu,\kappa \right)  \left \vert \xi \right \rangle
=\frac{1}{\mu}\left \vert \frac{\xi}{\mu}\right \rangle e^{\frac{i}{\mu}\left(
\xi_{1}\kappa_{2}-\xi_{2}\kappa_{1}\right)  }, \label{16.19}%
\end{equation}
it then follows%
\begin{align}
&  \int \frac{d^{2}\kappa}{\pi}U_{2}^{\dagger}\left(  \mu,\kappa \right)
\left \vert \xi^{\prime}\right \rangle \left \langle \xi \right \vert U_{2}\left(
\mu,\kappa \right) \nonumber \\
&  =\frac{1}{\mu^{2}}\int \frac{d^{2}\kappa}{\pi}e^{\frac{i}{\mu}\left[
\left(  \xi_{1}^{\prime}-\xi_{1}\right)  \kappa_{2}+\left(  \xi_{2}-\xi
_{2}^{\prime}\right)  \kappa_{1}\right]  }\left \vert \frac{\xi^{\prime}}{\mu
}\right \rangle \left \langle \frac{\xi}{\mu}\right \vert \nonumber \\
&  =4\pi \left \vert \frac{\xi}{\mu}\right \rangle \left \langle \frac{\xi}{\mu
}\right \vert \delta \left(  \xi_{1}^{\prime}-\xi_{1}\right)  \delta \left(
\xi_{2}-\xi_{2}^{\prime}\right)  . \label{16.20}%
\end{align}
Using the completeness of $\left \vert \xi \right \rangle $ and (\ref{16.20}) the
left-hand side (LHS) of (\ref{16.18}) can be reformed as%
\begin{align}
&  \text{LHS of Eq.(\ref{16.18})}\nonumber \\
&  =\int_{0}^{\infty}\frac{d\mu}{\mu^{3}}\int \frac{d^{2}\kappa d^{2}\xi
d^{2}\xi^{\prime}}{\pi^{3}}\left \langle \psi \right \vert \left.  \xi
\right \rangle \nonumber \\
&  \times \left \langle \xi \right \vert U_{2}\left(  \mu,\kappa \right)
\left \vert g_{1}\right \rangle \left \langle g_{2}\right \vert U_{2}^{\dagger
}\left(  \mu,\kappa \right)  \left \vert \xi^{\prime}\right \rangle \left \langle
\xi^{\prime}\right \vert \left.  \psi \right \rangle \nonumber \\
&  =4\int_{0}^{\infty}\frac{d\mu}{\mu^{3}}\int \frac{d^{2}\xi}{\pi}\left \vert
\psi \left(  \xi \right)  \right \vert ^{2}\left \langle g_{2}\right.  \left \vert
\frac{\xi}{\mu}\right \rangle \left \langle \frac{\xi}{\mu}\right.  \left \vert
g_{1}\right \rangle \nonumber \\
&  =\int \frac{d^{2}\xi}{\pi}\left \{  4\int_{0}^{\infty}\frac{d\mu}{\mu
}\left \vert \psi \left(  \mu \xi \right)  \right \vert ^{2}\right \}  \left \langle
g_{2}\right.  \left \vert \xi \right \rangle \left \langle \xi \right.  \left \vert
g_{1}\right \rangle , \label{16.21}%
\end{align}
where the integration value in $\{..\}$ is actually $\xi-$independent. Noting
that the mother wavelet $\psi \left(  \eta \right)  $ in Eq.(\ref{16.7}) is just
the function of $\left \vert \eta \right \vert ,$ so $\psi \left(  \xi \right)  $
is also the function of $\left \vert \xi \right \vert .$ In fact, using
Eqs.(\ref{16.7}) and (\ref{3.20}), we have
\begin{equation}
\psi \left(  \xi \right)  =e^{-1/2\left \vert \xi \right \vert ^{2}}\sum
_{n=0}^{\infty}K_{n,n}H_{n,n}\left(  \left \vert \xi \right \vert ,\left \vert
\xi \right \vert \right)  , \label{16.22}%
\end{equation}
where we have used the integral formula%
\begin{equation}
\int \frac{d^{2}z}{\pi}e^{\zeta \left \vert z\right \vert ^{2}+\xi z+\eta z^{\ast
}}=-\frac{1}{\zeta}e^{-\frac{\xi \eta}{\zeta}},\text{Re}\left(  \zeta \right)
<0. \label{16.23}%
\end{equation}
So we can rewrite (\ref{16.21}) as
\begin{equation}
\text{LHS of (\ref{16.18})}=C_{\psi}^{\prime}\int \frac{d^{2}\xi}{\pi
}\left \langle g_{2}\right.  \left \vert \xi \right \rangle \left \langle
\xi \right.  \left \vert g_{1}\right \rangle =C_{\psi}^{\prime}\left \langle
g_{2}\right.  \left \vert g_{1}\right \rangle , \label{16.24}%
\end{equation}
where%
\begin{equation}
C_{\psi}^{\prime}=4\int_{0}^{\infty}\frac{d\mu}{\mu}\left \vert \psi \left(
\mu \xi \right)  \right \vert ^{2}=4\int_{0}^{\infty}\frac{d\left \vert
\xi \right \vert }{\left \vert \xi \right \vert }\left \vert \psi \left(  \xi \right)
\right \vert ^{2}. \label{16.25}%
\end{equation}
Then we have completed the proof of the Parseval theorem for CWT in
(\ref{16.18}). Here, we should emphasize that (\ref{16.18}) is not only
different from the product of two 1D WTs, but also different from the usual WT
in 2D.

When $\left \vert g_{2}\right \rangle =\left \vert \eta \right \rangle ,$ by using
(\ref{16.14}) we see $\left \langle \eta \right \vert U_{2}^{\dagger}\left(
\mu,\kappa \right)  \left \vert \psi \right \rangle =\frac{1}{\mu}\psi \left(
\frac{\eta-\kappa}{\mu}\right)  ,$ then substituting it into (\ref{16.18})
yields%
\begin{equation}
g_{1}\left(  \eta \right)  =\frac{1}{C_{\psi}^{\prime}}\int_{0}^{\infty}%
\frac{d\mu}{\mu^{3}}\int \frac{d^{2}\kappa}{\pi \mu}W_{\psi}g_{1}\left(
\mu,\kappa \right)  \psi \left(  \frac{\eta-\kappa}{\mu}\right)  , \label{16.26}%
\end{equation}
which is just the inverse transform of the CWT. Especially, when $\left \vert
g_{1}\right \rangle =$ $\left \vert g_{2}\right \rangle ,$ Eq. (\ref{16.18})
reduces to%
\begin{align}
\int_{0}^{\infty}\frac{d\mu}{\mu^{3}}\int \frac{d^{2}\kappa}{\pi}\left \vert
W_{\psi}g_{1}\left(  \mu,\kappa \right)  \right \vert ^{2}  &  =C_{\psi}%
^{\prime}\int \frac{d^{2}\eta}{\pi}\left \vert g_{1}\left(  \eta \right)
\right \vert ^{2},\nonumber \\
\text{or }\int_{0}^{\infty}\frac{d\mu}{\mu^{3}}\int \frac{d^{2}\kappa}{\pi
}\left \vert \left \langle \psi \right \vert U_{2}\left(  \mu,\kappa \right)
\left \vert g_{1}\right \rangle \right \vert ^{2}  &  =C_{\psi}^{\prime
}\left \langle g_{1}\right.  \left \vert g_{1}\right \rangle , \label{16.27}%
\end{align}
which is named isometry of energy.

\subsection{Orthogonal property of mother wavelet in parameter space}

On the other hand, when $\left \vert g_{1}\right \rangle =\left \vert
\eta \right \rangle ,$ $\left \vert g_{2}\right \rangle =\left \vert \eta^{\prime
}\right \rangle $, Eq.(\ref{16.18}) becomes%
\begin{equation}
\frac{1}{C_{\psi}^{\prime}}\int_{0}^{\infty}\frac{d\mu}{\mu^{5}}\int
\frac{d^{2}\kappa}{\pi}\psi \left(  \frac{\eta^{\prime}-\kappa}{\mu}\right)
\psi^{\ast}\left(  \frac{\eta-\kappa}{\mu}\right)  =\pi \delta^{(2)}\left(
\eta-\eta^{\prime}\right)  , \label{16.28}%
\end{equation}
which is a new orthogonal property of mother wavelet in parameter space
spanned by $\left(  \mu,\kappa \right)  $. In a similar way, we take
$\left \vert g_{1}\right \rangle =\left \vert g_{2}\right \rangle =\left \vert
m,n\right \rangle ,$ a two-mode number state, since $\left \langle m,n\right.
\left \vert m,n\right \rangle =1,$ then we have
\begin{equation}
\int_{0}^{\infty}\frac{d\mu}{\mu^{3}}\int \frac{d^{2}\kappa}{\pi}\left \vert
\left \langle \psi \right \vert U_{2}\left(  \mu,\kappa \right)  \left \vert
m,n\right \rangle \right \vert ^{2}=C_{\psi}^{\prime}, \label{16.29}%
\end{equation}
or take $\left \vert g_{1}\right \rangle =\left \vert g_{2}\right \rangle
=\left \vert z_{1},z_{2}\right \rangle ,$ $\left \vert z\right \rangle
=\exp \left(  -\left \vert z\right \vert ^{2}/2+za^{\dagger}\right)  \left \vert
0\right \rangle $ is the coherent state, then
\begin{equation}
\int_{0}^{\infty}\frac{d\mu}{\mu^{3}}\int \frac{d^{2}\kappa}{\pi}\left \vert
\left \langle \psi \right \vert U_{2}\left(  \mu,\kappa \right)  \left \vert
z_{1},z_{2}\right \rangle \right \vert ^{2}=C_{\psi}^{\prime}. \label{16.30}%
\end{equation}

Next we examine a special example. When the mother wavelet is $\psi_{1}\left(
\eta \right)  $ in (\ref{16.9}), using (\ref{3.20}) we have $\psi \left(
\xi \right)  =\frac{1}{2}\left \vert \xi \right \vert ^{2}e^{-\frac{1}%
{2}\left \vert \xi \right \vert ^{2}},$ which leads to $C_{\psi}^{\prime}%
=\int_{0}^{\infty}\left \vert \xi \right \vert ^{3}e^{-\left \vert \xi \right \vert
^{2}}d\left \vert \xi \right \vert =\frac{1}{2}.$ Thus for $\psi_{1}\left(
\eta \right)  $, we see%
\begin{equation}
2\int_{0}^{\infty}\frac{d\mu}{\mu^{5}}\int \frac{d^{2}\kappa}{\pi}\psi
_{1}\left(  \frac{\eta^{\prime}-\kappa}{\mu}\right)  \psi_{1}^{\ast}\left(
\frac{\eta-\kappa}{\mu}\right)  =\pi \delta^{(2)}\left(  \eta-\eta^{\prime
}\right)  . \label{16.31}%
\end{equation}
Eq. (\ref{16.31}) can be checked as follows. Using (\ref{16.9}) and the
integral formula%
\begin{align}
&  \int_{0}^{\infty}u\left(  1-\frac{ux^{2}}{2}\right)  \left(  1-\frac
{uy^{2}}{2}\right)  e^{-u\frac{x^{2}+y^{2}}{2}}du\nonumber \\
&  =-\frac{4(x^{4}-4x^{2}y^{2}+y^{4})}{(x^{2}+y^{2})^{4}},\text{
}\operatorname{Re}\left(  x^{2}+y^{2}\right)  >0, \label{16.32}%
\end{align}
we can put the left-hand side (LHS) of (\ref{16.31}) into
\begin{equation}
\text{LHS of (\ref{16.31})}=-\int \frac{d^{2}\kappa}{\pi}\frac{4(x^{4}%
-4x^{2}y^{2}+y^{4})}{(x^{2}+y^{2})^{4}}, \label{16.33}%
\end{equation}
where $x^{2}=\left \vert \eta^{\prime}-\kappa \right \vert ^{2},$ $y^{2}%
=\left \vert \eta-\kappa \right \vert ^{2}.$

When $\eta^{\prime}=\eta,$ $x^{2}=y^{2},$%
\begin{equation}
\text{LHS of (\ref{16.31})}=\int \allowbreak \frac{d^{2}\kappa}{2\pi \left \vert
\kappa-\eta \right \vert ^{4}}=\int_{0}^{\infty}\allowbreak \int_{0}^{2\pi}%
\frac{drd\theta}{2\pi r^{3}}\rightarrow \infty. \label{16.34}%
\end{equation}
On the other hand, when $\eta \neq \eta^{\prime}$ and noticing that
\begin{align}
x^{2}  &  =\left(  \eta_{1}^{\prime}-\kappa_{1}\right)  ^{2}+\left(  \eta
_{2}^{\prime}-\kappa_{2}\right)  ^{2},\nonumber \\
y^{2}  &  =\left(  \eta_{1}-\kappa_{1}\right)  ^{2}+\left(  \eta_{2}%
-\kappa_{2}\right)  ^{2}, \label{16.35}%
\end{align}
which leads to $\mathtt{d}x^{2}\mathtt{d}y^{2}=4\left \vert J\right \vert
\mathtt{d}\kappa_{1}\mathtt{d}\kappa_{2}$, where $J\left(  x,y\right)
=\left \vert
\begin{array}
[c]{cc}%
\kappa_{1}-\eta_{1}^{\prime} & \kappa_{2}-\eta_{2}^{\prime}\\
\kappa_{1}-\eta_{1} & \kappa_{2}-\eta_{2}%
\end{array}
\right \vert $. As a result of (\ref{16.35}), (\ref{16.33}) reduces to
\begin{equation}
\text{LHS of (\ref{16.31})}=-4\int_{-\infty}^{\infty}\frac{dxdy}{\pi}%
\frac{xy(x^{4}-4x^{2}y^{2}+y^{4})}{\left \vert J\right \vert (x^{2}+y^{2})^{4}%
}=0\text{,} \label{16.36}%
\end{equation}
where we have noticed that $J\left(  x,y\right)  $ is the funtion of $\left(
x^{2},y^{2}\right)  .$ Thus we have%
\begin{equation}
\text{LHS of (\ref{16.31})}=\left \{
\begin{array}
[c]{cc}%
\infty, & \eta=\eta^{\prime},\\
0, & \eta \neq \eta^{\prime}.
\end{array}
\right.  =\text{RHS of (\ref{16.31}).} \label{16.37}%
\end{equation}

\subsection{CWT and Entangled Husimi distribution}

Recalling that in Ref.\cite{r41}, the so-called entangled Husimi operator
$\Delta_{h}\left(  \sigma,\gamma,\kappa \right)  $ has been introduced, which
is endowed with definite physical meaning, and it is found that the two-mode
squeezed coherent state $\left \vert \sigma,\gamma \right \rangle _{\kappa}$
representation of $\Delta_{h}\left(  \sigma,\gamma,\kappa \right)  ,$
$\Delta_{h}\left(  \sigma,\gamma,\kappa \right)  =$ $\left \vert \sigma
,\gamma,\kappa \right \rangle \left \langle \sigma,\gamma,\kappa \right \vert $.
The entangled Husimi operator $\Delta_{h}\left(  \sigma,\gamma,\kappa \right)
$ and the entangled Husimi distribution $F_{h}\left(  \sigma,\gamma
,\kappa \right)  $ of quantum state $\left \vert \psi \right \rangle $ are given
by%
\begin{equation}
\Delta_{h}\left(  \sigma,\gamma,\kappa \right)  =4\int d^{2}\sigma^{\prime
}d^{2}\gamma^{\prime}\Delta_{w}\left(  \sigma^{\prime},\gamma^{\prime}\right)
\exp \left \{  -\kappa \left \vert \sigma^{\prime}-\sigma \right \vert ^{2}-\frac
{1}{\kappa}\left \vert \gamma^{\prime}-\gamma \right \vert ^{2}\right \}  ,
\label{e4}%
\end{equation}
and
\begin{equation}
F_{h}\left(  \sigma,\gamma,\kappa \right)  =4\int d^{2}\sigma^{\prime}%
d^{2}\gamma^{\prime}F_{w}\left(  \sigma^{\prime},\gamma^{\prime}\right)
\exp \left \{  -\kappa \left \vert \sigma^{\prime}-\sigma \right \vert ^{2}-\frac
{1}{\kappa}\left \vert \gamma^{\prime}-\gamma \right \vert ^{2}\right \}  ,
\label{e5}%
\end{equation}
respectively, where $F_{w}\left(  \sigma^{\prime},\gamma^{\prime}\right)
=\left \langle \psi \right \vert \Delta_{w}\left(  \sigma^{\prime},\gamma
^{\prime}\right)  \left \vert \psi \right \rangle $ is two-mode Wigner function,
with $\Delta_{w}\left(  \sigma^{\prime},\gamma^{\prime}\right)  $\ being the
two-mode Wigner operator. Thus we are naturally led to studying the entangled
Husimi distribution function from the viewpoint of wavelet transformation. In
this subsection, we shall extend the relation between wavelet transformation
and Wigner-Husimi distribution function to the entangled case, that is to say,
we employ the CWT to investigate the entangled Husimi distribution function
(EHDF) by bridging the relation between CWT and EHDF. This is a convenient
approach for calculating various entangled Husimi distribution functions of
miscellaneous two-mode quantum states.

\subsubsection{CWT and its quantum mechanical version}

In Ref.\cite{r39}, the CWT has been proposed, i.e., the CWT of a complex
signal function $g\left(  \eta \right)  $ by $\psi$ is defined by
\begin{equation}
W_{\psi}g\left(  \mu,z\right)  =\frac{1}{\mu}\int \frac{d^{2}\eta}{\pi}g\left(
\eta \right)  \psi^{\ast}\left(  \frac{\eta-z}{\mu}\right)  , \label{e6}%
\end{equation}
whose admissibility condition for mother wavelets, $\int \frac{d^{2}\eta}{2\pi
}\psi \left(  \eta \right)  =0,$ is examined in the entangled state
representations $\left \langle \eta \right \vert $ and a family of new mother
wavelets (named the Laguerre--Gaussian wavelets) are found to match the CWT
\cite{r39}. In fact, by introducing the bipartite entangled state
representation $\left \langle \eta=\eta_{1}+\mathtt{i}\eta_{2}\right \vert ,$we
can treat (\ref{e5}) quantum mechanically,
\begin{equation}
W_{\psi}g\left(  \mu,z\right)  =\frac{1}{\mu}\int \frac{d^{2}\eta}{\pi
}\left \langle \psi \right \vert \left.  \frac{\eta-z}{\mu}\right \rangle
\left \langle \eta \right \vert \left.  g\right \rangle =\left \langle
\psi \right \vert U_{2}\left(  \mu,z\right)  \left \vert g\right \rangle ,
\label{e8}%
\end{equation}
where $z=z_{1}+iz_{2}\in C,$ $0<\mu \in R,$ $g\left(  \eta \right)
\equiv \left \langle \eta \right \vert \left.  g\right \rangle \ $and $\psi \left(
\eta \right)  =\left \langle \eta \right \vert \left.  \psi \right \rangle $ are the
wavefunction of state vector $\left \vert g\right \rangle $ and the mother
wavelet state vector $\left \vert \psi \right \rangle $ in $\left \langle
\eta \right \vert $ representation, respectively, and
\begin{equation}
U_{2}\left(  \mu,z\right)  \equiv \frac{1}{\mu}\int \frac{d^{2}\eta}{\pi
}\left \vert \frac{\eta-z}{\mu}\right \rangle \left \langle \eta \right \vert ,\;
\mu=e^{\lambda}, \label{e9}%
\end{equation}
is the two-mode squeezing-displacing operator. Noticing that the two-mode
squeezing operator has its natural expression in $\left \langle \eta \right \vert
$ representation (\ref{3.24}), which is different from the direct product of
two single-mode squeezing (dilation) operators, and the two-mode squeezed
state is simultaneously an entangled state, thus we can put Eq.(\ref{e9}) into
the following form,%
\begin{equation}
U_{2}\left(  \mu,z\right)  =S_{2}\left(  \mu \right)  \mathfrak{D}\left(
z\right)  , \label{e11}%
\end{equation}
where $\mathfrak{D}\left(  z\right)  $ is a two-mode displacement operator,
$\mathfrak{D}\left(  z\right)  \left \vert \eta \right \rangle =\left \vert
\eta-z\right \rangle $ and
\begin{align}
\mathfrak{D}\left(  z\right)   &  =\int \frac{d^{2}\eta}{\pi}\left \vert
\eta-z\right \rangle \left \langle \eta \right \vert \nonumber \\
&  =\exp \left[  iz_{1}\frac{P_{1}-P_{2}}{\sqrt{2}}-iz_{2}\frac{Q_{1}+Q_{2}%
}{\sqrt{2}}\right] \nonumber \\
&  =D_{1}\left(  -z/2\right)  D_{2}\left(  z^{\ast}/2\right)  . \label{e12}%
\end{align}
It the follows the quantum mechanical version of CWT is%
\begin{equation}
W_{\psi}g\left(  \mu,\zeta \right)  =\left \langle \psi \right \vert S_{2}\left(
\mu \right)  \mathfrak{D}\left(  z\right)  \left \vert g\right \rangle
=\left \langle \psi \right \vert S_{2}\left(  \mu \right)  D_{1}\left(
-z/2\right)  D_{2}\left(  z^{\ast}/2\right)  \left \vert g\right \rangle .
\label{e13}%
\end{equation}
Eq.(\ref{e13}) indicates that the CWT can be put into a matrix element in the
$\left \langle \eta \right \vert $ representation of the two-mode displacing and
the two-mode squeezing operators in Eq.(\ref{e10}) between the mother wavelet
state vector $\left \vert \psi \right \rangle $ and the state vector $\left \vert
g\right \rangle $ to be transformed. Once the state vector $\left \langle
\psi \right \vert $ as mother wavelet is chosen, for any state $\left \vert
g\right \rangle $ the matrix element $\left \langle \psi \right \vert U_{2}\left(
\mu,z\right)  \left \vert g\right \rangle $ is just the wavelet transform of
$g(\eta)$ with respect to $\left \langle \psi \right \vert .$ Therefore, various
quantum optical field states can then be analyzed by their wavelet transforms.

\subsubsection{Relation between CWT and EHDF}

In the following we shall show that the EHDF of a quantum state $\left \vert
\psi \right \rangle $ can be obtained by making a complex wavelet transform of
the Gaussian function $e^{-\left \vert \eta \right \vert ^{2}/2},$ i.e.,
\begin{equation}
\left \langle \psi \right \vert \Delta_{h}\left(  \sigma,\gamma,\kappa \right)
\left \vert \psi \right \rangle =e^{-\frac{1}{\kappa}\left \vert \gamma \right \vert
^{2}}\left \vert \int \frac{d^{2}\eta}{\sqrt{\kappa}\pi}e^{-\left \vert
\eta \right \vert ^{2}/2}\psi^{\ast}\left(  \frac{\eta-z}{\sqrt{\kappa}}\right)
\right \vert ^{2}, \label{e14}%
\end{equation}
where $\mu=e^{\lambda}=\sqrt{\kappa},$ $z=z_{1}+iz_{2},$ and
\begin{align}
z_{1}  &  =\frac{\cosh \lambda}{1+\kappa}\left[  \gamma^{\ast}-\gamma
-\kappa \left(  \sigma^{\ast}+\sigma \right)  \right]  ,\label{e15}\\
z_{2}  &  =\frac{i\cosh \lambda}{1+\kappa}\left[  \gamma+\gamma^{\ast}%
+\kappa \left(  \sigma-\sigma^{\ast}\right)  \right]  , \label{e16}%
\end{align}
$\Delta_{h}\left(  \sigma,\gamma,\kappa \right)  $ is named the entangled
Husimi operator by us,
\begin{align}
\Delta_{h}\left(  \sigma,\gamma,\kappa \right)   &  =\frac{4\kappa}{\left(
1+\kappa \right)  ^{2}}\colon \exp \left \{  -\frac{\left(  a_{1}+a_{2}^{\dag
}-\gamma \right)  \left(  a_{1}^{\dag}+a_{2}-\gamma^{\ast}\right)  }{1+\kappa
}\right. \nonumber \\
&  -\left.  \frac{\kappa \left(  a_{1}-a_{2}^{\dag}-\sigma \right)  \left(
a_{1}^{\dag}-a_{2}-\sigma^{\ast}\right)  }{1+\kappa}\right \}  \colon.
\label{e17}%
\end{align}
$\left \langle \psi \right \vert \Delta_{h}\left(  \sigma,\gamma,\kappa \right)
\left \vert \psi \right \rangle $ is the Husimi distribution function.

\textbf{Proof of Eq.(\ref{e14}).}

When the state to be transformed is $\left \vert g\right \rangle =\left \vert
00\right \rangle $ (the two-mode vacuum state), by noticing that $\left \langle
\eta \right.  \left \vert 00\right \rangle =e^{-\left \vert \eta \right \vert
^{2}/2},$ we can express Eq.(\ref{e8}) as%
\begin{equation}
\frac{1}{\mu}\int \frac{d^{2}\eta}{\pi}e^{-\left \vert \eta \right \vert ^{2}%
/2}\psi^{\ast}\left(  \frac{\eta-z}{\mu}\right)  =\left \langle \psi \right \vert
U_{2}\left(  \mu,z\right)  \left \vert 00\right \rangle . \label{e18}%
\end{equation}
To combine the CWTs with transforms of quantum states more tightly and
clearly, using the IWOP technique we can directly perform the integral in
Eq.(\ref{e9}) \cite{r42}%
\begin{align}
U_{2}\left(  \mu,z\right)   &  =\operatorname{sech}\lambda \exp \left[
-\frac{1}{2\left(  1+\mu^{2}\right)  }\left \vert z\right \vert ^{2}%
+a_{1}^{\dagger}a_{2}^{\dagger}\tanh \lambda+\frac{1}{2}\left(  z^{\ast}%
a_{2}^{\dagger}-za_{1}^{\dagger}\right)  \operatorname{sech}\lambda \right]
\nonumber \\
&  \times \exp \left[  \left(  a_{1}^{\dagger}a_{1}+a_{2}^{\dagger}a_{2}\right)
\ln \operatorname{sech}\lambda \right]  \exp \left(  \frac{z^{\ast}a_{1}-za_{2}%
}{1+\mu^{2}}-a_{1}a_{2}\tanh \lambda \right)  . \label{e19}%
\end{align}
where we have set $\mu=e^{\lambda}$, $\operatorname{sech}\lambda=\frac{2\mu
}{1+\mu^{2}}$, $\tanh \lambda=\frac{\mu^{2}-1}{\mu^{2}+1}$, and we have used
the operator identity $e^{ga^{\dagger}a}=\colon \exp \left[  \left(
e^{g}-1\right)  a^{\dagger}a\right]  \colon$. In particular, when $z=0,$
$U_{2}\left(  \mu,z=0\right)  $ becomes to the usual normally ordered two-mode
squeezing operator $S_{2}\left(  \mu \right)  $. From Eq.(\ref{e19}) it then
follows that
\begin{align}
U_{2}\left(  \mu,z\right)  \left \vert 00\right \rangle  &  =\operatorname{sech}%
\lambda \exp \left \{  -\frac{\left(  z_{1}-iz_{2}\right)  \left(  z_{1}%
+iz_{2}\right)  }{2\left(  1+\mu^{2}\right)  }+a_{1}^{\dagger}a_{2}^{\dagger
}\tanh \lambda \right. \nonumber \\
&  \left.  +\frac{1}{2}\left[  \left(  z_{1}-iz_{2}\right)  a_{2}^{\dagger
}-\left(  z_{1}+iz_{2}\right)  a_{1}^{\dagger}\right]  \operatorname{sech}%
\lambda \right \}  \left \vert 00\right \rangle . \label{e20}%
\end{align}
Substituting Eqs.(\ref{e15}), (\ref{e16}) and $\tanh \lambda=\frac{\kappa
-1}{\kappa+1},$ $\cosh \lambda=\frac{1+\kappa}{2\sqrt{\kappa}}$ into
Eq.(\ref{e20}) yields%
\begin{align}
&  e^{-\frac{1}{2\kappa}\left \vert \gamma \right \vert ^{2}-\frac{\sigma
\gamma^{\ast}-\gamma \sigma^{\ast}}{2\left(  \kappa+1\right)  }}U_{2}\left(
\mu,z_{1},z_{2}\right)  \left \vert 00\right \rangle \nonumber \\
&  =\frac{2\sqrt{\kappa}}{1+\kappa}\exp \left \{  -\frac{\left \vert
\gamma \right \vert ^{2}+\kappa \left \vert \sigma \right \vert ^{2}}{2\left(
\kappa+1\right)  }+\frac{\kappa \sigma+\gamma}{1+\kappa}a_{1}^{\dagger}%
+\frac{\gamma^{\ast}-\kappa \sigma^{\ast}}{1+\kappa}a_{2}^{\dagger}%
+a_{1}^{\dagger}a_{2}^{\dagger}\frac{\kappa-1}{\kappa+1}\right \}
\allowbreak \left \vert 00\right \rangle \left.  \equiv \right.  \left \vert
\sigma,\gamma \right \rangle _{\kappa}, \label{e21}%
\end{align}
then the CWT of Eq.(\ref{e18}) can be further expressed as%
\begin{equation}
e^{-\frac{1}{2\kappa}\left \vert \gamma \right \vert ^{2}-\frac{\sigma
\gamma^{\ast}-\gamma \sigma^{\ast}}{2\left(  \kappa+1\right)  }}\int \frac
{d^{2}\eta}{\mu \pi}e^{-\left \vert \eta \right \vert ^{2}/2}\psi^{\ast}\left(
\frac{\eta-z_{1}-iz_{2}}{\mu}\right)  =\left \langle \psi \right.  \left \vert
\sigma,\gamma \right \rangle _{\kappa}. \label{e22}%
\end{equation}

Using normally ordered form of the vacuum state projector $\left \vert
00\right \rangle \left \langle 00\right \vert =\colon e^{-a_{1}^{\dagger}%
a_{1}-a_{2}^{\dagger}a_{2}}\colon,$ and the IWOP method as well as
Eq.(\ref{e21}) we have%
\begin{align}
\left \vert \sigma,\gamma \right \rangle _{\kappa \kappa}\left \langle
\sigma,\gamma \right \vert  &  =\frac{4\kappa}{\left(  1+\kappa \right)  ^{2}%
}\colon \exp \left[  -\frac{\left \vert \gamma \right \vert ^{2}+\kappa \left \vert
\sigma \right \vert ^{2}}{\kappa+1}+\frac{\kappa \sigma+\gamma}{1+\kappa}%
a_{1}^{\dagger}+\frac{\gamma^{\ast}-\kappa \sigma^{\ast}}{1+\kappa}%
a_{2}^{\dagger}\right. \nonumber \\
&  \left.  +\frac{\kappa \sigma^{\ast}+\gamma^{\ast}}{1+\kappa}a_{1}%
+\frac{\gamma-\kappa \sigma}{1+\kappa}a_{2}+\frac{\kappa-1}{\kappa+1}\left(
a_{1}^{\dagger}a_{2}^{\dagger}+a_{1}a_{2}\right)  -a_{1}^{\dagger}a_{1}%
-a_{2}^{\dagger}a_{2}\right]  \colon \nonumber \\
&  =\frac{4\kappa}{\left(  1+\kappa \right)  ^{2}}\colon \exp \left \{
-\frac{\left(  a_{1}+a_{2}^{\dag}-\gamma \right)  \left(  a_{1}^{\dag}%
+a_{2}-\gamma^{\ast}\right)  }{1+\kappa}\right. \nonumber \\
&  -\left.  \frac{\kappa \left(  a_{1}-a_{2}^{\dag}-\sigma \right)  \left(
a_{1}^{\dag}-a_{2}-\sigma^{\ast}\right)  }{1+\kappa}\right \}  \colon \left.
=\right.  \Delta_{h}\left(  \sigma,\gamma,\kappa \right)  . \label{e23}%
\end{align}
Now we explain why $\Delta_{h}\left(  \sigma,\gamma,\kappa \right)  $ is the
entangled Husimi operator. Using the formula for converting an operator $A$
into its Weyl ordering form \cite{Weyl}%
\begin{equation}
A=4\int \frac{d^{2}\alpha d^{2}\beta}{\pi^{2}}\left \langle -\alpha
,-\beta \right \vert A\left \vert \alpha,\beta \right \rangle
\genfrac{}{}{0pt}{}{:}{:}%
\exp \{2\left(  \alpha^{\ast}a_{1}-a_{1}^{\dagger}\alpha+\beta^{\ast}%
a_{2}-a_{2}^{\dagger}\beta+a_{1}^{\dagger}a_{1}+a_{2}^{\dagger}a_{2}\right)
\}%
\genfrac{}{}{0pt}{}{:}{:}%
, \label{e24}%
\end{equation}
where the symbol $%
\genfrac{}{}{0pt}{}{:}{:}%
\genfrac{}{}{0pt}{}{:}{:}%
$ denotes the Weyl ordering, $\left \vert \beta \right \rangle $ is the usual
coherent state, substituting\ Eq.(\ref{e23}) into Eq.(\ref{e24}) and
performing the integration by virtue of the technique of integration within a
Weyl ordered product of operators, we obtain%
\begin{align}
\left \vert \sigma,\gamma \right \rangle _{\kappa \kappa}\left \langle
\sigma,\gamma \right \vert  &  =\frac{16\kappa}{\left(  1+\kappa \right)  ^{2}%
}\int \frac{d^{2}\alpha d^{2}\beta}{\pi^{2}}\left \langle -\alpha,-\beta
\right \vert \colon \exp \left \{  -\frac{\left(  a_{1}+a_{2}^{\dag}%
-\gamma \right)  \left(  a_{1}^{\dag}+a_{2}-\gamma^{\ast}\right)  }{1+\kappa
}\right. \nonumber \\
&  \left.  -\frac{\kappa \left(  a_{1}-a_{2}^{\dag}-\sigma \right)  \left(
a_{1}^{\dag}-a_{2}-\sigma^{\ast}\right)  }{1+\kappa}\right \}  \colon \left \vert
\alpha,\beta \right \rangle \nonumber \\
&  \times%
\genfrac{}{}{0pt}{}{:}{:}%
\exp \{2\left(  \alpha^{\ast}a_{1}-a_{1}^{\dagger}\alpha+\beta^{\ast}%
a_{2}-a_{2}^{\dagger}\beta+a_{1}^{\dagger}a_{1}+a_{2}^{\dagger}a_{2}\right)
\}%
\genfrac{}{}{0pt}{}{:}{:}%
\nonumber \\
&  =4%
\genfrac{}{}{0pt}{}{:}{:}%
\exp \left \{  -\kappa \left(  a_{1}-a_{2}^{\dag}-\sigma \right)  \left(
a_{1}^{\dag}-a_{2}-\sigma^{\ast}\right)  -\frac{1}{\kappa}\left(  a_{1}%
+a_{2}^{\dag}-\gamma \right)  \left(  a_{1}^{\dag}+a_{2}-\gamma^{\ast}\right)
\right \}
\genfrac{}{}{0pt}{}{:}{:}%
, \label{e25}%
\end{align}
where we have used the integral formula%
\begin{equation}
\int \frac{d^{2}z}{\pi}\exp \left(  \zeta \left \vert z\right \vert ^{2}+\xi z+\eta
z^{\ast}\right)  =-\frac{1}{\zeta}e^{-\frac{\xi \eta}{\zeta}},\text{Re}\left(
\zeta \right)  <0. \label{e26}%
\end{equation}
Eq.(\ref{e25}) is the Weyl ordering form of $\left \vert \sigma,\gamma
\right \rangle _{\kappa \kappa}\left \langle \sigma,\gamma \right \vert .$ Then
according to Weyl quantization scheme we know the Weyl ordering form of
two-mode Wigner operator is given by%
\begin{equation}
\Delta_{w}\left(  \sigma,\gamma \right)  =%
\genfrac{}{}{0pt}{}{:}{:}%
\delta \left(  a_{1}-a_{2}^{\dag}-\sigma \right)  \delta \left(  a_{1}^{\dag
}-a_{2}-\sigma^{\ast}\right)  \delta \left(  a_{1}+a_{2}^{\dag}-\gamma \right)
\delta \left(  a_{1}^{\dag}+a_{2}-\gamma^{\ast}\right)
\genfrac{}{}{0pt}{}{:}{:}%
, \label{e27}%
\end{equation}
thus the classical corresponding function of a Weyl ordered operator is
obtained by just replacing $a_{1}-a_{2}^{\dag}\rightarrow \sigma^{\prime}%
,a_{1}+a_{2}^{\dag}\rightarrow \gamma^{\prime},$ i.e.,%
\begin{align}
&  4%
\genfrac{}{}{0pt}{}{:}{:}%
\exp \left \{  -\kappa \left(  a_{1}-a_{2}^{\dag}-\sigma \right)  \left(
a_{1}^{\dag}-a_{2}-\sigma^{\ast}\right)  -\frac{1}{\kappa}\left(  a_{1}%
+a_{2}^{\dag}-\gamma \right)  \left(  a_{1}^{\dag}+a_{2}-\gamma^{\ast}\right)
\right \}
\genfrac{}{}{0pt}{}{:}{:}%
\nonumber \\
&  \rightarrow4\exp \left \{  -\kappa \left \vert \sigma^{\prime}-\sigma
\right \vert ^{2}-\frac{1}{\kappa}\left \vert \gamma^{\prime}-\gamma \right \vert
^{2}\right \}  , \label{e28}%
\end{align}
and in this case the Weyl rule is expressed as%
\begin{align}
\left \vert \sigma,\gamma \right \rangle _{\kappa \kappa}\left \langle
\sigma,\gamma \right \vert  &  =4\int d^{2}\sigma^{\prime}d^{2}\gamma^{\prime}%
\genfrac{}{}{0pt}{}{:}{:}%
\delta \left(  a_{1}-a_{2}^{\dag}-\sigma \right)  \delta \left(  a_{1}^{\dag
}-a_{2}-\sigma^{\ast}\right)  \delta \left(  a_{1}+a_{2}^{\dag}-\gamma \right)
\nonumber \\
&  \times \delta \left(  a_{1}^{\dag}+a_{2}-\gamma^{\ast}\right)
\genfrac{}{}{0pt}{}{:}{:}%
\exp \left \{  -\kappa \left \vert \sigma^{\prime}-\sigma \right \vert ^{2}-\frac
{1}{\kappa}\left \vert \gamma^{\prime}-\gamma \right \vert ^{2}\right \}
\nonumber \\
&  =4\int d^{2}\sigma^{\prime}d^{2}\gamma^{\prime}\Delta_{w}\left(
\sigma^{\prime},\gamma^{\prime}\right)  \exp \left \{  -\kappa \left \vert
\sigma^{\prime}-\sigma \right \vert ^{2}-\frac{1}{\kappa}\left \vert
\gamma^{\prime}-\gamma \right \vert ^{2}\right \}  . \label{e29}%
\end{align}
In reference to Eq.(\ref{e5}) in which the relation between the entangled
Husimi function and the two-mode Wigner function is shown, we know that the
right-hand side of Eq. (\ref{e29}) should be just the entangled Husimi
operator, i.e.
\begin{equation}
\left \vert \sigma,\gamma \right \rangle _{\kappa \kappa}\left \langle
\sigma,\gamma \right \vert =4\int d^{2}\sigma^{\prime}d^{2}\gamma^{\prime}%
\Delta_{w}\left(  \sigma^{\prime},\gamma^{\prime}\right)  \exp \left \{
-\kappa \left \vert \sigma^{\prime}-\sigma \right \vert ^{2}-\frac{1}{\kappa
}\left \vert \gamma^{\prime}-\gamma \right \vert ^{2}\right \}  =\Delta_{h}\left(
\sigma,\gamma,\kappa \right)  , \label{e30}%
\end{equation}
thus Eq. (\ref{e14}) is proved by combining Eqs.(\ref{e30}) and (\ref{e22}).

Thus we have further extended the relation between wavelet transformation and
Wigner-Husimi distribution function to the entangled case. That is to say, we
prove that the entangled Husimi distribution function of a two-mode quantum
state $\left \vert \psi \right \rangle $\ is just the modulus square of the
complex wavelet transform of $e^{-\left \vert \eta \right \vert ^{2}/2}$\ with
$\psi \left(  \eta \right)  $\ being the mother wavelet up to a Gaussian
function, i.e., $\left \langle \psi \right \vert \Delta_{h}\left(  \sigma
,\gamma,\kappa \right)  \left \vert \psi \right \rangle =e^{-\frac{1}{\kappa
}\left \vert \gamma \right \vert ^{2}}\left \vert \int \frac{d^{2}\eta}%
{\sqrt{\kappa}\pi}e^{-\left \vert \eta \right \vert ^{2}/2}\psi^{\ast}\left(
\left(  \eta-z\right)  /\sqrt{\kappa}\right)  \right \vert ^{2}$. Thus is a
convenient approach for calculating various entangled Husimi distribution
functions of miscellaneous quantum states.

\section{Symplectic Wavelet transformation (SWT)}

In this section we shall generalize the usual wavelet transform to symplectic
wavelet transformation (SWT) by using the coherent state representation
\cite{r43}.

\subsection{Single-mode SWT}

First we are motivated to generalize the usual wavelet transform, which
concerns about dilation, to optical Fresnel transform (we will explain this in
detail in section below), i.e. we shall use the
symplectic-transformed---translated versions of the mother wavelet
\begin{equation}
\psi_{r,s;\kappa}\left(  z\right)  =\sqrt{s^{\ast}}\psi \left[  s\left(
z-\kappa \right)  -r\left(  z^{\ast}-\kappa^{\ast}\right)  \right]
\label{17.1}%
\end{equation}
as a weighting function to synthesize the original complex signal $f\left(
z\right)  $,
\begin{align}
W_{\psi}f\left(  r,s;\kappa \right)   &  =\int \frac{d^{2}z}{\pi}f\left(
z\right)  \psi_{r,s;\kappa}^{\ast}\left(  z\right)  ,\text{ }\label{17.2}\\
d^{2}z  &  =dxdy,\text{ }z=x+iy,\nonumber
\end{align}
this is named the symplectic-transformed---translated wavelet transform. One
can see that the mother wavelet $\psi$ generates the other wavelets of the
family $\psi^{\ast}\left[  s\left(  z-\kappa \right)  -r\left(  z^{\ast}%
-\kappa^{\ast}\right)  \right]  $ through a translating transform followed by
a symplectic transform, ($r,s$ are the symplectic transform parameter,
$|s|^{2}-|r|^{2}=1,$ $\kappa$ is a translation parameter, $s$, $r$ and
$\kappa \in \mathrm{C}$)\textrm{,} this can be seen more clearly by writing the
second transform in matrix form
\begin{equation}
\left(
\begin{array}
[c]{c}%
z-\kappa \\
z^{\ast}-\kappa^{\ast}%
\end{array}
\right)  \rightarrow M\left(
\begin{array}
[c]{c}%
z-\kappa \\
z^{\ast}-\kappa^{\ast}%
\end{array}
\right)  ,\text{ }M\equiv \left(
\begin{array}
[c]{cc}%
s & -r\\
-r^{\ast} & s^{\ast}%
\end{array}
\right)  , \label{17.3}%
\end{equation}
where $M$ is a symplectic matrix satisfies $M^{T}JM=J$, $J=\left(
\begin{array}
[c]{cc}%
0 & I\\
-I & 0
\end{array}
\right)  $.

Symplectic matrices in Hamiltonian dynamics correspond to canonical
transformations and keep the Poisson bracket invariant, while in matrix optics
they represent ray transfer matrices of optical instruments, such as lenses
and fibers.

\subsubsection{Properties of symplectic-transformed---translated WT}

It is straightforward to evaluate this transform and its reciprocal transform
when $f\left(  z\right)  $ is the complex Fourier exponentials, $f\left(
z\right)  =\exp \left(  z\beta^{\ast}-z^{\ast}\beta \right)  ,$ (note that
$z\beta^{\ast}-z^{\ast}\beta$ is pure imaginary):%
\begin{align}
W_{\psi}f  &  =\sqrt{s}\int \frac{d^{2}z}{\pi}\exp \left(  z\beta^{\ast}%
-z^{\ast}\beta \right)  \psi^{\ast}\left[  s\left(  z-\kappa \right)  -r\left(
z^{\ast}-\kappa^{\ast}\right)  \right] \nonumber \\
&  =\sqrt{s}\int \frac{d^{2}z}{\pi}\exp \left[  \left(  z+\kappa \right)
\beta^{\ast}-\left(  z^{\ast}+\kappa^{\ast}\right)  \beta \right]  \psi^{\ast
}\left(  sz-rz^{\ast}\right) \nonumber \\
&  =\sqrt{s}\int \frac{d^{2}z^{\prime}}{\pi}\exp \left[  \left(  s^{\ast
}z^{\prime}+rz^{\prime \ast}+\kappa \right)  \beta^{\ast}-\left(  sz^{\prime
\ast}+r^{\ast}z^{\prime}+\kappa^{\ast}\right)  \beta \right]  \psi^{\ast
}\left(  z^{\prime}\right) \nonumber \\
&  =\exp \left[  \kappa \beta^{\ast}-\kappa^{\ast}\beta \right]  \sqrt{s}%
\int \frac{d^{2}z^{\prime}}{\pi}\exp \left[  z^{\prime}\left(  s^{\ast}%
\beta^{\ast}-r^{\ast}\beta \right)  -z^{\prime \ast}\left(  s\beta-r\beta^{\ast
}\right)  \right]  \psi^{\ast}\left(  z^{\prime}\right) \nonumber \\
&  =\sqrt{s}\exp \left[  \kappa \beta^{\ast}-\kappa^{\ast}\beta \right]
\Phi \left(  s^{\ast}\beta^{\ast}-r^{\ast}\beta \right)  , \label{17.4}%
\end{align}
where $\Phi$ is the complex Fourier transform of $\psi^{\ast}.$ Then we form
the adjoint operation%
\begin{align}
W_{\psi}^{\ast}\left(  W_{\psi}f\right)  \left(  z\right)   &  =\sqrt{s^{\ast
}}\int \frac{d^{2}\kappa}{\pi}\left(  W_{\psi}f\right)  \left(  r,s;\kappa
\right)  \psi \left[  s\left(  z-\kappa \right)  -r\left(  z^{\ast}-\kappa
^{\ast}\right)  \right] \nonumber \\
&  =\left \vert s\right \vert \Phi \left(  s^{\ast}\beta^{\ast}-r^{\ast}%
\beta \right)  \int \frac{d^{2}\kappa}{\pi}\exp \left[  \left(  \kappa+z\right)
\beta^{\ast}-\left(  \kappa^{\ast}+z^{\ast}\right)  \beta \right]  \psi \left[
-s\kappa+r\kappa^{\ast}\right] \nonumber \\
&  =\left \vert s\right \vert \exp \left(  z\beta^{\ast}-z^{\ast}\beta \right)
\Phi \left(  s^{\ast}\beta^{\ast}-r^{\ast}\beta \right) \nonumber \\
&  \times \int \frac{d^{2}\kappa^{\prime}}{\pi}\exp \left[  -\kappa^{\prime
}\left(  s^{\ast}\beta^{\ast}-r^{\ast}\beta \right)  +\kappa^{\prime \ast
}\left(  s\beta-r\beta^{\ast}\right)  \right]  \psi \left(  \kappa^{\prime
}\right) \nonumber \\
&  =\left \vert s\right \vert \exp \left(  z\beta^{\ast}-z^{\ast}\beta \right)
\left \vert \Phi \left(  s^{\ast}\beta^{\ast}-r^{\ast}\beta \right)  \right \vert
^{2}, \label{17.5}%
\end{align}
from which we have%
\begin{equation}
\int \frac{W_{\psi}^{\ast}\left(  W_{\psi}f\right)  \left(  z\right)  d^{2}%
s}{\left \vert s\right \vert ^{2}}=\exp \left(  z\beta^{\ast}-z^{\ast}%
\beta \right)  \int d^{2}s\frac{\left \vert \Phi \left(  s^{\ast}\beta^{\ast
}-r^{\ast}\beta \right)  \right \vert ^{2}}{\left \vert s\right \vert }
\label{17.6}%
\end{equation}
so we get the inversion formula%
\begin{equation}
f\left(  z\right)  =\exp \left(  z\beta^{\ast}-z^{\ast}\beta \right)
=\frac{\int d^{2}sW_{\psi}^{\ast}\left(  W_{\psi}f\right)  \left(  z\right)
/\left \vert s\right \vert ^{2}}{\int d^{2}s\left \vert \Phi \left(  s^{\ast}%
\beta^{\ast}-r^{\ast}\beta \right)  \right \vert ^{2}/\left \vert s\right \vert }.
\label{17.7}%
\end{equation}
Eq.(\ref{17.7}) leads us to impose the normalization
\begin{equation}
\int d^{2}s\left \vert \Phi \left(  s^{\ast}\beta^{\ast}-r^{\ast}\beta \right)
\right \vert ^{2}/\left \vert s\right \vert =1, \label{17.8}%
\end{equation}
in order to get the wavelet representation%
\begin{equation}
f\left(  z\right)  =\int d^{2}sW_{\psi}^{\ast}\left(  W_{\psi}f\right)
\left(  z\right)  /\left \vert s\right \vert ^{2}. \label{17.9}%
\end{equation}
Then we can have a form of Parseval's theorem for this new wavelet transform:

\textbf{\ Preposition: }\ For any $f$ and $f^{\prime}$ we have%
\begin{equation}
\int \int W_{\psi}f\left(  r,s;\kappa \right)  W_{\psi}f^{\prime \ast}\left(
r,s;\kappa \right)  \frac{d^{2}\kappa d^{2}s}{\left \vert s\right \vert ^{2}%
}=\int \frac{d^{2}z}{2\pi}f\left(  z\right)  f^{\prime \ast}\left(  z\right)  .
\label{17.10}%
\end{equation}
Proof: Let us assume $F(\beta)$ and $F^{\prime}(\beta)$ are the complex
Fourier transform of $f\left(  z\right)  $ and $f^{\prime}\left(  z\right)  $
respectively,
\begin{equation}
F(\beta)=\int \frac{d^{2}z}{2\pi}f\left(  z\right)  \exp \left(  z\beta^{\ast
}-z^{\ast}\beta \right)  \label{17.11}%
\end{equation}
recall the convolution theorem defined on complex Fourier transform,
\begin{align}
&  \int d^{2}zf\left(  \alpha-z,\alpha^{\ast}-z^{\ast}\right)  f^{\prime
}\left(  z\right) \nonumber \\
&  =\int d^{2}z\int \frac{d^{2}\beta}{2\pi}F(\beta)e^{\left(  \alpha^{\ast
}-z^{\ast}\right)  \beta-\left(  \alpha-z\right)  \beta^{\ast}}\int \frac
{d^{2}\beta^{\prime}}{2\pi}F^{\prime}(\beta^{\prime})\exp \left(  z^{\ast}%
\beta^{\prime}-z\beta^{\prime \ast}\right) \nonumber \\
&  =\int \int d^{2}\beta d^{2}\beta^{\prime}F^{\prime}(\beta^{\prime}%
)F(\beta)e^{\alpha^{\ast}\beta-\alpha \beta^{\ast}}\delta \left(  \beta
-\beta^{\prime}\right)  \delta \left(  \beta^{\ast}-\beta^{\prime \ast}\right)
\nonumber \\
&  =\int d^{2}\beta F(\beta)F^{\prime}(\beta)e^{\alpha^{\ast}\beta-\alpha
\beta^{\ast}} \label{17.12}%
\end{align}
so from (\ref{17.12}) and (\ref{17.1}), (\ref{17.2}) we see that $W_{\psi
}f\left(  r,s;\kappa \right)  =\int \frac{d^{2}z}{\pi}f\left(  z\right)
\psi_{r,s;\kappa}^{\ast}\left(  z\right)  $ can be considered as a convolution
in the form
\begin{align}
&  \int d^{2}zf\left(  z\right)  \psi^{\ast}\left[  s\left(  z-\kappa \right)
-r\left(  z^{\ast}-\kappa^{\ast}\right)  \right] \nonumber \\
&  =\int d^{2}\beta F(\beta)\Phi^{\ast}\left(  s\beta-r\beta^{\ast}\right)
\exp \left(  \kappa \beta^{\ast}-\kappa^{\ast}\beta \right)  \label{17.13}%
\end{align}
It then follows from (\ref{17.12}) that
\begin{align}
&  \int W_{\psi}f\left(  r,s;\kappa \right)  W_{\psi}f^{\prime \ast}\left(
r,s;\kappa \right)  d^{2}\kappa \nonumber \\
&  =\left \vert s\right \vert \int \frac{d^{2}\beta d^{2}\beta^{\prime}}{2\pi
}F(\beta)\Phi^{\ast}\left(  s\beta-r\beta^{\ast}\right) \nonumber \\
&  \times F^{\prime \ast}(\beta^{\prime})\Phi^{\prime}\left(  s\beta^{\prime
}-r\beta^{\prime \ast}\right)  \delta \left(  \beta-\beta^{\prime}\right)
\delta \left(  \beta^{\ast}-\beta^{\prime \ast}\right) \nonumber \\
&  =\left \vert s\right \vert \int \frac{d^{2}\beta}{2\pi}F(\beta)F^{\prime \ast
}(\beta)\left \vert \Phi \left(  s\beta-r\beta^{\ast}\right)  \right \vert ^{2},
\label{17.14}%
\end{align}
Therefore, using (\ref{17.8}) we see that the further integration yields%
\begin{align}
&  \int \frac{d^{2}s}{\left \vert s\right \vert ^{2}}\int W_{\psi}f\left(
r,s;\kappa \right)  W_{\psi}f^{\prime \ast}\left(  r,s;\kappa \right)
d^{2}\kappa \nonumber \\
&  =\int \frac{d^{2}\beta}{2\pi}F(\beta)F^{\prime \ast}(\beta)\int \frac{d^{2}%
s}{\left \vert s\right \vert }\left \vert \Phi \left(  s\beta-r\beta^{\ast
}\right)  \right \vert ^{2}\nonumber \\
&  =\int \frac{d^{2}\beta}{2\pi}F(\beta)F^{\prime \ast}(\beta)=\int \frac{d^{2}%
z}{2\pi}f\left(  z\right)  f^{\prime \ast}\left(  z\right)  , \label{17.15}%
\end{align}
which completes the proof.

\textbf{Theorem}: From the Proposition (\ref{17.10}) we have
\begin{equation}
\int \int W_{\psi}f\left(  r,s;\kappa \right)  \psi_{r,s;\kappa}\left(
z\right)  \frac{d^{2}\kappa d^{2}s}{\left \vert s\right \vert ^{2}}=f\left(
z\right)  , \label{17.16}%
\end{equation}
that is, there exists an inversion formula for arbitrary function $f\left(
z\right)  $. In fact, in Eq. (\ref{17.2}) when we take $f\left(  z\right)
=\delta \left(  z-z^{\prime}\right)  ,$ then
\begin{equation}
W_{\psi}f\left(  r,s;\kappa \right)  =\int \frac{d^{2}z}{\pi}f\left(  z\right)
\psi_{r,s;\kappa}^{\ast}\left(  z\right)  =\psi_{r,s;\kappa}^{\ast}\left(
z^{\prime}\right)  . \label{17.17}%
\end{equation}
Substituting (\ref{17.17}) into (\ref{17.15}) we obtain (\ref{17.16}).

\subsubsection{Relation between $W_{\psi}f\left(  r,s;\kappa \right)  $ and
optical Fresnel transform}

Now we explain why the idea of $W_{\psi}f\left(  r,s;\kappa \right)  $ is
originated from the optical Fresnel transform. We can visualize the
symplectic-transformed---translated wavelet transform in the context of
quantum mechanics, letting $f\left(  z\right)  \equiv \left \langle z\right \vert
\left.  f\right \rangle $, $\left \langle z\right \vert $ is the coherent state,
$\left \vert z\right \rangle =\exp \left[  za^{\dagger}-z^{\ast}a\right]
\equiv \left \vert \left(
\begin{array}
[c]{c}%
z\\
z^{\ast}%
\end{array}
\right)  \right \rangle $, $|0\rangle$ is the vacuum state in Fock space, then
Eq. (\ref{17.1}) can be expressed as%
\begin{align}
W_{\psi}f\left(  r,s;\kappa \right)   &  =\sqrt{s}\int \frac{d^{2}z}{\pi}%
\psi^{\ast}\left[  s\left(  z-\kappa \right)  -r\left(  z^{\ast}-\kappa^{\ast
}\right)  \right]  f\left(  z\right) \nonumber \\
&  =\sqrt{s}\int \frac{d^{2}z}{\pi}\left \langle \psi \right \vert \left.  \left(
\begin{array}
[c]{cc}%
s & -r\\
-r^{\ast} & s^{\ast}%
\end{array}
\right)  \left(
\begin{array}
[c]{c}%
z-\kappa \\
z^{\ast}-\kappa^{\ast}%
\end{array}
\right)  \right \rangle \left \langle z\right \vert \left.  f\right \rangle
\nonumber \\
&  =\left \langle \psi \right \vert F_{1}\left(  r,s,\kappa \right)  \left \vert
f\right \rangle , \label{17.18}%
\end{align}
where $F^{\left(  r,s,\kappa \right)  }$ is defined as
\begin{equation}
F_{1}\left(  r,s,\kappa \right)  =\sqrt{s}\int \frac{d^{2}z}{\pi}\left \vert
sz-rz^{\ast}\right \rangle \left \langle z+\kappa \right \vert ,\; \;
\label{17.19}%
\end{equation}
and $\left \vert sz-rz^{\ast}\right \rangle \equiv \left \vert \left(
\begin{array}
[c]{cc}%
s & -r\\
-r^{\ast} & s^{\ast}%
\end{array}
\right)  \left(
\begin{array}
[c]{c}%
z\\
z^{\ast}%
\end{array}
\right)  \right \rangle .$To know the explicit form of $F_{1}\left(
r,s,\kappa \right)  $, we employ the normal ordering of the vacuum projector
$\left \vert 0\right \rangle \left \langle 0\right \vert =:\exp \left(
-a^{\dagger}a\right)  :$ and the IWOP technique to perform the integration in
(\ref{17.19}), which leads to%
\begin{align}
F_{1}\left(  r,s,\kappa \right)   &  =\exp(\frac{1}{4}\left \vert \kappa
\right \vert ^{2}+\frac{r}{8s}\kappa^{\ast2}+\frac{r^{\ast}\kappa^{2}}%
{8s^{\ast}})\exp \left(  -\frac{r}{2s^{\ast}}a^{\dagger2}-\frac{\kappa
}{2s^{\ast}}\left(  \left \vert s\right \vert ^{2}+\left \vert r\right \vert
^{2}\right)  a^{\dagger}\right) \nonumber \\
&  \exp \left[  \left(  a^{\dagger}a+\frac{1}{2}\right)  \ln \frac{1}{s^{\ast}%
}\right]  \exp \left(  \frac{r^{\ast}}{2s^{\ast}}a^{2}-\frac{1}{2s^{\ast}%
}\left(  s^{\ast}\kappa^{\ast}+r^{\ast}\kappa \right)  a\right)  .
\label{17.20}%
\end{align}
The transformation matrix element of $F_{1}\left(  r,s,\kappa=0\right)  $ in
the coordinate representation $\left \vert x\right \rangle $ is just the kernel
of optical diffraction integration (\ref{5.10}) (Fresnel transform), this
explains our motivation to introduce $W_{\psi}f\left(  r,s;\kappa \right)  $.
In particular, when $r^{\ast}=r\equiv \sinh \lambda$, $s=s^{\ast}\equiv
\cosh \lambda,$ $F_{1}\left(  r,s,\kappa=0\right)  $ reduces to it the
well-known single-mode squeezing operator $\exp[\frac{\lambda}{2}\left(
a^{2}-a^{\dagger2}\right)  ]$ which corresponds to dilation in the usual WT.

\subsection{Entangled SWT}

In the above subsection, the mother wavelet is gained through a translating
transform followed by a symplectic transform. This motivation arises from the
consideration that symplectic transforms are more general than the dilated
transform, and are useful in Fresnel transform of Fourier optics, e.g. ray
transfer matrices of optical instruments, such as lenses and fibers in matrix
optics, while in quantum optics symplectic transforms correspond to
single-mode Fresnel operator (or generalized SU(1,1) squeezing operator).

Recalling that in section 9 we have introduced the 2-mode entangled Fresnel
operator which is a mapping of classical mixed transformation $\left(
z,z^{\prime}\right)  \rightarrow \left(  sz+rz^{\prime \ast},sz^{\prime
}+rz^{\ast}\right)  $ in 2-mode coherent state $\left \vert z,z^{\prime
}\right \rangle $ representation onto quantum operator $F_{2}\left(
r,s\right)  $, thus we are naturally led to develop the SWT in (\ref{17.2}) to
the so-called entangled SWT (ESWT) \cite{r44} for signals $g\left(
z,z^{\prime}\right)  $ defined in two complex planes,
\begin{equation}
W_{\phi}g\left(  r,s;k,k^{\prime}\right)  =%
{\displaystyle \iint}
\frac{d^{2}zd^{2}z^{\prime}}{\pi^{2}}g\left(  z,z^{\prime}\right)
\phi_{r,s;k,k^{\prime}}^{\ast}\left(  z,z^{\prime}\right)  , \label{17.21}%
\end{equation}
here
\begin{equation}
\phi_{r,s;k.k^{\prime}}\left(  z,z^{\prime}\right)  \equiv s^{\ast}\phi \left[
s\left(  z-k\right)  +r\left(  z^{\prime \ast}-k^{\prime \ast}\right)  ,s\left(
z^{\prime}-k^{\prime}\right)  +r\left(  z^{\ast}-k^{\ast}\right)  \right]  ,
\label{17.22}%
\end{equation}
is used as a weighting function to synthesize the signal $g\left(
z,z^{\prime}\right)  $ regarding to two complex planes. One can see that the
mother wavelet $\phi$ generates the family $\phi^{\ast}\left[  s\left(
z-k\right)  +r\left(  z^{\prime \ast}-k^{\prime \ast}\right)  ,s\left(
z^{\prime}-k^{\prime}\right)  +r\left(  z^{\ast}-k^{\ast}\right)  \right]  $
through a translating transform followed by an entangled symplectic transform.
We emphasize that this transform mixes the two complex planes, which is
different from the tensor product of two independent transforms $\left(
z,z^{\prime}\right)  \rightarrow \left[  s\left(  z-k\right)  -r\left(
z^{\ast}-k^{\ast}\right)  ,s\left(  z^{\prime}-k^{\prime}\right)  -r\left(
z^{\prime \ast}-k^{\prime \ast}\right)  \right]  $ given by (\ref{17.18}). The
new symplectic transform can be seen more clearly by writing it in matrix
form:%
\begin{equation}
\left(
\begin{array}
[c]{c}%
z-k\\
z^{\ast}-k^{\ast}\\
z^{\prime}-k^{\prime}\\
z^{\prime \ast}-k^{^{\prime}\ast}%
\end{array}
\right)  \longrightarrow \mathcal{M}\left(
\begin{array}
[c]{c}%
z-k\\
z^{\ast}-k^{\ast}\\
z^{\prime}-k^{\prime}\\
z^{\prime \ast}-k^{^{\prime}\ast}%
\end{array}
\right)  \text{, }\mathcal{M}=\left[
\begin{array}
[c]{cccc}%
s & 0 & 0 & r\\
0 & s^{\ast} & r^{\ast} & 0\\
0 & r & s & 0\\
r^{\ast} & 0 & 0 & s^{\ast}%
\end{array}
\right]  \label{17.23}%
\end{equation}
where $\mathcal{M}$ is symplectic satisfying $\mathcal{M}^{T}\mathcal{J}%
\mathcal{M}=\mathcal{J}$, $\mathcal{J}=\left[
\begin{array}
[c]{cc}%
0 & \mathcal{I}\\
-\mathcal{I} & 0
\end{array}
\right]  $ , $\mathcal{I}$ is the $2\times2$ unit matrix.

For Eq. (\ref{17.21}) being qualified as a new wavelet transform we must prove
that it possesses fundamental properties of the usual wavelet transforms, such
as the \textit{admissibility condition, }the Parseval's theorem and the
inversion formula. When $g\left(  z,z^{\prime}\right)  $ is the complex
Fourier exponential,
\begin{equation}
g_{1}\left(  z,z^{\prime}\right)  =\exp \left(  z\beta^{\ast}-z^{\ast}%
\beta+z^{\prime}\gamma^{\ast}-z^{\prime \ast}\gamma \right)  , \label{17.24}%
\end{equation}
according to (\ref{17.21})-(\ref{17.22}) we evaluate its ESWT
\begin{align}
W_{\phi}g_{1}  &  =%
{\displaystyle \iint}
\frac{d^{2}zd^{2}z^{\prime}}{\pi^{2}}\exp \left(  z\beta^{\ast}-z^{\ast}%
\beta+z^{\prime}\gamma^{\ast}-z^{\prime \ast}\gamma \right)  \phi
_{r,s;k.k^{\prime}}^{\ast}\left(  z,z^{\prime}\right) \nonumber \\
&  =s%
{\displaystyle \iint}
\frac{d^{2}zd^{2}z^{\prime}}{\pi^{2}}\phi^{\ast}\left[  sz+rz^{\prime \ast
},sz^{\prime}+rz^{\ast}\right] \nonumber \\
&  \times \exp \left[  \left(  z+k\right)  \beta^{\ast}-\left(  z^{\ast}%
+k^{\ast}\right)  \beta+\left(  z^{\prime}+k^{\prime}\right)  \gamma^{\ast
}-\left(  z^{\prime \ast}+k^{\prime \ast}\right)  \gamma \right]  . \label{17.25}%
\end{align}
Making the integration variables transform $sz+rz^{\prime \ast}\rightarrow w,$
$sz^{\prime}+rz^{\ast}\rightarrow w^{\prime},$ Eq. (\ref{17.25}) becomes%
\begin{align}
W_{\phi}g_{1}  &  =s\exp \left(  k\beta^{\ast}-k^{\ast}\beta+k^{\prime}%
\gamma^{\ast}-k^{\prime \ast}\gamma \right)
{\displaystyle \iint}
\frac{d^{2}wd^{2}w^{\prime}}{\pi^{2}}\phi^{\ast}\left(  w,w^{\prime}\right)
\nonumber \\
&  \times \exp \left[  w\left(  s^{\ast}\beta^{\ast}+r^{\ast}\gamma \right)
-w^{\ast}\left(  s\beta+r\gamma^{\ast}\right)  +w^{\prime}\left(  s^{\ast
}\gamma^{\ast}+r^{\ast}\beta \right)  -w^{\prime \ast}\left(  s\gamma
+r\beta^{\ast}\right)  \right]  , \label{17.26}%
\end{align}
the last integration is just the complex Fourier transform (CFT) of
$\phi^{\ast},$ denoting it as $\Phi^{\ast},$ we have
\begin{equation}
W_{\phi}g_{1}=s\exp \left(  k\beta^{\ast}-k^{\ast}\beta+k^{\prime}\gamma^{\ast
}-k^{\prime \ast}\gamma \right)  \Phi^{\ast}\left(  s^{\ast}\beta^{\ast}%
+r^{\ast}\gamma,\text{ }s^{\ast}\gamma^{\ast}+r^{\ast}\beta \right)  .
\label{17.27}%
\end{equation}
Then we form the adjoint operation of (\ref{17.27}),
\begin{align}
&  W_{\phi}^{\ast}\left(  W_{\phi}g_{1}\right)  \left(  z,z^{\prime}\right)
\nonumber \\
&  =s^{\ast}%
{\displaystyle \iint}
\frac{d^{2}kd^{2}k^{\prime}}{\pi^{2}}\{ \left(  W_{\phi}g\right)  \left(
r,s;k,k^{\prime}\right)  \} \phi \left[  s\left(  z-k\right)  +r\left(
z^{\prime \ast}-k^{\prime \ast}\right)  ,s\left(  z^{\prime}-k^{\prime}\right)
+r\left(  z^{\ast}-k^{\ast}\right)  \right] \nonumber \\
&  =\left \vert s\right \vert ^{2}\Phi^{\ast}\left(  s^{\ast}\beta^{\ast
}+r^{\ast}\gamma,\text{ }s^{\ast}\gamma^{\ast}+r^{\ast}\beta \right)
{\displaystyle \iint}
\frac{d^{2}kd^{2}k^{\prime}}{\pi^{2}}\exp \left(  k\beta^{\ast}-k^{\ast}%
\beta+k^{\prime}\gamma^{\ast}-k^{\prime \ast}\gamma \right) \nonumber \\
&  \times \phi \left[  s\left(  z-k\right)  +r\left(  z^{\prime \ast}%
-k^{\prime \ast}\right)  ,s\left(  z^{\prime}-k^{\prime}\right)  +r\left(
z^{\ast}-k^{\ast}\right)  \right] \nonumber \\
&  =\left \vert s\right \vert ^{2}\Phi^{\ast}\left(  s^{\ast}\beta^{\ast
}+r^{\ast}\gamma,\text{ }s^{\ast}\gamma^{\ast}+r^{\ast}\beta \right)
{\displaystyle \iint}
\frac{d^{2}kd^{2}k^{\prime}}{\pi^{2}}\phi \left[  -sk-rk^{\prime \ast
},-sk^{\prime}-rk^{\ast}\right] \nonumber \\
&  \times \exp \left[  \left(  k+z\right)  \beta^{\ast}-\left(  k^{\ast}%
+z^{\ast}\right)  \beta+\left(  k^{\prime}+z^{\prime}\right)  \gamma^{\ast
}-\left(  k^{\prime \ast}+z^{\prime \ast}\right)  \gamma \right] \nonumber \\
&  =\left \vert s\right \vert ^{2}\Phi^{\ast}\left(  s^{\ast}\beta^{\ast
}+r^{\ast}\gamma,\text{ }s^{\ast}\gamma^{\ast}+r^{\ast}\beta \right)
\exp \left(  z\beta^{\ast}-z^{\ast}\beta+z^{\prime}\gamma^{\ast}-z^{\prime \ast
}\gamma \right)
{\displaystyle \iint}
\frac{d^{2}vd^{2}v^{\prime}}{\pi^{2}}\phi \left(  v,v^{\prime}\right)
\nonumber \\
&  \times \exp \left[  -v\left(  s^{\ast}\beta^{\ast}+r^{\ast}\gamma \right)
+v^{\ast}\left(  s\beta+r\gamma^{\ast}\right)  -v^{\prime}\left(  s^{\ast
}\gamma^{\ast}+r^{\ast}\beta \right)  +v^{\prime \ast}\left(  s\gamma
+r\beta^{\ast}\right)  \right]  , \label{17.28}%
\end{align}
where the integration in the last line is just the CFT of $\phi$ (comparing
with (\ref{17.26})), thus (\ref{17.28}) leads to
\begin{equation}
W_{\phi}^{\ast}\left(  W_{\phi}g_{1}\right)  \left(  z,z^{\prime}\right)
=\left \vert s\right \vert ^{2}\exp \left(  z\beta^{\ast}-z^{\ast}\beta
+z^{\prime}\gamma^{\ast}-z^{\prime \ast}\gamma \right)  \left \vert \Phi \left(
s^{\ast}\beta^{\ast}+r^{\ast}\gamma,\text{ }s^{\ast}\gamma^{\ast}+r^{\ast
}\beta \right)  \right \vert ^{2}. \label{17.29}%
\end{equation}
From Eq. (\ref{17.29}) we have%
\begin{align}
&  \int d^{2}sW_{\phi}^{\ast}\left(  W_{\phi}g_{1}\right)  \left(
z,z^{\prime}\right)  /\left \vert s\right \vert ^{4}\nonumber \\
&  =\exp \left(  z\beta^{\ast}-z^{\ast}\beta+z^{\prime}\gamma^{\ast}%
-z^{\prime \ast}\gamma \right) \nonumber \\
&  \times \int d^{2}s\left \vert \Phi \left(  s^{\ast}\beta^{\ast}+r^{\ast}%
\gamma,\text{ }s^{\ast}\gamma^{\ast}+r^{\ast}\beta \right)  \right \vert
^{2}/\left \vert s\right \vert ^{2}, \label{17.30}%
\end{align}
which together with (\ref{17.24}) lead to
\begin{equation}
g_{1}\left(  z,z^{\prime}\right)  =\frac{\int d^{2}sW_{\phi}^{\ast}\left(
W_{\phi}g_{1}\right)  \left(  z,z^{\prime}\right)  /\left \vert s\right \vert
^{4}}{\int d^{2}s\left \vert \Phi \left(  s^{\ast}\beta^{\ast}+r^{\ast}%
\gamma,\text{ }s^{\ast}\gamma^{\ast}+r^{\ast}\beta \right)  \right \vert
^{2}/\left \vert s\right \vert ^{2}}. \label{17.31}%
\end{equation}
Eq. (\ref{17.31}) implies that we should impose the normalization
\begin{equation}
\int d^{2}s\left \vert \Phi \left(  s^{\ast}\beta^{\ast}+r^{\ast}\gamma,\text{
}s^{\ast}\gamma^{\ast}+r^{\ast}\beta \right)  \right \vert ^{2}/\left \vert
s\right \vert ^{2}=1, \label{17.32}%
\end{equation}
such that the reproducing process\textit{\ }exists
\begin{equation}
g_{1}\left(  z,z^{\prime}\right)  =\int d^{2}sW_{\phi}^{\ast}\left(  W_{\phi
}g_{1}\right)  \left(  z,z^{\prime}\right)  /\left \vert s\right \vert ^{4}.
\label{17.33}%
\end{equation}
(\ref{17.32}) may be named the generalized admissibility condition. Now we can
have the corresponding \emph{Parseval theorem}: For any $g$ and $g^{\prime}$
we have%
\begin{equation}%
{\displaystyle \iiint}
W_{\phi}g\left(  r,s;k,k^{\prime}\right)  W_{\phi}^{\ast}g^{\prime}\left(
r,s;k,k^{\prime}\right)  \frac{d^{2}kd^{2}k^{\prime}d^{2}s}{\left \vert
s\right \vert ^{4}}=%
{\displaystyle \iint}
d^{2}zd^{2}z^{\prime}g\left(  z,z^{\prime}\right)  g^{\prime \ast}\left(
z,z^{\prime}\right)  . \label{17.34}%
\end{equation}
\emph{Proof:}\textbf{\ }Assuming $F\left(  \beta,\gamma \right)  $ and
$F^{\prime}\left(  \beta,\gamma \right)  $ be CFT of $g\left(  z,z^{\prime
}\right)  $ and $g^{\prime}\left(  z,z^{\prime}\right)  $, respectively,%
\begin{equation}
F\left(  \beta,\gamma \right)  =%
{\displaystyle \iint}
\frac{d^{2}zd^{2}z^{\prime}}{\pi^{2}}g\left(  z,z^{\prime}\right)  \exp \left(
z\beta^{\ast}-z^{\ast}\beta+z^{\prime}\gamma^{\ast}-z^{\prime \ast}%
\gamma \right)  , \label{17.35}%
\end{equation}
recalling the corresponding convolution theorem
\begin{align}
&
{\displaystyle \iint}
d^{2}zd^{2}z^{\prime}g\left(  \alpha-z,\alpha^{\ast}-z^{\ast};\alpha^{\prime
}-z^{\prime},\alpha^{\prime \ast}-z^{\prime \ast}\right)  g^{\prime}\left(
z,z^{\prime}\right) \nonumber \\
&  =%
{\displaystyle \iint}
d^{2}\beta d^{2}\gamma F\left(  \beta,\gamma \right)  F^{\prime}\left(
\beta,\gamma \right)  \exp \left(  \alpha^{\ast}\beta-\alpha \beta^{\ast}%
+\alpha^{\prime \ast}\gamma-\alpha^{\prime}\gamma^{\ast}\right)  ,
\label{17.36}%
\end{align}
so from Eqs. (\ref{17.21}) and (\ref{17.35})-(\ref{17.36}) we see that
$W_{\phi}g\left(  r,s;k,k^{\prime}\right)  =%
{\displaystyle \iint}
\frac{d^{2}zd^{2}z\prime}{\pi^{2}}g\left(  z,z^{\prime}\right)  \phi
_{r,s;k,k^{\prime}}^{\ast}\left(  z,z^{\prime}\right)  $ can be considered as
a convolution in the form (noting that the CFT of $\phi^{\ast}$ is $\Phi
^{\ast},$ see (\ref{17.26})-(\ref{17.27}))%
\begin{align}
&
{\displaystyle \iint}
d^{2}zd^{2}z^{\prime}g\left(  z,z^{\prime}\right)  \phi^{\ast}\left[  s\left(
z-k\right)  +r\left(  z^{\prime \ast}-k^{\prime \ast}\right)  ,s\left(
z^{\prime}-k^{\prime}\right)  +r\left(  z^{\ast}-k^{\ast}\right)  \right]
\nonumber \\
&  =%
{\displaystyle \iint}
d^{2}\beta d^{2}\gamma F\left(  \beta,\gamma \right)  \Phi^{\ast}\left(
s^{\ast}\beta^{\ast}+r^{\ast}\gamma,\text{ }s^{\ast}\gamma^{\ast}+r^{\ast
}\beta \right)  \exp \left(  k\beta^{\ast}-k^{\ast}\beta+k^{\prime}\gamma^{\ast
}-k^{\prime \ast}\gamma \right)  . \label{17.37}%
\end{align}
Using Eq. (\ref{17.37}) we calculate%
\begin{align}
&
{\displaystyle \iint}
W_{\phi}g\left(  r,s;k,k^{\prime}\right)  W_{\phi}^{\ast}g^{\prime}\left(
r,s;k,k^{\prime}\right)  d^{2}kd^{2}k^{\prime}\nonumber \\
&  =\left \vert s\right \vert ^{2}%
{\displaystyle \iiiint}
d^{2}\beta d^{2}\gamma d^{2}\beta^{\prime}d^{2}\gamma^{\prime}F\left(
\beta,\gamma \right)  \Phi^{\ast}\left(  s^{\ast}\beta^{\ast}+r^{\ast}%
\gamma,\text{ }s^{\ast}\gamma^{\ast}+r^{\ast}\beta \right) \nonumber \\
&  \times F^{\prime^{\ast}}\left(  \beta^{\prime},\gamma^{\prime}\right)
\Phi \left(  s^{\ast}\beta^{\ast}+r^{\ast}\gamma,\text{ }s^{\ast}\gamma^{\ast
}+r^{\ast}\beta \right)  \delta \left(  \beta-\beta^{\prime}\right)
\delta \left(  \beta^{\ast}-\beta^{\prime \ast}\right)  \delta \left(
\gamma-\gamma^{\prime}\right)  \delta \left(  \gamma^{\ast}-\gamma^{\prime \ast
}\right) \nonumber \\
&  =\left \vert s\right \vert ^{2}%
{\displaystyle \iint}
d^{2}\beta d^{2}\gamma F\left(  \beta,\gamma \right)  F^{\prime \ast}\left(
\beta,\gamma \right)  \left \vert \Phi \left(  s^{\ast}\beta^{\ast}+r^{\ast
}\gamma,\text{ }s^{\ast}\gamma^{\ast}+r^{\ast}\beta \right)  \right \vert ^{2}.
\label{17.38}%
\end{align}
As a consequence of (\ref{17.32}) and (\ref{17.38}) the further integration
yields%
\begin{align}
&  \int \frac{d^{2}s}{\left \vert s\right \vert ^{4}}%
{\displaystyle \iint}
W_{\phi}g\left(  r,s;k,k^{\prime}\right)  W_{\phi}^{\ast}g^{\prime}\left(
r,s;k,k^{\prime}\right)  d^{2}kd^{2}k^{\prime}\nonumber \\
&  =%
{\displaystyle \iint}
d^{2}\beta d^{2}\gamma F\left(  \beta,\gamma \right)  F^{\prime \ast}\left(
\beta,\gamma \right)  \int d^{2}s\left \vert \Phi \left(  s^{\ast}\beta^{\ast
}+r^{\ast}\gamma,\text{ }s^{\ast}\gamma^{\ast}+r^{\ast}\beta \right)
\right \vert ^{2}/\left \vert s\right \vert ^{2}\nonumber \\
&  =%
{\displaystyle \iint}
d^{2}\beta d^{2}\gamma F\left(  \beta,\gamma \right)  F^{\prime \ast}\left(
\beta,\gamma \right)  =%
{\displaystyle \iint}
d^{2}zd^{2}z^{\prime}g\left(  z,z^{\prime}\right)  g^{\prime \ast}\left(
z,z^{\prime}\right)  , \label{17.39}%
\end{align}
which completes the proof.

\emph{Inversion Formula}:\emph{\ }From Eq. (\ref{17.34}) we have
\begin{equation}
g\left(  z,z^{\prime}\right)  =%
{\displaystyle \iiint}
W_{\phi}g\left(  r,s;k,k^{\prime}\right)  \phi_{r,s;k,k^{\prime}}\left(
z,z^{\prime}\right)  \frac{d^{2}kd^{2}k^{\prime}d^{2}s}{\pi^{2}\left \vert
s\right \vert ^{4}}, \label{17.40}%
\end{equation}
that is, there exists an inversion formula for $g\left(  z,z^{\prime}\right)
$ which represents the original signal $g\left(  z,z^{\prime}\right)  $ as a
superposition of wavelet functions $\phi_{r,s;k,k^{\prime}},$ with the value
of entangled wavelet transform $W_{\phi}g\left(  r,s;k,k^{\prime}\right)  $
serving as coefficients. In fact, in Eq. (\ref{17.21}) when we take
\begin{equation}
g\left(  z,z^{\prime}\right)  =\delta \left(  z-u\right)  \delta \left(
z^{\ast}-u^{\ast}\right)  \delta \left(  z^{\prime}-u^{\prime}\right)
\delta \left(  z^{\prime \ast}-u^{\prime \ast}\right)  , \label{17.41}%
\end{equation}
then
\begin{equation}
W_{\phi}g\left(  r,s;k,k^{\prime}\right)  =\frac{1}{\pi^{2}}\phi
_{r,s;k,k^{\prime}}^{\ast}\left(  u,u^{\prime}\right)  . \label{17.42}%
\end{equation}
Substituting (\ref{17.41})-(\ref{17.42}) into (\ref{17.39}), we obtain
(\ref{17.40}). We can visualize the ESWT in the context of quantum mechanics,
letting $g\left(  z,z^{\prime}\right)  =\left \langle z,z^{\prime}\right \vert
\left.  g\right \rangle \ $and using Eqs. (\ref{17.22})-(\ref{17.23}), Eq.
(\ref{17.21}) is expressed as%
\begin{align}
W_{\phi}g\left(  r,s;k,k^{\prime}\right)   &  =s%
{\displaystyle \iint}
\frac{d^{2}zd^{2}z^{\prime}}{\pi^{2}}\phi^{\ast}\left[  s\left(  z-k\right)
+r\left(  z^{\prime \ast}-k^{\prime \ast}\right)  ,s\left(  z^{\prime}%
-k^{\prime}\right)  +r\left(  z^{\ast}-k^{\ast}\right)  \right]  g\left(
z,z^{\prime}\right) \nonumber \\
&  =s%
{\displaystyle \iint}
\frac{d^{2}zd^{2}z^{\prime}}{\pi^{2}}\left \langle \phi \right.  \left \vert
\mathcal{M}\left(
\begin{array}
[c]{c}%
z-k\\
z^{\ast}-k^{\ast}\\
z^{\prime}-k^{\prime}\\
z^{\prime \ast}-k^{\prime \ast}%
\end{array}
\right)  \right \rangle \left \langle z,z^{\prime}\right \vert \left.
g\right \rangle =\left \langle \phi \right \vert F_{2}\left(  r,s;k,k^{\prime
}\right)  \left \vert g\right \rangle \label{17.43}%
\end{align}
where $F_{2}\left(  r,s;k,k^{\prime}\right)  $ is defined as%
\begin{equation}%
\begin{array}
[c]{c}%
F_{2}\left(  r,s;k,k^{\prime}\right)  =s%
{\displaystyle \iint}
\frac{d^{2}zd^{2}z^{\prime}}{\pi^{2}}\left \vert sz+rz^{\prime \ast},sz^{\prime
}+rz^{\ast}\right \rangle \left \langle z+k,z^{\prime}+k^{\prime}\right \vert ,\\
\left \vert sz+rz^{\prime \ast},sz^{\prime}+rz^{\ast}\right \rangle =\left \vert
sz+rz^{\prime \ast}\right \rangle _{1}\otimes \left \vert sz^{\prime}+rz^{\ast
}\right \rangle _{2}.
\end{array}
\label{17.44}%
\end{equation}
When $k=0$ and $k^{\prime}=0$, $F_{2}\left(  r,s;k=k^{\prime}=0\right)  $ is
just the 2-mode Fresnel operator.

Thus, we have extended the SWT of signals in one complex plane to ESWT of
signals defined in two complex planes, the latter is not the tensor product of
two independent SWTs, this generalization is inevitable, since it resembles
the extending from the single-mode squeezing transform (or Fresnel operator)
to the two-mode squeezing transform (or entangled Fresnel operator) in quantum optics.

\subsection{Symplectic-dilation mixed WT}

Next we shall introduce a new kind of WT, i.e., symplectic-dilation mixed WT
\cite{r45}. Recalling that in Ref. \cite{r46} we have constructed a new
entangled-coherent state representation (ECSR) $\left \vert \alpha
,x\right \rangle $,%
\begin{align}
\left \vert \alpha,x\right \rangle  &  =\exp \left[  -\frac{1}{2}x^{2}-\frac
{1}{4}\left \vert \alpha \right \vert ^{2}+(x+\frac{\alpha}{2})a_{1}^{\dagger
}\right. \nonumber \\
&  \left.  +(x-\frac{\alpha}{2})a_{2}^{\dagger}-\frac{1}{4}(a_{1}^{\dagger
}+a_{2}^{\dagger})^{2}\right]  \left \vert 00\right \rangle , \label{17.45}%
\end{align}
which is the common eigenvector of the operator $\left(  X_{1}+X_{2}\right)
/2$ and $a_{1}-a_{2},$ i.e., $\left(  a_{1}-a_{2}\right)  \left \vert
\alpha,x\right \rangle =\alpha \left \vert \alpha,x\right \rangle $ and $\frac
{1}{2}(X_{1}+X_{2})\left \vert \alpha,x\right \rangle =\frac{1}{\sqrt{2}%
}x\left \vert \alpha,x\right \rangle ,$ where $X_{i}=\frac{1}{\sqrt{2}}%
(a_{i}+a_{i}^{\dagger})$ is the coordinate operator, ($i=1,2)$. $\left \vert
\alpha,x\right \rangle $ is complete,%
\begin{equation}
\int_{-\infty}^{\infty}\frac{\mathtt{d}x}{\sqrt{\pi}}\int \frac{\mathtt{d}%
^{2}\alpha}{2\pi}\left \vert \alpha,x\right \rangle \left \langle \alpha
,x\right \vert =1, \label{17.46}%
\end{equation}
and exhibits partly non-orthogonal property (for $\alpha)$ and orthonormal
property (for $x),$
\begin{align}
&  \left \langle \alpha^{\prime},x^{\prime}\right.  \left \vert \alpha
,x\right \rangle \nonumber \\
&  =\sqrt{\pi}\exp \left[  -\frac{1}{4}(\left \vert \alpha \right \vert
^{2}+\left \vert \alpha^{\prime}\right \vert ^{2})+\frac{1}{2}\alpha
\alpha^{\prime \ast}\right]  \delta \left(  x^{\prime}-x\right)  , \label{17.47}%
\end{align}
so $\left \vert \alpha,x\right \rangle $ possess behavior of both the coherent
state and the entangled state. An interesting question is: Can we introduce a
new kind of continuous WT for which the $\left \vert \alpha,x\right \rangle $
representation underlies? The answer is affirmative. Our motivation of this
issue comes from the mixed lens-Fresnel transform in classical optics
\cite{r47} (see (\ref{44}) below).

By synthesizing (\ref{15.3}) and (\ref{17.2}) and in reference to
(\ref{17.46}) we propose the mixed WT for $g\left(  \alpha,x\right)  $
($\alpha=\alpha_{1}+\mathtt{i}\alpha_{2}$):
\begin{equation}
W_{\psi}g\left(  s,r,\kappa;\mathrm{a},b\right)  \equiv \int_{-\infty}^{\infty
}\frac{\mathtt{d}x}{\sqrt{\pi}}\int \frac{\mathtt{d}^{2}\alpha}{2\pi}g\left(
\alpha,x\right)  \psi_{s,r,\kappa;\mathrm{a},b}^{\ast}\left(  \alpha,x\right)
. \label{17.48}%
\end{equation}
where $\mathtt{d}^{2}\alpha=\mathtt{d}\alpha_{1}\mathtt{d}\alpha_{2},$ the
family of mother wavelet $\psi$ involves both the the symplectic transform of
$\alpha$ and the dilation-transform\ of $x$,%
\begin{equation}
\psi_{s,r,\kappa;\mathrm{a},b}\left(  \alpha,x\right)  =\sqrt{\frac{s^{\ast}%
}{\left \vert \mathrm{a}\right \vert }}\psi \left[  s\left(  \alpha
-\kappa \right)  -r\left(  \alpha^{\ast}-\kappa^{\ast}\right)  ,\frac
{x-b}{\mathrm{a}}\right]  . \label{17.49}%
\end{equation}
Letting $g\left(  \alpha,x\right)  \equiv$ $\left \langle \alpha,x\right \vert
\left.  g\right \rangle ,$ then (\ref{17.48}) can be expressed as quantum
mechanical version%
\begin{equation}
W_{\psi}g\left(  s,r,\kappa;\mathrm{a},b\right)  =\left \langle \psi \right \vert
U\left(  s,r,\kappa;\mathrm{a},b\right)  \left \vert g\right \rangle ,
\label{17.50}%
\end{equation}
where $U\left(  s,r,\kappa;\mathrm{a},b\right)  $ is defined as%
\begin{align}
U\left(  s,r,\kappa;\mathrm{a},b\right)   &  =\sqrt{\frac{s}{\left \vert
\mathrm{a}\right \vert }}\int_{-\infty}^{\infty}\frac{\mathtt{d}x}{\sqrt{\pi}%
}\int \frac{\mathtt{d}^{2}\alpha}{2\pi}\nonumber \\
&  \times \left \vert s\alpha-r\alpha^{\ast},\frac{x-b}{\mathrm{a}}\right \rangle
\left \langle \alpha+\kappa,x\right \vert . \label{17.51}%
\end{align}
$U\left(  s,r,\kappa=0;\mathrm{a},b=0\right)  $ is just the generalized
squeezing operator, which causes a lens-Fresnel mixed transform.

For Eq. (\ref{17.48}) being qualified as a new WT we must prove that it
possesses fundamental properties of the usual WTs, such as the admissibility
condition, the Parseval theorem and the inversion formula. It is
straightforward to evaluate the transform (\ref{17.48}) and its reciprocal
transform when $g\left(  \alpha,x\right)  $\ is the exponential $g_{1}\left(
\alpha,x\right)  =\exp \left(  \alpha^{\ast}\beta-\alpha \beta^{\ast}%
-\mathtt{i}px\right)  ,$%
\begin{align}
W_{\psi}g_{1}  &  =\sqrt{\frac{s}{\left \vert \mathrm{a}\right \vert }}%
e^{\kappa^{\ast}\beta-\kappa \beta^{\ast}-\mathtt{i}pb}\int_{-\infty}^{\infty
}\frac{\mathtt{d}x}{\sqrt{\pi}}\int \frac{\mathtt{d}^{2}\alpha}{2\pi
}\nonumber \\
&  \times \psi^{\ast}(s\alpha-r\alpha^{\ast},\frac{x}{\mathrm{a}}%
)e^{\alpha^{\ast}\beta-\alpha \beta^{\ast}-\mathtt{i}px}. \label{17.52}%
\end{align}
Making the integration variables transform $s\alpha-r\alpha^{\ast}\rightarrow
w,\frac{x}{\mathrm{a}}\rightarrow x^{\prime},$ leading to $\mathtt{d}%
^{2}\alpha \rightarrow \mathtt{d}^{2}w$ and $\int_{-\infty}^{\infty}%
\mathtt{d}x\rightarrow \left \vert \mathrm{a}\right \vert \int_{-\infty}^{\infty
}\mathtt{d}x^{\prime}$, (\ref{17.52}) becomes%
\begin{equation}
W_{\psi}g_{1}=\sqrt{s\left \vert \mathrm{a}\right \vert }\Phi^{\ast}\left(
s^{\ast}\beta^{\ast}-r^{\ast}\beta,\text{ }\mathrm{a}p\right)  e^{\kappa
^{\ast}\beta-\kappa \beta^{\ast}-\mathtt{i}pb}, \label{17.53}%
\end{equation}
where $\Phi^{\ast}$ is just the Fourier transform of $\psi^{\ast},$%
\begin{align}
\Phi^{\ast}\left(  s^{\ast}\beta^{\ast}-r^{\ast}\beta,\text{ }\mathrm{a}%
p\right)   &  =\int_{-\infty}^{\infty}\frac{\mathtt{d}x^{\prime}}{\sqrt{\pi}%
}\int \frac{\mathtt{d}^{2}w}{2\pi}\psi^{\ast}\left(  w,x^{\prime}\right)
\nonumber \\
&  \times e^{w^{\ast}\left(  s\beta-r\beta^{\ast}\right)  -w\left(  s^{\ast
}\beta^{\ast}-r^{\ast}\beta \right)  -\mathtt{i}\mathrm{a}px^{\prime}}.
\label{17.54}%
\end{align}
Then we perform the adjoint WT of (\ref{17.48}), using (\ref{17.49})\ and
(\ref{17.53}) we see%
\begin{align}
&  W_{\psi}^{\ast}\left(  W_{\psi}g_{1}\right)  \left(  \alpha,x\right)
\nonumber \\
&  =\sqrt{\frac{s^{\ast}}{\left \vert \mathrm{a}\right \vert }}\int_{-\infty
}^{\infty}\frac{\mathtt{d}b}{\sqrt{\pi}}\int \frac{\mathtt{d}^{2}\kappa}{2\pi
}W_{\psi}g_{1}\nonumber \\
&  \times \psi \left[  s\left(  \alpha-\kappa \right)  -r\left(  \alpha^{\ast
}-\kappa^{\ast}\right)  ,\frac{x-b}{\mathrm{a}}\right] \nonumber \\
&  =\left \vert s\right \vert \left \vert \mathrm{a}\right \vert g_{1}\left(
\alpha,x\right)  \Phi^{\ast}\left(  s^{\ast}\beta^{\ast}-r^{\ast}\beta,\text{
}\mathrm{a}p\right)  \int_{-\infty}^{\infty}\frac{\mathtt{d}b^{\prime}}%
{\sqrt{\pi}}\nonumber \\
&  \times \int \frac{\mathtt{d}^{2}\kappa^{\prime}}{2\pi}e^{\kappa^{\prime}%
\beta^{\ast}-\kappa^{\prime \ast}\beta+\mathtt{i}\mathrm{a}pb^{\prime}}%
\psi \left(  s\kappa^{\prime}-r\kappa^{\prime \ast},b^{\prime}\right)
\nonumber \\
&  =\left \vert s\right \vert \left \vert \mathrm{a}\right \vert g_{1}\left(
\alpha,x\right)  \left \vert \Phi \left(  s^{\ast}\beta^{\ast}-r^{\ast}%
\beta,\text{ }\mathrm{a}p\right)  \right \vert ^{2}. \label{17.55}%
\end{align}
From Eq. (\ref{17.55}) we obtain%
\begin{align}
&  \int_{-\infty}^{\infty}\frac{\mathtt{d}\mathrm{a}}{\mathrm{a}^{2}}\int
\frac{\mathtt{d}^{2}s}{\left \vert s\right \vert ^{2}}W_{\psi}^{\ast}\left(
W_{\psi}g_{1}\right)  \left(  \alpha,x\right) \nonumber \\
&  =g_{1}\left(  \alpha,x\right)  \int_{-\infty}^{\infty}\frac{\mathtt{d}%
\mathrm{a}}{\left \vert \mathrm{a}\right \vert }\int \frac{\mathtt{d}^{2}%
s}{\left \vert s\right \vert }\left \vert \Phi \left(  s^{\ast}\beta^{\ast
}-r^{\ast}\beta,\text{ }\mathrm{a}p\right)  \right \vert ^{2}, \label{17.56}%
\end{align}
which leads to
\begin{equation}
g_{1}\left(  \alpha,x\right)  =\frac{\int_{-\infty}^{\infty}\frac
{\mathtt{d}\mathrm{a}}{\mathrm{a}^{2}}\int \frac{\mathtt{d}^{2}s}{\left \vert
s\right \vert ^{2}}W_{\psi}^{\ast}\left(  W_{\psi}g_{1}\right)  \left(
\alpha,x\right)  }{\int_{-\infty}^{\infty}\frac{\mathtt{d}\mathrm{a}%
}{\left \vert \mathrm{a}\right \vert }\int \frac{\mathtt{d}^{2}s}{\left \vert
s\right \vert }\left \vert \Phi \left(  s^{\ast}\beta^{\ast}-r^{\ast}\beta,\text{
}\mathrm{a}p\right)  \right \vert ^{2}}. \label{17.57}%
\end{equation}
Eq. (\ref{17.57}) implies that we should impose the normalization
\begin{equation}
\int_{-\infty}^{\infty}\frac{\mathtt{d}\mathrm{a}}{\left \vert \mathrm{a}%
\right \vert }\int \frac{\mathtt{d}^{2}s}{\left \vert s\right \vert }\left \vert
\Phi \left(  s^{\ast}\beta^{\ast}-r^{\ast}\beta,\text{ }\mathrm{a}p\right)
\right \vert ^{2}=1, \label{17.58}%
\end{equation}
such that the reproducing process\textit{\ }exists
\begin{equation}
g_{1}\left(  \alpha,x\right)  =\int_{-\infty}^{\infty}\frac{\mathtt{d}%
\mathrm{a}}{\mathrm{a}^{2}}\int \frac{\mathtt{d}^{2}s}{\left \vert s\right \vert
^{2}}W_{\psi}^{\ast}\left(  W_{\psi}g_{1}\right)  \left(  \alpha,x\right)  .
\label{17.59}%
\end{equation}
(\ref{17.58}) may be named the generalized \textit{admissibility condition}.
Now we can have the corresponding \emph{Parseval theorem}: For any $g$ and
$g^{\prime}$ we have%
\begin{align}
&  \int_{-\infty}^{\infty}\frac{\mathtt{d}\mathrm{a\mathtt{d}}b}%
{\mathrm{a}^{2}}\int \frac{\mathtt{d}^{2}\kappa \mathtt{d}^{2}s}{\left \vert
s\right \vert ^{2}}W_{\psi}g\left(  s,r,\kappa;\mathrm{a},b\right)  W_{\psi
}^{\ast}g^{\prime}\left(  s,r,\kappa;\mathrm{a},b\right) \nonumber \\
&  =\int_{-\infty}^{\infty}\mathtt{d}x\int \mathtt{d}^{2}\alpha g\left(
\alpha,x\right)  g^{\prime \ast}\left(  \alpha,x\right)  . \label{17.60}%
\end{align}
\emph{Proof:}\textbf{\ }Assuming $F\left(  \beta,p\right)  $ and $F^{\prime
}\left(  \beta,p\right)  $ be the Fourier transforms of $g\left(
\alpha,x\right)  $ and $g^{\prime}\left(  \alpha,x\right)  $, respectively,%
\begin{equation}
F\left(  \beta,p\right)  =\int_{-\infty}^{\infty}\frac{\mathtt{d}x}{\sqrt
{2\pi}}\int \frac{\mathtt{d}^{2}\alpha}{\pi}g\left(  \alpha,x\right)
e^{\alpha \beta^{\ast}-\alpha^{\ast}\beta+\mathtt{i}px}, \label{17.61}%
\end{equation}
In order to prove (\ref{17.60}), we first calculate $W_{\psi}g\left(
s,r,\kappa;\mathrm{a},b\right)  $. In similar to deriving Eq.(\ref{19}), using
(\ref{14}), (\ref{15}) and the inversion formula of (\ref{17.61}) we have%
\begin{align}
&  W_{\psi}g\left(  s,r,\kappa;\mathrm{a},b\right) \nonumber \\
&  =\sqrt{\frac{s}{\left \vert \mathrm{a}\right \vert }}\int_{-\infty}^{\infty
}\frac{\mathtt{d}p}{\sqrt{2\pi}}\int \frac{\mathtt{d}^{2}\beta}{\pi}F\left(
\beta,p\right)  \int_{-\infty}^{\infty}\frac{\mathtt{d}x}{\sqrt{\pi}}\int
\frac{\mathtt{d}^{2}\alpha}{2\pi}\nonumber \\
&  \times e^{\alpha^{\ast}\beta-\alpha \beta^{\ast}-\mathtt{i}px}\psi^{\ast
}\left[  s\left(  \alpha-\kappa \right)  -r\left(  \alpha^{\ast}-\kappa^{\ast
}\right)  ,\frac{x-b}{\mathrm{a}}\right] \nonumber \\
&  =\sqrt{s\left \vert \mathrm{a}\right \vert }\int_{-\infty}^{\infty}%
\frac{\mathtt{d}p}{\sqrt{2\pi}}\int \frac{\mathtt{d}^{2}\beta}{\pi}F\left(
\beta,p\right) \nonumber \\
&  \times \Phi^{\ast}\left(  s^{\ast}\beta^{\ast}-r^{\ast}\beta,\text{
}\mathrm{a}p\right)  e^{\kappa^{\ast}\beta-\kappa \beta^{\ast}-\mathtt{i}pb}.
\label{17.62}%
\end{align}
It then follows
\begin{align}
&  \int_{-\infty}^{\infty}\mathtt{d}b\int \mathtt{d}^{2}\kappa W_{\psi}g\left(
s,r,\kappa;\mathrm{a},b\right)  W_{\psi}^{\ast}g^{\prime}\left(
s,r,\kappa;\mathrm{a},b\right) \nonumber \\
&  =\left \vert \mathrm{a}s\right \vert \int_{-\infty}^{\infty}\mathtt{d}%
p\mathtt{d}p^{\prime}\int \mathtt{d}^{2}\beta \mathtt{d}^{2}\beta^{\prime
}F\left(  \beta,p\right)  F^{\prime \ast}\left(  \beta^{\prime},p^{\prime
}\right) \nonumber \\
&  \times \Phi^{\ast}\left(  s^{\ast}\beta^{\ast}-r^{\ast}\beta,\text{
}\mathrm{a}p\right)  \Phi \left(  s^{\ast}\beta^{\prime \ast}-r^{\ast}%
\beta^{\prime},\mathrm{a}p^{\prime}\right) \nonumber \\
&  \times \int_{-\infty}^{\infty}\frac{\mathtt{d}b}{2\pi}\int \frac
{\mathtt{d}^{2}\kappa}{\pi^{2}}e^{\kappa^{\ast}\left(  \beta-\beta^{\prime
}\right)  -\kappa \left(  \beta^{\ast}-\beta^{\prime \ast}\right)
+\mathtt{i}\left(  p^{\prime}-p\right)  b}\nonumber \\
&  =\left \vert \mathrm{a}s\right \vert \int_{-\infty}^{\infty}\mathtt{d}%
p\int \mathtt{d}^{2}\beta F\left(  \beta,p\right)  F^{\prime \ast}\left(
\beta,p\right)  \left \vert \Phi \left(  s^{\ast}\beta^{\ast}-r^{\ast}%
\beta,\mathrm{a}p\right)  \right \vert ^{2}. \label{17.63}%
\end{align}
Substituting (\ref{17.63}) into the left-hand side (LHS) of (\ref{17.60}) and
using (\ref{17.58}) we see%
\begin{align}
\text{LHS of (\ref{26})}  &  =\int_{-\infty}^{\infty}\mathtt{d}p\int
\mathtt{d}^{2}\beta F\left(  \beta,p\right)  F^{\prime \ast}\left(
\beta,p\right) \nonumber \\
&  \times \int_{-\infty}^{\infty}\frac{\mathtt{d}\mathrm{a}}{\left \vert
\mathrm{a}\right \vert }\int \frac{\mathtt{d}^{2}s}{\left \vert s\right \vert
}\left \vert \Phi^{\ast}\left(  s^{\ast}\beta^{\ast}-r^{\ast}\beta
,\mathrm{a}p\right)  \right \vert ^{2}\nonumber \\
&  =\int_{-\infty}^{\infty}\mathtt{d}p\int \mathtt{d}^{2}\beta F\left(
\beta,p\right)  F^{\prime \ast}\left(  \beta,p\right)  . \label{17.64}%
\end{align}
Thus we complete the proof of Eq.(\ref{17.60}).

\emph{Inversion Formula}:\emph{\ }From Eq. (\ref{17.60}) we have
\begin{equation}
g\left(  \alpha,x\right)  =\int_{-\infty}^{\infty}\frac{\mathtt{d\mathrm{a}%
d}b}{\sqrt{\pi}\mathrm{a}^{2}}\int \frac{\mathtt{d}^{2}\kappa \mathtt{d}^{2}%
s}{2\pi \left \vert s\right \vert ^{2}}W_{\psi}g\left(  s,r,\kappa;\mathrm{a}%
,b\right)  \psi_{s,r,\kappa;\mathrm{a},b}\left(  \alpha,x\right)  ,
\label{17.65}%
\end{equation}
that is the inversion formula for the original signal $g\left(  \alpha
,x\right)  $ expressed by a superposition of wavelet functions $\psi
_{s,r,\kappa;\mathrm{a},b}\left(  \alpha,x\right)  ,$ with the value of
continuous WT $W_{\psi}g\left(  s,r,\kappa;\mathrm{a},b\right)  $ serving as
coefficients. In fact, in Eq. (\ref{17.48}) when we take $g\left(
\alpha,x\right)  =\delta \left(  \alpha-\alpha^{\prime}\right)  \delta \left(
\alpha^{\ast}-\alpha^{\prime \ast}\right)  \delta \left(  x-x^{\prime}\right)
,$ then%
\begin{equation}
W_{\psi}g\left(  s,r,\kappa;\mathrm{a},b\right)  =\frac{1}{2\pi \sqrt{\pi}}%
\psi_{s,r,\kappa;\mathrm{a},b}^{\ast}\left(  \alpha^{\prime},x^{\prime
}\right)  . \label{17.66}%
\end{equation}
Substituting (\ref{17.66}) into (\ref{17.64}) yields (\ref{17.65}).

We can visualize the new WT $W_{\psi}g\left(  s,r,\kappa;\mathrm{a},b\right)
$ in the context of quantum optics. Noticing that the generalized squeezing
operator $U\left(  s,r,\kappa=0;\mathrm{a},b=0\right)  $ in (\ref{17.51}) is
an image of the combined mapping of\ the classical real dilation transform
$x\rightarrow$ $x/\mathrm{a}$ ($\mathrm{a}>0$) and the classical complex
symplectic transform $\left(  \alpha,\alpha^{\ast}\right)  \rightarrow \left(
s\alpha-r\alpha^{\ast},s^{\ast}\alpha^{\ast}-r^{\ast}\alpha \right)  $ in
$\left \vert \alpha,x\right \rangle $ representation, one can use the technique
of integration within normal product of operators to perform the integration
in (\ref{17.51}) to derive its explicit form (see Eq. (15) in Ref.
\cite{r12a}). The transform matrix element of $U\left(  s,r,\kappa
=0;\mathrm{a},b=0\right)  $ in the entangled state representation $\left \vert
\eta \right \rangle $ is%
\begin{align}
&  \left \langle \eta \right \vert U\left(  s,r,\kappa=0;\mathrm{a},b=0\right)
\left \vert \eta^{\prime}\right \rangle \nonumber \\
&  =\sqrt{\frac{s}{\mathrm{a}}}\int_{-\infty}^{\infty}\frac{\mathtt{d}x}%
{\sqrt{\pi}}\int \frac{\mathtt{d}^{2}\alpha}{2\pi}\left \langle \eta \right.
\left \vert s\alpha-r\alpha^{\ast},\frac{x}{\mathrm{a}}\right \rangle
\left \langle \alpha,x\right \vert \left.  \eta^{\prime}\right \rangle .
\label{17.67}%
\end{align}
In Fock space $\left \vert \eta=\eta_{1}+\mathtt{i}\eta_{2}\right \rangle $ is
two-mode EPR entangled state in (\ref{3.11}).

Then using (\ref{17.45}) and (\ref{3.11}), we obtain%

\begin{equation}
\left \langle \eta \right.  \left \vert \alpha,x\right \rangle =\frac{1}{\sqrt{2}%
}\exp \left[  -\frac{\alpha^{2}+\left \vert \alpha \right \vert ^{2}}{4}-\frac
{1}{2}\eta_{1}^{2}+\eta_{1}\alpha-\mathtt{i}\eta_{2}x\right]  . \label{17.68}%
\end{equation}
Substituting (\ref{17.68}) into (\ref{17.67}) and using (\ref{5.2})$,$ we
obtain%
\begin{align}
&  \left \langle \eta \right \vert U\left(  s,r,\kappa=0;\mathrm{a},b=0\right)
\left \vert \eta^{\prime}\right \rangle \nonumber \\
&  =\frac{\pi}{\sqrt{\mathrm{a}}}\delta \left(  \eta_{2}^{\prime}-\eta
_{2}/\mathrm{a}\right)  \frac{1}{\sqrt{2\mathtt{i}\pi B}}\nonumber \\
&  \times \exp \left[  \frac{\mathtt{i}}{2B}\left(  A\eta_{1}^{\prime2}%
-2\eta_{1}\eta_{1}^{\prime}+D\eta_{1}^{2}\right)  \right]  . \label{17.69}%
\end{align}
which is just the kernel of a mixed lens$-$Fresnel transform, i.e., the
variable $\eta_{1}$ of the object experiences a generalized Fresnel transform,
while $\eta_{2}$ undergoes a lens transformation. Thus, based on $\left \vert
\alpha,x\right \rangle $ we have introduced SDWT which involves both the real
variable dilation-transform\ and complex variable symplectic transform,
corresponding to the lens-Fresnel mixed transform in classical optics.

\section{Fresnel-Hadamard combinatorial transformation}

In the theoretical study of quantum computer, of great importance is the
Hadamard transform. This operation is $n$ Hadamard gates acting in parallel on
$n$ qubits. The Hadamard transform produces an equal superposition of all
computational basis states. From the point of view of Deutsch-Jozsa quantum
algorithm, the Hadamard transform is an example of the $N=2^{n}$ quantum
Fourier transform, which can be expressed as \cite{r48}%
\begin{equation}
\left \vert j\right \rangle =\frac{1}{\sqrt{2^{n}}}%
{\displaystyle \sum \limits_{k=0}^{2^{n}-1}}
e^{2\pi ijk/2^{n}}\left \vert k\right \rangle . \label{j18.1}%
\end{equation}
Now the continuous Hadamard transform, used to go from the coordinate basis
$\left \vert x\right \rangle $ to the momentum basis$,$ is defined as\cite{r49}
\begin{equation}
\mathfrak{F}\left \vert x\right \rangle =\frac{1}{\sqrt{\pi}\sigma}\int
_{-\infty}^{\infty}dy\exp \left(  2ixy/\sigma^{2}\right)  \left \vert
y\right \rangle , \label{j18.2}%
\end{equation}
where $\sigma$ is the scale length. $\mathfrak{F}$ is named Hadamard operator.
Using the completeness of $\int_{-\infty}^{\infty}dx\left \vert x\right \rangle
\left \langle x\right \vert =1,$ we have%
\begin{equation}
\mathfrak{F}=\frac{1}{\sqrt{\pi}\sigma}%
{\displaystyle \iint_{-\infty}^{\infty}}
dxdy\exp \left(  2ixy/\sigma^{2}\right)  \left \vert y\right \rangle \left \langle
x\right \vert . \label{j18.3}%
\end{equation}
The above two transforms (Fresnel transform and Hadmard transform) are
independent of each other, an interesting question thus naturally arises: can
we combine the two transforms together? To put it in another way, can we
construct a combinatorial operator which play the role of both Fresnel
transform and Hadmard transform for two independent optical modes? The answer
is affirmative, in this section we try to construct so-called Fresnel-Hadmard
combinatorial transform.

\subsection{The Hadamard-Fresnel combinatorial operator}

Based on the coherent-entangled representation $\left \vert \alpha
,x\right \rangle $, and enlightened by Eq. (\ref{5.4}) and (\ref{j18.2}) we now
construct the following ket-bra integration \cite{r50}%
\begin{equation}
U=\frac{\sqrt{s}}{\sqrt{\pi}\sigma}\int \frac{d^{2}\alpha}{\pi}%
{\displaystyle \iint}
dxdy\exp \left(  2ixy/\sigma^{2}\right)  \left \vert s\alpha-r\alpha^{\ast
},y\right \rangle \left \langle \alpha,x\right \vert , \label{j18.4}%
\end{equation}
we name $U$ the Hadamard-Fresnel combinatorial operator.

Substituting Eq.(\ref{j18.2}) into Eq.(\ref{j18.4}), and using the two-mode
vacuum projector's normally ordered form $\left \vert 00\right \rangle
\left \langle 00\right \vert =\colon \exp \left[  -a_{1}^{+}a_{1}-a_{2}^{+}%
a_{2}\right]  \colon$ as well as the IWOP technique we get%
\begin{equation}
U=\frac{\sqrt{s}}{\sqrt{\pi}\sigma}\colon \int \frac{d^{2}z}{\pi}A\left(
z,z^{\ast}\right)
{\displaystyle \iint}
dxdyB(x,y)e^{C}\colon,
\end{equation}
where%
\[
C\equiv-\frac{\left(  a_{1}^{+}+a_{2}^{+}\right)  ^{2}+\left(  a_{1}%
+a_{2}\right)  ^{2}}{4}-a_{1}^{+}a_{1}-a_{2}^{+}a_{2},
\]%
\[
B(x,y)\equiv \exp \left[  -\frac{y^{2}+x^{2}}{2}+y\left(  a_{1}^{+}+a_{2}%
^{+}\right)  +x\left(  a_{1}+a_{2}\right)  +\frac{2ixy}{\sigma^{2}}\right]  ,
\]
and
\[
A\left(  z,z^{\ast}\right)  \equiv \exp \left[  -\frac{\left \vert sz-rz^{\ast
}\right \vert ^{2}+\left \vert z\right \vert ^{2}}{4}+\frac{sz-rz^{\ast}}%
{2}\left(  a_{1}^{\dagger}-a_{2}^{\dagger}\right)  +\frac{z^{\ast}\left(
a_{1}-a_{2}\right)  }{2}\right]  ,
\]
they are all within the normal ordering symbol $::$ $.$ Now performing the
integration over $dxdy$ within $::$ and remembering that all creation
operators are commute with all annihilation operators (the essence of the IWOP
technique) so that they can be considered c-number during the integration, we
can finally obtain%
\begin{align}
U  &  =\frac{1}{\sqrt{s^{\ast}}}\frac{4\sqrt{\pi}\sigma}{\sqrt{\sigma^{4}+4}%
}\colon \exp \left \{  -\frac{1}{2}\frac{r}{s^{\ast}}\left(  \frac{a_{1}^{\dag
}-a_{2}^{\dag}}{\sqrt{2}}\right)  ^{2}+\frac{\sigma^{4}-4}{2\left(  \sigma
^{4}+4\right)  }\left(  \frac{a_{1}^{\dag}+a_{2}^{\dag}}{\sqrt{2}}\right)
^{2}\right. \nonumber \\
&  +\left(  \frac{1}{s^{\ast}}-1\right)  \frac{a_{1}^{\dag}-a_{2}^{\dag}%
}{\sqrt{2}}\frac{a_{1}-a_{2}}{\sqrt{2}}+\left(  \frac{4i\sigma^{2}}{\sigma
^{4}+4}-1\right)  \frac{a_{1}^{\dag}+a_{2}^{\dag}}{\sqrt{2}}\frac{a_{1}+a_{2}%
}{\sqrt{2}}\nonumber \\
&  +\left.  \frac{1}{2}\frac{r^{\ast}}{s^{\ast}}\left(  \frac{a_{1}-a_{2}%
}{\sqrt{2}}\right)  ^{2}+\frac{\sigma^{4}-4}{2\left(  \sigma^{4}+4\right)
}\left(  \frac{a_{1}+a_{2}}{\sqrt{2}}\right)  ^{2}\right \}  \colon,
\label{j18.5}%
\end{align}
which is the normally ordered form of Hadamard-Fresnel combinatorial operator.

\subsection{The properties of Hadamard-Fresnel operator}

Note
\begin{equation}
\left[  \frac{a_{1}-a_{2}}{\sqrt{2}},\frac{a_{1}^{\dag}+a_{2}^{\dag}}{\sqrt
{2}}\right]  =0, \label{j18.6}%
\end{equation}
\ and
\begin{equation}
\left[  \frac{a_{1}-a_{2}}{\sqrt{2}},\frac{a_{1}^{\dag}-a_{2}^{\dag}}{\sqrt
{2}}\right]  =1,\text{ \ }\left[  \frac{a_{1}+a_{2}}{\sqrt{2}},\frac
{a_{1}^{\dag}+a_{2}^{\dag}}{\sqrt{2}}\right]  =1, \label{j18.7}%
\end{equation}
$\frac{a_{1}-a_{2}}{\sqrt{2}}$ can be considered a mode independent of another
mode $\frac{a_{1}^{\dag}+a_{2}^{\dag}}{\sqrt{2}},$ thus we have the operator
identity
\begin{equation}
\text{exp}\left[  f\left(  a_{1}^{\dag}\pm a_{2}^{\dag}\right)  \left(
a_{1}\pm a_{2}\right)  \right]  =\colon \exp[\frac{1}{2}\left(  e^{2f}%
-1\right)  \left(  a_{1}^{\dag}\pm a_{2}^{\dag}\right)  \left(  a_{1}\pm
a_{2}\right)  ]\colon. \label{j18.8}%
\end{equation}
Using (\ref{j18.8}) we can rewrite Eq.(\ref{j18.7}) as%
\begin{equation}
U=U_{2}U_{1}=U_{1}U_{2}, \label{j18.9}%
\end{equation}
where%
\begin{align}
U_{1}  &  =\frac{4\sqrt{\pi}\sigma}{\sqrt{\sigma^{4}+4}}\exp \left[
\frac{\sigma^{4}-4}{2\left(  \sigma^{4}+4\right)  }\left(  \frac{a_{1}^{\dag
}+a_{2}^{\dag}}{\sqrt{2}}\right)  ^{2}\right] \nonumber \\
&  \exp \left[  \frac{a_{1}^{\dag}+a_{2}^{\dag}}{\sqrt{2}}\frac{a_{1}+a_{2}%
}{\sqrt{2}}\ln \frac{4i\sigma^{2}}{\left(  \sigma^{4}+4\right)  }\right]
\exp \left[  \frac{\sigma^{4}-4}{2\left(  \sigma^{4}+4\right)  }\left(
\frac{a_{1}+a_{2}}{\sqrt{2}}\right)  ^{2}\right]  \label{j18.10}%
\end{align}
and%
\begin{align}
U_{2}  &  =\exp \left[  -\frac{r}{2s^{\ast}}\left(  \frac{a_{1}^{\dag}%
-a_{2}^{\dag}}{\sqrt{2}}\right)  ^{2}\right]  \exp \left[  \left(  \frac
{a_{1}^{\dag}-a_{2}^{\dag}}{\sqrt{2}}\frac{a_{1}-a_{2}}{\sqrt{2}}+\frac{1}%
{2}\right)  \ln \frac{1}{s^{\ast}}\right] \nonumber \\
&  \exp \left[  \frac{r^{\ast}}{2s^{\ast}}\left(  \frac{a_{1}-a_{2}}{\sqrt{2}%
}\right)  ^{2}\right]  , \label{j18.11}%
\end{align}
while $U_{2}$ is the Fresnel operator for mode $\frac{a_{1}-a_{2}}{\sqrt{2}},$
$U_{1}$ is named the Hadamard operator for mode $\frac{a_{1}+a_{2}}{\sqrt{2}%
}.$

It then follows
\begin{align}
U\frac{a_{1}-a_{2}}{\sqrt{2}}U^{-1}  &  =U_{2}\frac{a_{1}-a_{2}}{\sqrt{2}%
}U_{2}^{-1}=s^{\ast}\frac{a_{1}-a_{2}}{\sqrt{2}}+r\frac{a_{1}^{\dag}%
-a_{2}^{\dag}}{\sqrt{2}},\nonumber \\
U\frac{a_{1}^{\dag}-a_{2}^{\dag}}{\sqrt{2}}U^{-1}  &  =U_{2}\frac{a_{1}^{\dag
}-a_{2}^{\dag}}{\sqrt{2}}U_{2}^{-1}=r^{\ast}\frac{a_{1}-a_{2}}{\sqrt{2}%
}+s\frac{a_{1}^{\dag}-a_{2}^{\dag}}{\sqrt{2}}, \label{j18.12}%
\end{align}
from which we see the Hadamard-Fresnel combinatorial operator can play the
role of Fresnel transformation for $\frac{a_{1}-a_{2}}{\sqrt{2}}.$ Physically,
$\frac{a_{1}-a_{2}}{\sqrt{2}}$ and $\frac{a_{1}+a_{2}}{\sqrt{2}}$ can be two
output fields of a beamsplitter.

In a similar way, we have
\begin{align}
U\frac{a_{1}+a_{2}}{\sqrt{2}}U^{-1}  &  =U_{1}\frac{a_{1}+a_{2}}{\sqrt{2}%
}U_{1}^{-1}=\frac{1}{4i\sigma^{2}}\left[  \left(  \sigma^{4}+4\right)
\frac{a_{1}+a_{2}}{\sqrt{2}}-\left(  \sigma^{4}-4\right)  \frac{a_{1}^{\dag
}+a_{2}^{\dag}}{\sqrt{2}}\right]  ,\nonumber \\
U\frac{a_{1}^{\dag}+a_{2}^{\dag}}{\sqrt{2}}U^{-1}  &  =U_{1}\frac{a_{1}^{\dag
}+a_{2}^{\dag}}{\sqrt{2}}U_{1}^{-1}=\frac{1}{4i\sigma^{2}}\left[  -\left(
\sigma^{4}+4\right)  \frac{a_{1}^{\dag}+a_{2}^{\dag}}{\sqrt{2}}+\left(
\sigma^{4}-4\right)  \frac{a_{1}+a_{2}}{\sqrt{2}}\right]  \label{j18.13}%
\end{align}
which for the quadrature $X_{i}=\left(  a_{i}+a_{i}^{\dag}\right)  /\sqrt{2},$
\ $P_{i}=\left(  a_{i}-a_{i}^{\dag}\right)  /\sqrt{2}i,$ $\left(
i=1,2\right)  ,$ leads to
\begin{equation}
U\frac{X_{1}+X_{2}}{2}U^{-1}=\frac{\sigma^{2}}{4}\left(  P_{1}+P_{2}\right)
,\text{ \  \ }U\left(  P_{1}+P_{2}\right)  U^{-1}=-\frac{4}{\sigma^{2}}%
\frac{X_{1}+X_{2}}{2}, \label{j18.14}%
\end{equation}
from which we see that the Hadamard-Fresnel combinatorial operator also plays
the role of exchanging the total momentum---average position followed by a
squeezing transform, with the squeezing parameter being $\frac{\sigma^{2}}%
{4}.$

The mutual transform in (\ref{j18.14}) can be realized by
\begin{align}
e^{i\frac{\pi}{2}\left(  a_{1}^{\dag}a_{1}+a_{2}^{\dag}a_{2}\right)  }\left(
X_{1}+X_{2}\right)  e^{-i\frac{\pi}{2}\left(  a_{1}^{\dag}a_{1}+a_{2}^{\dag
}a_{2}\right)  }  &  =P_{1}+P_{2},\text{ }\label{j18.15}\\
\text{\ }e^{i\frac{\pi}{2}\left(  a_{1}^{\dag}a_{1}+a_{2}^{\dag}a_{2}\right)
}\left(  P_{1}+P_{2}\right)  e^{-i\frac{\pi}{2}\left(  a_{1}^{\dag}a_{1}%
+a_{2}^{\dag}a_{2}\right)  }  &  =-\left(  X_{1}+X_{2}\right)  \text{\ }
\label{j18.16}%
\end{align}
while the two-mode squeezing operator is $S_{2}=\exp \left[  \ln \frac{2}%
{\sigma^{2}}\left(  a_{1}^{\dag}a_{2}^{\dag}-a_{1}a_{2}\right)  \right]  ,$
therefore%
\begin{equation}
U_{1}=S_{2}^{-1}e^{i\frac{\pi}{2}\left(  a_{1}^{\dag}a_{1}+a_{2}^{\dag}%
a_{2}\right)  }. \label{j18.17}%
\end{equation}
From Eq.(\ref{j18.9}) and Eq.(\ref{j18.17}), we see that the Hadamard-Fresnel
combinatorial operator can be decomposed as
\begin{equation}
U=U_{2}S_{2}^{-1}e^{i\frac{\pi}{2}\left(  a_{1}^{\dag}a_{1}+a_{2}^{\dag}%
a_{2}\right)  }=S_{2}^{-1}e^{i\frac{\pi}{2}\left(  a_{1}^{\dag}a_{1}%
+a_{2}^{\dag}a_{2}\right)  }U_{2}. \label{j18.18}%
\end{equation}
It can be also seen that $U$ is unitary, $U^{+}U=UU^{+}=1$.

In this section, we have introduced the Fresnel-Hadamard combinatorial
operator by virtue of the IWOP technique. This unitary operator plays the role
of both Fresnel transformation for mode $\frac{a_{1}-a_{2}}{\sqrt{2}}$ and
Hadamard transformation for mode $\frac{a_{1}+a_{2}}{\sqrt{2}},$ respectively,
and the two transformations are combinatorial. We have shown that the two
transformations are concisely expressed in the coherent-entangled state
representation as a projective operator in integration form. We also found
that the Fresnel-Hadamard operator can be decomposed as $U_{2}S_{2}%
^{-1}e^{i\frac{\pi}{2}\left(  a_{1}^{\dag}a_{1}+a_{2}^{\dag}a_{2}\right)  },$
a Fresnel operator $U_{2}$, a two-mode squeezing operator $S_{2}^{-1}$ and the
total momentum-average position exchanging operator. Physically, $\frac
{a_{1}-a_{2}}{\sqrt{2}}$ and $\frac{a_{1}+a_{2}}{\sqrt{2}}$ can be two output
fields of a beamsplitter. If an optical device can be designed for
Fresnel-Hadamard combinatorial transform, then it can be directly applied to
these two output fields of the beamsplitter. \bigskip

In summary, although quantum optics and classical optics are so different, no
matter in the mathematical tools they employed or in a conceptual view
(quantum optics concerning the wave-particle duality of optical field with an
emphasis on its nonclassical properties, whereas classical optics works on the
distribution aqnd propagation of the light waves), that it is a new
exploration to link them systematically. However, In this review, via the
route of developing Dirac's symbolic method we have revealed some links
between them by mapping classical symplectic transformation in the coherent
state representation onto quantum unitary operators (GFO), throughout our
discussion the IWOP technique is indispensable for the derivation. We have
resorted to the quantum optical interpretation of various classical optical
transformations by adopting quantum optics concepts such as the coherent
states, squeezed states, and entangled states, etc. Remarkably, we have
endowed complex fractional Fourier transform, Hankel transform with quantum
optical representation-transform interpretation. Our formalism, starting from
quantum optics theory, not only provides quantum mechanical account of various
classical optical transformations, but also have found their way back to some
new classical transformations, e.g. entangled Fresnel transform,
Fresnel-wavelet transform, etc, which may have realistic optical
interpretation in the future. As Dirac predicted, functions that have been
applied in classical optical problems may be translated in an operator
language in quantum mechanics, and vice-versa. We expect that the content of
this work may play some role in quantum states engineering, i.e., optical
field states' preparation and design.

Once the correspondence in this respect between the two distinct fields is
established, the power of Dirac's symbolic method can be fully displayed to
solve some new problems in classical optics, e.g., to find new eigen-modes of
some optical transforms; to extend the research region of classical optics
theoretically by introducing new transforms (for example, the entangled
Fresnel transforms), which may bring attention of experimentalists, who may
get new ideas to implement these new classical optical transformations.

\textbf{Acknowledgement:} This work supported by the National Natural Science
Foundation of China, Grant No. 10775097 and 10874174, and a grant from the Key
Programs Foundation of Ministry of Education of China (No. 210115), and the
Research Foundation of the Education Department of Jiangxi Province of China
(grant no. GJJ10097).

\end{document}